\documentclass[times,twocolumn,final]{elsarticle}

\usepackage{medima}
\usepackage{framed,multirow}

\usepackage{amssymb}
\usepackage{latexsym}

\usepackage{url}
\usepackage{xcolor}
\usepackage{longtable}
\usepackage{subfigure}
\usepackage{amsmath,amssymb,amsfonts}
\usepackage{algorithmic}
\usepackage{graphicx}
\usepackage{gensymb}
\usepackage{multirow}
\usepackage{textcomp}
\usepackage{adjustbox}
\usepackage[figuresright]{rotating}
\usepackage{pdflscape}
\usepackage{threeparttable}
\usepackage{makecell}

\usepackage{hyperref}
\usepackage[hyphenbreaks]{breakurl}
\usepackage{booktabs}
\usepackage{breakurl}
\usepackage{url}

\definecolor{newcolor}{rgb}{.8,.349,.1}

\begin{document}

\verso{Tao Li \textit{et~al.}}

\begin{frontmatter}

\title{Applications of Deep Learning in Fundus Images: A Review}

\author[1]{Tao \snm{Li}}
\author[1]{Wang \snm{Bo}}
\author[1]{Chunyu \snm{Hu}}
\author[1]{Hong \snm{Kang}}
\author[2]{Hanruo \snm{Liu}}
\author[1]{Kai \snm{Wang}\corref{cor1}}
\cortext[cor1]{Corresponding author:
  Tel.: +0-000-000-0000;
  fax: +0-000-000-0000;}
\author[3]{Huazhu \snm{Fu}}
\address[1]{College of Computer Science, Nankai University, Tianjin 300350, China}
\address[2]{Beijing Tongren Hospital, Capital Medical University, Address, Beijing 100730 China}
\address[3]{Inception Institute of Artificial Intelligence (IIAI), Abu Dhabi, UAE}

\received{11 Nov 2020}
\finalform{11 Nov 2020}
\accepted{12 Jan 2021}
\availableonline{20 Jan 2021}
\communicated{S. Sarkar}

\begin{abstract}
The use of fundus images for the early screening of eye diseases is of great clinical importance.
Due to its powerful performance, deep learning is becoming more and more popular in related applications, such as lesion segmentation, biomarkers segmentation, disease diagnosis and image synthesis.
Therefore, it is very necessary to summarize the recent developments in deep learning for fundus images with a review paper.
In this review, we introduce 143 application papers with a carefully designed hierarchy.
Moreover, 33 publicly available datasets are presented.
Summaries and analyses are provided for each task.
Finally, limitations common to all tasks are revealed and possible solutions are given.
We will also release and regularly update the state-of-the-art results and newly-released datasets at https://github.com/nkicsl/Fundus\_Review to adapt to the rapid development of this field.

\end{abstract}


\end{frontmatter}


\section{Introduction}
\label{sec1}
According to the World Vision Report\footnote{https://www.who.int/publications/i/item/world-report-on-vision} released by the World Health Organization in October 2019, more than 418 million people worldwide suffer from glaucoma,  diabetic retinopathy (DR),  age-related macular degeneration (AMD) or other eye diseases which can cause blindness.
Patients with eye diseases are often unaware of the aggravation of asymptomatic conditions \citep{article}, so early screening and treatment of eye diseases is particularly important.

\begin{figure}
\centering
\includegraphics[height=5cm]{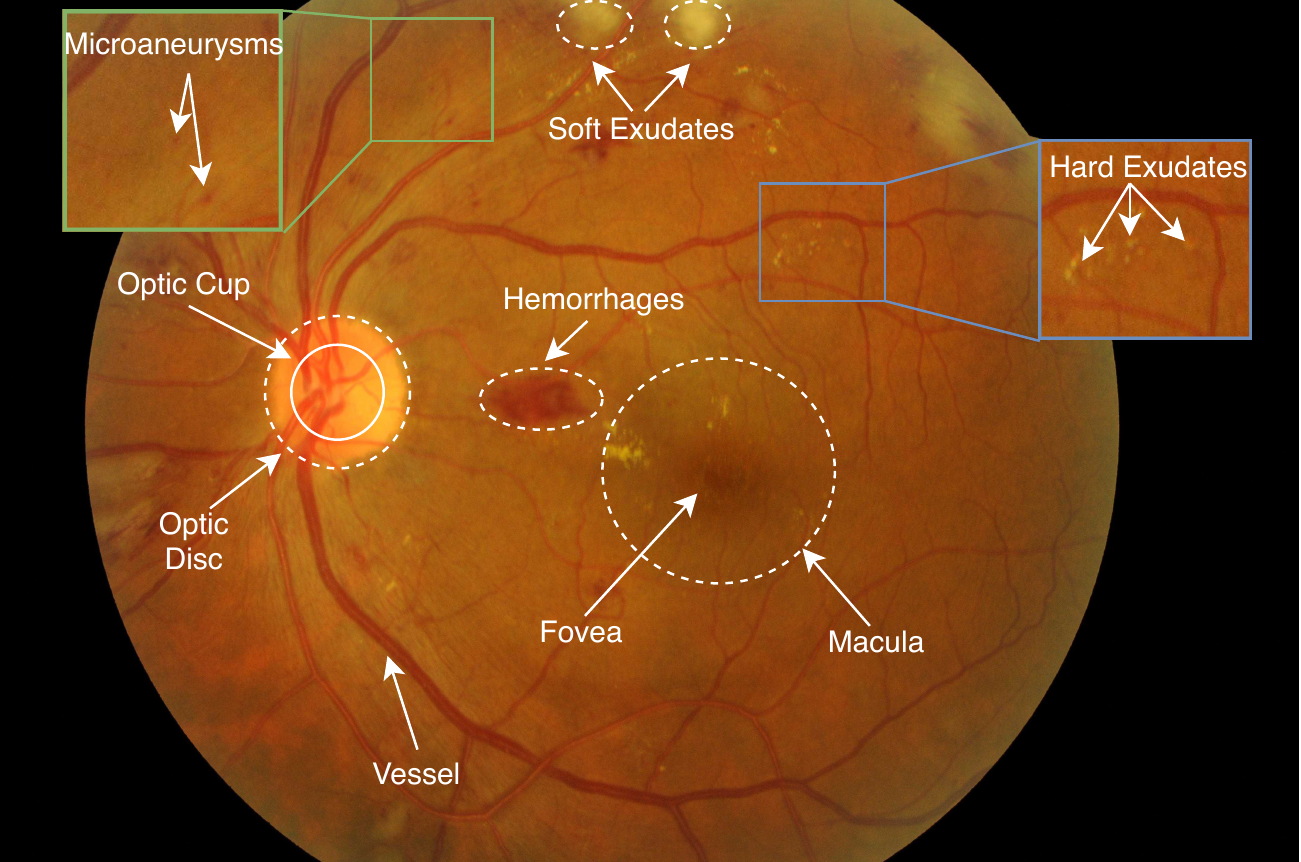}\\
\caption{A fundus image from the IDRiD dataset illustrating important biomarkers and lesions.}
\label{fig:fundus_sample}
\end{figure}

A fundus image is a projection of the fundus captured by a monocular camera on a 2D plane.
Unlike other eye scans, such as OCT images and angiographs, fundus images can be acquired in a non-invasive and cost-effective manner, making them more suitable for large-scale screening \citep{DBLP:conf/icip/EdupugantiCK18}.
An example of a fundus image is presented in Fig. \ref{fig:fundus_sample}.

Many important biomarkers can be seen in the fundus image, such as optic disc (OD), optic cup (OC), macula, fovea, blood vessel, and some DR related lesions, such as microaneurysms (MAs), hemorrhages (HEs), hard exudates (EXs), and soft exudates (SEs).
Fundus images can be used to diagnose a variety of eye diseases, including glaucoma, DR, AMD, cataract, retinopathy of prematurity (ROP), and diabetic macular edema (DME).

Recently, data-driven deep learning has been widely applied to ophthalmic disease diagnosis based on fundus images.
Compared to traditional methods that use manually designed features, deep learning models can achieve better performance by automatically optimizing the features in an end-to-end manner.
Most applications of deep learning in fundus images can be coarsely divided into classification, segmentation and synthesis tasks.
For brevity, we only list widely used backbones in fundus image applications.
Diagnosis and grading of ophthalmic diseases are two examples of classification tasks, 
VGG-Net \citep{DBLP:journals/corr/SimonyanZ14a}, Inception \citep{DBLP:conf/cvpr/SzegedyLJSRAEVR15,DBLP:conf/cvpr/SzegedyVISW16,DBLP:conf/aaai/SzegedyIVA17}, ResNet \citep{DBLP:conf/cvpr/HeZRS16} and DenseNet \citep{DBLP:conf/cvpr/HuangLMW17} are the most widely used classification backbone networks. 
In terms of segmentation tasks, identifying lesions and biomarkers is of great importance in the diagnosis of diseases. 
In addition to those used for classification, other networks widely used for segmentation in fundus images include 
FCN \citep{DBLP:conf/cvpr/LongSD15}, SegNet \citep{DBLP:journals/pami/BadrinarayananK17}, U-Net \citep{DBLP:conf/miccai/RonnebergerFB15}, MaskRCNN \citep{DBLP:conf/iccv/HeGDG17} and DeeplabV3+ \citep{DBLP:conf/eccv/ChenZPSA18}.
Finally, in the field of fundus image synthesis, generative adversarial network (GANs) \citep{DBLP:conf/nips/GoodfellowPMXWOCB14} are the dominant architecture.

\begin{figure}
\centering
\includegraphics[height=5.5cm]{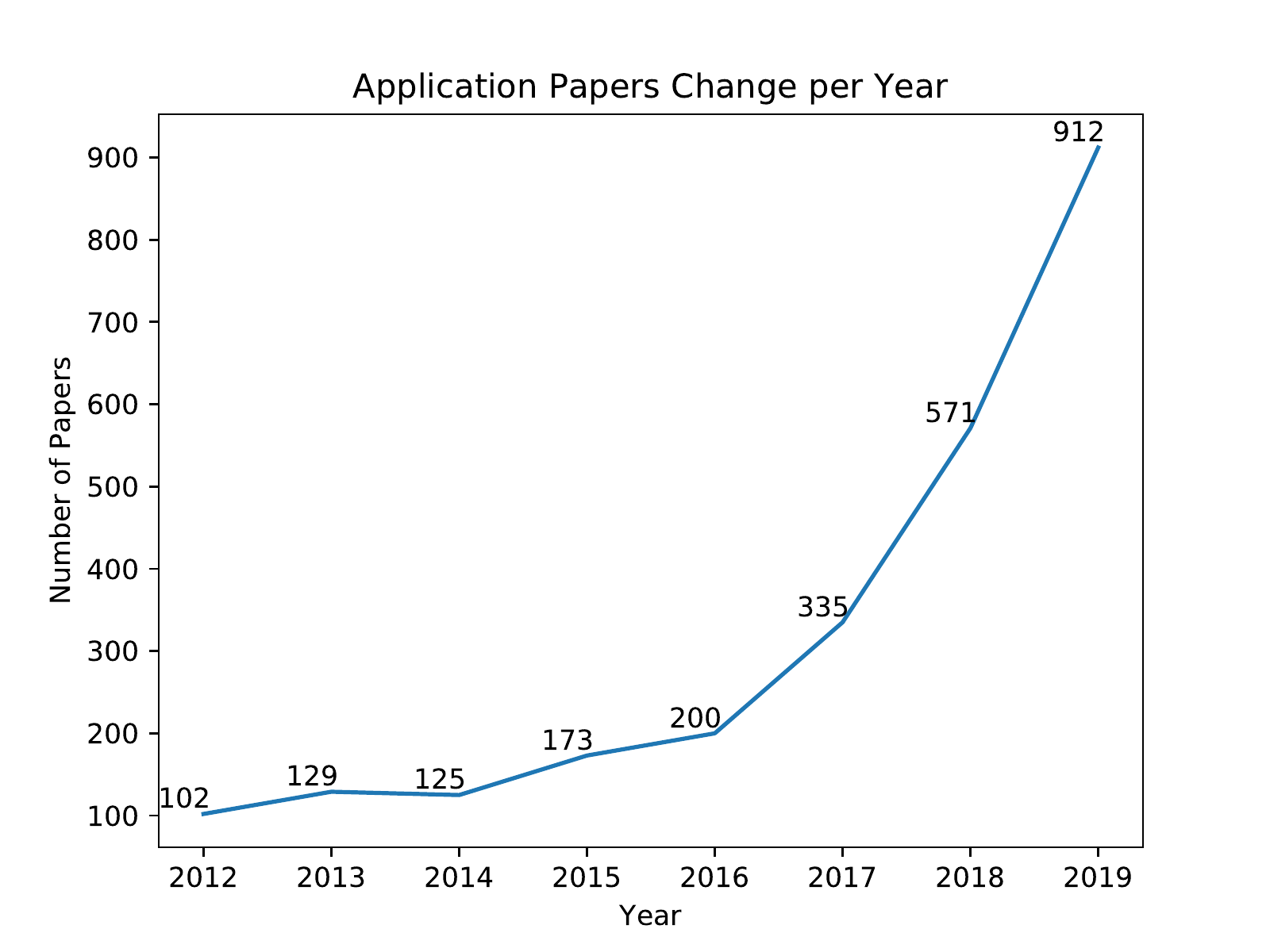}\\
\caption{Number of papers on fundus images and deep learning in recent years.}
\label{fig:paper_number}
\end{figure}

\textbf{Motivation.} The results in Fig. \ref{fig:paper_number} show that the number of papers on fundus images and deep learning are increasing year by year.
While several review papers already exist for these, they are all different from our review. 
For instance, \cite{abramoff2010retinal} and
\cite{DBLP:journals/midm/ZhangSLCDWKW014} only focus on classical machine learning methods;
\cite{DBLP:journals/artmed/SalamatMR19}, \cite{DBLP:journals/jois/KanseY19} and \cite{DBLP:journals/cmpb/MocciaMHM18} only consider specific individual diseases, such as DR or glaucoma;
and \cite{SCHMIDTERFURTH20181}, \cite{Rahimy2018Deep} and \cite{Artificial2018,Ting2018Artificial,TING2019100759} do not discuss specific deep learning methods or structures, but instead simply use ``a deep learning method'' and similar terms to refer to all methods.
Therefore, it is necessary to provide a high-quality review that analyzes the trends and highlights the future directions for the applications of deep learning in fundus images.

\textbf{Data.} In this review, we focus on the successful application of deep learning methods in fundus images from January 2016 to August 2020.
We collected 143 papers from the
\emph{DBLP}\footnote{https://dblp.uni-trier.de/db/}, \emph{ScienceDirect}\footnote{https://www.sciencedirect.com/}, \emph{JAMA Network} \footnote{https://jamanetwork.com/},
\emph{Investigative Ophthalmology \& Visual Science}\footnote{http://iovs.arvojournals.org/} and \emph{Web of Science}\footnote{http://apps.webofknowledge.com/} databases using the following keywords:
\emph{retina},
\emph{fundus},
\emph{diabetes retinopathy},
\emph{glaucoma},
\emph{age-related macular degeneration},
\emph{cataract},
\emph{retinal vessel},
\emph{optic disc / disk / cup},
\emph{fundus / retinal + lesion / abnormal},
\emph{hemorrhage},
\emph{microaneurysm},
\emph{exudate},
\emph{neovascularization},
\emph{drusen},
\emph{fundus / retinal + synthesis / enhance},
\emph{fundus / retinal + hypertension / stroke},
\emph{fundus / retinal + kidney / brain / heart},
and \emph{fundus / retinal + cardiovascular / cerebrovascular}.

\begin{figure}
\centering
\includegraphics[height=5.5cm]{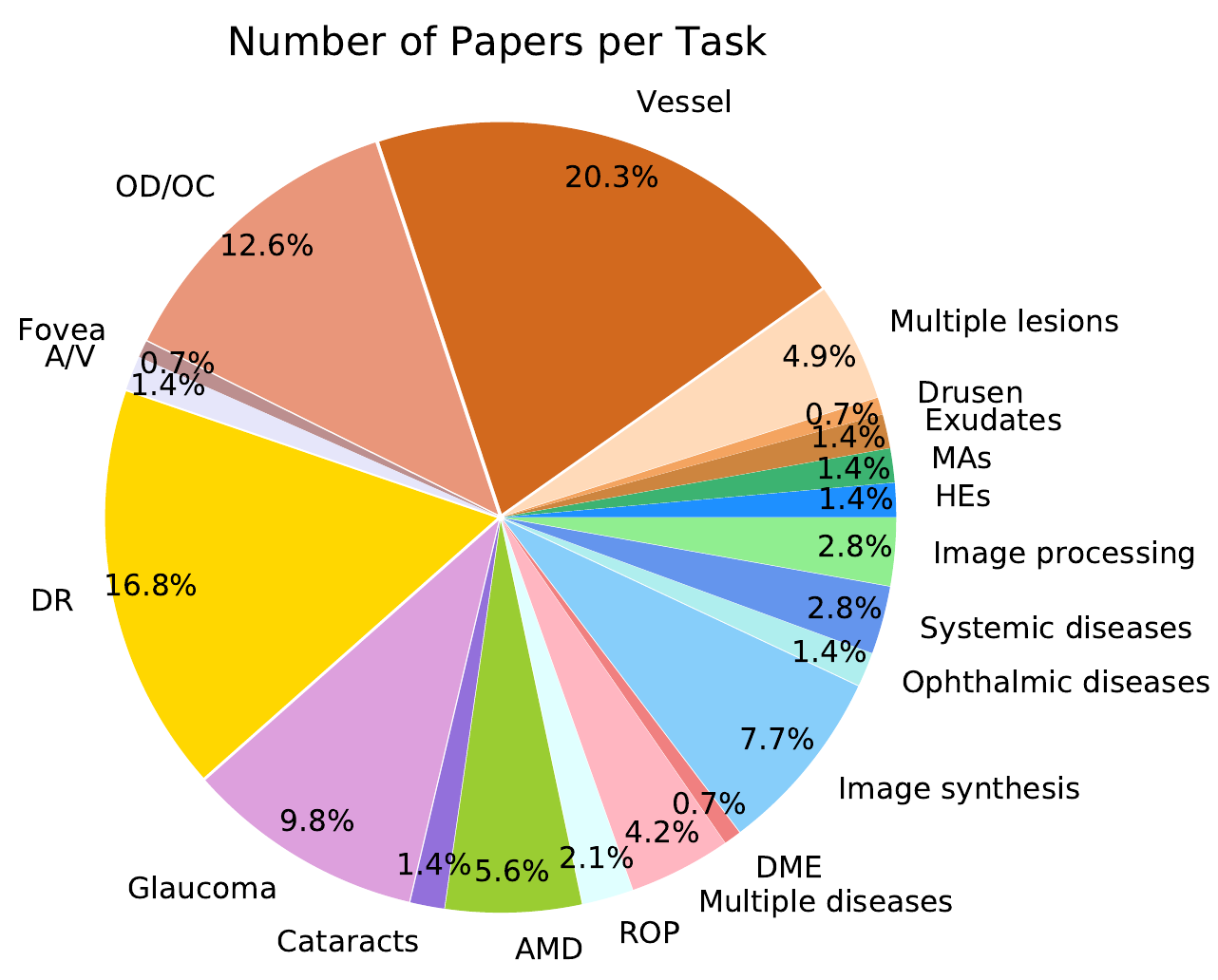}\\
\caption{The distribution of papers per task.}
\label{fig:paper_per_source}
\end{figure}

\begin{figure}[!t]
\centering
\subfigure[Papers per year]{\includegraphics[width=4.1cm]{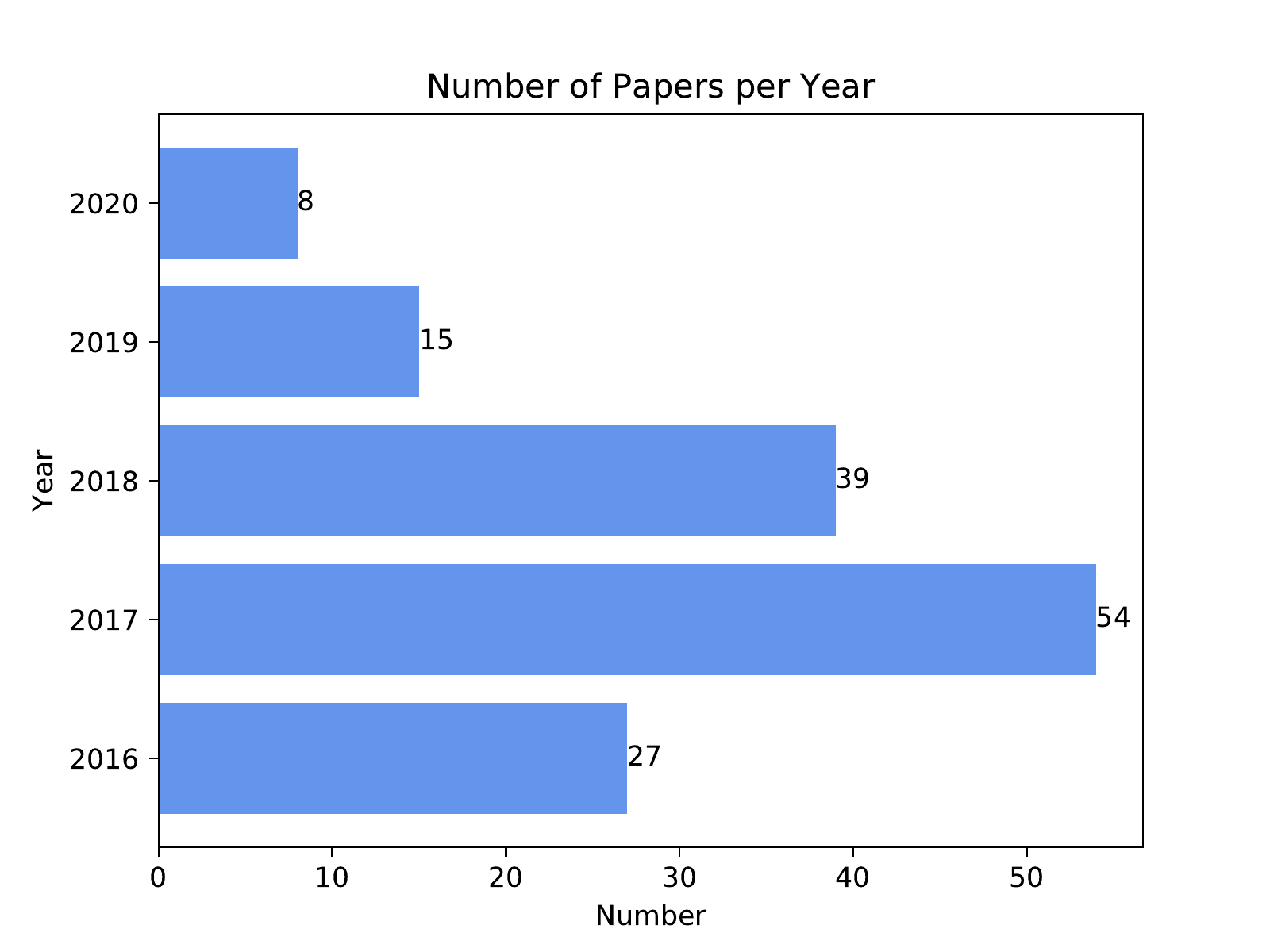}}
\subfigure[Papers per source]{\includegraphics[width=4.6cm]{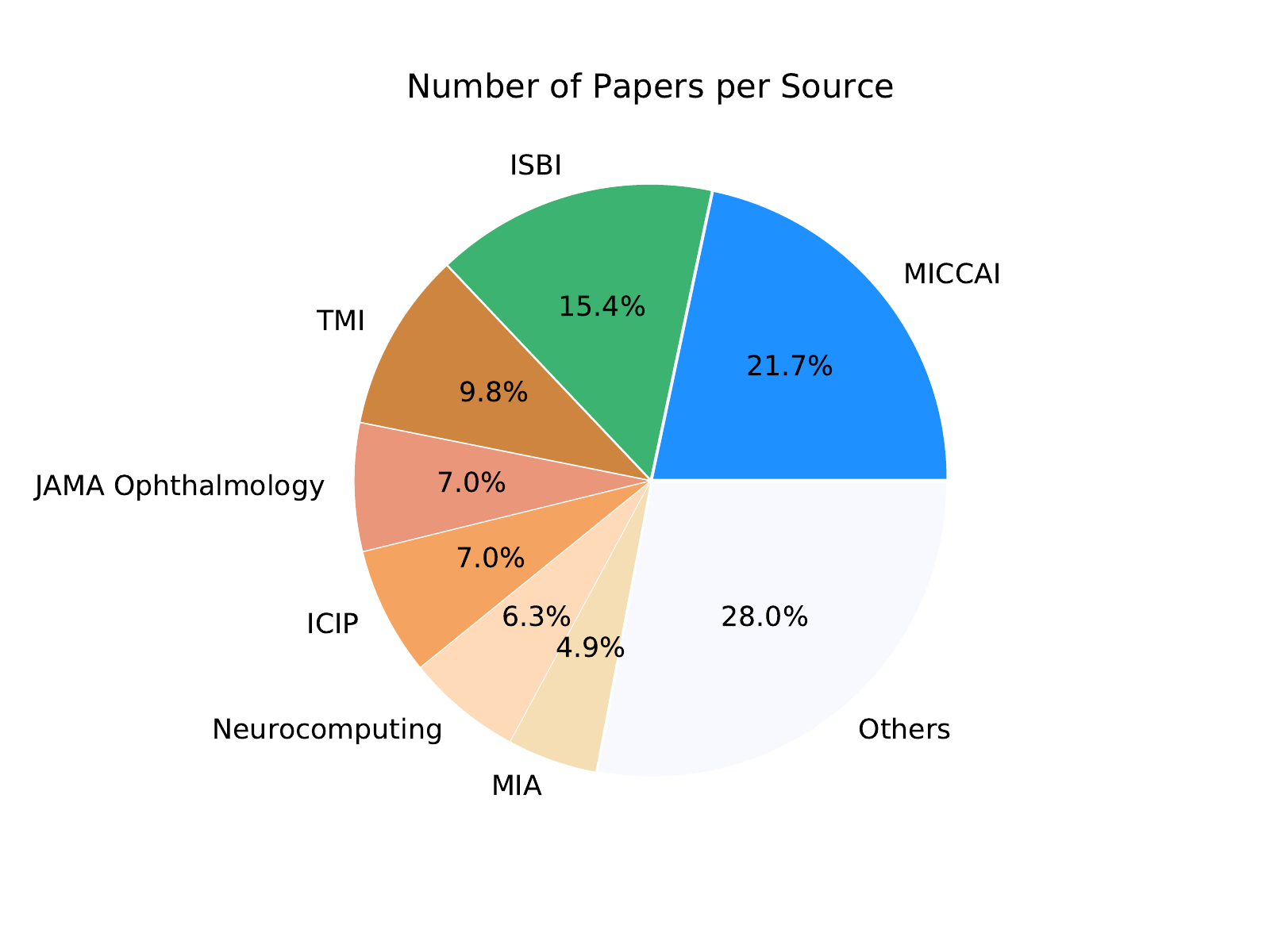}}
\caption{The distributions of papers per year and source.}
\label{fig:papers_stat}
\end{figure}

The conference sources include \emph{CVPR}, \emph{AAAI}, \emph{MICCAI}, \emph{ISBI}, \emph{ICMLA}, and \emph{ICIP},
and the journal sources include \emph{IEEE TIP}, \emph{IEEE TMI}, \emph{MIA}, \emph{JAMA}, \emph{JAMA Ophthalmology}, \emph{Ophthalmology}, \emph{Investigative Ophthalmology \& Visual Science}, \emph{Diabetes Care}, \emph{Nature Biomedical Engineering}, \emph{IEEE TBME}, \emph{IEEE JBHI}, \emph{Pattern Recognition}, \emph{Neural Networks}, \emph{Information Science}, \emph{Knowledge Based Systems}, \emph{Expert Systems with Applications}, \emph{Neurocomputing} and \emph{Future Generation Computer Systems}. 
The distribution of papers per task is shown in Fig. \ref{fig:paper_per_source}.
Distributions of papers per year/source are shown in Fig. \ref{fig:papers_stat}.

\begin{figure}[!t]
\centering
\includegraphics[height=5.3cm]{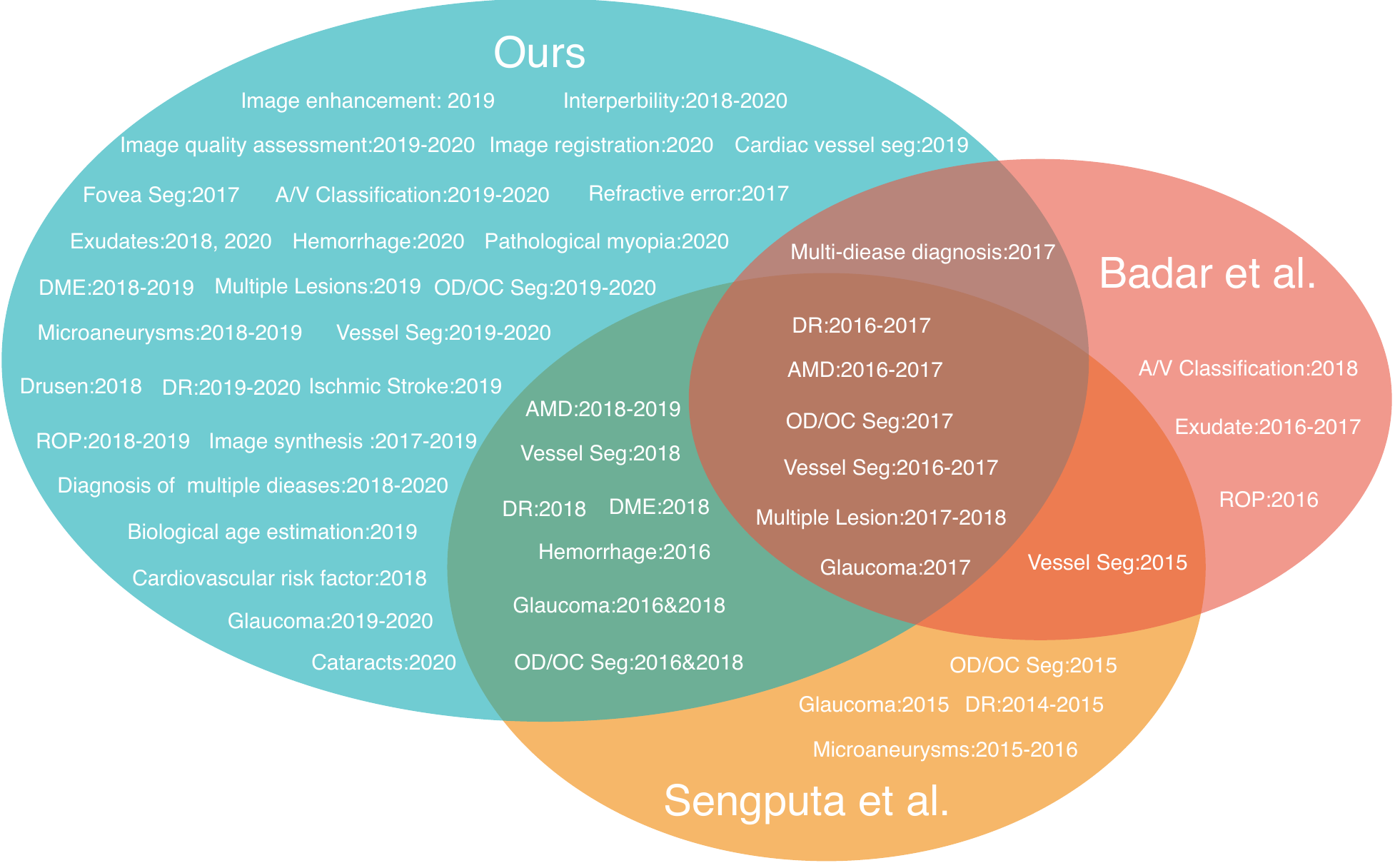}\\
\caption{Comparison with two recent reviews similar in view, style and year.}
\label{fig:compare_with_other_reviews}
\end{figure}

\begin{figure*}
\centering
\includegraphics[scale=0.44,angle=90]{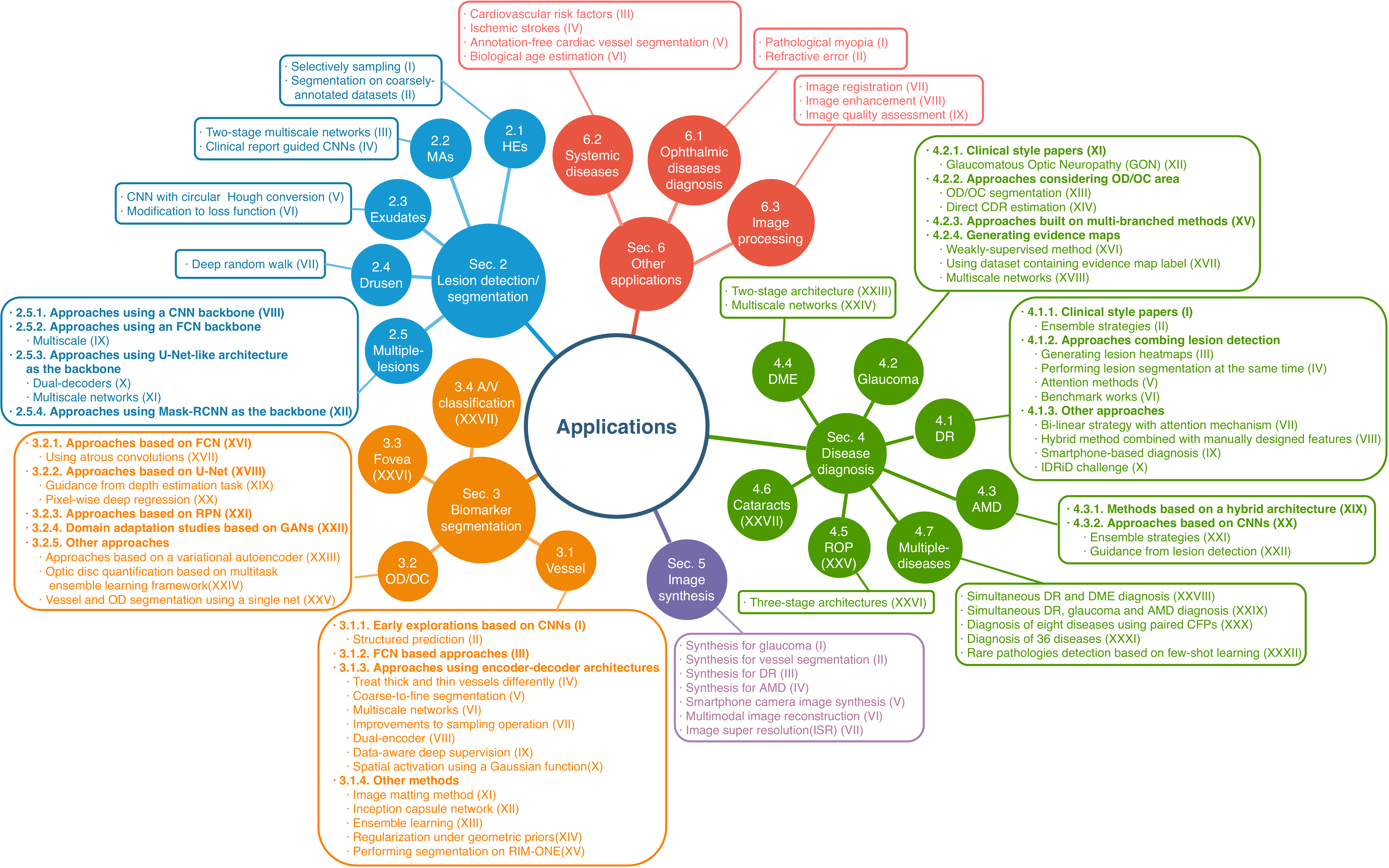}\\
\caption{The knowledge graph is summarized in this review. The lookup table for the references in the graph can be found in the \emph{Appendix}.}
\label{fig:knowledge_graph}
\end{figure*}

Two recent reviews, \cite{DBLP:journals/csr/BadarHF20} and \cite{DBLP:journals/artmed/SenguptaSLGL20}, are similar to ours in terms of view (fundus image, not ophthalmology), style (technical, not clinical), and method (deep learning, not artificial intelligence or machine learning).
However, only 34 papers and 62 papers are reviewed in each, respectively, while we review 143 papers.
Further, as shown in Fig. \ref{fig:compare_with_other_reviews}, the scopes are also different.
Finally, as shown in Fig. \ref{fig:knowledge_graph}, this review utilizes a carefully designed multi-layer hierarchy to organize the related works in a more intuitive manner.

\textbf{Contributions.} \textsl{First}, we give a comprehensive review of the applications of deep learning in fundus images.
Compared to recent works, this review covers more recent papers, more eye diseases and more challenging tasks, especially including image synthesis and several interesting applications in Section \ref{sec:Other_application}.
\textsl{Second}, we carefully design the taxonomy of our paper. 
A knowledge graph is summarized in Fig. \ref{fig:knowledge_graph}.
The lookup table for the references in the knowledge graph is presented in the \emph{Appendix}.
This can help readers to quickly find content of interest. \textsl{Third}, summaries and analyses are provided for each task.
Limitations that are common to current approaches are also described and possible solutions given in Section \ref{sec_conclu}.
This may provide inspiring ideas for researchers in this field.

\section{Lesion detection/segmentation}
\label{sec:leision_detection_segmentation}
In this section we will review how deep learning methods have been applied to lesion detection/segmentation.
The widely used datasets for this task are shown in Tab. \ref{tab:dataset_DR}. The \emph{availability} column is linkable in the soft copy.
Because of the correlation between lesion detection/segmentation and DR diagnosis, there is an overlap between datasets used for the two. 
\begin{table*}

\small
\caption{Widely used datasets of lesion detection/segmentation and DR diagnosis/grading}
\begin{center}

\resizebox{\textwidth}{!}{

\begin{tabular}{p{3cm} p{3.3cm} p{1.8cm} p{5cm} p{3.1cm}}
\hline
Dataset name & Number of images & Resolution & Camera & Availability\\
\hline
DIARETDB0 & 130 (110 DR, 20 Normal) & - & digital fundus cameras with unknown camera settings, FVO 50\degree & \href{https://www.it.lut.fi/project/imageret/}{available online}\tnote{1}$^{\rm 1}$ \\
\hline
DIARETDB1 & 89 (84 DR, 5 Normal) & 1500$\times$1152 & ZEISS FF 450plus fundus camera with Nikon F5 digital camera, FOV 50\degree & \href{https://www.it.lut.fi/project/imageret/}{available online} \tnote{2}$^{\rm 2}$ \\
\hline
Retinopathy Online Challenge & 100 & - & a Topcon NW 100, a Topcon NW 200, or a CanonCR5-45NM, 2 differently shaped FOVs & \href{http://webeye.ophth.uiowa.edu/ROC/}{available on registration}$^{\rm 3}$ \\
\hline
RC-RGB-MA & 250 & 2595$\times$1944 & a DRS non-mydriatic fundus camera, FOV45\degree & \href{http://www.retinacheck.org/datasets}{available online}$^{\rm 4}$\\
\hline
RC-SLO-MA & 58 & 1024$\times$1024 & an EasyScan camera (i-Optics Inc., the Netherlands), FOV45\degree & \href{http://www.retinacheck.org/datasets}{available online}$^{\rm 5}$\\
\hline
IDRiD & 516 & 4288$\times$2848 & a Kowa VX-10 alpha digital fundus camera, FOV 50\degree & \href{https://ieee-dataport.org/open-access/indian-diabetic-retinopathy-image-dataset-idrid}{available online}$^{\rm 6}$\\
\hline
Messidor & 1200 & 1440$\times$960, 2240$\times$1488, 2304$\times$1536 & a color video 3CCD camera on a Topcon TRC NW6 non-mydriatic retinograph with FOV 45\degree & \href{http://www.adcis.net/en/third-party/messidor/}{available on registration}$^{\rm 7}$\\
\hline
Messidor-2 & 1748 & 1440$\times$960, 2240$\times$1488, 2304$\times$1536&a Topcon TRC NW6 non-mydriatic fundus camera with FOV 45\degree & \href{http://www.adcis.net/en/third-party/messidor2/}{available on registration}$^{\rm 8}$\\
\hline
e-ophtha EX & 47 with 12,278 exudates, 35 healthy & ranging from 1440$\times$960 to 2544$\times$1696 & - & \href{http://www.adcis.net/en/third-party/e-ophtha/}{available on registration}$^{\rm 9}$\\
\hline
e-ophtha MA & 148 with 1306 MA, 233 healthy & ranging from 1440$\times$960 to 2544$\times$1696 & - & \href{http://www.adcis.net/en/third-party/e-ophtha/}{available on registration}$^{\rm 10}$\\
\hline
DDR & 13,673  & mixed & 42 types of fundus cameras with a 45\degree FOV & \href{https://github.com/nkicsl/DDR-dataset}{available online}$^{\rm 11}$\\
\hline
Kaggle/EyePACS & 35,126 train, 53,576 test &-&multiple fundus cameras and different ﬁelds of views & \href{https://www.kaggle.com/c/diabetic-retinopathy-detection/data}{available on registration}$^{\rm 12}$\\
\hline
CLEOPATRA & 298 & - & multiple fundus cameras & not available online\\
\hline

\end{tabular}
}


\end{center} 
\begin{tablenotes}
     \item[1]$^{\rm 1}$https://www.it.lut.fi/project/imageret/
     \item[2]$^{\rm 2}$https://www.it.lut.fi/project/imageret/
     \item[3]$^{\rm 3}$http://webeye.ophth.uiowa.edu/ROC/
     \item[4]$^{\rm 4}$http://www.retinacheck.org/datasets
     \item[5]$^{\rm 5}$http://www.retinacheck.org/datasets
     \item[6]$^{\rm 6}$https://ieee-dataport.org/open-access/indian-diabetic-retinopathy-image-dataset-idrid
     \item[7]$^{\rm 7}$http://www.adcis.net/en/third-party/messidor/
     \item[8]$^{\rm 8}$http://www.adcis.net/en/third-party/messidor2/
     \item[9]$^{\rm 9}$http://www.adcis.net/en/third-party/e-ophtha/
     \item[10]$^{\rm 10}$http://www.adcis.net/en/third-party/e-ophtha/
     \item[11]$^{\rm 11}$https://github.com/nkicsl/DDR-dataset
     \item[12]$^{\rm 12}$https://www.kaggle.com/c/diabetic-retinopathy-detection/data
     
\end{tablenotes}


\label{tab:dataset_DR}
\end{table*}
The DIARETDB0 dataset \citep{Kalesnykiene_diaretdb0:evaluation} consists of 130 images, of which 110 contain signs of DR (EXs, SEs, MAs, HEs and neovascularization) and 20 are normal.
DIARETDB1 \citep{DBLP:conf/bmvc/KauppiKKLSRVUKP07} consists of 89 images, of which 84 contain at least mild non-proliferative signs of DR (MA) and five are normal.
The RC-RGB-MA dataset \citep{DBLP:journals/tip/DashtbozorgZHR18} contains 250 images.
MAs were annotated by two experts at bounding-box level.
Images in the RC-SLO-MA dataset \citep{DBLP:journals/tip/DashtbozorgZHR18} were captured using scanning laser ophthalmoscopy (SLO).
The dataset contains 58 images with MA labels.
The Retinopathy Online Challenge (ROC) dataset \citep{DBLP:journals/tmi/NiemeijerGCMQSZHLMWCYMLHCRKGFA10} consists of 100 images that have been divided into a training set and a test set, both containing 50 images.
Center locations of MAs are labeled by experts.
The e-ophtha dataset can be divided into two subsets, namely e-ophtha EX and e-ophtha MA.
E-ophtha EX \citep{decenciere2013teleophta:} provides pixel-level labels for EX segmentation.
It consists of 47 images with exudates and 35 with no lesions.
E-ophtha MA \citep{decenciere2013teleophta:} consists of 148 images with MAs or small HEs and 233 healthy images.
The Messidor dataset \citep{decenciere2014feedback} consists of 1,200 images obtained from three ophthalmologic departments.
540 images are normal and 660 images are abnormal.
Messidor is divided into three sets, one per department, with different resolutions.
800 images were acquired with pupil dilation and 400 without.
Messidor-2 \citep{10.1001/jamaophthalmol.2013.1743} extended Messidor to 1,748 images.
Unlike Messidor, images of Messidor-2 are all in pairs.
The CLEOPATRA dataset \citep{2014A} consists of 298 images obtained from 15 hospitals in the UK.
It was acquired by different fundus cameras.
Therefore, the images have different resolutions.
Two experts were invited to annotate the ground truths for EXs, HEs and MAs.
The first expert marked all images and the second marked 135 images.
CLEOPATRA is not available online.
The two names Kaggle and EyePACS \citep{EyePacs} refer to the same dataset which was provided by EyePACS and used in the “Diabetic Retinopathy Detection--Identify signs of diabetic retinopathy in eye images" Kaggle competition. 
Kaggle dataset consists of 35,126 training images graded into five DR stages and 53,576 test images of undisclosed stages.
Images in the Kaggle dataset were obtained using multiple fundus cameras with differnt fileds of view.
The IDRiD dataset \citep{DBLP:journals/data/PorwalPKKDSM18} was used in the “Diabetic Retinopathy: Segmentation and Grading Challenge" held by ISBI in 2018.
It consists of three tasks, namely segmentation, disease grading and localization, with official training and test sets provided.
The segmentation task consists of 81 images with ground truths provided for lesions (MAs, HEs, EXs, SEs) and OD areas.
The disease grading task consists of 516 images with severity grade for DR and DME.
The localization task also consists of 516 images, with annotations for OD and fovea center localization.
Note that images in IDRiD have relatively high resolution.
The DDR dataset \citep{DBLP:journals/isci/LiGWGLK19} consists of 13,673 images which were obtained from 147 hospitals, covering 23 provinces in China.
Image level annotations with five classes of DR severity are provided for all images.
In addition, 757 images are provided with pixel-level and bounding-box-level annotations for lesions (MAs, EXs, SEs and HEs).

Experimental results for lesion segmentation on the various datasets introduced in this section are provided in Tab. \ref{tab:result_lesion_IDRiD}, \ref{tab:result_lesion_Eophtha}, \ref{tab:result_lesion_DiaretDB1} and \ref{tab:result_lesion_other}.
\begin{table*}
\small
\caption{Summary of several results for lesion detection/segmentation on IDRiD dataset}
\begin{center}

\resizebox{\textwidth}{!}{

\begin{tabular}{p{2.4cm} p{1.7cm} p{4.1cm} p{0.6cm} p{0.5cm} p{0.5cm} p{0.7cm} p{0.8cm} p{0.7cm} p{0.9cm} }
\hline
Reference & Backbone & Loss & PR/\% & SE/\% & SP/\% & ACC/\% & AUPR/\% & AUC/\% & F1/\% \\
\hline
\multicolumn{10}{c}{Hemorrhage detection/segmentation}\\
\hline
\cite{DBLP:journals/ijon/GuoLKLZW19} & FCN & Top-k loss, Bin loss &- &- &- &- & -& \textbf{67.34} &- \\
\cite{DBLP:conf/isbi/YanHWQXC19}  & U-Net & weighted CE & -& -& -&- & \textbf{70.3} &- &-\\
\hline
\multicolumn{10}{c}{Microaneurysms detection/segmentation}\\
\hline
\cite{DBLP:conf/miccai/SarhanAYNE19}(geometric) & FCN & Dice loss, CE and Triplet loss & \textbf{61.128} & 28.07 &  -& -& 41.96 & -& \textbf{38.4877}\\
\cite{DBLP:journals/ijon/GuoLKLZW19} & FCN & Top-k loss, Bin loss & -& -&- &- &- & \textbf{46.27} &- \\
\cite{DBLP:conf/isbi/YanHWQXC19} & U-Net & weighted CE & -& -& -& -& \textbf{52.5} &- &-\\
\cite{DBLP:journals/kbs/XueYQQQZCLLL19} & Mask-RCNN & log loss, regression loss, CE loss& - & \textbf{76.4} & \textbf{99.8} & \textbf{99.7} &- & -&-\\
\hline
\multicolumn{10}{c}{Hard exudate detection/segmentation}\\
\hline
\cite{DBLP:journals/ijon/GuoWKLGL20} & HED & Top-k loss, Bin loss &- & \textbf{95.74} &- &- &- & \textbf{98.71} & \textbf{95.57} \\
\cite{DBLP:journals/ijon/GuoLKLZW19} & FCN & Top-k loss, Bin loss & -& -& -& -&- & 79.45 &-\\
\cite{DBLP:conf/isbi/YanHWQXC19} & U-Net & weighted CE & -& -& -&- & \textbf{88.9} & -&-\\
\cite{DBLP:journals/kbs/XueYQQQZCLLL19} & Mask-RCNN & log loss, regression loss, CE loss& - & 77.9 & \textbf{99.6}  & \textbf{99.2} &- &- &-\\
\hline
\multicolumn{10}{c}{Soft exudate detection/segmentation}\\
\hline
\cite{DBLP:journals/ijon/GuoLKLZW19} & FCN & Top-k loss, Bin loss &- &- &- &- & -& \textbf{71.13}  &-\\
\cite{DBLP:conf/isbi/YanHWQXC19} & U-Net & weighted CE &- &- &- &- & \textbf{67.9} & -&-\\
\hline 
\end{tabular}
}

\end{center} 
\label{tab:result_lesion_IDRiD}
\end{table*}

\begin{table*}
\small
\caption{Summary of several results for lesion detection/segmentation on E-ophtha dataset}
\begin{center}

\resizebox{\textwidth}{!}{

\begin{tabular}{p{2.6cm} p{3.3cm} p{1.7cm} p{2.6cm} p{0.6cm} p{0.6cm} p{0.6cm} p{0.7cm} p{0.9cm} p{0.7cm} p{0.6cm}}
\hline
Reference & Task & Backbone & Loss & PR/\% & SE/\% & SP/\% & ACC/\% & AUPR/\% & AUC/\% & F1/\% \\
\hline
\cite{2018Retinal} & MA classification & CNN & -& -&- &- &- & \textbf{86} & \textbf{94} &- \\
\hline
\cite{DBLP:journals/ijon/GuoLKLZW19} & MA segmentation & FCN & Top-k loss, Bin loss & -& -& -& -& -& \textbf{16.87} &-\\
\cite{DBLP:journals/kbs/XueYQQQZCLLL19} & MA segmentation & Mask-RCNN & log loss, regression loss and CE loss& - & \textbf{67.2} & \textbf{99.8} & \textbf{99.7} &- & -&-\\
\hline
\cite{2018Retinal} & Exudates classification & CNN &- &- &- &- &- & \textbf{64} & \textbf{95} &- \\
\hline
\cite{DBLP:journals/ijon/GuoWKLGL20} & EX detection & HED & Top-k loss, Bin loss &- & \textbf{86.44} &- &- & -& \textbf{91.84} & \textbf{87.01} \\
\cite{DBLP:journals/ijon/GuoLKLZW19} & EX segmentation & FCN & Top-k loss, Bin loss &- &- & -&- &- & 41.71 &- \\
\cite{DBLP:journals/kbs/XueYQQQZCLLL19} & EX segmentation & Mask-RCNN & log loss, regression loss and CE loss& - & 84.6 & \textbf{98.8} & \textbf{98.4} &- & -&-\\
\hline
\cite{DBLP:journals/tmi/PlayoutDC19} & Bright Lesion segmentation & U-Net & loss based on Cohen’s coefficient & \textbf{78.50} & \textbf{80.02} & \textbf{99.88} & \textbf{99.77} &- &- & \textbf{79.25}\\
\hline
\cite{DBLP:journals/tmi/PlayoutDC19} & Red Lesion segmentation & U-Net & loss based on Cohen’s coefficient &\textbf{75.26} & \textbf{75.62} & \textbf{99.99} & \textbf{99.88} &- &- & \textbf{75.44}\\
\hline 
\end{tabular}
}
\end{center} 
\label{tab:result_lesion_Eophtha}
\end{table*}

\begin{table*}
\small
\caption{Summary of several results for lesion detection/segmentation on DiaretDB1 dataset}
\begin{center}

\resizebox{\textwidth}{!}{

\begin{tabular}{p{2.6cm}p{3.4cm}p{1.1cm}p{4.2cm}p{0.6cm}p{0.6cm}p{0.6cm}p{0.7cm}p{0.7cm}p{0.6cm}}
\hline
Reference & Task & Backbone & Loss & PR/\% & SE/\% & SP/\% & ACC/\% & AUC/\% & F1/\%\\
\hline
\cite{DBLP:journals/tmi/DaiFLHSWJ18} & MA detection & CNN & -& \textbf{99.7} & \textbf{87.8} &- & \textbf{96.1} & \textbf{93.4}&-\\
\hline
\cite{DBLP:journals/eswa/Adem18}& Exudate detection & CNN &- & -& \textbf{99.2} & \textbf{97.97} &- & -&-\\
\hline
\cite{DBLP:conf/miccai/PlayoutDC18} & Bright lesion segmentation & U-Net &loss based on Cohen’s coefficient &-  & 75.35 & 99.86 &- &-&- \\

\cite{DBLP:journals/tmi/PlayoutDC19} & Bright lesion segmentation & U-Net & loss based on Cohen’s coefficient & \textbf{81.70} & \textbf{88.29} & \textbf{99.93} & \textbf{99.89} & -& \textbf{84.87}\\
\hline
\cite{DBLP:conf/miccai/PlayoutDC18} & Red lesion segmentation & U-Net &loss based on Cohen’s coefficient &-  & 66.91 & 99.82 &- &- &-\\

\cite{DBLP:journals/tmi/PlayoutDC19} & Red lesion segmentation & U-Net & loss based on Cohen’s coefficient & \textbf{78.96} & \textbf{85.18} & \textbf{99.89} & \textbf{99.83} &- & \textbf{81.95}\\
\hline 
\end{tabular}

}

\end{center} 
\label{tab:result_lesion_DiaretDB1}
\end{table*}

\begin{table*}
\small
\caption{Summary of several results for lesion detection/segmentation on other datasets}
\begin{center}

\resizebox{\textwidth}{!}{

\begin{tabular}{p{2.5cm} p{2.4cm} p{1.6cm} p{2.2cm} p{3cm} p{0.6cm} p{0.6cm}  p{0.8cm} p{0.9cm}}
\hline
Reference & Task & Dataset & Backbone & Loss & SE/\% & SP/\% & AUC/\% & mAP/\%\\
\hline
\cite{DBLP:journals/tmi/GrinsvenGHTS16} & HE detection & Kaggle & CNN & CE & \textbf{83.7} & \textbf{85.1} & \textbf{89.4} & -\\
\hline
\cite{DBLP:journals/tmi/GrinsvenGHTS16} & HE detection & Messidor & CNN & CE  & \textbf{91.9} & \textbf{91.4}  & \textbf{97.2}& -\\
\hline
\cite{DBLP:conf/isbi/HuangLLWCWYT20} & HE segmentation & private & CNN & MSE, IoU, GIoU & -& -  & -  & \textbf{52.20}\\
\hline
\cite{DBLP:conf/miccai/YanCWLLWYZ18} & Drusen segmentation & STARE, DRIVE & Encoder-decoder Network & -& \textbf{92.02} & \textbf{97.30}  &- &-\\
\hline
\cite{DBLP:journals/eswa/Adem18}& Exudate detection & DiaretDB0 & CNN  &- & \textbf{100} & \textbf{98.41} &- & - \\
\hline
\cite{DBLP:journals/eswa/Adem18}& Exudate detection & DrimDB & CNN  & -& \textbf{100} & \textbf{98.44} &- &- \\
\hline
\cite{DBLP:journals/isci/TanFSBRCA17} & EX detection & CLEOPATRA & CNN & log-likelihood function  & \textbf{87.58} & \textbf{98.73} &- & -\\
\hline
\cite{DBLP:journals/isci/TanFSBRCA17} & HE detection & CLEOPATRA & CNN & log-likelihood function  &  \textbf{62.57} & \textbf{98.93} & -& -\\
\hline
\cite{DBLP:journals/isci/TanFSBRCA17} & MA detection & CLEOPATRA & CNN & log-likelihood function  & \textbf{46.06} & \textbf{97.99} &- & -\\
\hline
\cite{DBLP:journals/ijon/GuoLKLZW19} & EX segmentation & DDR & FCN & Top-k loss, Bin loss  & -&- & \textbf{55.46}&- \\
\hline
\cite{DBLP:journals/ijon/GuoLKLZW19} & SE segmentation & DDR & FCN & Top-k loss, Bin loss  & -& -& \textbf{26.48} &- \\
\hline
\cite{DBLP:journals/ijon/GuoLKLZW19} & HE segmentation & DDR & FCN & Top-k loss, Bin loss  & -&- & \textbf{35.86}&- \\
\hline
\cite{DBLP:journals/ijon/GuoLKLZW19} & MA segmentation & DDR & FCN & Top-k loss, Bin loss  &- &- & \textbf{10.52}&- \\
\hline 
\end{tabular}

}

\end{center} 
\label{tab:result_lesion_other}
\end{table*}

\subsection{Hemorrhages}
Hemorrhages (HEs) are one of the visible pathological signs of DR.
Accurate detection or segmentation of HEs is important for DR diagnosis.
In the task of lesion detection/segmentation, patch-based methods are quite popular because of the limited number of images in datasets and the need to reduce computational costs.
Patch-based methods can generate tens of thousands of patches with only dozens of images, which can help improve performance and alleviate the problem of overfitting.
However, HEs (as well as other lesions) are typically relatively small in size, with their pixels only making up a small proportion of the whole image.
This leads to an imbalance problem,
where only a few patches contain lesions and a large number do not contribute much to the lesion detection/segmentation task.
Imbalance is also common in other lesion detection/segmentation tasks in this section, details of which will not be repeated for brevity.
There are two main directions in the improvement of hemorrhage detection/segmentation; namely selective sampling and performing segmentation on coarsely-annotated datasets.

\textbf{Selective sampling.} \cite{DBLP:journals/tmi/GrinsvenGHTS16} proposed a method called selective sampling to reduce the use of redundant data and speed up CNN training.
They invited three experts to relabel the Messidor dataset and a subset of Kaggle.
During the training process, weights of samples were dynamically adjusted according to the current iteration's classification results, so that the informative samples were more likely to be included in the next training iteration.
Inspired by VGG, they designed a nine-layer CNN as the classifier.
On the Kaggle competition and  Messidor datasets, experimental results showed that the CNN with selective sampling (SeS) outperformed the CNN without selective sampling (NSeS), and SeS reduced the number of training epochs from 170 to 60.

\textbf{Segmentation on coarsely-annotated datasets.} \cite{DBLP:conf/isbi/HuangLLWCWYT20} proposed a bounding box refining network (BBR-Net) which can generate more accurate bounding box annotations for coarsely-annotated data.
Then they utilized a RetinaNet \citep{DBLP:conf/iccv/LinGGHD17} to detect hemorrhage.
Rather than using the finely annotated IDRiD dataset, they performed hemorrhage detection on a private dataset with coarsely annotated bounding box.
They first established a dataset containing image pairs.
For each pair, one image was taken from IDRiD and the other was obtained by simulating coarsely-annotated bounding boxes.
BBR-Net took coarsely annotated patches as input and finely annotated patches as target.
After training, the authors introduced their private data to obtain more accurate bounding box annotations, and then sent the results to the RetinaNet for hemorrhage detection.

\textbf{Discussion.}
The selective sampling method alleviates the problem of data imbalance.
Selective sampling is also used in other applications, which will be introduced in the following sections.
The explorations made by \cite{DBLP:conf/isbi/HuangLLWCWYT20} also offer a promising direction.
The generation of more accurate bounding box annotations can be seen as image synthesis, which will be discusses in more detail in Section \ref{sec:image_synthesis}.

However, there are still some limitations in the current HEs detection applications.
First, the imbalance problem needs to be further studied.
Second, compared to other lesions, less research has focused on HEs.
More attention needs to be paid to this area for its importance in DR diagnosis.
Third, pixel-level segmentation and detection are required.
More datasets that provide pixel-level labels for HEs, like DDR, still need to be explored.

\subsection{Microaneurysms}
MAs are the earliest clinical sign of DR and have thus captured more research interests.
There are several barriers affecting the segmentation of MAs, including the existence of other lesions with similar color, extremely low contrast, and variation in image lighting, clarity and background texture.
Two-stage multiscale architectures and guidance from clinical reports are some successful strategies for MAs detection.

\textbf{Two-stage multiscale networks.} \cite{DBLP:conf/miccai/SarhanAYNE19} proposed a two-stage deep learning approach embedding a triplet loss for microaneurysm segmentation.
The first stage is called the hypothesis generation network (HGN), in which multiscale FCNs are employed to generate a region of interest (ROI).
The second stage is known as the patch-wise refinement network (PRN), in which patches extracted from around ROIs are passed to a modified ResNet-50 for classification.
The authors introduced the triplet loss into the PRN to extract discriminative features.
Further, the previously mentioned selective sampling method \citep{DBLP:journals/tmi/GrinsvenGHTS16} is utilized to reduce the computational cost and solve the data imbalance problem.

\textbf{Clinical report guided CNNs.} \cite{DBLP:journals/tmi/DaiFLHSWJ18} proposed a clinical report guided multi-sieving convolutional neural network (MS-CNN) for the detection of MAs.
They first trained a weak image-to-text model from clinical reports and fundus images to generate a rough segmentation of microaneurysms.
Then the proposed MS-CNN was used to generate final high-quality segmentation using the rough segmentation as guidance.
In order to tackle the data imbalance problem, MS-CNN adopts a method similar to boosting. 
Specifically, MS-CNN is composed of multiple CNNs, where the false positives from the previous CNN are fed into the following CNN as negative examples.

\textbf{Discussion.} Several effective methods have been employed in MA segmentation, including multiscale networks, guidance from clinical reports and utilization of the triplet loss.
The extraction of ROIs and cascaded architecture adopted in MS-CNN alleviate the imbalance problem.
However, the two-stage architecture and cascaded architecture of MS-CNN lack efficiency.
They use multiple base networks, leading to a huge number of parameters to be trained.
Thus, one promising direction in MA segmentation would be to reduce complexity of the networks while maintaining high performance.

\subsection{Exudates}
Soft and hard exudates are usually the basis for the diagnosis of DR.
Accurate detection of SEs and EXs are thus crucial for timely treatment.
Like other lesion detection/segmentation tasks, there are several challenges.
The barriers include low contrast, varied sizes and similarity to other lesions.
There are several approaches for exudate detection, most of which can be divided into CNN with circular Hough conversion and modifications to the loss function.

\textbf{CNN with circular Hough conversion.} \cite{DBLP:journals/eswa/Adem18} introduced a three-layer CNN architecture for the binary classification of exudated and exudate-free fundus images.
During pre-processing, the OD region was removed by applying several methods, including adaptive histogram equalization, Canny edge detection and circular Hough conversion.

\textbf{Modification to loss function.} \cite{DBLP:journals/ijon/GuoWKLGL20} proposed a top-k loss and a bin loss to enhance performance for exudate segmentation.
The class balanced cross entropy (CBCE) loss \citep{7410521} solved the class imbalance problem to some extent.
However, this introduced the new problem of loss imbalance, where background similar to exudate tends to be misclassified.
The main reason is that with the different weights for background and foreground pixels in CBCE loss, the loss for misclassifying a background pixel is much smaller than that for misclassifying a hard exudate pixel.
To solve this loss imbalance problem, top-k loss is proposed, which considers all hard exudate pixels but only top-k background pixels with the larger loss.
They also proposed a fast version of top-k loss named bin loss with consideration of efficiency.

\textbf{Discussion.}
In exudate detection, some works have focused on modifying the loss function.
The top-k loss and bin loss solved the loss imbalance problem caused by the use of CBCE.
However, the misclassification problem still remains.
Moreover, only baseline models have been tested and no innovative architecture has been proposed.
\cite{DBLP:journals/eswa/Adem18}'s work is based on a CNN.
Several more recent models, such as encoder-decoder networks, need to be utilized in exudate detection.

\subsection{Drusen}
Drusen, the main manifestations of the disease, can be used to assist in the diagnosis of AMD.
There are four main challenges to drusen segmentation: their yellowish-white color is similar to the fundus image and OD; uneven brightness and interference from other biomarkers, such as blood vessels, is common; the drusen often have irregular shapes; and boundaries may be blurred.

\textbf{Deep random walk.} \cite{DBLP:conf/miccai/YanCWLLWYZ18} proposed a deep random walk method to successfully segment drusen from fundus images. 
The proposed architecture is composed of three main parts.
Fundus images are first passed into a deep feature extraction module, which consists of two branches: a SegNet-like network capturing deep semantic features and a three-layer CNN capturing low-level features.
Then the captured features are fused together and passed into a named affinity learning module to obtain pixel-pixel affinities for formulating the transition matrix of the random walk.
Finally, a deep random walk module is applied to propagate manual labels.
This model achieved state-of-the-art performance on the STARE and DRIVE datasets.

\textbf{Discussion.}
As can be seen, only one effective approach has been introduced so far.
Other architectures and methods need to be explored.
Further, the segmentation of drusen is closely related to the diagnosis of AMD.
Therefore, one of the future works can be extending the original drusen segmentation to serve as an evidence for AMD diagnosis.

\subsection{Multiple lesions}
Most previous works only segment/detect one type of lesion or treat all lesions as a single group (usually red lesions or bright lesions).
However, segmenting multiple lesions simultaneously is of more practical value.
More and more researchers are thus focusing on multi-lesion segmentation/detection.
The challenges found in the individual lesion detection/segmentation tasks, including imbalance, contrast, illumination, etc, still exist.
Further the inter-class similarity for different lesions, such as HEs and MAs becomes more prominent.
All these factors make multi-lesion segmentation a challenging task.

\subsubsection{Approaches using a CNN backbone} 
\cite{DBLP:journals/isci/TanFSBRCA17} conducted the first work to segment multiple lesions, including  exudates, haemorrhages and microaneurysms, automatically and simultaneously using a 10-layer CNN, with the outputs evaluated at the pixel level.
Their work demonstrated that it is possible to segment several lesions simultaneously using a single CNN architecture.
\cite{2018Retinal} used a CNN to perform five-class classification on image patches.
The five classes consist of 1) normal, 2) microaneurysms, 3) dot-blot hemorrhages, 4) exudates or cotton wool spots, and 5) high-risk lesions such as neovascularization, venous beading, scarring, and so forth.
They invited two ophthalmologists to verify and relabel a subset of the Kaggle dataset containing 243 images.
The image patching method was used, proving that good performance can be obtained using such a method, even with limited training samples.

\subsubsection{Approaches using an FCN backbone} 
\textbf{Multiscale} networks are important models that have been applied to many fields. 
\cite{DBLP:journals/ijon/GuoLKLZW19} proposed a small object segmentation network (L-Seg) which can segment four kinds of lesions, including microaneurysms, soft exudates, hard exudates and hemorrhages, simultaneously.
The backbone network is a VGG-16, which has five groups of convolution layers and three fully connected (FC) layers.
They removed all the FC layers and the fifth pooling layer and added a side extraction layer which consists of a 1$\times$1 conv and upsampling to every conv group (except the first one) with deep supervision.
The final output is obtained by multiscale weighted fusion of the side extraction layers, instead of simple element-wise sum.
The bin loss \citep{DBLP:journals/ijon/GuoWKLGL20} was also used to solve the problem of class imbalance and loss imbalance.

\subsubsection{Approaches using U-Net-like architecture as the backbone}
There are two main directions explored in this section, namely dual-decoders and multiscale networks.

\textbf{Dual-decoders.} \cite{DBLP:conf/miccai/PlayoutDC18} proposed an extension to U-Net which is capable of segmenting red and bright lesions simultaneously.
They are the first to use fully convolutional approaches for joint lesion segmentation.
Several novel developments were used in their decoder, including residual connections, global convolutions and mixed-pooling.
They used two identical decoders, each specialized for one lesion category.
Near the end of training, they also added two fully connected conditional random fields (CRFs) \citep{DBLP:conf/nips/KrahenbuhlK11}.
In their subsequent work, \cite{DBLP:journals/tmi/PlayoutDC19}, made several modifications.
They proposed a novel unsupervised method to enhance segmentation performance by training the network at image-level labels when pixel-level annotations are limited.
They introduced an exchange layer which aims to share parameters between two decoders softly, instead of employing hard parameter sharing as previously \citep{DBLP:conf/miccai/PlayoutDC18}.

\textbf{Multiscale networks.} \cite{DBLP:conf/isbi/YanHWQXC19} combined local and global features to segment microaneurysms, soft exudates, hard exudates and hemorrhages.
A GlobalNet was used to capture more context features, taking a downsampled version of original images as input.
They also employed a LocalNet which takes cropped image patches as input, aiming to capture more detailed information.
GlobalNet and LocalNet both use a U-Net-like encoder-decoder architecture as their backbone.

\subsubsection{Approaches using Mask-RCNN as the backbone}  \cite{DBLP:journals/kbs/XueYQQQZCLLL19} proposed a deep membrane system for simultaneous MAs, EXs and OD segmentation.
A hybrid structure, consisting of a dynamic membrane system and communication channels between cells, was designed.
Three types of rules, i.e. T-rules, G-rules and D-rules were proposed for the computation and communication of the system, solving complex real applications in parallel.
Mask-RCNN served as the computational cell of the membrane system. 

\subsubsection{Discussion}
In this subsection, we have seen that various base networks have been applied to multi-lesion detection, including recent models like U-Net and Mask-RCNN.
Multiscale methods have also been explored, and have been proven quite suitable for this task.
Architectures modifications have also been introduced, with the dual-decoder being one notable example.
Such a framework may work well on other similar scenarios.
The proposed deep membrane system is quite innovative in fundus image analysis and is expected to be further explored.

However, there are still some limitations.
First, compared to other segmentation and detection tasks like blood vessel segmentation and OD/OC segmentation, the performance, and in particular sensitivity of lesion segmentation/detection needs to be further improved. 
Second, there are still several works that focus on red or bright lesion segmentation instead of individual lesions.
However, the specific segmentation and detection of individual lesions is more practical.
Third, pixel-wise segmentation should be emphasized, and datasets with pixel-level lesion annotations deserve more attention.

\section{Biomarker segmentation}
\label{sec:biomarker_segmentation}
\subsection{Vessel segmentation}
Segmentation of retinal blood vessels is of paramount importance in the diagnosis of various ophthalmic diseases including diabetic retinopathy and glaucoma \citep{5660089}.
With the use of powerful deep learning techniques such as CNNs, FCNs and recently U-Net, excellent performance has been achieved.
However, there still remain some factors making retinal blood vessel segmentation a challenging task.
These factors include varying contrast and intensity among different datasets, inter-vessel differences between thick and thin vessels, the presence of optic disc and lesions, limited annotated data and so on.
We will discuss how these problems were addressed in the following subsections.

The most commonly used datasets in retinal blood vessel segmentation include DRIVE, STARE, CHASE\_DB1 and HRF. 
The DRIVE dataset \citep{DBLP:journals/tmi/StaalANVG04} consists of 40 images, seven of which show signs of mild early DR.
DRIVE is officially divided into a training set and a test set, both containing 20 images.
A single manual segmentation of vessels is provided in the training set and two manual segmentations are provided in the test set.
Border masks are also available for all images.
The STARE dataset \citep{DBLP:journals/tmi/HooverKG00} consists of 400 images, 20 of which have two manual blood vessel segmentations annotated by two experts.
Ten of the images contain pathologies.
Coarsely annotated centerline-level artery/vein labels of 10 images are also provided.
The CHASE\_DB1 dataset \citep{owen2009measuring} consists of 28 images, obtained from both eyes of 14 multi-ethnic school children.
The HRF dataset \citep{DBLP:journals/ijbi/BudaiBMHM13} consists of 45 images, of which 15 are healthy, 15 have DR and 15 are glaucomatous.
Compared to the other three datasets, images from HRF have a higher resolution (3504$\times$2336).
More details are shown in Tab. \ref{tab:dataset_vessel}. 
Experimental results on different datasets are shown in Tab. \ref{tab:result_vessel_drive}, \ref{tab:result_vessel_stare}, \ref{tab:result_vessel_chase} and \ref{tab:result_vessel_hrf}.

\begin{table*}
\caption{Widely used datasets for vessel segmentation}
\small
\begin{center}

\resizebox{\textwidth}{!}{

\begin{tabular}{p{1.8cm}p{4.4cm}p{2cm}p{4cm}p{3cm}}
\hline
Dataset name & Number of images & Resolution & Camera & Availability\\
\hline
DRIVE &40 (33 healthy, 7 mild early DR) & 768$\times$584&a Canon CR5 non-mydriatic 3CCD camera, FOV 45\degree & \href{https://drive.grand-challenge.org/Download/}{available on registration}$^{\rm 1}$\\
\hline
STARE & 400 (vessel segmentation labeling of 40 , A/V labeling of 10) &700 $\times$ 605&a TopCon TRV-50 fundus camera, FOV35\degree & \href{http://cecas.clemson.edu/~ahoover/stare/}{available online}$^{\rm 2}$\\
\hline
CHASE\_DB1&28 &1280$\times$ 960& - & \href{https://blogs.kingston.ac.uk/retinal/chasedb1/}{available online}$^{\rm 3}$\\
\hline
HRF&45, 15 each of healthy, DR and glaucomatous &3504 $\times$ 2336& a Canon CR-1 fundus camera with FOV 45\degree & \href{http://www5.cs.fau.de/research/data/fundus-images/}{available online}$^{\rm 4}$\\
\hline
\end{tabular}

}

\end{center} 
\begin{tablenotes}
     \item[1]$^{\rm 1}$https://drive.grand-challenge.org/Download/
     \item[2]$^{\rm 2}$http://cecas.clemson.edu/~ahoover/stare/
     \item[3]$^{\rm 3}$https://blogs.kingston.ac.uk/retinal/chasedb1/
     \item[4]$^{\rm 4}$http://www5.cs.fau.de/research/data/fundus-images/
\end{tablenotes}
\label{tab:dataset_vessel}
\end{table*}

\begin{table*}
\small
\caption{Summary of several results for vessel segmentation on DRIVE dataset}
\begin{center}

\resizebox{\textwidth}{!}{

\begin{tabular}{p{3.2cm} p{2.3cm} p{3.6cm} p{0.9cm} p{0.9cm} p{0.9cm} p{0.9cm} p{0.9cm}}
\hline
Reference & Backbone & Loss & SE/\% & SP/\%  & ACC/\%   & AUC/\%  & F1/\% \\
\hline
\cite{DBLP:conf/icip/KhalafYF16} & CNN & - & 83.97 & 95.62 & 94.56 & - & -\\
\hline
\cite{DBLP:journals/tmi/LiskowskiK16}  & CNN & CE & \textbf{91.60} & 92.41 & 92.30 & 97.38& -\\
\hline
\cite{DBLP:journals/ijon/YuQZDQC20} & CNN & - & 76.43 & 98.03 & 95.24 & 97.23 & -\\
\hline
\cite{DBLP:conf/miccai/FuXLW016}  & FCN & CBCE & 76.03 &- & 95.23 &-  &-\\
\hline
\cite{DBLP:conf/isbi/DasguptaS17}  & FCN & CE & 76.91 & 98.01 & 95.33 & 97.44 &- \\
\hline
\cite{DBLP:conf/icip/Feng0Y17} & FCN & CBCE & 78.11 & 98.39 & 95.60 & 97.92 &- \\
\hline
\cite{DBLP:journals/eswa/OliveiraPS18} & FCN & categorical CE & 80.39 & 98.04 & 95.76 & 98.21 &-\\
\hline
\cite{DBLP:conf/miccai/ZhangC18} & U-Net & CE & 87.23 & 96.18 & 95.04  & 97.99& \\
\hline 
\cite{DBLP:conf/icip/HeZZLF018} & U-Net & Focal loss & 77.61& 97.92 & 95.19&-  & 81.29\\
\hline
\cite{DBLP:journals/tbe/YanYC18} & U-Net & Proposed segment-level loss & 76.53 & 98.18 & 95.42& 97.52 &-\\
\hline
\cite{DBLP:journals/titb/YanYC19} & U-Net & CE & 76.31 & 98.20& 95.38 & 97.50& -\\
\hline
\cite{DBLP:conf/miccai/WuXSZC18}& U-Net & CE & 78.44& 98.19& 95.67& 98.07 &-\\
\hline
\cite{DBLP:journals/nn/WuXSZC20} & U-Net & CE & 79.96 & 98.13 & 95.82 & 98.30& -\\
\hline
\cite{DBLP:conf/isbi/00210HWC20} & U-Net & CE & 78.49 & 98.13 & 95.67 & 97.88 & 82.41\\
\hline
\cite{DBLP:journals/ijon/HuZNZCXG18}  & FCN & improved CE & 77.72 & 97.93 & 95.33 & 97.59 &- \\
\hline
\cite{DBLP:conf/miccai/WuX0ZLZC19} & U-Net & CE & 80.38 & 98.02 & 95.78 & 98.21 &-\\
\hline
\cite{DBLP:journals/eswa/SoomroAGHZP19} & SegNet & CBCE & 87 & 98.5 & 95.6 & 98.6 &- \\
\hline
\cite{DBLP:conf/miccai/ZhangFYZWYTX19} & U-Net & - & 81.00 & 98.48 & 96.92 & 98.56 &- \\
\hline
\cite{DBLP:conf/miccai/Wang0H19} & U-Net & CE and Jaccard loss & 79.40 & 98.16 & 95.67 & 97.72 & \textbf{82.70}\\
\hline
\cite{DBLP:conf/miccai/MaYMWDZ19} & U-Net & CE & 79.16 & 98.11 & 95.70  & 98.10 &-\\
\hline
\cite{DBLP:journals/pr/ZhaoLC20}  & Dense U-Net & global pixel loss, local matting loss & 83.29& 97.67&- &- & 82.29\\
\hline
\cite{DBLP:conf/isbi/MishraCH20} & U-Net & CE & 89.16& 96.01 & 95.40 & 97.24&-\\
\hline
\cite{DBLP:journals/ijon/FengZPT20} & FCN & MSE & 76.25 & 98.09 & 95.28 & 96.78 &-\\
\hline
\cite{DBLP:journals/tip/CherukuriGBM20} & Residual FCN & MSE & 84.25 & \textbf{98.49} & \textbf{97.23}& \textbf{98.70}&- \\
\hline
\cite{DBLP:conf/isbi/KrommR20} & CapsNet & margin loss & 76.51 & 98.18 & 95.47 & 97.50 &-\\
\hline
\cite{DBLP:conf/miccai/Liu0019} & No-reference net & MSE & 80.72 & 97.80 & 95.59 &97.79& 82.25 \\
\hline 
\label{tab:result_vessel_drive}
\end{tabular}

}

\end{center} 
\end{table*}

\begin{table*}
\small
\caption{Summary of several results for vessel segmentation on STARE dataset}
\begin{center}


\begin{tabular}{p{3.2cm} p{2.3cm} p{3.6cm} p{0.9cm} p{0.9cm} p{0.9cm} p{0.9cm} p{0.9cm}}
\hline
Reference & Backbone & Loss & SE/\% & SP/\% & ACC/\%  & AUC/\% & F1/\%\\
\hline
\cite{DBLP:journals/tmi/LiskowskiK16} & CNN & CE & \textbf{93.07} & 93.04 & 93.09 & 98.20& -\\
\hline
\cite{DBLP:journals/ijon/YuQZDQC20} & CNN & - & 78.37 & 98.22 & 96.13 & 97.87& - \\
\hline
\cite{DBLP:conf/miccai/FuXLW016}  & FCN & CBCE & 74.12 &- & 95.85 &-  &-\\
\hline
\cite{DBLP:journals/eswa/OliveiraPS18} & FCN & categorical CE & 83.15 & 98.58 & 96.94 & 99.05 &-\\
\hline
\cite{DBLP:conf/miccai/ZhangC18} & U-Net & CE & 76.73 & 99.01 & 97.12 & 98.82 & -\\
\hline 
\cite{DBLP:conf/icip/HeZZLF018} & U-Net & Focal loss & 81.20 & 98.95 & 97.04 & -& 85.53\\
\hline
\cite{DBLP:journals/tbe/YanYC18}  & U-Net & Proposed segment-level loss & 75.81  & 98.46 & 96.12 & 98.01 &-\\
\hline
\cite{DBLP:journals/titb/YanYC19} & U-Net & CE & 77.35 & 98.57 & 96.38 & 98.33& -\\
\hline
\cite{DBLP:journals/nn/WuXSZC20} & U-Net & CE & 79.63 & 98.63 & 96.72 & 98.75& -\\
\hline
\cite{DBLP:conf/isbi/00210HWC20} & U-Net & CE & 90.24 & \textbf{99.34} & \textbf{98.49} & \textbf{99.60} & \textbf{91.84}\\
\hline
\cite{DBLP:journals/ijon/HuZNZCXG18}  & FCN & improved CE & 75.43 & 98.14 & 96.32 & 97.51 &- \\
\hline
\cite{DBLP:journals/ijon/FengZPT20} & FCN & MSE &  77.09 & 98.48 & 96.33 & 97 &-\\
\hline
\cite{DBLP:journals/eswa/SoomroAGHZP19} & SegNet & CBCE & 84.8 & 98.6 & 96.8 & 98.8 &- \\
\hline
\cite{DBLP:journals/tip/CherukuriGBM20} & Residual FCN & MSE & 86.64 & 98.95 & 98.03& 99.35 &- \\
\hline
\cite{DBLP:journals/pr/ZhaoLC20} & Dense U-Net & global pixel loss, local matting loss &  84.33 &  98.57 &- &- & 83.51\\
\hline
\cite{DBLP:conf/isbi/MishraCH20} & U-Net & CE & 87.71 & 96.34 & 95.71 & 97.42&-\\
\hline
\cite{DBLP:conf/miccai/Liu0019} & No-reference net & MSE & 77.71& 98.43 & 96.23 & 97.93 & 80.36 \\
\hline 
\label{tab:result_vessel_stare}
\end{tabular}

\end{center} 
\end{table*}

\begin{table*}
\small
\caption{Summary of several results for vessel segmentation on CHASE\_DB1 dataset}
\begin{center}

\resizebox{\textwidth}{!}{

\begin{tabular}{p{3.2cm} p{2.3cm} p{3.6cm} p{0.9cm} p{0.9cm} p{0.9cm} p{0.9cm} p{0.9cm}}
\hline
Reference & Backbone & Loss & SE/\% & SP/\% & ACC/\%  & AUC/\% & F1/\%\\
\hline
\cite{DBLP:conf/miccai/FuXLW016} & FCN & CBCE & 71.30 & -& 94.89 & - &-\\
\hline
\cite{DBLP:journals/eswa/OliveiraPS18} & FCN & categorical CE & 77.79 & 98.64 & 96.53 & 98.55 &-\\
\hline
\cite{DBLP:conf/miccai/ZhangC18}  & U-Net & CE & 76.70 & \textbf{99.09} & 97.70 & 99.00 &- \\
\hline
\cite{DBLP:journals/tbe/YanYC18} & U-Net & Proposed segment-level loss & 76.33 & 98.09 & 96.10  & 97.81 &-\\
\hline
\cite{DBLP:journals/titb/YanYC19} & U-Net & CE & 76.41 & 98.06 & 96.07 & 97.76& -\\
\hline
\cite{DBLP:conf/miccai/WuXSZC18} & U-Net & CE & 75.38 & 98.47 & 96.37& 98.25 &-\\
\hline
\cite{DBLP:journals/nn/WuXSZC20} & U-Net & CE & 80.03 & 98.80 & 96.88 & 98.94& -\\
\hline
\cite{DBLP:conf/isbi/00210HWC20} & U-Net & CE & 79.48 & 98.42 & 96.48 & 98.47 & 82.20\\
\hline
\cite{DBLP:conf/miccai/WuX0ZLZC19} & U-Net & CE & 81.32 & 98.14 & 96.61 & 98.60 &-\\
\hline
\cite{DBLP:journals/eswa/SoomroAGHZP19} & SegNet & CBCE & \textbf{88.6} & 98.2 & 97.6 & 98.5 & -\\
\hline
\cite{DBLP:conf/miccai/ZhangFYZWYTX19}  & U-Net & & 81.86 & 98.48 & 97.43 & 98.63 &- \\
\hline
\cite{DBLP:journals/tip/CherukuriGBM20} & Residual FCN & MSE & 80.17 & 99.08 & \textbf{97.88} &  98.64 &- \\
\hline
\cite{DBLP:conf/miccai/Wang0H19} & U-Net & CE and Jaccard loss & 80.74 & 98.21 & 96.61  & 98.12 & 80.37\\
\hline
\cite{DBLP:conf/isbi/MishraCH20} & U-Net & CE & 88.05 & 96.51 & 96.01 & 97.63 &-\\
\hline
\cite{DBLP:conf/miccai/Liu0019}  & No-reference net & MSE & 87.69 & 98.43 & 97.42  & \textbf{99.05} & \textbf{85.98} \\
\hline 
\label{tab:result_vessel_chase}
\end{tabular}
}

\end{center} 
\end{table*}

\begin{table*}
\small
\caption{Summary of several results for vessel segmentation on HRF dataset}
\begin{center}

\resizebox{\textwidth}{!}{

\begin{tabular}{p{2.7cm} p{1.7cm} p{4.6cm} p{0.9cm} p{0.9cm} p{0.9cm} p{0.9cm} p{0.9cm}}
\hline
Reference & Backbone & Loss & SE/\% & SP/\% & ACC/\%  & AUC/\% & F1/\%\\
\hline
\cite{DBLP:journals/eswa/SoomroAGHZP19} & SegNet & CBCE & \textbf{82.9} & 96.1 & \textbf{96.2} & \textbf{98.5} &- \\
\hline
\cite{DBLP:journals/pr/ZhaoLC20} & Dense U-Net & global pixel loss, local matting loss & 78.09 & \textbf{98.18} &- & -& \textbf{78.13}\\
\hline 
\label{tab:result_vessel_hrf}
\end{tabular}
}

\end{center} 
\end{table*}

\subsubsection{Early explorations based on CNNs} Before fully convolutional networks were widely used, vessel segmentation was regarded as a pixel-by-pixel classification task and structured prediction was still a problem to be solved.
The usual approach was to crop the images into patches as input, and uses CNNs whose last few layers are fully connected to predict the label of the center pixel of each patch.
The approach proposed by \cite{DBLP:conf/icip/KhalafYF16} is a typical one.
They used a CNN containing three conv layers to perform vessel segmentation.
The last FC layer of the proposed CNN contains three neurons, representing the probability of central pixels being large vessels, small vessels or background, respectively.
 \cite{DBLP:journals/ijon/YuQZDQC20} also used a CNN whose last layers are FC layers for vessel segmentation.
And they conducted further research based on the segmentation results.
They first extracted vascular trees from the segmented vessels using a graph-based method.
Then two algorithms were proposed for the hierarchical division of retinal vascular networks.

\textbf{Structured prediction} has been explored by several researchers.
\cite{DBLP:journals/tmi/LiskowskiK16} used a CNN for segmentation.
Their FC layer comprises two neurons representing the vessel and background.
They also explored a structured prediction scheme, which can simultaneously predict labels of all pixels in an s$\times$s window of an n$\times$n patch.
Their approach was to set the number of the last FC layer's neurons to s$\times$s. 
Each neuron represents one pixel in the window, and the output is a set of two-dimensional vectors instead of scalars.

\subsubsection{FCN based approaches} Fully convolutional networks provide an end-to-end solution, addressing the issue of structured prediction.
Hence, they were quickly applied to vessel segmentation.
\cite{DBLP:conf/miccai/FuXLW016} proposed a fully convolutional network called DeepVessel.
They employed a side-output layer to help the network learn multiscale features.
At the end of the net, a CRF layer was used to further model non-local pixel correlations.
\cite{DBLP:conf/isbi/DasguptaS17} also proposed a fully convolutional network.
Their network contains six conv layers, one downsampling layer and one upsampling layer.
\cite{DBLP:conf/icip/Feng0Y17} proposed a fully convolutional network which can be considered as a simplified version of U-Net.
Their network only upsamples and downsamples twice.
In order to solve the class imbalance between background and blood vessels, they defined an entropy which measures the proportion of vessel pixels in the patch. 
During the training process, half of the patches are selected from the patches with the highest entropy, and the other half are randomly selected.
\cite{DBLP:journals/eswa/OliveiraPS18} proposed an FCN architecture which is similar to that of \cite{DBLP:conf/icip/Feng0Y17}.
In the pre-processing phase, they utilized a stationary wavelet transform (SWT) to obtain additional channels for the input images.
\cite{DBLP:journals/ijon/HuZNZCXG18} proposed a multiscale network inspired by RCF \citep{DBLP:conf/cvpr/LiuCHWB17}, which merges feature maps of every middle layer with the output.
Similar to \cite{DBLP:journals/ijon/GuoLKLZW19}, they also removed all the FC layers of VGG-16 as their backbone network.
At the end of their net, fully connected CRFs are employed. 
An improved cross-entropy loss was also proposed to focus on hard examples.

\subsubsection{Approaches using encoder-decoder architectures} Because of their excellent ability to extract features and extraordinary performance in practice, encoder-decoder architectures, especially U-Net, are still the most popular segmentation frameworks applied to fundus images up to now.
There are many directions for improvement in this area, as will be discussed next.

\textbf{Treating thick and thin vessels differently.} In order to improve performance on capillaries, one possible solution is to treat thick and thin vessels differently.
\cite{DBLP:conf/miccai/ZhangC18} proposed a multi-label architecture.
They used opening and dilation operations to expand the original vessel and background into five classes, namely 0 (other background pixels), 1 (background near thick vessels), 2 (background near thin vessels), 3 (thick vessels) and 4 (thin vessels).
The proposed architecture uses a U-Net with residual connection as the backbone.
A side-output layer was also introduced to capture multiscale features.
\cite{DBLP:conf/icip/HeZZLF018} introduced an operation named local de-regression (LODESS) to get additional labels.
After the LODESS, the original binary labels (vessel and background) were further divided into five classes, specifically 0 (the center of big vessels), 1 (the edge of big vessels), 2 (the center and edge of small vessels), 3 (the center of background) and 4 (edge of background).
\cite{DBLP:journals/tbe/YanYC18} introduced a segment-level loss which assigns different weights to different segments according to their thickness.
They first obtained vessel segments from the whole vessel tree based on skeletonization and then estimated the relative thickness of each segment.
Then a weight assigning strategy was designed to give thinner segments higher weights.
\cite{DBLP:journals/titb/YanYC19} proposed a three-stage model for vessel segmentation.
They first applied a skeletonization method to extract the skeletons.
For each skeleton pixel, the diameter of the maximum inscribed circle that is completely covered by vessel pixels is considered the thickness.
A ThickSegmenter and a ThinSegmenter were utilized for thick and thin vessel segmentation respectively.
Note that, when calculating the loss, only thick vessel pixels were counted for the ThickSegmenter and thin vessels for the ThinSegmenter.
Finally, the results of the two segmenters were passed to a FusionSegmenter to get the final result.

\textbf{Coarse-to-fine segmentation.} This is another approach that employs two branches: the first takes fundus images as input to get a preliminary result and the second further refines it.
\cite{DBLP:conf/miccai/WuXSZC18} proposed a multiscale network followed network (MS-NFN) to improve performance on capillaries.
Input images are passed into two different branches, namely the `up-pool' NFN and `pool-up' NFN.
The two branches both have identical U-Net-like structures. 
The first network converts input patches into a probability map, and the second performs further refinement.
The difference between the two NFNs is that the `up-pool' NFN upsamples before downsampling, and `pool-up' is the opposite. 
Finally, probability maps of the two NFNs are averaged to generate the final prediction.
In their subsequent work \citep{DBLP:journals/nn/WuXSZC20}, they added some modifications to NFN to form a new network named NFN+.
Compared to NFN, the main extensions include: introducing inter-network connections between the preceding and following networks; replacing the `up-pool' and `pool-up' networks with an identical U-Net-like architecture; and removing the ensemble operation.
\cite{DBLP:conf/isbi/00210HWC20} proposed a coarse-to-fine supervision network (CTF-Net) for vessel segmentation.
Their CTF-Net consists of two U-shaped networks, namely the coarse segNet producing preliminary predicted map and the fine segNet further enhancing performance.
They also proposed a feature augmentation module (FAM-residual block) to improve the ability of the network to extract features.

\textbf{Mutliscale networks.} This is another important direction that has been explored.
\cite{DBLP:conf/miccai/WuX0ZLZC19} proposed Vessel-Net, which is based on the multiscale method.
They first implemented an Inception-Residual (IR) block inspired by Inception and ResNet that can be embedded into U-Net.
Four supervision paths were introduced to the net, including: a traditional supervision path; a richer feature supervision path, which resizes all stages of the encoder's output to the same size as the input patches (48$\times$48) and then concatenates them; and two multiscale supervision paths, where feature maps generated by the encoder with size 12$\times$12 and 24$\times$24 are passed into a 1$\times$1 conv layer with Relu and softmax.
\cite{DBLP:journals/ijon/FengZPT20} proposed a cross-connected convolutional neural network (CcNet) for vessel segmentation, which also utilizes the multiscale method.
The CcNet had two paths.
The first is the primary path, which has more convolutional kernels than the other path to extract more features.
The other is called the secondary path.
Each conv layer of the primary path is connected to all conv layers of the secondary path to learn multiscale features.

\textbf{Improvements to sampling operation.} The downsampling and upsampling operations will change the resolution of the feature maps, which is not ideal for the segmentation task.
Several works thus tried to improve or replace these two operations.
\cite{DBLP:journals/eswa/SoomroAGHZP19} proposed a strided-CNN model to improve the sensitivity.
They first performed pre-processing including morphological mappings and principal component analysis (PCA).
Then the processed images were passed to a SegNet-like encoder-decoder architecture.
The pooling operation was replaced with a strided-conv inspired by \cite{DBLP:journals/corr/SpringenbergDBR14}.
\cite{DBLP:conf/miccai/ZhangFYZWYTX19} proposed the Attention Guided Network (AG-Net) for vessel segmentation.
An attention guided filter inspired by \cite{DBLP:journals/pami/He0T13} was proposed.
Specifically, it takes high-resolution feature maps from the encoder and low-resolution feature maps from the lower stage of the decoder as input and produces high-resolution feature maps as output.
The attention guided filter can preserve edge and structural information.
Note that AG-Net can also perform OD/OC segmentation.

\textbf{Dual-encoder.} \cite{DBLP:conf/miccai/Wang0H19} proposed the Dual Encoding U-Net (DEU-Net).
DEU-Net consists of two encoders.
The first, inspired by the global convolutional network \citep{DBLP:conf/cvpr/PengZYLS17}, has a spatial path with larger kernels to capture more spatial information.
The second, inpired by Inception, has a context path with multiple kernels to get more context features.
A feature fusion module was proposed to fuse the features extracted by the two encoders at the top stage.
Channel attention was used to replace the skip-connection in the original U-Net.

\textbf{Data-aware deep supervision.} \cite{DBLP:conf/isbi/MishraCH20} added a data-aware deep supervision path to a U-Net-like network.
Based on the concept of effective receptive field (EFT) proposed by \cite{DBLP:conf/nips/LuoLUZ16}, where the output-affecting region is actually smaller than the theoretical receptive field (RF), they proposed the concept of layer-wise effective receptive fields (LERFs), which are calculated by the gradient of the loss function using back-propagation.
The average vessel width was taken as the target object size.
The convolutional layer with the smallest absolute difference between its LERF and vessel width was selected as the target layer, and considered as the preeminent layer.
Deep supervision was used in the target layer.

\textbf{Spatial activation using a Gaussian function.} \cite{DBLP:conf/miccai/MaYMWDZ19} proposed a multitask network which can perform vessel segmentation and A/V classification simultaneously.
In view of the observation that the value of capillary vessels and boundary vessels in a probability map is close to 0.5, they proposed a spatial activation module that assigns higher weights to the thin vessels by a Gaussian function.
Deep supervision was also utilized.

\subsubsection{Other methods} Several other methods have been proposed to improve model performance.
They not only achieve good experimental performance, but are also inspiring.

\textbf{Image matting method.} \cite{DBLP:journals/pr/ZhaoLC20} transformed the segmentation problem to a related matting problem.
A trimap was first obtained using a bi-level thresholding of the score map.
Then the retinal images and corresponding trimaps were sent to an end-to-end matting network to get the foreground matte.
They proposed a local matting loss together with a global pixel loss for training.
The final segmentation map was obtained by applying a threshold to all pixels of the matte.

\textbf{Inception capsule network.} \cite{DBLP:conf/isbi/KrommR20} combined the Capsule network \citep{DBLP:conf/nips/SabourFH17} with the Inception architecture for vessel segmentation and centerline extraction.
Their Inception Capsule network has a shallow architecture with fewer parameters and does not need data augmentation.

\textbf{Ensemble learning.} \cite{DBLP:conf/miccai/Liu0019} proposed a novel and simple unsupervised ensemble strategy for vessel segmentation. 
They multiplied the output results of the best performing recent networks by the weights to obtain a result.
The weights of results were then trained and they finally obtained better results than a single network.

\textbf{Regularization under geometric priors.} \cite{DBLP:journals/tip/CherukuriGBM20} proposed a domain enriched deep network for vessel segmentation.
A representation network was first employed.
Two geometrical regularizers, including an orientation diversity regularizer and a data adaptive noise regularizer, were added to the loss function to learn specific geometric features.
After that they introduced a network containing residual blocks with no downsampling/upsampling steps, instead of using U-Net-like most other works.

\textbf{Performing segmentation on RIM-ONE.} \cite{DBLP:conf/isbi/NaserySG20} performed vessel segmentation on the RIM-ONE dataset.
Compared to the DRIVE dataset, RIM-ONE is of a lower quality and does not have any vessel annotations.
Instead of performing image synthesis to get high-quality images, they transformed high-quality images with expert labels from the DRIVE dataset to resemble poor-quality target images.
To accomplish this, substantial vignetting masks were used.
Then a U-Net was trained using the resulting images and their corresponding labels.
Once trained, the net could be used to obtain vessel masks of images from the RIM-ONE dataset.

\subsubsection{Discussion}
From the model discussed in this section, we can see the development of base networks used, for vessel segmentation, from CNNs to FCNs to U-Net-like architectures.
The use of CNNs and FCNs has become less common recently, while U-Net-like architectures are very popular.
However, the feature extraction ability of U-Net is inadequate.
U-Net only has 10 convolution operations in the encoder, which is even less than that in VGG-16.
Therefore, several works have focused on how to improve the feature extraction ability.
Alternatives include Dense U-Net, Residual U-Net, and a dual-encoder network.
Another disadvantage of U-Net is that there are four paired sampling operations (downsampling and upsampling), which is not ideal for the segmentation task.
Several studies have tried to alleviate this problem by, for instance, using a shallower version of U-Net that has two or three paired sampling operations.
Multiscale methods can also be utilized to improve the performance of the segmentation task.
Low-level spatial features and high-level semantic features can both be focused on by the network.
In order to solve the problem of poor performance on thin and edge vessels, method for treating thick and thin vessels differently have been explored.
It is worth noting that there are several other inspiring and also interesting studies, including data-aware deep supervision, spatial activation, image matting, using a Inception capsule network, ensemble learning and performing segmentation on RIM-ONE.

However, there are still several limitations.
First, there is still room for improvement in thin and edge vessels segmentation.
Specifically, sensitivity and accuracy need to be further improved while maintaining specificity and AUC.
Second, there are only three commonly used datasets for vessel segmentation, namely DRIVE, STARE and CHASE\_DB1.
And they contain fewer images than datasets for other tasks.
On the one hand, more experiments need to be carried on the high-resolution HRF dataset.
On the other hand, more images should be collected and annotated.
Researchers can also employ image synthesis methods.
Third, the imbalance problem also exists in the vessel segmentation task and it is even more challenging to solve than in other tasks like lesion and OD/OC segmentation for the irregular shape of blood vessels.
The typical approach is to use a class-balancing loss.
It is worth noting that a selective sampling method based on entropy could be effective.
More attention should thus be paid to the imbalance problem.

\subsection{OD/OC Segmentation}
\label{sec:od}
Cup-to-Disc ratio (CDR) is a widely accepted and used standard for the diagnosis of glaucoma.
It is calculated as the ratio of vertical cup diameter (VCD) and vertical disc diameter (VDD) \citep{PHENE20191627}.
The segmentation of the optic cup (OC) and optic disc (OD) is therefore very important for the diagnosis of glaucoma.
Compared to OD segmentation, OC segmentation is a more challenging task for its subtle boundaries. 
Further, there is an imbalance problem for OC, as the OC region only accounts for a low proportion of extracted ROIs.

The datasets used in this field are shown in Tab. \ref{tab:dataset_glaucoma}.
Similar to lesion segmentation/detection, there is also an overlap between datasets used in OD/OC segmentation and glaucoma diagnosis.
The ONHSD dataset \citep{DBLP:journals/tmi/LowellHSBRFK04} consists of 99 images obtained from 50 patients of various ethnic backgrounds.
Further, 96 images have discernable ONH.
The Drions-DB dataset \citep{DBLP:journals/artmed/CarmonaRGM08} consists of 110 images, belonging to 55 patients with glaucoma (23.1\%) and eye hypertension (76.9\%). 
The images were obtained from a hospital in Spain.
The ORIGA dataset \citep{5626137} consists of 650 images, of which 168 are glaucomatous and 482 are normal.
The boundaries of OD and OC, CDR value and a label indicating whether glaucoma exists or not are provided for each image.
The RIM-ONE-r3 dataset \citep{DBLP:conf/cbms/FumeroASSG11} consists of 169 ONH images, of which 118 are normal, 11 have ocular hypertension (OHT) and 40 are glaucomatous.
Five-class labels were provided by five experts.
The ACHIKO-K dataset \citep{6566371} consists of 258 images, which were obtained from 67 glaucomatous patients from Korea.
144 images are of glaucomatous eyes and 114 are normal.
The Drishti-GS dataset \citep{DBLP:conf/isbi/SivaswamyKJJT14} contains 101 images, which are officially divided into 50 training images and 51 test images.
Images were obtained from a hospital in India.
The SCES dataset \citep{baskaran2015the} consists of 1,676 images, each from a single subject, which only provide clinical diagnoses.
46 images of SCES are glaucomatous.
The RIGA dataset \citep{10.1117/12.2293584} is made up of three parts, namely 460 images from MESSIDOR, 195 images from the Bin Rushed Ophthalmic Center and 95 images from the Magrabi Eye Center, making the total number of images 750.
Each image was manually annotated by six ophthalmologists.
The LAG dataset \citep{DBLP:journals/tmi/LiXLLWJWFW20} contains 11,760 images, of which 6,882 do not have glaucoma and 4,878 are suspicious.
5,824 images were further annotated with attention labels, in which 2,392 display glaucoma and the remaining 3,432 do not.

Experimental results are shown in Tab. \ref{tab:result_OD_drishtigs},  \ref{tab:result_OD_origa}, \ref{tab:result_OD_rimone}, \ref{tab:result_OD_refuge} and \ref{tab:result_OD_other}.
``A" in the tables means balanced accuracy,``E" is the widely used overlapping error, and $\delta$ denotes absolute CDR error.

\begin{table*}
\small
\caption{Widely used datasets for OD/OC segmentation and glaucoma diagnosis/grading}\label{tab:resized}
\begin{center}

\resizebox{\textwidth}{!}{

\begin{tabular}{ p{2.3cm} p{3.1cm} p{3.2cm} p{4.3cm} p{2.5cm}}
\hline
Dataset name & Number of images & Resolution & Camera & Availability\\
\hline
ONHSD & 100 & 640$\times$480 & a Canon CR6 45MNf fundus camera, FOV 45\degree & \href{http://www.aldiri.info/Image\%20Datasets/ONHSD.aspx}{available online}$^{\rm 1}$\\
\hline
Drishti-GS & 101  & 2896$\times$1944 & a fundus camera with FOV 30\degree & \href{http://cvit.iiit.ac.in/projects/mip/drishti-gs/mip-dataset2/Home.php}{available online}$^{\rm 2}$\\
\hline
Drions-DB&110 &600$\times$400& a colour analogical fundus camera & \href{https://www.researchgate.net/publication/326460478_Glaucoma_dataset_-_DRIONS-DB}{available online}$^{\rm 3}$\\
\hline
ORIGA & 650 (168 glaucomatous, 482 normal)& 3072$\times$2048 & - & not available online\\
\hline
RIGA& 750 & ranging from 2240$\times$1488 to 2743$\times$1936 & multiple fundus cameras with different FOV & \href{https://deepblue.lib.umich.edu/data/concern/data_sets/3b591905z}{available online}$^{\rm 4}$\\
\hline
RIM-ONE & 169 ONH & - &a fundus camera Nidek AFC-210 with a body of a Canon EOS 5D Mark II of 21.1 megapixels & not available online\\
\hline
ACHIKO-K & 258 (144 glaucomatic) & 640$\times$480; 2144$\times$1424; 3216$\times$2136, etc & NIKON D80, NIKON D90 & \href{https://oar.a-star.\%20edu.sg/jspui/handle/123456789/1080?mode=full}{available online}$^{\rm 5}$\\
\hline
SEED & 235 (43 glaucoma) & - & - & not available online\\
\hline
REFUGE &1200 & 2124$\times$2056, 1634$\times$1634 & a Zeiss Visucam 500 fundus camera and a Canon CR-2 device& \href{https://refuge.grand-challenge.org/}{available online}$^{\rm 6}$\\
\hline
SCES & 1676 & 3072$\times$2048 & - & not available online\\
\hline
SINDI & 5783 & 3072$\times$2048 & - & not available online\\
\hline
LAG  & 11,760 (6882 glaucoma) & ranging from 582$\times$597 to 3456$\times$5184 & 3 types of devices: Topcon, Canon and Carl Zeiss & \href{https://github.com/smilell/AG-CNN}{available online}$^{\rm 7}$\\
\hline
\end{tabular}
}

\end{center} 
\begin{tablenotes}
     \item[1]$^{\rm 1}$http://www.aldiri.info/Image\%20Datasets/ONHSD.aspx
     \item[2]$^{\rm 2}$http://cvit.iiit.ac.in/projects/mip/drishti-gs/mip-dataset2/Home.php
     \item[3]$^{\rm 3}$https://www.researchgate.net/publication/326460478\_Glaucoma\_dataset\_-\_DRIONS-DB
     \item[4]$^{\rm 4}$https://deepblue.lib.umich.edu/data/concern/data\_sets/3b591905z/
     \item[5]$^{\rm 5}$https://oar.a-star.\%20edu.sg/jspui/handle/123456789/1080?mode=full
     \item[6]$^{\rm 6}$https://refuge.grand-challenge.org/
     \item[7]$^{\rm 7}$https://github.com/smilell/AG-CNN
\end{tablenotes}
\label{tab:dataset_glaucoma}
\end{table*}

\begin{table*}
\small
\caption{Summary of several results for OD/OC segmentation on Drishiti-GS dataset}
\begin{center}

\resizebox{\textwidth}{!}{

\begin{tabular}{p{3.5cm} p{2.7cm} p{3.9cm} p{0.9cm} p{0.9cm} p{0.9cm} p{0.9cm} p{0.9cm}}
\hline
Reference & Backbone & Loss & \multicolumn{2}{c}{OD} & \multicolumn{2}{c}{OC} & $\delta$  \\
\hline
 &  &  & Dice/\% & IoU/\% & Dice/\% & IoU/\% & \\
\hline
\cite{DBLP:conf/icip/EdupugantiCK18} & FCN & weighted CE & -& 69.58 &- & 81.22 & - \\

\cite{DBLP:conf/icip/MohanKS18} & FCN & bootstrapped CE and Dice loss & 96.4 & - & - & - &- \\

\cite{DBLP:conf/icip/MohanKS19} & FCN & bootstrapped CE and Dice loss & 97.13 &- &- & - &-\\

\cite{DBLP:journals/ijon/LiuHLCZZ19} & FCN & spatial-aware error function & \textbf{98} & - & 89 &  - &-\\

\cite{DBLP:journals/titb/Shankaranarayana19} & Encoder-decoder net & multi-class CE & 96.3 & - & 84.8 & -& 0.1045\\

\cite{DBLP:conf/isbi/ShahKP19}(PSBN) & U-Net & logarithmic dice loss & 95 &91 &88 & \textbf{80} & - \\

\cite{DBLP:conf/isbi/ShahKP19}(WRoIM) & U-Net & logarithmic dice loss & 96 &\textbf{93}& 89& \textbf{80} & - \\

\cite{DBLP:journals/tmi/WangYYFH19} & Deeplab, GAN & dice coefficient loss, smoothness loss and adversarial loss & 97.4 & -& \textbf{90.1} & - &\textbf{0.048} \\ 

\cite{DBLP:conf/miccai/WangYLYFH19} & DeeplabV3+, GAN & CE, MSE, Adversarial loss &96.1&- &86.2 & - &- \\
\hline
\label{tab:result_OD_drishtigs}
\end{tabular}
}

\end{center} 
\end{table*}

\begin{table*}
\small
\caption{Summary of several results for OD/OC segmentation on ORIGA dataset}
\begin{center}

\resizebox{\textwidth}{!}{

\begin{tabular}{p{3cm} p{2.8cm} p{3.5cm} p{0.6cm} p{0.6cm} p{0.6cm} p{0.6cm}p{0.6cm} p{0.9cm}p{0.9cm}}
\hline
Reference & Backbone & loss &\multicolumn{2}{c}{OD} & \multicolumn{2}{c}{OC} & \multicolumn{2}{c}{Rim} & $\delta$\\
\hline
&  &  & A/\% & E& A/\% & E & A/\% & E & \\
\hline
\cite{DBLP:journals/ijon/LiuHLCZZ19}  & FCN & spatial-aware error function & -& 0.059 & -& \textbf{0.208} & -& \textbf{0.215} &-\\

\cite{DBLP:journals/tmi/FuCXWLC18} & U-Net & proposed multi-label loss & 98.3 & 0.071 & 93.0 &  0.230 & 94.1 & 0.233 & 0.071 \\ 

\cite{DBLP:journals/titb/Shankaranarayana19} & Encoder-decoder net & multi-class CE & 97.4 & \textbf{0.051} & 92.8 & 0.212 & - & - & 0.067 \\

\cite{DBLP:conf/miccai/YinWXMYZT19} & RPN & Multi-label CE & \textbf{98.6} & 0.066 & \textbf{94.2} & \textbf{0.208}& \textbf{94.9} & 0.224& \textbf{0.065} \\

\cite{DBLP:journals/tbe/JiangDCGXFLL20} & atrous CNN and RPN & Smooth L$_1$ loss and BCE & -& 0.063 & -& 0.209 & - & - & 0.068 \\
\hline
\label{tab:result_OD_origa}
\end{tabular}
}

\end{center} 
\end{table*}

\begin{table*}
\small
\caption{Summary of several results for OD/OC segmentation on RIM-ONE-r3 dataset}
\begin{center}

\resizebox{\textwidth}{!}{

\begin{tabular}{p{2cm} p{2.7cm} p{3.7cm} p{0.6cm} p{0.6cm} p{0.7cm} p{0.6cm} p{0.6cm} p{0.6cm} p{0.7cm} p{0.6cm} p{0.6cm}}
\hline
Reference & Backbone & loss & \multicolumn{4}{c}{OD} & \multicolumn{4}{c}{OC} & $\delta$ \\
\hline
 & & & A/\% & E & Dice/\% & IoU/\% & A/\% & E & Dice/\% & IoU/\% \\
\hline
\cite{DBLP:journals/titb/Shankaranarayana19} & Encoder-decoder net & multi-class CE & \textbf{97.5} & \textbf{0.058} & \textbf{97.0} &- &\textbf{92.0} & \textbf{0.284} & \textbf{87.6} & - & 0.066 \\

\cite{DBLP:conf/isbi/ShahKP19}(PSBN) & U-Net & logarithmic dice loss & -&- & 91 & 84 & - &- & 75 & 60 & -  \\

\cite{DBLP:conf/isbi/ShahKP19}(WRoIM) & U-Net & logarithmic dice loss & -& - & 94 & \textbf{90}& -& - & 82 & \textbf{71} &-\\

\cite{DBLP:journals/tmi/WangYYFH19} & Deeplab, GAN & dice coefficient loss, smoothness loss, adversarial loss &- &- & 96.8 & -&- &-  & 85.6  & -& \textbf{0.049}\\ 

\cite{DBLP:conf/miccai/WangYLYFH19} & DeeplabV3+, GAN & CE, MSE, Adversarial loss & - &- &  89.8 & - & -&-  & 81.0 & - &- \\
\hline 
\label{tab:result_OD_rimone}
\end{tabular}
}

\end{center} 
\end{table*}

\begin{table*}
\small
\caption{Summary of several results for OD/OC segmentation on REFUGE dataset}
\begin{center}

\resizebox{\textwidth}{!}{

\begin{tabular}{p{2.5cm} p{1.2cm} p{3.7cm} p{0.6cm} p{0.6cm} p{0.7cm} p{0.6cm} p{0.6cm} p{0.7cm} p{0.6cm} p{0.6cm} p{0.6cm}}
\hline
Reference & Backbone & Loss & \multicolumn{3}{c}{OD} & \multicolumn{3}{c}{OC} & \multicolumn{2}{c}{Rim} & $\delta$ \\
\hline
 &  &  & A/\% & E & Dice/\% & A/\% & E & Dice/\% & A/\% & E & \\
\hline
\cite{DBLP:conf/isbi/WangDR0XX19}  & RPN & Weighted CE, regression loss & -& - & 95.3 & -& -& 87.2 &- & -& 0.047\\
\hline
\cite{DBLP:conf/miccai/YinWXMYZT19}  & RPN & Multi-label CE & \textbf{97.9} & \textbf{0.088} &- & \textbf{98.0} & \textbf{0.223} &-  & \textbf{93.6} & \textbf{0.204} & 0.048  \\
\hline
\cite{DBLP:journals/tmi/WangYYFH19}  & Deeplab, GAN & dice coefficient loss, smoothness loss and adversarial loss & -& - & \textbf{96.02} & -& - & \textbf{88.26}  & -& -& \textbf{0.0450}\\ 
\hline 
\cite{DBLP:conf/miccai/LiuKLZF19}  & GAN & dice segmentation loss, adversarial loss and MSE loss &- &- & 94.16  &- &-  & 86.27 &- &- & 0.0481\\
\hline 
\label{tab:result_OD_refuge}
\end{tabular}
}

\end{center} 
\end{table*}

\begin{table*}
\small
\caption{Summary of several results for OD/OC segmentation on other datasets}
\begin{center}

\resizebox{\textwidth}{!}{

\begin{tabular}{p{2.5cm}p{1.4cm}p{2.3cm}p{3.8cm}p{0.6cm}p{0.7cm}p{0.6cm}p{0.7cm}p{0.6cm}}
\hline
Reference & Dataset & Backbone & Loss& \multicolumn{2}{c}{OD} & \multicolumn{2}{c}{OC} & $\delta$\\
\hline
 &  & &  & E & Dice/\% & E & Dice/\% & \\
\hline
\cite{DBLP:conf/icip/MohanKS18} & DrionsDB & FCN & bootstrapped CE, Dice loss &- & 95.5 &- & - &-\\

\cite{DBLP:conf/icip/MohanKS19} & DrionsDB & FCN & bootstrapped CE, Dice loss & - & \textbf{96.6}&- &- &-\\
\hline
\cite{DBLP:conf/icip/MohanKS18} & MESSIDOR & FCN & bootstrapped CE, Dice loss  &-  & 95.7  &-&-&-\\

\cite{DBLP:conf/icip/MohanKS19} & MESSIDOR& FCN & bootstrapped CE, Dice loss &- & \textbf{96.8} &- &- &-\\

\hline
\cite{DBLP:journals/tbe/JiangDCGXFLL20} & SCES & atrous CNN, RPN & Smooth L$_1$ loss, BCE & \textbf{0.063} &- &\textbf{0.209} &- & \textbf{0.068} \\

\hline
\cite{DBLP:conf/miccai/SedaiMHMG17} & EyePACS & VAE & negative KL-divergence, BCE &- &- &- &- & \textbf{0.80} \\
\hline 
\label{tab:result_OD_other}
\end{tabular}
}

\end{center} 
\end{table*}

\subsubsection{Approaches based on FCN}Similar to blood vessel segmentation, fully convolutional networks were widely used in early OD/OC segmentation.
\cite{DBLP:conf/icip/EdupugantiCK18} used FCN-8s to perform OD/OC segmentation.
They also explored various strategies, such as assigning higher weights to edges in the loss function, for further improvement.

\textbf{Using atrous convolutions} is quite common in this field. \cite{DBLP:conf/icip/MohanKS18} proposed a structure named Fine-Net which has a symmetrical encoder-decoder architecture.
Inspired by full-resolution residual networks (FRRNs) \citep{DBLP:conf/cvpr/PohlenHML17}, they used several full-resolution residual units (FRRU) in Fine-Net.
An atrous convolution was also introduced to alleviate the memory cost while ensuring reliable performance at the same time.
In their subsequent work, \cite{DBLP:conf/icip/MohanKS19} proposed P-Net.
P-Net is used to obtain preliminary segmentation results by taking downscaled images as input.
The output of P-Net is upscaled and then sent to Fine-Net as guidance for further segmentation.
DenseBlock and atrous convolution are combined as the dense atrous block (DB) to formulate P-Net.
DBs with different dilation rates are used to capture multiscale features, inspired by atrous spatial pyramid pooling (ASPP) \citep{DBLP:journals/corr/ChenPSA17}.
\cite{DBLP:journals/ijon/LiuHLCZZ19} proposed a spatial-aware neural network which adopts a multiscale method.
First, an atrous CNN model is used to extract spatially denser feature maps.
Then, the extracted features are fed to a pyramid filtering module to obtain multiscale features.
Finally, the multiscale features are passed to a spatial-aware segmentation network to get the final result.

\subsubsection{Approaches based on U-Net}There are also several works using U-Net as the baseline.
\cite{DBLP:journals/tmi/FuCXWLC18} proposed M-Net for OD/OC segmentation.
M-Net contains a multiscale input layer. 
Images are down-sampled to form an image pyramid as U-Net's input, to obtain a multi-level receptive field. 
In order to segment OD and OC simultaneously, the authors proposed a multi-label loss based on the dice loss, which is intended to solve the problem of multi-label and data imbalance.
Moreover, a polar transformation was introduced to obtain spatial consistency and increase the proportion of the OD/OC in a patch. 
Their approach greatly inspired subsequent work.

\cite{DBLP:conf/isbi/ShahKP19} proposed two different methods named the Parameter-Shared Branched Network (PSBN) and Weak Region of Interest Model-based segmentation (WRoIM).
With fewer parameters, they obtained comparable performance to state-of-the-art approaches.
PSBN has two branches, which are used to generate masks for the OD and OC, respectively. 
Encoders of the two branches share parameters, and the OC branch uses cropped activations from the OD branch.
WRoIM first obtains a coarse OD area through a small U-Net structure (one conv block for downsampling and one conv block for upsampling), and uses the extracted ROI as another U-Net's input to perform fine segmentation.

\textbf{Guidance from depth estimation task} can boost performance of OD/OC segmentation. \cite{DBLP:journals/titb/Shankaranarayana19} proposed a fully 
convolutional network for retinal depth estimation and used the results to guide OD/OC segmentation.
They proposed a Dilated Residual Inception (DRI) module utilizing convolution kernels with different dilation rates in the way of an Inception Block to extract multiscale features.
In order to ensure that the retinal depth estimation branch guides OD/OC segmentation, they proposed a multi-modal feature fusion module to fuse feature maps from the depth estimation branch and the OD/OC segmentation branch.

\textbf{Pixel-wise deep regression} is a novel direction under exploration.  \cite{DBLP:conf/miccai/MeyerGMC18} reformulated the segmentation task as a pixel-wise regression task to perform OD and fovea detection simultaneously.
A bi-distance map is first obtained, illustrating the distance between every pixel and its nearest landmarker, namely OD and fovea.
Then a U-Net-like deep network is utilized for distance regression and obtaining a globally consistent prediction map.

\subsubsection{Approaches based on RPN}The RPN method from Faster-RCNN \citep{DBLP:conf/nips/RenHGS15} and Mask R-CNN \citep{DBLP:conf/iccv/HeGDG17} has also been applied in the segmentation of OD/OC.
Inspired by RPN, \cite{DBLP:conf/isbi/WangDR0XX19} proposed the Ellipse Proposal Network (EPN) to detect ellipse regions.
They used \{$X_{0}$, $Y_{0}$, $F_{1}$, $F_{2}$\} as parameters of elliptical anchors, where ($X_{0}$, $Y_{0}$) denotes the center coordinate of the ellipse and $F_{1}$, $F_{2}$ denote the major and minor axis, respectively.
Two EPN branches were used to detect OD and OC.
In addition, a spatial attention path was introduced between the OD to OC branches to guide the detection of OC.
\cite{DBLP:conf/miccai/YinWXMYZT19} proposed a PM-Net, which is also inspired by RPN.
They introduced a segmentation branch to RPN to provide more accurate proposals for localization of the optic nerve head (ONH) area.
A pyramid RoIAlign module was also proposed to capture multiscale features.
\cite{DBLP:journals/tbe/JiangDCGXFLL20} proposed JointRCNN for OD and OC segmentation.
An atrous-VGG16 is first used for feature extraction.
The extracted features are passed to two parallel branches, i.e., a disc proposal network (DPN) and a cup proposal network (CPN), for OD and OC region proposal.
A disc attention module is employed between the DPN and CPN to determine where the OC is located based on the result of the DPN.

\subsubsection{Domain adaptation studies based on GANs}
\label{sec:OD_seg_domain_adapt}
Domain adaptation is another direction that has been explored.
\cite{DBLP:journals/tmi/WangYYFH19} proposed the patch-based Output Space Adversarial Learning (pOSAL) framework.
Images from the source and target domains are first passed into a lightweight network to extract ROIs.
The extraction network is based on Deeplabv3+ \citep{DBLP:conf/eccv/ChenZPSA18} but utilizes MobileNetV2 \citep{DBLP:journals/corr/abs-1801-04381} as the backbone.
The authors also designed a morphology-aware segmentation loss for the network.
Then the ROIs extracted are passed to a patch discriminator (Patch GAN) for adversarial learning, which can learn abstract spatial and shape features from the label distribution of the source domain.
In their following work, \cite{DBLP:conf/miccai/WangYLYFH19} proposed an unsupervised domain adaptation network named Boundary and Entropy-Driven Adversarial Learning (BEAL).
Based on the observation that predictions in the target domain made by the network trained on the source domain tend to contain ambiguous and inaccurate boundaries and the corresponding entropy map is noisy with high-entropy outputs, they introduced two segmentation branches focused on the boundary and entropy map, respectively.
An adversarial learning method was introduced to encourage predictions of the two branches to be domain-invariant.
\cite{DBLP:conf/miccai/LiuKLZF19} proposed an unsupervised domain adaptation architecture named Collaborative Feature Ensembling Adaptation (CFEA).
Their framework consists of three parts; namely, the source domain network (SN), which learns from the source domain with labels, the target domain student network (TSN) and the target domain teacher network (TTN), which learn from target domains without labels.
Adversarial learning was introduced between SN and TSN, where the supervised SN enables the segmentation network obtain more precise prediction, and the unsupervised TSN introduces a perturbation to the training of the network.
The MSE between TSN and TTN was calculated to help the student network's learning.

\subsubsection{Other approaches}
There are serveral other approaches for addressing vessel and OD segmentation using a variational autoencoder (VAE) as the backbone, tackling a novel task named optic disc quantification and using a single net.

\textbf{Approaches based on a variational autoencoder.} Note that some works were built on VAE.
\cite{DBLP:conf/miccai/SedaiMHMG17} proposed a semi-supervised method to perform OC segmentation using limited labeled training samples.
First, a generative variational autoencoder (GVAE) is trained using a large amout of unlabeled data to learn the feature embedding.
Then a segmentation variational autoencoder (SVAE) is used to predict the OC mask by leveraging the feature embedding provided by the GVAE.

\textbf{OD quantification based on multitask ensemble learning framework.} 
Unlike traditional OD/OC segmentation, optic disk quantification refers to the simultaneous quantification of a series of medical indicators, namely two vertical diameters (OD, OC), two complete regions (Disc, Rim), and 16 local regions \citep{Garway1998Quantitative}.
Accurate optic disc quantification provides effective help in the diagnosis and treatment of many eye diseases such as chronic glaucoma \citep{DBLP:conf/miccai/ManinisPAG16}.
\cite{DBLP:conf/miccai/ZhaoCLZ019} proposed a multitask ensemble learning framework (DMTFs) to perform optic disc quantification.
To the best of our knowledge, they were the first to use deep learning to accomplish this task. 
In their following work \citep{DBLP:journals/mia/ZhaoL20}, they made several modifications to their original model, including incorporating a conduct feature interaction module for highly correlated tasks.

\textbf{Vessel and OD segmentation using a single net.} \cite{DBLP:conf/miccai/ManinisPAG16} proposed Deep Retinal Image Understanding (DRIU) for vessel and OD segmentation.
They used VGG-16 as “base network" and removed its FC layers.
The feature maps of different levels from the base network are resized and fused to form two task-specific “specialized" layers, which are used to perform vessel segmentation and OD segmentation simultaneously.

\subsubsection{Discussion}
Compared to other segmentation tasks like vessel or lesion segmentation, the segmentation of the OD/OC is more similar to natural image segmentation due to its ellipse shape.
Therefore, several architectures taken from natural image segmentation are used in this task, including Deeplabv3+ and Mask-RCNN.
Such networks are rarely seen in vessel segmentation or lesion segmentation.
Further, compared to the other two segmentation tasks, the research on OD/OC segmentation is the most complete.
Architectures of FCN, U-Net, Deeplabv3$+$ and Mask-RCNN have been used,
and methods like multiscale and polar transformations have been tried.

Future work on OD/OC segmentation may lie in the following directions.
First, the more accurate task of OD quantification may be promising.
Second, the problems of domain shift and poor generation performance still exist.
Therefore, domain adaption should be paid more attention to.
It is also worth noting that the REFUGE2 challenge is ongoing.
Researchers can follow the competition to see the latest methods and research directions.

\subsection{Fovea segmentation}
The fovea is one of the most significant landmarks in fundus images.
Segmenting the fovea can help define the risk of a particular lesion in retinal diagnosis.
However, due to the lack of publicly-available datasets, few works focus on fovea segmentation.

\cite{DBLP:conf/isbi/SedaiTRCG17} proposed a two-stage framework for fovea segmentation, using a subset of EyePACS as training data.
The first stage is a coarse network, which performs coarse segmentation to localize the fovea region.
The authors discarded the FC layers of VGG-16 to make it a fully convolutional network.
Feature maps at different levels are upsampled to the same size as the input images and fused together to get the output.
The second stage is a fine network, which takes the ROI regions obtained by the coarse network to generate the final result.
The only difference between the fine network and coarse network is that the fine network only uses the last two blocks of VGG-16 to get the segmentation output.

It is clear that only one remarkable work has been introduced for fovea segmentation.
More researches are thus called for to obtain better performance on fovea segmentation.
Moreover, the framework used by \cite{DBLP:conf/isbi/SedaiTRCG17} is a two-stage architecture.
More architectures need to be explored.

\subsection{A/V classification}
Subdividing the blood vessels in fundus images into arteries and veins is of vital importance for the early diagnosis of many diseases. For example, a low ratio between arteriolar and venular width (AVR) can predict diabetes and many cardiovascular diseases \citep{DBLP:journals/tmi/NiemeijerXDGGFA11}.
The widely used datasets for this task are as follows. Compared to the DRIVE dataset used in vessel segmentation, the DRIVE-AV dataset \citep{DBLP:conf/miccai/HuAG13} further provides pixel-level artery/vein labels.
As mentioned before, it has 20 images in the training set and 20 images in the test set.
DRIVE-AV is also called RITE in some papers.
The LES-AV dataset\footnote{https://ignaciorlando.github.io/\#publications\_selected} \citep{DBLP:conf/miccai/OrlandoBKBBB18} consists of 22 images with pixel-level labels.
The INSIPRE-AVR dataset\footnote{http://www.retinacheck.org/datasets} \citep{DBLP:journals/tip/DashtbozorgMC14} consists of 40 images with only centerline-level annotation.
The private IOSTAR dataset \citep{DBLP:conf/iciar/Abbasi-Sureshjani15} consists of 24 images annotated by two experts.

\cite{DBLP:conf/isbi/GaldranM0MC19} regarded the A/V classification task as a four-class segmentation problem, with categories including background, artery, vein and uncertain.
They used a U-Net-like structure to classify the arteries and veins directly without segmenting the vessel tree first.
To the best of our knowledge, they are the first to focus on pixel-level uncertainty in the task of vascular segmentation and classification.
\cite{DBLP:conf/isbi/RajMKS20} proposed an Artery-Vein Net (AV-Net) for A/V classification.
The backbone network is ResNet-50 and squeeze-excitation (SE) blocks are used.
Feature maps of different scales are upsampled and fused to the same size as the input image to get the segmentation map.
AV-Net does need a segmented vasculature map as input, instead only requiring a single wavelength, color fundus image.
Finally, as introduced previously, \cite{DBLP:conf/miccai/MaYMWDZ19} performed A/V classification while segmenting blood vessels.

\textbf{Discussion.} From the above methods we can see that A/V classification is a promising direction.
The general tendency is to directly perform A/V classification without performing vessel segmentation first.
However, A/V classification is an even more challenging task than vessel segmentation.
Further in existing works, there is still the problem of arteries and veins appearing in a single vessel segment, which is not common in reality.

\section{Disease diagnosis/grading}
\label{sec:disease_diagnosis_grading}
\subsection{Diabetic retinopathy}
Diabetic retinopathy (DR) is a vascular disease that affects normal blood vessels in the eye and is the leading cause of preventable blindness worldwide \citep{Wilkinson2003Proposed}.
There is a unified standard for DR classification, namely the International Clinical Diabetic Retinopathy Scale (ICDRS).
According to this standard, the severity of DR can be graded into five classes, namely 0 (no apparent DR), 1 (mild DR), 2 (moderate DR), 3 (severe DR), 4 (poliferative DR).
The most commonly used datasets are shown in Tab. \ref{tab:dataset_DR}.
All of them have been introduced in Section \ref{sec:leision_detection_segmentation}.
The experimental results are shown in Tab. \ref{tab:result_DR}.

\begin{table*}
\small
\caption{Summary of several results for DR diagnosis/grading}
\begin{center}

\resizebox{\textwidth}{!}{

\begin{tabular}{p{2.7cm} p{1.7cm} p{2cm} p{2cm} p{2.9cm} p{0.6cm} p{0.6cm} p{0.7cm} p{0.9cm} }
\hline
Reference & Dataset & Category & Backbone & Loss & SE/\% & SP/\%  & AUC/\%  & Kappa/\% \\
\hline
\cite{David2016Improved} & Messidor-2 & 4 & CNN &- & \textbf{96.8} & \textbf{87.0} & \textbf{98.0} &- \\

\cite{10.1001/jama.2016.17216} & Messidor-2 &  2 & Inception-v3 &-  & 87.0 &  \textbf{98.5} & \textbf{99.0} &- \\

\cite{Gargeya2017Automated} & Messidor-2 & 2 & CNN & 2-class categorical CE &  \textbf{93} &  87  & 94 &- \\
\hline
\cite{DBLP:conf/miccai/WangYSFLW17} & Messidor & 5 & CNN & -& - & - &  95.7 &-  \\

\cite{DBLP:conf/miccai/LinGWWCWCW18} & Messidor & 5& CNN & - & -& -& \textbf{96.8} &- \\
\hline
\cite{10.1001/jama.2016.17216} & EyePACS &  2 & Inception-v3 & - & 90.3 & \textbf{98.1} & \textbf{99.1} & -\\

\cite{Gargeya2017Automated} & EyePACS & 2 & CNN & 2-class categorical CE & \textbf{94} & 98 & 97 &- \\
\hline
\cite{Gargeya2017Automated} & E-Ophtha & 2 & CNN & 2-class categorical CE & \textbf{90} & \textbf{94} & \textbf{95} & -\\

\cite{DBLP:journals/mia/QuellecCBCL17} & E-Ophtha & 2 & CNN & -& -&- & 94.9 &- \\
\hline
\cite{DBLP:conf/miccai/WangYSFLW17} & Kaggle & 5 & CNN &- & - & - & 85.4& - \\

\cite{DBLP:conf/miccai/LinGWWCWCW18} & Kaggle & 5& CNN & - & -&- &- &  85.9 \\

\cite{DBLP:conf/isbi/RoyTCSMMG17} & Kaggle & 5 & CNN & - & -& -&- & \textbf{86} \\

\cite{DBLP:conf/miccai/YangLLWFZ17} & Kaggle & 4 & CNN &- &- & - &\textbf{95.90} &-  \\

\cite{DBLP:journals/mia/QuellecCBCL17} & Kaggle & 2 & CNN & -&- & -& \textbf{95.5} & -\\
\hline
\cite{DBLP:conf/icip/GondalKGFH17} & DiaretDB1 & 2 & CNN & -& \textbf{93.6} & \textbf{97.6} & \textbf{95.4} &-\\
\hline
\cite{DBLP:conf/aaai/FooHLLW20} & SiDRP14-15 & 5(No DR here) & U-Net, VGG16 & binary CE &- & -& \textbf{78.56} & -\\
\hline
\cite{DBLP:conf/aaai/FooHLLW20} & IDRiD & 5(No DR here) & U-Net, VGG16 & binary CE &- & -&  \textbf{99.00} &- \\

\hline
\cite{DBLP:conf/miccai/LinGWWCWCW18} & private & 5& CNN & - & -&- &- &  \textbf{87.5} \\
\hline
 \cite{Krause2017Grader} & private & 5 (moderate or worse DR here) & Inception-v4 &- & \textbf{97.1} & \textbf{92.3}  & \textbf{98.6} & \textbf{84} \\
\hline
\cite{Lidc180147} & private & 2 & Inception-v3 &-& \textbf{92.5} & \textbf{98.5} & \textbf{95.5}  &-\\
\hline
 \cite{DBLP:journals/kbs/ZhangZYGHCY19} & private & 2 & CNN & CE & \textbf{97.5} & \textbf{97.7} & \textbf{97.7} &- \\

 \cite{DBLP:journals/kbs/ZhangZYGHCY19} & private & 4 & CNN & CE & \textbf{98.1} & \textbf{98.9} & - &- \\
\hline
\cite{10.1001/jamaophthalmol.2019.2004} & hospital in Sankara & 2 & CNN &- & \textbf{92.1} &  \textbf{95.2} &  \textbf{98.0} & -\\
\hline
\cite{10.1001/jamaophthalmol.2019.2004} & hospitals in Aravind & 2 & CNN &- & \textbf{88.9} & \textbf{92.2} & \textbf{96.3}& -\\
\hline 
\end{tabular}
}

\end{center} 
\label{tab:result_DR}
\end{table*}

\subsubsection{Clinical style papers}There are several clinical style papers, usually found in clinical journals such as JAMA and Diabetic Care.
These papers usually pay more attention to actual clinical meaning rather than network architecture improvements.
Most of the training datasets were collected by the authors, rather than using public datasets.

\cite{David2016Improved} proposed a system that can automatically detect DR, called IDx-DR X2.1.
It applies a set of CNN-based detectors for each image in the detection.
Their CNN structure is inspired by AlexNet and VGG and is able to predict four labels, namely negative (no or mild DR), referable DR (rDR), vision-threatening DR (vtDR), and low exam quality (protocol errors or low-quality images).
CNN-based anatomy detectors can further detect hemorrhages, exudates, and other lesions.
\cite{Gargeya2017Automated} also used a CNN to perform DR binary classification.
They used a CNN containing five residual blocks to extract image features.
The features extracted by the deep CNN and metadata information were fed into a decision tree model for binary classification.
\cite{Lidc180147} used a deep learning algorithm (DLA) for the detection of referable DR.
For the training and validation set, they collected 71,043 images from a website named LabelMe and invited 27 ophthalmologists to annotate them.
They used four Inception-v3 networks for different tasks, namely 1) classification of vision-threatening referable DR, 2) classification of DME, 3) evaluation of image quality for DR, and 4) assessment of image quality and of the availability of the macular region of DME. 

\textbf{Ensemble strategies} are commonly used in this area. \cite{10.1001/jama.2016.17216} used a CNN for binary classification of with/without DR.
They used a dataset of 128,175 images, which were annotated three to seven times by 54 experts.
The specific network uses the structure of Inception-v3, and an ensemble of ten networks trained with the same data.
The final result is the average of all network outputs.
\cite{Krause2017Grader} used a CNN for the five-class classification of DR.
Their improvements over \cite{10.1001/jama.2016.17216} include: using Inception-v4 instead of Inception-v3, using a larger dataset during training, and using higher-resolution input images.
Their network structure is also an ensemble of ten networks.
\cite{DBLP:journals/kbs/ZhangZYGHCY19} established a high-quality labeled dataset, and adopted an ensemble strategy to perform two-class and four-class classifications.
Features extracted from different CNN models are passed through the corresponding SDNN modules, which are defined as component classifiers.
Then, the features are fused and fed into a FC layer to generate the final results.

\subsubsection{Approaches combining lesion detection}
Considering the internal correlation between the diagnosis of DR and detection of hemorrhages, exudates and other lesions, many works also generate heatmaps of lesions while performing DR diagnostic grading.
These methods consist of: generating lesion heatmaps, lesion segmentation, attention method and two benchmark works.

\textbf{Generating lesion heatmaps.} \cite{DBLP:conf/miccai/YangLLWFZ17} proposed a two-stage DCNN that can simultaneously delineate the lesions and perform DR severity grading.
The first stage is a local network, which extracts local features for lesion detection.
The second stage is a global network for the grading of DR.
A weighted lesion map is obtained from the local network and the original fundus images.
An imbalanced weighting scheme was introduced to pay more attention to lesion patches while performing DR grading.
\cite{DBLP:conf/icip/GondalKGFH17} adopted unsupervised learning to perform DR grading and generate lesion heatmaps using only image-level labels.
Their main network uses the o\_O solution \citep{oO-Solution}, replacing the last dense layer with a global average pooling (GAP) layer.
Their way of generating heatmaps was mainly inspired by \cite{DBLP:conf/cvpr/ZhouKLOT16}.
\cite{DBLP:journals/mia/QuellecCBCL17} proposed a solution to generate a heatmap that shows what roles the pixels in an image play in image-level prediction.
They can detect both image-level referable DR and pixel-level biomarkers.
Their network's baseline is also the o\_O solution, and a method called backward-forward propagation was proposed to optimize the parameters.

\textbf{Performing lesion segmentation at the same time.} \cite{DBLP:conf/aaai/FooHLLW20} used an encoder-decoder network for DR grading and lesion segmentation.
They replaced the encoder of U-Net with VGG-16, which has five groups of conv layers.
Correspondingly, the decoder is modified as a mirror of the encoder.
This architecture can perform lesion segmentation naturally.
Then, for DR grading, they attached a GAP layer to the saddle layer of the network for classification.
They further proposed a semi-supervised approach to increase the number of training images.

\textbf{Attention methods.} Attention mechanisms are also commonly used in DR diagnosis and grading.
\cite{DBLP:conf/miccai/WangYSFLW17} proposed a Zoom-in-Net that can simultaneously perform five-class DR grading and generate attention maps highlighting lesions.
Zoom-in-Net consists of three parts, namely a main net (M-Net) using Inception-Resnet as the backbone, which aims to extract features and can output diagnostic results; an A-Net, which can generate attention maps using only image-level supervision; and a C-Net, which simulates the zoom-in operation when clinicians examine images.
\cite{DBLP:conf/miccai/LinGWWCWCW18} proposed a framework based on anti-noise detection and attention-based fusion which can perform five-class DR grading.
They first extract the features using a CNN, then feed them into a designed center-sample detector to generate lesion maps.
Lesion maps and original images are sent to the proposed attention fusion network (AFN), which can learn the weights of the original images and lesion maps to reduce the influence of unnecessary information.

\textbf{Benchmark works.} Although, according to priori knowledge, detecting related lesions is helpful for the diagnosis/grading of DR, lesion detection is actually a complex and difficult task, and there exists a trade-off between lesion detection and DR grading.

\cite{DBLP:journals/isci/LiGWGLK19} built a dataset called DDR.
DDR is the only dataset considering both DR and lesion detection; it is the largest dataset for lesion detection and second largest for DR grading.
The authors evaluated ten state-of-the-art deep learning models on this dataset, including five classification models, two segmentation models, and three detection models.
Although these methods achieved a maximum acc of 0.8284 in DR grading, their performance in lesion segmentation and detection was particularly poor, indicating that the detection or segmentation of lesions is a very challenging task.
\cite{DBLP:conf/isbi/AhmadKP19} performed a benchmark work on Messidor-2.
They evaluated eight state-of-the-art deep learning classification models and generated class activation maps (CAMs) of lesions at the same time.
The results showed that there is a trade-off between classification and localization.
As the networks' depth and parameters increased, they performed better in classification, while performing worse in localization.

\subsubsection{Other approaches}
There are several other approaches for DR grading, including a bi-linear strategy, a hybrid method and the IDRiD challenge.

\textbf{Bi-linear strategy with attention mechanism.} \cite{DBLP:conf/icip/ZhaoZHTCCX19} proposed a BiRA-Net to perform DR grading.
In the introduced RA-Net, features extracted from ResNet were fed to a proposed attention net to pay more attention to the decisive areas for grading.
A bilinear strategy was adopted to train two RA-Nets for more fine-grained classification.

\textbf{Hybrid method combined with manually designed features.} \cite{DBLP:conf/isbi/RoyTCSMMG17} proposed a strategy that combines CNN and dictionary-based strategies for DR severity assessment. 
The activation value of the second fully connected layer (FC2) of the CNN was converted to a discriminative pathology histogram (DPH) and generative pathology histogram (GPH), which consist of manually designed features with specific concerns. 
The two histogram feature vectors and the original-size image were fused with the CNN's FC2 response, and finally a decision tree classifier was used to obtain the final result.

\textbf{Smartphone-based diagnosis.} \cite{10.1001/jamaophthalmol.2019.2923} proposed an offline DR screening system on a smartphone to detect referable DR.
Users can download the app and get DR diagnosis results instantly.
It is unbelievable to imagine that the diagnosis of DR can be performed by such a low-cost device in such a convenient way.
Such an offline system is of great significance to areas with limited medical resources.

\textbf{IDRiD challenge.} \cite{DBLP:journals/mia/PorwalPKDSBLWLG20} described the IDRiD dataset and outlined the setup of the challenge ``Diabetic Retinopathy Segmentation and Grading'' held at ISBI2018.
They also discussed a variety of deep learning models that were outstanding in the competition, as well as lessons learned from analyzing of the submissions.

\subsubsection{Discussion}
The diagnosis/grading of DR has been widely studied.
There are several clinical style papers in this field.
In these works, a large number of images were typically collected and labeled and the significance of using deep learning in actual clinical diagnosis was assessed.
From a technical point of view, the diagnosis/grading of DR is a classification task.
All we need to do is to predict a number indicating the stage of DR.
However, only providing a single number may confuse clinicians.
They also need to know why the network makes certain decisions, and what are deemed decisive regions.
Therefore, many works have focused on generating heatmaps or performing lesion segmentation at the same time.
Other effective methods like attention mechanisms and hybrid methods have also been explored.
It is also worth noting that a remarkable smartphone-based offline diagnosis system has been created.

However, the existing researches still face several shortcomings.
Rather than fine segmentation, the heatmap is typically generated coarsely and cannot provide lesion labels.
Therefore, performing lesion segmentation and DR diagnosis/grading at the same time is a promising direction.
However, as discussed previously, there is a trade-off between DR diagnosis/grading and lesion segmentation.
This is mainly because the high-level semantic features needed for the classification task tend to lack the spatial information required for segmentation.
Reaching this trade-off is important for this multi-task problem.

\subsection{Glaucoma}
Glaucoma is one of the major causes of blindness worldwide.
The number of glaucoma infections is expected to grow to 112 million by 2040.
Because of its irreversibility, early screening of glaucoma is extremely important \citep{BOURNE2013e339, tham2014global}.
The datasets used in this field are shown in Tab. \ref{tab:dataset_glaucoma}.
The SINDI, REFUGE and SEED datasets were not introduced in Section \ref{sec:od}.
The SINDI dataset \citep{DBLP:journals/tmi/FuCXZWLC18} was established to assess the risk factors of visual impairment in the Singapore-Indian community.
It consists of 5,783 images, of which 5,670 are normal and 113 are glaucomatous.
The REFUGE dataset \citep{DBLP:journals/mia/OrlandoFBKBDFHK20} was used in the REFUGE challenge.
It contains 1,200 fundus images with segmentation ground truth and clinical glaucoma labels.
The SEED study \citep{zheng2013how} was conducted in southwestern Singapore between 2004 and 2011.
The population included 3,353 Chinese, 3,280 Malays and 3,400 Indian adults aged 40 and older.
Experimental results are shown in Tab. \ref{tab:result_glaucoma}.

\begin{table*}
\small
\caption{Summary of several results for glaucoma diagnosis/grading}
\begin{center}

\resizebox{\textwidth}{!}{

\begin{tabular}{p{2.6cm} p{1.8cm} p{3.4cm} p{2.2cm} p{0.6cm} p{0.6cm} p{0.7cm} p{0.9cm} p{0.9cm} }
\hline
Reference & Dataset & Backbone & Loss & SE/\% & SP/\% & ACC/\% & BACC/\% & AUC/\% \\
\hline
\cite{DBLP:conf/cvpr/LiXWJL19}/\cite{DBLP:journals/tmi/LiXLLWJWFW20} & RIM-ONE & CNN & K-L divergence function and CE & 84.8 & 85.5 & 85.2 & -& 91.6 \\

\cite{DBLP:journals/eswa/FerreiraFSSG18} & RIM-ONE, DRISHTI-GS & U-Net, CNN &- &\textbf{100} & \textbf{100} & \textbf{100} &- & \textbf{100} \\
\hline
\cite{DBLP:conf/aaai/ZhaoLZC019} & ORIGA & CNN & contrastive loss and hinge loss & -&- & -& -& \textbf{92} \\

\cite{DBLP:journals/titb/LiaoZZCHZ20} & ORIGA & CNN & -& - & -& -& -& 88 \\
\hline
\cite{DBLP:conf/cvpr/LiXWJL19} & LAG & CNN & K-L divergence function and CE & \textbf{95.4} & 95.2 & 95.3 & -& 97.5 \\

\cite{DBLP:journals/tmi/LiXLLWJWFW20} & LAG & CNN & K-L divergence function and CE& \textbf{95.4}& \textbf{96.7} & \textbf{96.2} & -& \textbf{98.3} \\
\hline
\cite{DBLP:conf/icip/PalMS18} & DRIONS-DB & Encoder-decoder network & Reconstruction loss and CE & - & -& -& -& \textbf{92.3} \\
\hline
\cite{DBLP:journals/tmi/FuCXZWLC18} & SCES & U-Net, ResNet50 & Dice coefficient loss and CE & \textbf{84.78} & \textbf{83.80} &- & \textbf{84.29} & \textbf{91.83} \\
\hline
\cite{DBLP:journals/tmi/FuCXZWLC18} & SINDI & U-Net, ResNet50 & Dice coefficient loss and CE & \textbf{78.76} & \textbf{71.15} & -& \textbf{74.95} & \textbf{81.73} \\
\hline
\cite{DBLP:journals/isci/RaghavendraFBGT18} & Private & CNN & -& \textbf{98.00} & \textbf{98.30} & \textbf{98.13} & -& -\\
\hline
\cite{Li2018Efficacy} & Private & Inception-v3 &- & \textbf{95.6} & \textbf{92.0} &- & -& \textbf{98.6} \\
\hline
\cite{PHENE20191627} & Private & Inception-v3 &- & -&- &- &- & \textbf{94.5} \\
\hline
\cite{DBLP:journals/kbs/ChaiLX18} & Private & FCN, CNN, Faster-RCNN & CE & \textbf{92.33} &\textbf{90.90} & \textbf{91.51} & -&-  \\
\hline
\cite{10.1001/jamaophthalmol.2019.3501} & Private FIGD & ResNet & CE & \textbf{96.2} & \textbf{97.7} & -& -& \textbf{99.6} \\
\hline 
\end{tabular}
}

\end{center} 
\label{tab:result_glaucoma}
\end{table*}

\subsubsection{Clinical style papers}
\label{sec_glaucoma_clnical}
Similar to DR, there are several clinical applications of glaucoma diagnosis.
\cite{DBLP:journals/isci/RaghavendraFBGT18} utilized an 18-layer CNN containing five conv layers to perform glaucoma classification.
They obtained 1,426 images from Kasturba Medical College, Manipal, India.
\cite{10.1001/jamaophthalmol.2019.3501} proposed a deep learning system (DLS) named Glaucoma Diagnosis with Convoluted Neural Networks (GD-CNN), based on ResNet.
They established a dataset named FIGD consisting of 241,032 images.
They further proposed an online deep learning (ODL) system to improve the generalization ability of GD-CNN.

\textbf{Glaucomatous Optic Neuropathy (GON)} diagnosis is also widely studied. \cite{Li2018Efficacy} used deep learning to perform  binary classification of GON.
They downloaded 70,000 fundus images from the online dataset LabelMe\footnote{http://www.labelme.org/}, and selected 48,116 images for annotation.
They invited 27 qualified ophthalmologists for labeling and used Inception-v3 as the classification network.
\cite{PHENE20191627} collected 86,618 images from several sources, including EyePACS, Inoveon\footnote{http://www.inoveon.com/}, AREDS, UK Biobank\footnote{https://www.ukbiobank.ac.uk/aboutbiobankuk} and three hospitals in India.
They invited 43 graders to perform image-level and feature-level labeling.
They also used Inception-v3 and trained an ensemble of 10 networks.
Their network can predict referable GON and the presence/absence of various ONH features at the same time.

\subsubsection{Approaches considering OD/OC area}
\label{sec:glaucoma_od}
Based on the priori knowledge that the OD and OC area can be helpful to diagnose glaucoma, many methods have paid attention to these two areas.
Applications can be subdivided into OD/OC segmentation and direct CDR estimation.

\textbf{OD/OC segmentation.} \cite{DBLP:journals/eswa/FerreiraFSSG18} designed a texture descriptor with a CNN to diagnose glaucoma.
They first used a U-Net to segment the OD area, then inspired by the domain knowledge of biology, designed a module called phylogenetic diversity indexes to extract semantic features, and finally used a CNN-based classifier for the diagnosis of glaucoma.
\cite{DBLP:conf/icip/PalMS18} designed a G-EyeNet for the classification of glaucoma, which performs particularly well when the dataset is small.
They first perform OD segmentation, then the extracted ROIs are fed to a U-Net-like architecture for image reconstruction.
Finally, an FC layer followed by a softmax classifier were incorporated into to the encoder for glaucoma classification.

\textbf{Direct CDR estimation.} \cite{DBLP:journals/titb/ZhaoCLCGL20} abandoned the intermediate step of segmenting the OD and OC area and decided to directly estimate CDR from fundus images.
In the proposed MFPPNet, fundus images are passed through three DenseBlocks, and then the extracted features go through a feature pyramid pooling module and a fully connected feature fusion module to learn and fuse multiscale features.
Finally, random forest regression is used to perform CDR regression.

\subsubsection{Approaches built on multi-branched methods}
\label{sec:glaucoma_branch}
Multi-branched methods are also widely explored for glaucoma diagnosis.
The results of multiple networks with different focuses are fused together to achieve higher accuracy.
\cite{DBLP:journals/tmi/FuCXZWLC18} proposed the Disc-aware Ensemble Network (DENet) which contains four branches.
The global image stream learns the image-level global features, employing a ResNet-50 as the backbone and using the original images as input.
The second stream is a segmentation-guided network using a U-Net to segment the OD area as guidance for the other two branches.
FC layers are connected to the saddle layer of U-Net to output classification results.
The local disc region stream and the disc polar transformation stream both take ResNet-50 as the classifier, with the former taking the disc region crop as input and the other take polar transformed version.
\cite{DBLP:journals/kbs/ChaiLX18} designed a multi-branch neural network (MB-NN) combining domain knowledge.
MB-NN takes three branches as input.
The first is a set of original images.
The second branch is the optical disc region generated by Faster-RCNN.
The third branch contains domain knowledge features, which include image features such as CDR and PPA size, and non-image features such as age, intraocular pressure and eye sight.

\subsubsection{Generating evidence maps}
\label{sec:glaucoma_heatmap}
There are also some inspiring studies that generate evidence maps when performing glaucoma diagnosis.
The approaches include a weakly supervised method, using a LAG dataset containing evidence label and a multiscale method.

\textbf{Weakly supervised method.} \cite{DBLP:conf/aaai/ZhaoLZC019} proposed a weakly-supervised multi-task Learning method (WSMTL) to perform accurate evidence identification, optic disc segmentation and automated glaucoma diagnosis simultaneously.
First a skip and densely connected CNN is used to capture multiscale features.
Then, the extracted features are fed to the proposed pyramid integration structure to generate high-resolution evidence maps.
These evidence maps are passed to a constrained clustering branch which clusters pixels with relational constraints. 
The evidence maps are also fed to a fully-connected discriminator to diagnose glaucoma.

\textbf{Using dataset containing evidence map label.} \cite{DBLP:conf/cvpr/LiXWJL19} established a large-scale attention-based glaucoma (LAG) dataset.
LAG contains 5,824 fundus images and attention maps provided by ophthalmologists.
They also proposed an AG-CNN to diagnose glaucoma.
First, an attention prediction subnet was introduced to generate attention maps.
In this subnet, multiscale and channel attention methods are utilized.
Then, a pathological area localization subnet was designed to locate the pathological area, in which attention maps are embedded to feature maps at each stage.
Finally, the located pathological areas and predicted attention maps are concatenated together and fed to a glaucoma classiﬁcation subnet to predict the binary label of glaucoma.
In their subsequent work \citep{DBLP:journals/tmi/LiXLLWJWFW20}, they extended their LAG dataset to 11,760 fundus images.
They also proposed a weakly supervised learning strategy for AG-CNN.

\textbf{Multiscale networks.} \cite{DBLP:journals/titb/LiaoZZCHZ20} introduced a clinically interpretable ConvNet architecture (EAMNet) for glaucoma diagnosis.
They first used a CNN as backbone network with several residual blocks to extract useful features.
Then a method named Multi-Layers Average Pooling (M-LAP) was proposed to bridge the gap between low-level localization information and high-level semantic information.
Moreover, evidence activation maps (EAMs) were obtained by weighted summation of feature maps.

\subsubsection{Discussion}
Like in DR grading, there are several glaucoma diagnosis papers that care more about clinical applications, as discussed in Section \ref{sec_glaucoma_clnical}.
Further, Sections \ref{sec:glaucoma_od}, \ref{sec:glaucoma_branch}, and \ref{sec:glaucoma_heatmap} can all be regarded as focusing on the OC area from different aspects.
Section \ref{sec:glaucoma_od} describes methods that perform OD/OC segmentation or CDR estimation and glaucoma diagnosis simultaneously.
In Section \ref{sec:glaucoma_branch}, OD/OC segmentation serves as a branch to guide the glaucoma diagnosis task.
Finally, in Section \ref{sec:glaucoma_heatmap}, heatmaps are generated to highlight decisive regions for glaucoma diagnosis.

However, there is still room for improvement in the diagnosis of glaucoma.
First, just like in DR diagnosis, while heatmaps can be used to provide guidance for diagnosis,
the more accurate task of OD/OC segmentation should be emphasized.
Second, the diagnosis of glaucoma does not only lie in CDR estimation;
there are several other factors that can affect the result, such as age, race and family history.
However, few works focus on these factors.

\subsection{AMD}
Age-related macular degeneration (AMD) is the leading cause of vision loss among people aged 50 and above.
6.2 million people worldwide suffered from AMD in 2015 \citep{20161545}.
The datasets used in this field are shown in Tab. \ref{tab:dataset_AMD}.
AREDS is widely used in AMD diagnosis.
The AREDS dataset \citep{1999573} consists of over 206,500 images acquired from 5,208 participants.
iChallenge-AMD was used as the dataset of the iChallenge competition.
It consists of 1,200 images, of which ~77\% are from non-AMD subjects and ~23\% are from AMD patients.
Labels for AMD/non-AMD, disc boundaries and fovea locations and lesion boundaries are provided.
The KORA dataset \citep{brandl2016features} was acquired from 2,840 individuals aged 25 to 74 years old from South Germany.
Experimental results are shown in Tab. \ref{tab:result_AMD}.
\begin{table*}
\small
\caption{Widely used datasets for AMD diagnosis/grading}
\begin{center}

\resizebox{\textwidth}{!}{

\begin{tabular}{p{3cm} p{4.6cm} p{2cm} p{2cm} p{3.8cm}}
\hline
Dataset name & Number of images & Resolution & Camera & Availability\\
\hline
AREDS & Over 206,500 images & - & - & \href{https://www.ncbi.nlm.nih.gov/projects/gap/cgi-bin/study.cgi?study_id=phs000001.v3.p1}{available online}$^{\rm 1}$\\
\hline
iChallenge-AMD & 1200 & - & - & \href{http://ai.baidu.com/broad/introduction?dataset=amd}{available on registration}$^{\rm 2}$\\
\hline
KORA & images from 2840 individuals & - & - & \href{https://epi.helmholtz-muenchen.de/}{available online}$^{\rm 3}$ \\
\hline
\end{tabular}
}

\end{center} 
\begin{tablenotes}
     \item[1]$^{\rm 1}$https://www.ncbi.nlm.nih.gov/projects/gap/cgi-bin/study.cgi?study\_id=phs000001.v3.p1
     \item[2]$^{\rm 2}$http://ai.baidu.com/broad/introduction?dataset=amd
     \item[3]$^{\rm 3}$https://epi.helmholtz-muenchen.de/
\end{tablenotes}
\label{tab:dataset_AMD}
\end{table*}

\begin{table*}
\small
\caption{Summary of several results for AMD diagnosis/grading}
\begin{center}

\resizebox{\textwidth}{!}{

\begin{tabular}{p{3.1cm} p{1.1cm} p{2.1cm} p{2.3cm} p{1.2cm} p{0.9cm} p{0.9cm} p{1.1cm} p{0.9cm} p{0.9cm}}
\hline
Reference & Dataset & Backbone & Loss & Category & SE/\% & SP/\% & ACC/\% & AUC/\% & Kappa/\%\\
\hline
\cite{DBLP:conf/isbi/BurlinaFJWB16} & AREDS & CNN with SVM & - & 2({1}vs.{3,4})& \textbf{93.4} & \textbf{95.6} & \textbf{95.0} &- &-\\

\cite{10.1001/jamaophthalmol.2017.3782} & AREDS & CNN with SVM & -& 2 & - & -&  88.4$\sim$91.6 & \textbf{94$\sim$96} &-\\

\cite{DBLP:conf/icmla/HortaJPPKBFB17} & AREDS & CNN with RF & -&2& 66.34 & 88.95 & 79.04 & 84.76 &-\\

\cite{DBLP:conf/isbi/GovindaiahHSB18} & AREDS & CNN & -&2& -&- & 92.5
&- &-\\

\cite{DBLP:conf/isbi/GovindaiahHSB18} & AREDS & CNN & -&  & -&- & 
\textbf{83}&- &-\\

\cite{10.1001/jamaophthalmol.2018.4118} & AREDS & ResNet50 &Regression loss & 4 & -&- & 75.7&- &-\\

\cite{Peng2018DeepSeeNet} & AREDS & Inception-v3 &-& 6 &\textbf{59.0} &\textbf{93.0} & \textbf{67.1} &- & \textbf{55.8}\\

\cite{10.1001/jamaophthalmol.2018.4118} & AREDS & ResNet50 &Regression loss & 9 & -&- & \textbf{59.1} &- &-\\

\cite{Grassmann2018A} & AREDS, KORA & CNN & weighted k metric& 13 & -& -& \textbf{63.3} &- &-\\

\hline
\cite{DBLP:journals/fgcs/TanBSHBRRRSGCA18} & Collected & CNN & - &2& \textbf{96.43} & \textbf{93.75} & \textbf{95.45} &- &- \\
\hline 
\end{tabular}
}

\end{center} 
\label{tab:result_AMD}
\end{table*}

\subsubsection{Methods based on a hybrid architecture}
\cite{DBLP:conf/isbi/BurlinaFJWB16} are one of the very first to use deep learning for AMD diagnosis.
They used a pre-trained OverFeat DCNN to map original images into a 4,096-dimensional feature vector.
Then the vectors were passed through a linear SVM classifier to output accurate AMD binary classification results, namely disease-free/early stages and referable intermediate/advanced stages.
In their following work \citep{10.1001/jamaophthalmol.2017.3782}, they expanded the previous method by using datasets about 10 to 20 times larger.
\cite{DBLP:conf/icmla/HortaJPPKBFB17} made several modifications to \cite{DBLP:conf/isbi/BurlinaFJWB16}.
They added some side channel features such as sunlight, education and gender.
Further, they used two inscribed rectangles of fundus images at different scales as input, and thus obtained 8,192-dimensional feature vectors.
However, the dimension of the side channel features was much smaller than 8,192 dimensions.
In order to alleviate this imbalance, they used PCA for dimension reduction.
These features were then fused together to train a random forest classifier for final AMD classification.

\subsubsection{Approaches based on CNNs}
\cite{DBLP:conf/isbi/GovindaiahHSB18} evaluated the performance of deep learning networks in two-class (no or early AMD and intermediate or advanced AMD) and four-class (no AMD, early AMD, intermediate AMD and advanced AMD) classification for AMD.
The networks evaluated include VGG-16 with transfer learning, VGG-16 without transfer learning, and ResNet-50.
The experimental results showed that, whether on two-class or four-class classifications, VGG-16 without transfer learning performs best.
\cite{DBLP:journals/fgcs/TanBSHBRRRSGCA18} designed a 14-layer CNN for early AMD diagnosis.
Their data was obtained from the Ophthalmology Department of Kasturba Medical College (KMC), included 402 normal images, 583 retinal images with early, intermediate AMD, or GA and 125 retinal images with evidence of wet AMD.
\cite{10.1001/jamaophthalmol.2018.4118} used deep learning for detailed severity characterization and estimation of five-year risk among patients with AMD.
The classification network used was ResNet-50.
For AMD severity scales, they employed four-step and nine-step scales, respectively.
For the estimation of five-year risk of progression to advanced AMD, they evaluated three deep learning-based strategies, namely Soft Prediction, Hard Prediction and Regressed Prediction.

\textbf{Ensemble strategies} have also been used in AMD diagnosis. \cite{Grassmann2018A} used an ensemble network for 13-class AMD classification.
The pre-processed images were independently trained using six different CNNs (AlexNet, VGG-16, GoogLeNet, Inception-v3, ResNet, Inception-ResNet-v2). The results of the six networks were fused using random forests.

\textbf{Guidance from lesion detection} is helpful to AMD diagnosis. \cite{Peng2018DeepSeeNet} proposed a DeepSeeNet to grade the severity of AMD (0-5).
Their network consists of three parts, Drusen Net (D-Net) for detecting drusen in 3 sizes (none/small, medium and large), Pigment-Net (P-Net) for detecting pigment abnormalities (hypopigmentary or hyperpigmentary) and Late AMD-Net (LA-Net) to detect the presence of late AMD (neovascular AMD or central GA). 
The three subnetworks all use an Inception-v3 structure.

\subsubsection{Discussion}
In this section, we have discussed approaches for AMD diagnosis.
All of them are based on a CNN architecture.
Hybrid methods, ensemble strategies and guidance from lesion detection have also been explored.
However, there are still several limitations.
First, the attention paid to AMD does not match its prevalence and severity.
There is much less research on AMD diagnosis than on DR and glaucoma.
Second, the datasets and the number of images used for AMD diagnosis are far fewer than those for DR and glaucoma.
Finally, the actual amount of data in the website is inconsistent with what is claimed in the original paper.

\subsection{DME}Diabetic macular edema (DME) is the most common complication of DR and may cause severe vision loss \citep{Ciulla2003Diabetic}.
Two approaches addressing this task use a two-stage architecture and multiscale method respectively.

\textbf{Two-stage architecture.} \cite{DBLP:journals/ijon/MoZF18} proposed the cascaded deep residual networks for DME diagnosis.
The datasets used include e-ophtha EX and Hamilton Eye Institute Macular Edema Dataset (HEI-MED).
The HEI-MED dataset\footnote{http://www.vibot.u-bourgogne.fr/luca/heimed.php} \citep{GIANCARDO2012216} consists of 169 images, of which 115 are healthy and 54 contain exudates.
Their framework consists of two stages.
The first stage is an exudate segmentation network adopting a deep fully convolutional residual network (FCRN).
Then a fixed-size region is cropped, where the center pixel has the maximal probability value.
The cropped region is fed to the second stage, which is a deep residual network performing binary classification.

\textbf{Multiscale networks.} \cite{DBLP:conf/miccai/HeZWC019} proposed a DME-Net based on the multiscale method for DME classification.
They used IDRiD and MESSIDOR as their datasets.
They first passed fundus images through a U-Net to generate fovea and hard exudate region masks.
Then a multiscale feature extraction module using VGG-16 as the backbone was designed, in which a GAP operation was applied to feature maps of each stage.
Then the features were concatenated to obtain multiscale features.
They passed the original fundus images, obtained fovea and hard exudate region masks, and the macular region cropped from fundus images through the proposed multiscale feature extraction modules respectively.
The features were fused together and fed to a XGBoost classifier \citep{DBLP:conf/kdd/ChenG16} to output the final results.

\textbf{Discussion.} The two approaches introduced above both detect exudates as guidance for the DME diagnosis task.
Therefore, the architectures used are both multi-stage.
Possible future work for this task may be to design a more lightweight architecture and decrease the parameters to be trained.
Another promising direction is to detect DME and DR at the same time.
Such works can be seen in Section \ref{sec:Other_app_multu_diease}.

\subsection{ROP}
Retinopathy of prematurity (ROP) is an eye disease that often occurs in infants with low birth weight or premature birth.
It is the main cause of childhood blindness \citep{Tasman2006Retinopathy}.
\cite{10.1001/jamaophthalmol.2018.1934} used deep CNNs to diagnose plus disease in ROP.
The plus disease is defined as arterial tortuosity and venous dilation of the posterior retinal vessels that is greater than or equal to that found in a standard published retinal photograph \citep{ROPPlus}.
The presence of plus disease is the most critical feature of severe, treatment-requiring ROP.
Their training dataset contains 5,511 images, and formed part of the multicenter Imaging and Informatics in Retinopathy of Prematurity (i-ROP) cohort
study.
They first used a U-Net for preprocessing, then an Inception-v1 architecture was employed to diagnose plus disease.
Experimental results showed that the deep CNNs outperformed six of eight ROP experts invited.
\cite{10.1001/jamaophthalmol.2019.2433} used deep learning to objectively monitor ROP progression.
Their data was also from the i-ROP study.
In their work, a quantitative ROP vascular severity score was developed using previous work \citep{10.1001/jamaophthalmol.2018.1934}.
Tracking the quantitative severity score may be an effective method for identifying patients at risk of disease progression.

\textbf{Three-stage architectures} have been shown to be suitable frameworks for ROP diagnosis.
\cite{DBLP:journals/tmi/HuCZJY19} used deep learning to classify ROP.
To solve the problem of insufficient labeled data, they collected 2,668 examinations obtained from 720 infants.
Each examination consists of several fundus images from different views.
Because only a few of the n images in one ROP examination may contain features that can diagnose ROP, it is necessary to extract features from all the images in one ROP examination and fuse these features.
To this end, they designed a three-stage network which is divided into three parts: feature extraction, feature fusion and classification.
In the feature extraction stage, there are a total of n CNN networks with identical structures, which are used to extract features from the n images in one examination.
In the feature fusion stage, they considered two ways of max and mean, to fuse the n h$\times$w$\times$c features from the feature extraction stage into one h$\times$w$\times$c feature map.
The final classification stage has a convolutional layer and a GAP layer for the final binary classification of ROP.
For the network structure, they considered VGG-16, Inception-v2 and ResNet-50, and also experimented with different image resolutions.

\textbf{Discussion.} From \cite{DBLP:journals/tmi/HuCZJY19}'s work we can see that only a few images of a ROP examination contain useful features.
This situation is similar to MRI and CT image processing.
Researchers can thus borrow some inspiration from these tasks.
However, publicly available datasets are still limited in number.
More finely-annotated datasets are called for.

\subsection{Cataracts}
Cataracts can cause severe vision loss and are one of the most serious eye diseases that can cause blindness.
\cite{DBLP:journals/tmi/ZhouLL20} used deep learning for cataract diagnosis (non-cataract/cataract) and cataract grading (non-cataract, mild cataract, moderate cataract and severe cataract).
They first established a dataset containing 1,335 images from Beijing Tongren hospital, China.
433 images of the dataset are non-cataract and 922 images have cataracts.
They proposed discrete state transition (DST) and empirical DST (EDST) strategies.
In the DST strategy, the weights and activation values were restricted in a unified discrete space, while in EDST they were restricted in an exponential discrete space.
DST and EDST can reduce the networks' energy consumption and prevent overfitting.
When using priori knowledge, they extracted the improved Haar wavelet features and visible structure features from fundus image as the input of DST-MLP and EDST-MLP.
When not using priori knowledge, they used DST-ResNet and EDST-ResNet as classification networks.
\cite{DBLP:journals/titb/XuZLGZ20} introduced a hybrid global-local representation CNN for cataract grading.
They established a dataset consisting of 8,030 images, which were manually annotated by ophthalmologists.
They first used AlexNet to learn global features and then used a deconvolutional network (DN) in each CNN layer to analyze which pixel contributes most to the classification, and explain the misclassification cases.
Then a hybrid model which is an ensemble of several AlexNets, was employed combining global and local features.

\textbf{Discussion.} Cataract diagnosis using fundus images is a promising direction.
However, there are several limitations remaining.
First, there are no publicly available datasets.
Second, there is no unified grading standard like in DR diagnosis.
There limitations make it difficult to compare different works.

\subsection{Diagnosis of multiple diseases}
\label{sec:Other_app_multu_diease}
The diagnoses of different eye diseases may affect each other.
For example, for a patient who has both glaucoma and cataracts, it may be difficult to diagnose the glaucoma because of the unclear biomarkers caused by the cataracts.
Therefore, the diagnosis of multiple diseases may be a possible solution to this problem.
Moreover, the diagnosis of multiple diseases simultaneously is more convenient and helpful to clinicians.
The diagnosis of multiple diseases can be divided into simultaneous DR and DME diagnosis, simultaneous DR, glaucoma and AMD diagnosis, the diagnosis of eight diseases using paired CFPs, the diagnosis of 36 diseases and rare pathologies detection.

\textbf{Simultaneous DR and DME diagnosis.} \cite{DBLP:journals/tmi/LiHYZFH20} proposed a cross-disease attention network (CANet), which can simultaneously diagnose DR and DME.
CANet contains two different types of attention modules.
Disease-specific attention module is used to selectively learn useful features for diagnosing the specific disease.
Disease-dependent attention module can further learn the internal relationship between the two diseases.
Features extracted from ResNet-50 are passed through two disease-specific attention modules and two disease-dependent attention modules successively.
They used IDRiD and MESSIDOR as their datasets.
The diagnosis of the two diseases can be mutually enhanced.
\cite{DBLP:conf/isbi/TuGZCFGCYL20} proposed a multi-task network named feature Separation and Union Network (SUNet) for simultaneous DR and DME grading.
Experiments were carried on the IDRiD dataset.
They first used ResNet-34 to extract features for all tasks.
Then a feature blending block was proposed, which contains a sequence of feature separation and feature union layers.
The feature separation layers learn task-specific features, i.e., diagnosis features for the multi-disease diagnosis block (MD-Block) and lesion features for lesion regularize net (LR-Net).
The feature union layers are able to learn useful union features for both branches.
Finally, the MD-Block is used to predict results of DR and DME grading.

\textbf{Simultaneous DR, glaucoma, and AMD diagnosis.} \cite{10.1001/jama.2017.18152} used a CNN network to perform referable DR, vision-threatening DR, possible referable glaucoma and referable AMD diagnosis simultaneously.
Their dataset contains 494,661 retinal images, which were obtained from the ongoing Singapore National Diabetic Retinopathy Screening Program (SIDRP) \citep{2016Cost}.
The backbone network used is VGG-Net.

\textbf{Diagnosis of eight diseases using paired CFPs.} \cite{DBLP:conf/isbi/LiYHW0G20} proposed a Dense Correlation Network (DCNet) to diagnose eight diseases using paired color fundus photographs (CTF) from the ODIR dataset\footnote{https://github.com/nkicsl/OIA-ODIR}.
The ODIR dataset consists of 10,000 paired images from 5,000 Chinese patients.
Eight kinds of labels denoting the stages of specific diseases are provided for each image.
Here, paired refers to images of the left eye and the right eye from the same patient.
DCNet consists of a shared CNN feature extractor for paired CFPs, ResNet in this case, a spatial correlation module (SCM) and a final classifier.
The SCM is utilized to capture dense correlations between extracted features and fuse relevant ones.

\textbf{Diagnosis of 36 diseases.} \cite{DBLP:conf/miccai/WangJZG19} used multi-task learning to diagnose 36 diseases simultaneously.
To achieve this, they collected and relabeled 200,817 images with 36 categories, of which 17,385 images have more than one label.
Their proposed network structure is divided into two stages.
The first stage has a modified YOLO-v3 \citep{DBLP:journals/corr/abs-1804-02767} as the main structure, which is used to detect the macula and the OD/OC area.
The second level has three branches, namely the general task stream, macular task stream, and optic-disc task stream, which use original images, the macula area, and the OD and OC area as inputs, respectively.
The general task stream uses Inception-ResNet-v2 as the backbone network to detect general retinal diseases, fusing features from the other two streams.
The macular task stream is used to detect macular diseases.
It uses Inception-v3 as the backbone network and fuses features from the general task stream.
The optic-disc task stream also uses Inception-v3 as the backbone network, but it is independent and does not fuse features from the other branches.

\textbf{Rare pathologies detection based on few-shot learning.} \cite{DBLP:journals/mia/QuellecLCMC20} used few-shot learning to perform rare pathologies detection.
They used the OPHDIAT dataset for training.
This dataset \citep{MASSIN2008227} consists of 763,848 images acquired from the Ile-de-France area.
DR grading is provided for every image.
Moreover, the ophthalmologists also indicated his or her findings in free-form text.
The images contain 41 conditions, some of which are rare pathologies.
Based on the observation that CNNs trained to detect frequent conditions, such as DR, also cluster many other unrelated conditions in the feature space, the authors trained a CNN classifier and derived several simple probabilistic models from its feature space to detect rare conditions, solving the few-shot learning problem.

\textbf{Discussion.}
The diagnosis of multiple diseases is very significance in clinical practice.
In this subsection, we first discussed the simultaneous diagnosis of DR and DME and the diagnosis of three different diseases.
Then we introduced three works that focus on multiple disease diagnosis for up to 41 classes.
The methods used vary but are all inspiring.
In conclusion, this developing direction is very promising and deserves more attention,
since more comprehensive systems are needed in practice.
Therefore, more experiments should be carried on newly built datasets such as ODIR.  

\section{Image synthesis}
\label{sec:image_synthesis}
As mentioned before, the training datasets for medical imaging often consist of a fewer number of images than in other deep learning tasks.
Further, high-quality annotated datasets are often costly to obtain.
One possible solution is image synthesis.
Image synthesis can increase the number of fundus images, help us to better understand the images and improve model performance.

\textbf{Synthesis for glaucoma.} \cite{DBLP:conf/isbi/DeshmukhS19} proposed a deep learning based method to synthesize the ONH region of fundus images.
Given the OD, OC, and blood vessel segmentation masks from arbitrary fundus images, their method can generate high-quality images with vessels bending at the edges of the OC, like in real images.
The generator contains four U-Nets.
Three parallel branches take an OC, OD and vessel mask as input, respectively, and the outputs are jointly passed through another U-Net to generate RGB images.
The synthetic and original images along with their corresponding OC, OD and vessel masks compose the input of the discriminator, which employs a five-layer FCN as the backbone.
For the datasets, they used Drishti-GS and DRIVE.
\cite{DBLP:journals/tmi/Diaz-PintoCNMXF19} used deep convolutional generative adversarial networks (DCGANs) to obtain a fundus image synthesizer and used a semi-supervised method for glaucoma assessment.
Their system can not only generate synthetic images, but also provide labels automatically.
They collected 14 datasets as the training set, including an unprecedented 86,926 images.
Their two systems, DCGAN and SS-DCGAN, have similar structures, and both contain a generator and a discriminator.
The difference between them is that the DCGAN only performs image synthesis, while the SS-DCGAN can predict glaucoma by changing the final output layer of the discriminator of DCGAN.
\cite{DBLP:conf/miccai/WangXLWG19} proposed a pathology-aware visualization strategy for glaucoma classification and a pathology-based GAN (Patho-GAN) for image synthesis.
For brevity, the pathology-aware visualization net will not be described here.
Different from the usual GAN network, the synthetic images generated by the generator and the original images are passed through the pathology-aware visualization net and the pathological loss is calculated.
Specific pathological areas can be enhanced by optimizing the pathological loss.
They used LAG as their dataset.

\textbf{Synthesis for vessel segmentation.}
\cite{DBLP:journals/tmi/CostaGMNAMC18} utilized a GAN to perform retinal image synthesis.
An adversarial autoencoder is first trained to reconstruct vessel maps and learn a latent space associated with a normal distribution.
Then the generated vessel maps are passed to a GAN.
The generator of GAN is used to generate synthetic retinal images that can fool the discriminator.
Synthetic pairs and the real pairs are then passed to the discriminator.
Once trained, a synthetic retinal image can be generated using decoder of the adversarial autoencoder and generator of the GAN, using a normal distribution as input.
In their implementation, the vessel annotation-free dataset Messidor-1 and DRIVE dataset were used.
\cite{DBLP:journals/mia/ZhaoLMC18} proposed a Tub-GAN and a Tub-sGAN for retinal and neuronal image synthesis, which work well on small datasets.
They aimed to learn a mapping from a tubular structured annotation to a synthetic image.
In Tub-GAN, a GAN is employed with a vessel map ground truth and random noise as input.
In their Tub-sGAN, style transfer is incorporated using VGG-Net learning style features and content features.
In Zhao et al.'s following work \citep{DBLP:journals/tmi/ZhaoLMGDC19}, they proposed a R-sGAN to perform image synthesis for further segmentation of unannotated fundus images.
Their framework consists of two stages.
In the first stage, vessel maps from finely-annotated datasets and retinal images from unannotated images are passed to R-sGAN as input to generate retinal images that have the same style as the unannotated datasets.
R-sGAN is a non-linear variant of GRU \citep{DBLP:journals/corr/ChungGCB14}.
Then, image pairs of vessel maps and generated retinal images are passed to a segmentation network for training.
Once trained, the segmentation network can perform segmentation of unannotated images.
Both works used the DRIVE, STARE and HRF datasets.

\textbf{Synthesis for DR.}
\cite{DBLP:conf/miccai/ZhouHCZL019} proposed a diabetic retinopathy generative adversarial network (DR-GAN).
It can generate high-resolution images given arbitrary DR grading and lesion information.
They designed a generator conditioned on vessel and lesion masks to generate high-resolution images.
They introduced a fine-grained design which aims to learn better and more realistic local details.
A multiscale discriminators framework was also employed, containing three identical discriminators.
The only difference between the discriminators is the resolution of input images.
For the dataset, they used EyePACS, IDRiD and DRIVE.

\textbf{Synthesis for AMD.}
\cite{10.1001/jamaophthalmol.2018.6156} utilized a GAN to perform image synthesis for AMD.
The GAN model was trained using 133,821 fundus images from AREDS as input.
They invited two ophthalmologists to diagnose AMD on real images and synthetic images.
The results obtained are similar for real and synthetic images.
Moreover, a classification network trained only on the synthetic images showed similar performance to the network trained only on real images.

\textbf{Smartphone camera image synthesis.}
It is relatively easy as well as cost-effective to collect fundus images using a smartphone camera (SC).
However, the images are often low-quality, have uneven illumination and other problems.
\cite{DBLP:conf/isbi/VS19} proposed a ResCycleGAN for image synthesis using SC images.
Their modifications over CycleGAN \citep{8237506} are two-fold; namely, they introduced a residual connection and proposed a structure similarity based loss function.
The dataset used consists of 540 images acquired using an iPhone6.

\textbf{Multimodal image reconstruction.}
Different modalities provide complementary views of the same real-world object.
Image reconstruction from one modality to another is a self-supervised task.
On the one hand, important general features can be learned in the reconstruction process.
On the other hand, images which are obtained invasively in practice can be obtained using non-invasive modal images.
\cite{DBLP:conf/miccai/HervellaRNO18} employed U-Net to perform image reconstruction from retinography to angiography.
They used the publicly available Isfahan MISP dataset\footnote{https://misp.mui.ac.ir/data/eye-images.html} \citep{DBLP:journals/cmmm/AlipourRA12}, which contains 59 retinography/angiography pairs.

\textbf{Image super resolution (ISR).}
ISR takes low-resolution images as input and outputs super-resolved (SR) images.
This is useful to several downstream, such as small or blurred lesion and biomarker detection. 
\cite{DBLP:conf/miccai/MahapatraBHG17} proposed an ISR method based on GANs.
A local saliency map is obtained by combining abstraction, element distribution and uniqueness.
Then a local saliency loss is calculated and added to the cost function.
Entropy filtering is performed to highlight compact regions.
They used DRIVE, STARE and CHASE\_DB1 as their datasets.

\textbf{Discussion.}
Image synthesis based on deep learning is a relatively new task in fundus image processing.
Synthetic images can be used to help training, which can improve performance and alleviate overfitting.
In terms of architecture, nearly all approaches used GANs.
The latest powerful variant, CycleGAN, can also be seen in some applications.
Image synthesis will likely be a very popular direction in the near future.
It is hard to imagine what kinds of explorations could be done using GANs.

\section{Other applications}
\label{sec:Other_application}

\subsection{Ophthalmic disease diagnosis}
There are several works focusing on rare pathologies, such as pathological myopia and refractive error.

\textbf{Pathological myopia} is a common disease that can cause loss of vision.
\cite{DBLP:conf/isbi/GuoWZLWLLX20} introduced a lesion-aware segmentation network (LSN) to perform atrophy and detachment segmentation, which is related to pathological myopia.
The architecture is a U-Net-like encoder-decoder network.
The authors added a classification branch to the saddle layer to predict the existence of lesions.
A feature fusion module is used in the decoder, which is designed as a multiscale network.
To further boost the sensitivity to lesion edges, they added a loss function named edge overlap rate (EOR).
The training set used was taken from from PALM challenge\footnote{https://palm.grand-challenge.org/} in ISBI 2019, consisting of 400 images.

\textbf{Refractive error} is one of the leading causes of visual impairment.
\cite{DBLP:journals/corr/abs-1712-07798} used deep learning to diagnose this.
They used UK Biobank and AREDS as their datasets.
The architecture is a combination of ResNet and soft attention \citep{DBLP:conf/icml/XuBKCCSZB15}.
Their network can also generate attention maps.
Results showed that the foveal region is one of the most import regions for making predictions.

\textbf{Discussion.}
It is good to see that these diseases have received some attention, despite not being as prevalent as other diseases like DR, glaucoma, etc.
Diagnoses of rare pathologies are also important.
It is expected that the successful experiences in other diseases can easily be well extended to the rare pathologies diagnoses.
Current challenges lie in the lack of data.

\subsection{Systemic diseases}
As shown in previous studies, the condition of the retina may reflect other diseases.
In fact, many ophthalmologists have used ophthalmoscopes to diagnose systemic diseases such as hypertension, sarcoidosis and CMV infection. \citep{SCHMIDTERFURTH20181}. 
We thus discuss studies on systemic disease diagnosis as follows.

\textbf{Cardiovascular risk factors.}
\cite{Poplin2018PredictionOC} used a deep learning method to predict multiple cardiovascular risk factors including age, gender, smoking, systolic pressure (SBP) and so on.
Their training dataset includes the data of 284,335 patients, and was collected from the UK Biobank and EyePACS.
Inception-v3 was used as their classification network.
Moreover, to help clinicians better understand the decision process of the CNN network, they used a mechanism named soft attention to generate heatmaps, which can highlight decisive regions in the process of CNN classification.

\textbf{Ischemic strokes.}
\cite{DBLP:conf/aaai/LimLXTWLH19} utilized deep learning methods to predict strokes from fundus images.
Images positive for ischemic stroke were obtained from the MCRS study \citep{2009Retinal} while negative images were taken from five other fundus image datasets.
Their classification network is VGG-16.
They also adopted the feature isolation method.
First, a U-Net was used to segment the vessel tree from original images.
Then, vascular maps were used as the input of the classification network, providing additional information.

\textbf{Annotation-free cardiac vessel segmentation.}
\cite{DBLP:conf/miccai/Yu0GWLYDLZ19} proposed a knowledge transfer based shape-consistent generative adversarial network (SC-GAN) and a simpler Add U-Net for cardiac vessel segmentation.
In SC-GAN, an average fundus image and digital subtraction angiography (DSA) image were passed to a generator to obtain a synthetic image that had both retinal vessels and coronary arteries.
A shape-consistent loss was proposed to ensure shape-consistency.
A discriminator was then trained using synthetic images and real DSA images as input.
Finally, a U-Net was trained using synthetic DSA images with synthetic labels for cardiac vessel segmentation.
In Add U-Net, a U-Net was trained using an average fundus image and DSA image as input and a combination of fundus image annotations and the Frangi segmentation \citep{10.1007/BFb0056195} results of DSA images as labels.
They used DRIVE as the source domain and collected 1,092 coronary angiographies (DSA) with
no annotations.

\textbf{Biological age estimation.}
Biological age (BA) is a widely used aging biomarker.
\cite{DBLP:conf/miccai/LiuWLJHHMH19} developed a CNN classifier to estimate BA based on retinal images.
Two datasets named the Yangxi Dataset and Shenzhen Dataset were collected, containing 5,825 and 2,911 adults aged 50 years or older, respectively.
They employed a detail manipulation method to enhance the global details of the non-specific global anatomical and physiological features related to aging.
Then a VGG-19 network was used to estimate BA.
They also proposed a joint loss to boost performance.
Results showed that their method outperforms existing `brain age' models. 

\textbf{Discussion.}
The diagnosis of systemic diseases using fundus images is an encouraging direction.
With the successful studies using multi-disease and smartphone-based offline diagnosis systems,
there is promise of predicting systematic diseases in a remote, non-invasive, offline, convenient way.

\subsection{Image processing}
Here we discuss approaches for two aspects of image processing, namely image registration and image quality assessment.
Both of these are important for the processing and selection of images.

\textbf{Image registration.}
\cite{DBLP:journals/ijon/ZouHZZLL20} proposed an unsupervised architecture for non-rigid retinal image registration. 
They formulated the image registration task as a parameterized deformation function.
Thus, the aim is to regress the non-linear spatial correspondence between a pair of images.
For this regression task, they proposed the Structure-Driven Regression Network (SDRN) framework, which utilizes a multiscale method to focus on global and local features simultaneously.
They used the publicly available Fundus Image Registration (FIRE) dataset\footnote{https://projects.ics.forth.gr/cvrl/fire/} \citep{FIRE}, which consists of 129 retinal images forming 134 image pairs.

\textbf{Image enhancement.}
Image enhancement is another way to improve performance on existing datasets.
\cite{DBLP:conf/miccai/ZhaoYCL19} proposed a data-driven strategy to enhance blurred fundus images in a weakly supervised manner.
Their strategy uses two unpaired datasets for training.
Their method is the first end-to-end deep generative model for blurred retinal image enhancement.
They designed two generators with the same structure; one to enhance low-quality images to high-quality ones, and the other to convert high-quality images to low-quality ones for training reference.
Similarly, they designed two corresponding discriminators with the same structure.
They also introduced a dynamic retinal image feature limit to guide the generator to improve performance and avoid the over-enhancement of extremely blurred areas.
They used a private dataset which consists of 550 blurry images and 550 high-quality images for training and 60 blurry images for testing.
The blurry images are from cataract patients and the high-quality ones are from normal people.

\textbf{Image quality assessment.}
Retinal image quality assessment (RIQA) is important for ensuring the quality of images used by clinicians and deep learning systems.
\cite{DBLP:conf/miccai/FuWSCX0S19} proposed the Multiple Color-space Fusion Network (MCF-Net) for RIQA.
They first re-annotated an Eye-Quality (EyeQ) dataset with 28,792 images from EyePACS.
Compared with other datasets, they extended binary labels (`Accept' and `Reject') to ternary labels (`Good', `Usable' and `Reject').
For the model architecture, they first transferred the original RGB color-space to HSV and LAB color-space.
Then, retinal images with different color-spaces were passed to their corresponding base network to extract features.
Feature fusion was performed at a feature level and prediction level.
\cite{DBLP:journals/mia/ShenSFLDSQJS20} proposed a domain-invariant interpretable fundus IQA system.
In order to improve the interpretability, they added three clinically accepted aspects (artifact, clarity and field definition) to the output and a visual feedback.
A coarse-to-fine architecture was introduced to locate landmarks, including OD and fovea, for robustness.
In order to generalize well on different datasets, they adopted a semi-tied adversarial discriminative domain adaption model.
They collected their dataset from patients who participated in the Shanghai Diabetic Retinopathy Screening Program (SDRSP).

\textbf{Discussion.}
The above three directions are all significant in supporting other tasks.
And the use of deep learning make these image processings effective.
\cite{DBLP:journals/mia/ShenSFLDSQJS20}'s work is particularly inspiring.
Domain invariance and visualization, which are useful for clinicians, are also goals common to almost all other tasks.
It is clear that deep learning is on the frontier of research, and more architectures and methods should be explored.

\section{Conclusions and Discussions}
\label{sec_conclu}
\subsection{Conclusions}
As demonstrated in Sections \ref{sec:leision_detection_segmentation} to \ref{sec:image_synthesis}, the performance of deep learning for fundus image diagnostic tasks is quite impressive.
In fact, deep learning methods have even achieved better performance than experienced humans in some cases.
Specifically, deep learning can provide helpful suggestions for ophthalmologists.
It can detect and segment important biomarkers, such as lesions, blood vessels, OD/OC, and provide evidence for physicians to diagnose specific diseases.
It can also directly predict whether a patient has an ophthalmic disease, and can serve as a powerful assistant for physicians in the screening of glaucoma, DR, AMD and other ophthalmic diseases.
\cite{10.1001/jamaophthalmol.2019.2004} compared the performance of deep learning with that of a human expert and a trained grader in DR diagnosis in two hospitals in India.
Their results showed that the deep learning system is well generalized to the actual data of India.
Further, for the first time in ophthalmology, it was verified in practice rather than on datasets that deep learning achieves comparable or even better performance than human experts.
The excellent performance of deep learning makes it a promising replacement for traditional computer-aided diagnostic systems (CADs).
In fact, IDx \citep{David2016Improved} has already been approved by the US FDA for practical use.
We believe that more deep learning methods will further be deployed as stable, efficient, and robust diagnostic systems for practical clinical diagnosis.

In terms of network structure, the classification backbone network has evolved from VGG and Inception-v1 to Inception-v2, Inception-v3, ResNet and DenseNet, while the segmentation backbone network has evolved from manually designed CNNs to FCNs and then to U-Net, Mask-RCNN, DeeplabV3+, etc.
However, using deep learning in fundus image analysis is more than simply applying the backbone networks to specific tasks.
There are many practical problems emerging.
For example, the number of pixels of biomarkers such as lesions, OD/OC, and blood vessels is much smaller than that of the background.
Further, the number of pixels of the unique curved structure of blood vessels, especially capillaries, makes it a hard sample problem.
To solve these problems, specific methods are required according to the characteristics of each task.
Advances in methods range from simple applications of deep learning to multi-branch, multiscale, and coarse-to-fine networks, as well as attention mechanisms, and so on.
We have summarized the approaches in Sections \ref{sec:leision_detection_segmentation} to \ref{sec:image_synthesis} according to tasks and methods, and provided an overall summary in Fig. \ref{fig:knowledge_graph}.

\subsection{Limitations and possible solutions}
Although the application of deep learning in the field of fundus image analysis has achieved gratifying performance, it is worth noting that it still has limitations in many other aspects.
The problems which restrict the performance have not been solved, and the inherent limitations of deep learning also remain unresolved.
Studying how to solve these limitations will be a key issue in the future of this field.
We have discussed limitations for specific tasks in each section.
Thus, here we will only list limitations which are common to all tasks, and provide possible solutions to these.
\subsubsection{Lack of high-quality labeled data}
As mentioned previously, deep learning is data-driven.
There are many large-scale datasets in the field of natural image processing.
For instance, ImageNet \citep{DBLP:conf/cvpr/DengDSLL009} has more than 14 million images.
Fundus image datasets, however, are quite limited, as with other medical fields.
Unlike natural images, the labeling of fundus images needs to be completed by experts, and is very difficult.
For example, because of the lack of depth information, an expert typically requires eight minutes to label a fundus image for OD/OC segmentation and glaucoma diagnosis \citep{DBLP:conf/ictai/LimCHL15}.
These limitations have resulted in a lack of high-quality labeled fundus images.
The smaller the size of the dataset, the more likely it will lead to lower accuracy and overfitting.

In addition to waiting for ophthalmologists to label more data, researchers can also seek measures to help alleviate this problem:

\textbf{Weakly supervised learning.}
Weakly supervised learning can solve the lack of high-quality labeled data to a certain extent.
We saw in Sections \ref{sec:leision_detection_segmentation} to \ref{sec:image_synthesis} that weakly supervised methods have already been applied to the field of fundus images. Weakly supervised learning can be divided into incomplete, inexact and inaccurate supervision.
Incomplete supervision means that part of the data is labeled, while the other part is not; that is, the labels are incomplete.
Active and semi-supervised learning can be used to solve this.
Inexact supervision means the granularity of the annotation does not match the problem to be solved.
Multi-instance learning can be used to address this.
Inaccurate supervision means that the annotation is not completely accurate, so there are samples with incorrect annotations.
One can consider learning with noisy labels to solve this.
For specific relevant methods please refer to \cite{zhou2018a}.

\textbf{Image synthesis and enhancement.}
Unlike weakly supervised learning, which addresses the lack of high-quality annotations, image enhancement can improve the quality of the image, while image synthesis can directly generate realistic images, even with labels.
The excellent performance of image generation is due to the use of powerful GAN models.
In fact, GANs have become mainstream for many image generation problems, such as style transfer, image inpainting, super resolution, and so on.
Section \ref{sec:image_synthesis} introduced several specific examples of GAN-based image synthesis.
We believe that as the quality of the generated images improves, there will be more approaches that use these for training, and there will be in turn more research to further improve the quality of the synthesized images.

\textbf{Federated learning.}
Creating a high-quality annotated dataset involves more than simply inviting ophthalmologists to annotate the data.
It is a complicated matter, and there are many other considerations, including data privacy, competition in various research institutes and hospitals, and relevant laws and regulations.
Note that many of the fundus image datasets are also private.
How to achieve data sharing while satisfying diverse research groups, complying with regulations and not infringing upon user privacy is an urgent problem to be solved.
In 2016, Google proposed federated learning to solve the ``data islands'' problem.
Federated learning can be divided into horizontal federated learning, vertical federated learning and federated transfer learning.
Horizontal federated learning means that the data features of two datasets overlap more and users overlap less.
Vertical federated learning means that the two datasets have more user overlap and less user feature overlap.
Federated transfer learning means that both user and data features overlap little.
Federated learning is still a relatively new field to be explored.
Readers can refer to \cite{DBLP:journals/corr/abs-1912-04977} for more information about federated learning.

\subsubsection{Imbalance}
The imbalance in fundus images is mainly an imbalance in foreground and background, or number of samples in different classes.
A large portion of the methods introduced in Sections \ref{sec:leision_detection_segmentation} to \ref{sec:image_synthesis} were proposed to solve this problem.
For instance, imbalance between foreground and background occurs in many fundus image tasks, with the number of pixels in lesions, blood vessels, OD and OC, being much smaller than the number of pixels in their respective backgrounds.
This imbalance will directly increase the difficulty of training.
For lesions and OD/OC, using the detection network to extract the ROI as the input of the network can increase the proportion of the foreground.
Such an approach can be seen in \cite{DBLP:journals/kbs/ChaiLX18}, \cite{DBLP:journals/tmi/FuCXZWLC18}, \cite{DBLP:conf/miccai/SarhanAYNE19}, \cite{DBLP:conf/isbi/ShahKP19} and \cite{DBLP:journals/tmi/WangYYFH19}.
Using spatial attention, as done by \cite{DBLP:conf/miccai/WangYSFLW17} and \cite{DBLP:conf/icip/ZhaoZHTCCX19}, is also very common and allows the network to focus on areas that are more decisive for solving tasks.
It is worth noting that \cite{DBLP:journals/tmi/FuCXWLC18} performed polar coordinate transformation to alleviate the imbalance between foreground and background based on the unique ellipse shape of the OD/OC region.
An imbalance in the number of samples in different classes is also very common in fundus images.
One solution is to use a class balance loss function, such as cross-entropy loss and focal loss.
Selective sampling is also a direction that has been explored.
It uses a carefully designed sampling strategy to maintain a certain proportion of samples in different classes during each training epoch, thereby avoiding imbalance.
This strategy was used in \cite{DBLP:journals/tmi/GrinsvenGHTS16}, \cite{DBLP:conf/icip/GondalKGFH17}, \cite{DBLP:journals/tmi/DaiFLHSWJ18} and \cite{DBLP:conf/miccai/SarhanAYNE19}.

\subsubsection{Poor generalization performance}
There are certain differences between the various fundus image datasets, including acquisition camera, resolution, light source intensity, parameter settings, and so on.
The differences between the datasets pose a challenge to the generalization performance of deep learning models.
In fact, even some state-of-the-art models only perform well on certain datasets and degrade on others.
This problem is mainly caused by the distribution difference between different datasets, that is, domain shift \citep{DBLP:conf/miccai/GhafoorianMKKMP17}.
Domain adaption, introduced in Section \ref{sec:OD_seg_domain_adapt}, can be used to enhance the model's performance on the target domain and solve the problems caused by the domain shift.
This strategy was used in \cite{DBLP:conf/miccai/WangYLYFH19}, \cite{DBLP:journals/tmi/WangYYFH19}, and \cite{DBLP:conf/miccai/LiuKLZF19} to enhance the generalization performance of the optic disc segmentation model.
Domain adaption is still an area being explored, and there are various diverse methods for it.
Readers can learn more from \cite{DBLP:journals/ijon/WangD18}.

\subsubsection{High consumption of deep learning}
The excellent performance of deep learning comes at the cost of very high consumption,
since the size of the parameters are much larger than those of traditional machine learning.
This not only means that the models require significant computational resources and time during training, but also prevents them from being deployed on portable devices, such as binocular indirect ophthalmoscopes \citep{DBLP:conf/icip/HajabdollahiENK18}.
One direction to solve the high-consumption problem of deep learning is to design more novel network structures and explore operations or layers with low computational load and low memory consumption.
However, this is mainly the work of basic network researchers.
Researchers in this field can directly apply or learn from mature models, for example using a lightweight network to decrease complexity.
For instance, MobileNet \citep{DBLP:journals/corr/abs-1801-04381} could be used instead of ResNet as the backbone.
The models could also be compressed by quantizing the network weight to reduce the complexity.
One approach is to quantize the weight and activation values in a CNN from a 32-bit float number to a low-bit number, or constrain the weights and activations to binary values and so on.
The model pruning method can also compress the models.
This strategy changes some parameters in the model to zero and skips the calculation.
In addition, there are methods such as Huffman coding for the weights of the model.
Some researchers have also explored how to reduce the consumption of fundus image network models specifically.
For example, \cite{DBLP:conf/icip/HajabdollahiENK18} used the method of quantifying weights and pruning to reduce the complexity of a blood vessel segmentation model.

\subsubsection{Lack of interpretability}
An important issue in the application of deep learning to actual medical systems is to what extent doctors accept its ``black box''.
This lack of interpretability is an inherent defect in deep learning.
Fortunately, several studies have focused on this issue.
Approaches for solving this can be divided into generating heatmaps, clinical meaning of heatmaps and other explorations.

\textbf{Generating heatmaps.} 
The basic idea of several studies is to merge the feature maps of each layer of the deep network to generate a heatmap, called a class activation map (CAM) or evidence map.
The generated heatmap shows which part of the image the deep network referred to when making its final judgment.
\cite{10.1001/jamaophthalmol.2018.6035} tried to generate heatmaps for DR and GON diagnosis systems.
They used a threshold strategy when generating the final probability map to visualize decisive regions for the prediction.
The application of this idea can also be seen in \cite{DBLP:conf/miccai/YangLLWFZ17}, \cite{DBLP:conf/icip/GondalKGFH17}, \cite{DBLP:journals/mia/QuellecCBCL17}, \cite{DBLP:conf/cvpr/LiXWJL19}, \cite{DBLP:conf/aaai/ZhaoLZC019} and \cite{DBLP:journals/tmi/LiXLLWJWFW20}.

\textbf{Clinical meaning of heatmaps.} The heatmap generated not only provides a cue that how deep learning makes decisions, but also provides guidance and assistance to the diagnostic process.
\cite{DBLP:conf/isbi/MengHS20} explored how the process of generating heatmaps can improve performance of disease diagnosis and explainability of the net.
They first generated heatmaps using a gradient-based classification activation map (Grad-CAM) \citep{DBLP:conf/iccv/SelvarajuCDVPB17}.
Then the network was fine-tuned by several designed losses and ophthalmologist intervention.
Experimental results on a private dataset showed performance improvement for the classification task.
\cite{SAYRES2019552} evaluated the role of deep learning in guiding diagnosis.
They invited 10 experts to grade DR in 796 fundus images.
Each image had three forms: the original image without auxiliary information, the image with only grading results and the images with grading results and heatmaps.
Images were randomly assigned to different ophthalmologists.
The results showed that the assistance of deep learning diagnostic results improves the accuracy and confidence of experts in diagnosing DR, especially with heatmaps.

\textbf{Other explorations.} Note that there are several other studies on the topic of interpretability. 
\cite{DBLP:journals/mia/AraujoAMPMCMC20} proposed a deep learning-based grading system named DR$|$GRADUATE.
In addition to the grading of DR, it can also estimate how uncertain the prediction is.
\cite{DBLP:journals/ijon/TorreVP20} proposed a deep learning-based interpretable classifier for DR grading.
In their classifier, a score similar to the concept of relevance was assigned to every point of the input and hidden spaces.
The scores indicated the contributions to the final prediction. 
\cite{DBLP:conf/aaai/NiuG0LWSZXDC19} explored interpretability in the diagnosis of DR and borrowed some ideas from Koch's law in infectious diseases.

\section*{Acknowledgments}
This work is partially supported by the National Natural Science Foundation (61872200),  the Natural Science Foundation of Tianjin (19JCZDJC31600, 18YFYZCG00060) and the Open Project Fund of State Key Laboratory of Computer Architecture, Institute of Computing Technology, Chinese Academy of Sciences No. CARCH201905 

\bibliographystyle{model2-names.bst}\biboptions{authoryear}
\bibliography{refs}

\begin{thebibliography}{242}
\expandafter\ifx\csname natexlab\endcsname\relax\def\natexlab#1{#1}\fi
\providecommand{\url}[1]{\texttt{#1}}
\providecommand{\href}[2]{#2}
\providecommand{\path}[1]{#1}
\providecommand{\DOIprefix}{doi:}
\providecommand{\ArXivprefix}{arXiv:}
\providecommand{\URLprefix}{URL: }
\providecommand{\Pubmedprefix}{pmid:}
\providecommand{\doi}[1]{\href{http://dx.doi.org/#1}{\path{#1}}}
\providecommand{\Pubmed}[1]{\href{pmid:#1}{\path{#1}}}
\providecommand{\bibinfo}[2]{#2}
\ifx\xfnm\relax \def\xfnm[#1]{\unskip,\space#1}\fi
\bibitem[{Abbasi{-}Sureshjani et~al.(2015)Abbasi{-}Sureshjani, Smit{-}Ockeloen,
  Zhang and ter Haar~Romeny}]{DBLP:conf/iciar/Abbasi-Sureshjani15}
\bibinfo{author}{Abbasi{-}Sureshjani, S.}, \bibinfo{author}{Smit{-}Ockeloen,
  I.}, \bibinfo{author}{Zhang, J.}, \bibinfo{author}{ter Haar~Romeny, B.M.},
  \bibinfo{year}{2015}.
\newblock \bibinfo{title}{Biologically-inspired supervised vasculature
  segmentation in {SLO} retinal fundus images}, in: \bibinfo{editor}{Kamel,
  M.}, \bibinfo{editor}{Campilho, A.J.C.} (Eds.), \bibinfo{booktitle}{Image
  Analysis and Recognition - 12th International Conference, {ICIAR} 2015,
  Niagara Falls, ON, Canada, July 22-24, 2015, Proceedings},
  \bibinfo{publisher}{Springer}. pp. \bibinfo{pages}{325--334}.
\newblock \URLprefix \url{https://doi.org/10.1007/978-3-319-20801-5\_35},
  \DOIprefix\doi{10.1007/978-3-319-20801-5\_35}.
\bibitem[{Abramoff et~al.(2010)Abramoff, Garvin and
  Sonka}]{abramoff2010retinal}
\bibinfo{author}{Abramoff, M.D.}, \bibinfo{author}{Garvin, M.K.},
  \bibinfo{author}{Sonka, M.}, \bibinfo{year}{2010}.
\newblock \bibinfo{title}{Retinal imaging and image analysis}.
\newblock \bibinfo{journal}{IEEE Reviews in Biomedical Engineering}
  \bibinfo{volume}{3}, \bibinfo{pages}{169--208}.
\bibitem[{Abràmoff et~al.(2013)Abràmoff, Folk, Han, Walker, Williams,
  Russell, Massin, Cochener, Gain, Tang, Lamard, Moga, Quellec and
  Niemeijer}]{10.1001/jamaophthalmol.2013.1743}
\bibinfo{author}{Abràmoff, M.D.}, \bibinfo{author}{Folk, J.C.},
  \bibinfo{author}{Han, D.P.}, \bibinfo{author}{Walker, J.D.},
  \bibinfo{author}{Williams, D.F.}, \bibinfo{author}{Russell, S.R.},
  \bibinfo{author}{Massin, P.}, \bibinfo{author}{Cochener, B.},
  \bibinfo{author}{Gain, P.}, \bibinfo{author}{Tang, L.},
  \bibinfo{author}{Lamard, M.}, \bibinfo{author}{Moga, D.C.},
  \bibinfo{author}{Quellec, G.}, \bibinfo{author}{Niemeijer, M.},
  \bibinfo{year}{2013}.
\newblock \bibinfo{title}{{Automated Analysis of Retinal Images for Detection
  of Referable Diabetic Retinopathy}}.
\newblock \bibinfo{journal}{JAMA Ophthalmology} \bibinfo{volume}{131},
  \bibinfo{pages}{351--357}.
\newblock \URLprefix \url{https://doi.org/10.1001/jamaophthalmol.2013.1743},
  \DOIprefix\doi{10.1001/jamaophthalmol.2013.1743}.
\bibitem[{{Abràmoff} et~al.(2010){Abràmoff}, {Garvin} and {Sonka}}]{5660089}
\bibinfo{author}{{Abràmoff}, M.D.}, \bibinfo{author}{{Garvin}, M.K.},
  \bibinfo{author}{{Sonka}, M.}, \bibinfo{year}{2010}.
\newblock \bibinfo{title}{Retinal imaging and image analysis}.
\newblock \bibinfo{journal}{IEEE Reviews in Biomedical Engineering}
  \bibinfo{volume}{3}, \bibinfo{pages}{169--208}.
\bibitem[{Adem(2018)}]{DBLP:journals/eswa/Adem18}
\bibinfo{author}{Adem, K.}, \bibinfo{year}{2018}.
\newblock \bibinfo{title}{Exudate detection for diabetic retinopathy with
  circular hough transformation and convolutional neural networks}.
\newblock \bibinfo{journal}{Expert Syst. Appl.} \bibinfo{volume}{114},
  \bibinfo{pages}{289--295}.
\newblock \URLprefix \url{https://doi.org/10.1016/j.eswa.2018.07.053},
  \DOIprefix\doi{10.1016/j.eswa.2018.07.053}.
\bibitem[{Ahmad et~al.(2019)Ahmad, Kasukurthi and
  Pande}]{DBLP:conf/isbi/AhmadKP19}
\bibinfo{author}{Ahmad, M.}, \bibinfo{author}{Kasukurthi, N.},
  \bibinfo{author}{Pande, H.}, \bibinfo{year}{2019}.
\newblock \bibinfo{title}{Deep learning for weak supervision of diabetic
  retinopathy abnormalities}, in: \bibinfo{booktitle}{16th {IEEE} International
  Symposium on Biomedical Imaging, {ISBI} 2019, Venice, Italy, April 8-11,
  2019}, \bibinfo{publisher}{{IEEE}}. pp. \bibinfo{pages}{573--577}.
\newblock \URLprefix \url{https://doi.org/10.1109/ISBI.2019.8759417},
  \DOIprefix\doi{10.1109/ISBI.2019.8759417}.
\bibitem[{Alipour et~al.(2012)Alipour, Rabbani and
  Akhlaghi}]{DBLP:journals/cmmm/AlipourRA12}
\bibinfo{author}{Alipour, S.H.M.}, \bibinfo{author}{Rabbani, H.},
  \bibinfo{author}{Akhlaghi, M.}, \bibinfo{year}{2012}.
\newblock \bibinfo{title}{Diabetic retinopathy grading by digital curvelet
  transform}.
\newblock \bibinfo{journal}{Comput. Math. Methods Medicine}
  \bibinfo{volume}{2012}, \bibinfo{pages}{761901:1--761901:11}.
\newblock \URLprefix \url{https://doi.org/10.1155/2012/761901},
  \DOIprefix\doi{10.1155/2012/761901}.
\bibitem[{Almazroa et~al.(2018)Almazroa, Alodhayb, Osman, Ramadan, Hummadi,
  Dlaim, Alkatee, Raahemifar and Lakshminarayanan}]{10.1117/12.2293584}
\bibinfo{author}{Almazroa, A.}, \bibinfo{author}{Alodhayb, S.},
  \bibinfo{author}{Osman, E.}, \bibinfo{author}{Ramadan, E.},
  \bibinfo{author}{Hummadi, M.}, \bibinfo{author}{Dlaim, M.},
  \bibinfo{author}{Alkatee, M.}, \bibinfo{author}{Raahemifar, K.},
  \bibinfo{author}{Lakshminarayanan, V.}, \bibinfo{year}{2018}.
\newblock \bibinfo{title}{{Retinal fundus images for glaucoma analysis: the
  RIGA dataset}}, in: \bibinfo{editor}{Zhang, J.}, \bibinfo{editor}{Chen, P.H.}
  (Eds.), \bibinfo{booktitle}{Medical Imaging 2018: Imaging Informatics for
  Healthcare, Research, and Applications}, \bibinfo{organization}{International
  Society for Optics and Photonics}. \bibinfo{publisher}{SPIE}. pp.
  \bibinfo{pages}{55 -- 62}.
\newblock \URLprefix \url{https://doi.org/10.1117/12.2293584},
  \DOIprefix\doi{10.1117/12.2293584}.
\bibitem[{Ara{\'{u}}jo et~al.(2020)Ara{\'{u}}jo, Aresta, Mendon{\c{c}}a, Penas,
  Maia, Carneiro, Mendon{\c{c}}a and
  Campilho}]{DBLP:journals/mia/AraujoAMPMCMC20}
\bibinfo{author}{Ara{\'{u}}jo, T.}, \bibinfo{author}{Aresta, G.},
  \bibinfo{author}{Mendon{\c{c}}a, L.}, \bibinfo{author}{Penas, S.},
  \bibinfo{author}{Maia, C.}, \bibinfo{author}{Carneiro, {\^{A}}.},
  \bibinfo{author}{Mendon{\c{c}}a, A.M.}, \bibinfo{author}{Campilho, A.},
  \bibinfo{year}{2020}.
\newblock \bibinfo{title}{Dr{\(\vert\)}graduate: Uncertainty-aware deep
  learning-based diabetic retinopathy grading in eye fundus images}.
\newblock \bibinfo{journal}{Medical Image Anal.} \bibinfo{volume}{63},
  \bibinfo{pages}{101715}.
\newblock \URLprefix \url{https://doi.org/10.1016/j.media.2020.101715},
  \DOIprefix\doi{10.1016/j.media.2020.101715}.
\bibitem[{Badar et~al.(2020)Badar, Haris and
  Fatima}]{DBLP:journals/csr/BadarHF20}
\bibinfo{author}{Badar, M.}, \bibinfo{author}{Haris, M.},
  \bibinfo{author}{Fatima, A.}, \bibinfo{year}{2020}.
\newblock \bibinfo{title}{Application of deep learning for retinal image
  analysis: {A} review}.
\newblock \bibinfo{journal}{Comput. Sci. Rev.} \bibinfo{volume}{35},
  \bibinfo{pages}{100203}.
\newblock \URLprefix \url{https://doi.org/10.1016/j.cosrev.2019.100203},
  \DOIprefix\doi{10.1016/j.cosrev.2019.100203}.
\bibitem[{Badrinarayanan et~al.(2017)Badrinarayanan, Kendall and
  Cipolla}]{DBLP:journals/pami/BadrinarayananK17}
\bibinfo{author}{Badrinarayanan, V.}, \bibinfo{author}{Kendall, A.},
  \bibinfo{author}{Cipolla, R.}, \bibinfo{year}{2017}.
\newblock \bibinfo{title}{Segnet: {A} deep convolutional encoder-decoder
  architecture for image segmentation}.
\newblock \bibinfo{journal}{{IEEE} Trans. Pattern Anal. Mach. Intell.}
  \bibinfo{volume}{39}, \bibinfo{pages}{2481--2495}.
\newblock \URLprefix \url{https://doi.org/10.1109/TPAMI.2016.2644615},
  \DOIprefix\doi{10.1109/TPAMI.2016.2644615}.
\bibitem[{Baskaran et~al.(2015)Baskaran, Foo, Cheng, Narayanaswamy, Zheng, Wu,
  Saw, Foster, Wong and Aung}]{baskaran2015the}
\bibinfo{author}{Baskaran, M.}, \bibinfo{author}{Foo, R.C.},
  \bibinfo{author}{Cheng, C.}, \bibinfo{author}{Narayanaswamy, A.},
  \bibinfo{author}{Zheng, Y.}, \bibinfo{author}{Wu, R.}, \bibinfo{author}{Saw,
  S.}, \bibinfo{author}{Foster, P.J.}, \bibinfo{author}{Wong, T.Y.},
  \bibinfo{author}{Aung, T.}, \bibinfo{year}{2015}.
\newblock \bibinfo{title}{The prevalence and types of glaucoma in an urban
  chinese population: The singapore chinese eye study}.
\newblock \bibinfo{journal}{JAMA Ophthalmology} \bibinfo{volume}{133},
  \bibinfo{pages}{874--880}.
\bibitem[{Bourne et~al.(2013)Bourne, Stevens, White, Smith, Flaxman, Price,
  Jonas, Keeffe, Leasher, Naidoo, Pesudovs, Resnikoff and
  Taylor}]{BOURNE2013e339}
\bibinfo{author}{Bourne, R.R.A.}, \bibinfo{author}{Stevens, G.A.},
  \bibinfo{author}{White, R.A.}, \bibinfo{author}{Smith, J.L.},
  \bibinfo{author}{Flaxman, S.R.}, \bibinfo{author}{Price, H.},
  \bibinfo{author}{Jonas, J.B.}, \bibinfo{author}{Keeffe, J.},
  \bibinfo{author}{Leasher, J.}, \bibinfo{author}{Naidoo, K.},
  \bibinfo{author}{Pesudovs, K.}, \bibinfo{author}{Resnikoff, S.},
  \bibinfo{author}{Taylor, H.R.}, \bibinfo{year}{2013}.
\newblock \bibinfo{title}{Causes of vision loss worldwide, 1990–2010: a
  systematic analysis}.
\newblock \bibinfo{journal}{The Lancet Global Health} \bibinfo{volume}{1},
  \bibinfo{pages}{e339 -- e349}.
\newblock \URLprefix
  \url{http://www.sciencedirect.com/science/article/pii/S2214109X1370113X},
  \DOIprefix\doi{https://doi.org/10.1016/S2214-109X(13)70113-X}.
\bibitem[{Brandl et~al.(2016)Brandl, Breinlich, Stark, Enzinger, Asenmacher,
  Olden, Grassmann, Graw, Heier, Peters et~al.}]{brandl2016features}
\bibinfo{author}{Brandl, C.}, \bibinfo{author}{Breinlich, V.A.},
  \bibinfo{author}{Stark, K.}, \bibinfo{author}{Enzinger, S.},
  \bibinfo{author}{Asenmacher, M.}, \bibinfo{author}{Olden, M.},
  \bibinfo{author}{Grassmann, F.}, \bibinfo{author}{Graw, J.},
  \bibinfo{author}{Heier, M.}, \bibinfo{author}{Peters, A.}, et~al.,
  \bibinfo{year}{2016}.
\newblock \bibinfo{title}{Features of age-related macular degeneration in the
  general adults and their dependency on age, sex, and smoking: Results from
  the german kora study}.
\newblock \bibinfo{journal}{PLOS ONE} \bibinfo{volume}{11}.
\bibitem[{Brown et~al.(2018)Brown, Campbell, Beers, Chang, Ostmo, Chan, Dy,
  Erdogmus, Ioannidis, Kalpathy-Cramer, Chiang, for~the Imaging and
  in~Retinopathy of Prematurity~(i ROP)
  Research~Consortium}]{10.1001/jamaophthalmol.2018.1934}
\bibinfo{author}{Brown, J.M.}, \bibinfo{author}{Campbell, J.P.},
  \bibinfo{author}{Beers, A.}, \bibinfo{author}{Chang, K.},
  \bibinfo{author}{Ostmo, S.}, \bibinfo{author}{Chan, R.V.P.},
  \bibinfo{author}{Dy, J.}, \bibinfo{author}{Erdogmus, D.},
  \bibinfo{author}{Ioannidis, S.}, \bibinfo{author}{Kalpathy-Cramer, J.},
  \bibinfo{author}{Chiang, M.F.}, \bibinfo{author}{for~the Imaging},
  \bibinfo{author}{in~Retinopathy of Prematurity~(i ROP) Research~Consortium,
  I.}, \bibinfo{year}{2018}.
\newblock \bibinfo{title}{{Automated Diagnosis of Plus Disease in Retinopathy
  of Prematurity Using Deep Convolutional Neural Networks}}.
\newblock \bibinfo{journal}{JAMA Ophthalmology} \bibinfo{volume}{136},
  \bibinfo{pages}{803--810}.
\newblock \URLprefix \url{https://doi.org/10.1001/jamaophthalmol.2018.1934},
  \DOIprefix\doi{10.1001/jamaophthalmol.2018.1934}.
\bibitem[{Budai et~al.(2013)Budai, Bock, Maier, Hornegger and
  Michelson}]{DBLP:journals/ijbi/BudaiBMHM13}
\bibinfo{author}{Budai, A.}, \bibinfo{author}{Bock, R.},
  \bibinfo{author}{Maier, A.K.}, \bibinfo{author}{Hornegger, J.},
  \bibinfo{author}{Michelson, G.}, \bibinfo{year}{2013}.
\newblock \bibinfo{title}{Robust vessel segmentation in fundus images}.
\newblock \bibinfo{journal}{Int. J. Biomed. Imaging} \bibinfo{volume}{2013},
  \bibinfo{pages}{154860:1--154860:11}.
\newblock \URLprefix \url{https://doi.org/10.1155/2013/154860},
  \DOIprefix\doi{10.1155/2013/154860}.
\bibitem[{Burlina et~al.(2016)Burlina, Freund, Joshi, Wolfson and
  Bressler}]{DBLP:conf/isbi/BurlinaFJWB16}
\bibinfo{author}{Burlina, P.}, \bibinfo{author}{Freund, D.E.},
  \bibinfo{author}{Joshi, N.}, \bibinfo{author}{Wolfson, Y.},
  \bibinfo{author}{Bressler, N.M.}, \bibinfo{year}{2016}.
\newblock \bibinfo{title}{Detection of age-related macular degeneration via
  deep learning}, in: \bibinfo{booktitle}{13th {IEEE} International Symposium
  on Biomedical Imaging, {ISBI} 2016, Prague, Czech Republic, April 13-16,
  2016}, \bibinfo{publisher}{{IEEE}}. pp. \bibinfo{pages}{184--188}.
\newblock \URLprefix \url{https://doi.org/10.1109/ISBI.2016.7493240},
  \DOIprefix\doi{10.1109/ISBI.2016.7493240}.
\bibitem[{Burlina et~al.(2018)Burlina, Joshi, Pacheco, Freund, Kong and
  Bressler}]{10.1001/jamaophthalmol.2018.4118}
\bibinfo{author}{Burlina, P.M.}, \bibinfo{author}{Joshi, N.},
  \bibinfo{author}{Pacheco, K.D.}, \bibinfo{author}{Freund, D.E.},
  \bibinfo{author}{Kong, J.}, \bibinfo{author}{Bressler, N.M.},
  \bibinfo{year}{2018}.
\newblock \bibinfo{title}{{Use of Deep Learning for Detailed Severity
  Characterization and Estimation of 5-Year Risk Among Patients With
  Age-Related Macular Degeneration}}.
\newblock \bibinfo{journal}{JAMA Ophthalmology} \bibinfo{volume}{136},
  \bibinfo{pages}{1359--1366}.
\newblock \URLprefix \url{https://doi.org/10.1001/jamaophthalmol.2018.4118},
  \DOIprefix\doi{10.1001/jamaophthalmol.2018.4118}.
\bibitem[{Burlina et~al.(2019)Burlina, Joshi, Pacheco, Liu and
  Bressler}]{10.1001/jamaophthalmol.2018.6156}
\bibinfo{author}{Burlina, P.M.}, \bibinfo{author}{Joshi, N.},
  \bibinfo{author}{Pacheco, K.D.}, \bibinfo{author}{Liu, T.Y.A.},
  \bibinfo{author}{Bressler, N.M.}, \bibinfo{year}{2019}.
\newblock \bibinfo{title}{{Assessment of Deep Generative Models for
  High-Resolution Synthetic Retinal Image Generation of Age-Related Macular
  Degeneration}}.
\newblock \bibinfo{journal}{JAMA Ophthalmology} \bibinfo{volume}{137},
  \bibinfo{pages}{258--264}.
\newblock \URLprefix \url{https://doi.org/10.1001/jamaophthalmol.2018.6156},
  \DOIprefix\doi{10.1001/jamaophthalmol.2018.6156}.
\bibitem[{Burlina et~al.(2017)Burlina, Joshi, Pekala, Pacheco, Freund and
  Bressler}]{10.1001/jamaophthalmol.2017.3782}
\bibinfo{author}{Burlina, P.M.}, \bibinfo{author}{Joshi, N.},
  \bibinfo{author}{Pekala, M.}, \bibinfo{author}{Pacheco, K.D.},
  \bibinfo{author}{Freund, D.E.}, \bibinfo{author}{Bressler, N.M.},
  \bibinfo{year}{2017}.
\newblock \bibinfo{title}{{Automated Grading of Age-Related Macular
  Degeneration From Color Fundus Images Using Deep Convolutional Neural
  Networks}}.
\newblock \bibinfo{journal}{JAMA Ophthalmology} \bibinfo{volume}{135},
  \bibinfo{pages}{1170--1176}.
\newblock \URLprefix \url{https://doi.org/10.1001/jamaophthalmol.2017.3782},
  \DOIprefix\doi{10.1001/jamaophthalmol.2017.3782}.
\bibitem[{{California Healthcare Foundation}(2015)}]{EyePacs}
\bibinfo{author}{{California Healthcare Foundation}}, \bibinfo{year}{2015}.
\newblock \bibinfo{title}{Diabetic retinopathy detection - identify signs of
  diabetic retinopathy in eye images}.
\newblock
  \bibinfo{howpublished}{\url{https://www.kaggle.com/c/diabetic-retinopathy-detection/overview}}.
\bibitem[{Carmona et~al.(2008)Carmona, Rinc{\'{o}}n,
  Garc{\'{\i}}a{-}Feijo{\'{o}} and
  Mart{\'{\i}}nez{-}de{-}la{-}Casa}]{DBLP:journals/artmed/CarmonaRGM08}
\bibinfo{author}{Carmona, E.J.}, \bibinfo{author}{Rinc{\'{o}}n, M.},
  \bibinfo{author}{Garc{\'{\i}}a{-}Feijo{\'{o}}, J.},
  \bibinfo{author}{Mart{\'{\i}}nez{-}de{-}la{-}Casa, J.M.},
  \bibinfo{year}{2008}.
\newblock \bibinfo{title}{Identification of the optic nerve head with genetic
  algorithms}.
\newblock \bibinfo{journal}{Artif. Intell. Medicine} \bibinfo{volume}{43},
  \bibinfo{pages}{243--259}.
\newblock \URLprefix \url{https://doi.org/10.1016/j.artmed.2008.04.005},
  \DOIprefix\doi{10.1016/j.artmed.2008.04.005}.
\bibitem[{Carson et~al.(2018)Carson, Caroline, Laura and Daniel}]{2018Retinal}
\bibinfo{author}{Carson, L.}, \bibinfo{author}{Caroline, Y.},
  \bibinfo{author}{Laura, H.}, \bibinfo{author}{Daniel, R.},
  \bibinfo{year}{2018}.
\newblock \bibinfo{title}{Retinal lesion detection with deep learning using
  image patches}.
\newblock \bibinfo{journal}{Investigative Ophthalmology \& Visual Science}
  \bibinfo{volume}{59}, \bibinfo{pages}{590--596}.
\bibitem[{Chai et~al.(2018)Chai, Liu and Xu}]{DBLP:journals/kbs/ChaiLX18}
\bibinfo{author}{Chai, Y.}, \bibinfo{author}{Liu, H.}, \bibinfo{author}{Xu,
  J.}, \bibinfo{year}{2018}.
\newblock \bibinfo{title}{Glaucoma diagnosis based on both hidden features and
  domain knowledge through deep learning models}.
\newblock \bibinfo{journal}{Knowl. Based Syst.} \bibinfo{volume}{161},
  \bibinfo{pages}{147--156}.
\newblock \URLprefix \url{https://doi.org/10.1016/j.knosys.2018.07.043},
  \DOIprefix\doi{10.1016/j.knosys.2018.07.043}.
\bibitem[{Chen et~al.(2017)Chen, Papandreou, Schroff and
  Adam}]{DBLP:journals/corr/ChenPSA17}
\bibinfo{author}{Chen, L.}, \bibinfo{author}{Papandreou, G.},
  \bibinfo{author}{Schroff, F.}, \bibinfo{author}{Adam, H.},
  \bibinfo{year}{2017}.
\newblock \bibinfo{title}{Rethinking atrous convolution for semantic image
  segmentation}.
\newblock \bibinfo{journal}{CoRR} \bibinfo{volume}{abs/1706.05587}.
\newblock \URLprefix \url{http://arxiv.org/abs/1706.05587}.
\bibitem[{Chen et~al.(2018)Chen, Zhu, Papandreou, Schroff and
  Adam}]{DBLP:conf/eccv/ChenZPSA18}
\bibinfo{author}{Chen, L.}, \bibinfo{author}{Zhu, Y.},
  \bibinfo{author}{Papandreou, G.}, \bibinfo{author}{Schroff, F.},
  \bibinfo{author}{Adam, H.}, \bibinfo{year}{2018}.
\newblock \bibinfo{title}{Encoder-decoder with atrous separable convolution for
  semantic image segmentation}, in: \bibinfo{editor}{Ferrari, V.},
  \bibinfo{editor}{Hebert, M.}, \bibinfo{editor}{Sminchisescu, C.},
  \bibinfo{editor}{Weiss, Y.} (Eds.), \bibinfo{booktitle}{Computer Vision -
  {ECCV} 2018 - 15th European Conference, Munich, Germany, September 8-14,
  2018, Proceedings, Part {VII}}, \bibinfo{publisher}{Springer}. pp.
  \bibinfo{pages}{833--851}.
\newblock \URLprefix \url{https://doi.org/10.1007/978-3-030-01234-2\_49},
  \DOIprefix\doi{10.1007/978-3-030-01234-2\_49}.
\bibitem[{Chen and Guestrin(2016)}]{DBLP:conf/kdd/ChenG16}
\bibinfo{author}{Chen, T.}, \bibinfo{author}{Guestrin, C.},
  \bibinfo{year}{2016}.
\newblock \bibinfo{title}{Xgboost: {A} scalable tree boosting system}, in:
  \bibinfo{editor}{Krishnapuram, B.}, \bibinfo{editor}{Shah, M.},
  \bibinfo{editor}{Smola, A.J.}, \bibinfo{editor}{Aggarwal, C.C.},
  \bibinfo{editor}{Shen, D.}, \bibinfo{editor}{Rastogi, R.} (Eds.),
  \bibinfo{booktitle}{Proceedings of the 22nd {ACM} {SIGKDD} International
  Conference on Knowledge Discovery and Data Mining, San Francisco, CA, USA,
  August 13-17, 2016}, \bibinfo{publisher}{{ACM}}. pp.
  \bibinfo{pages}{785--794}.
\newblock \URLprefix \url{https://doi.org/10.1145/2939672.2939785},
  \DOIprefix\doi{10.1145/2939672.2939785}.
\bibitem[{Cherukuri et~al.(2020)Cherukuri, G, Bala and
  Monga}]{DBLP:journals/tip/CherukuriGBM20}
\bibinfo{author}{Cherukuri, V.}, \bibinfo{author}{G, V.K.B.},
  \bibinfo{author}{Bala, R.}, \bibinfo{author}{Monga, V.},
  \bibinfo{year}{2020}.
\newblock \bibinfo{title}{Deep retinal image segmentation with regularization
  under geometric priors}.
\newblock \bibinfo{journal}{{IEEE} Trans. Image Process.} \bibinfo{volume}{29},
  \bibinfo{pages}{2552--2567}.
\newblock \URLprefix \url{https://doi.org/10.1109/TIP.2019.2946078},
  \DOIprefix\doi{10.1109/TIP.2019.2946078}.
\bibitem[{Chung et~al.(2014)Chung, G{\"{u}}l{\c{c}}ehre, Cho and
  Bengio}]{DBLP:journals/corr/ChungGCB14}
\bibinfo{author}{Chung, J.}, \bibinfo{author}{G{\"{u}}l{\c{c}}ehre, {\c{C}}.},
  \bibinfo{author}{Cho, K.}, \bibinfo{author}{Bengio, Y.},
  \bibinfo{year}{2014}.
\newblock \bibinfo{title}{Empirical evaluation of gated recurrent neural
  networks on sequence modeling}.
\newblock \bibinfo{journal}{CoRR} \bibinfo{volume}{abs/1412.3555}.
\newblock \URLprefix \url{http://arxiv.org/abs/1412.3555}.
\bibitem[{Ciulla et~al.(2003)Ciulla, Amador and Zinman}]{Ciulla2003Diabetic}
\bibinfo{author}{Ciulla, T.A.}, \bibinfo{author}{Amador, A.G.},
  \bibinfo{author}{Zinman, B.}, \bibinfo{year}{2003}.
\newblock \bibinfo{title}{Diabetic retinopathy and diabetic macular edema:
  Pathophysiology, screening, and novel therapies}.
\newblock \bibinfo{journal}{Diabetes Care} \bibinfo{volume}{26},
  \bibinfo{pages}{2653--2664}.
\bibitem[{Costa et~al.(2018)Costa, Galdran, Meyer, Niemeijer, Abramoff,
  Mendon{\c{c}}a and Campilho}]{DBLP:journals/tmi/CostaGMNAMC18}
\bibinfo{author}{Costa, P.}, \bibinfo{author}{Galdran, A.},
  \bibinfo{author}{Meyer, M.I.}, \bibinfo{author}{Niemeijer, M.},
  \bibinfo{author}{Abramoff, M.}, \bibinfo{author}{Mendon{\c{c}}a, A.M.},
  \bibinfo{author}{Campilho, A.J.C.}, \bibinfo{year}{2018}.
\newblock \bibinfo{title}{End-to-end adversarial retinal image synthesis}.
\newblock \bibinfo{journal}{{IEEE} Trans. Medical Imaging}
  \bibinfo{volume}{37}, \bibinfo{pages}{781--791}.
\newblock \URLprefix \url{https://doi.org/10.1109/TMI.2017.2759102},
  \DOIprefix\doi{10.1109/TMI.2017.2759102}.
\bibitem[{Dai et~al.(2018)Dai, Fang, Li, Hou, Sheng, Wu and
  Jia}]{DBLP:journals/tmi/DaiFLHSWJ18}
\bibinfo{author}{Dai, L.}, \bibinfo{author}{Fang, R.}, \bibinfo{author}{Li,
  H.}, \bibinfo{author}{Hou, X.}, \bibinfo{author}{Sheng, B.},
  \bibinfo{author}{Wu, Q.}, \bibinfo{author}{Jia, W.}, \bibinfo{year}{2018}.
\newblock \bibinfo{title}{Clinical report guided retinal microaneurysm
  detection with multi-sieving deep learning}.
\newblock \bibinfo{journal}{{IEEE} Trans. Med. Imaging} \bibinfo{volume}{37},
  \bibinfo{pages}{1149--1161}.
\newblock \URLprefix \url{https://doi.org/10.1109/TMI.2018.2794988},
  \DOIprefix\doi{10.1109/TMI.2018.2794988}.
\bibitem[{Dasgupta and Singh(2017)}]{DBLP:conf/isbi/DasguptaS17}
\bibinfo{author}{Dasgupta, A.}, \bibinfo{author}{Singh, S.},
  \bibinfo{year}{2017}.
\newblock \bibinfo{title}{A fully convolutional neural network based structured
  prediction approach towards the retinal vessel segmentation}, in:
  \bibinfo{booktitle}{14th {IEEE} International Symposium on Biomedical
  Imaging, {ISBI} 2017, Melbourne, Australia, April 18-21, 2017},
  \bibinfo{publisher}{{IEEE}}. pp. \bibinfo{pages}{248--251}.
\newblock \URLprefix \url{https://doi.org/10.1109/ISBI.2017.7950512},
  \DOIprefix\doi{10.1109/ISBI.2017.7950512}.
\bibitem[{Dashtbozorg et~al.(2014)Dashtbozorg, Mendon{\c{c}}a and
  Campilho}]{DBLP:journals/tip/DashtbozorgMC14}
\bibinfo{author}{Dashtbozorg, B.}, \bibinfo{author}{Mendon{\c{c}}a, A.M.},
  \bibinfo{author}{Campilho, A.J.C.}, \bibinfo{year}{2014}.
\newblock \bibinfo{title}{An automatic graph-based approach for artery/vein
  classification in retinal images}.
\newblock \bibinfo{journal}{{IEEE} Trans. Image Process.} \bibinfo{volume}{23},
  \bibinfo{pages}{1073--1083}.
\newblock \URLprefix \url{https://doi.org/10.1109/TIP.2013.2263809},
  \DOIprefix\doi{10.1109/TIP.2013.2263809}.
\bibitem[{Dashtbozorg et~al.(2018)Dashtbozorg, Zhang, Huang and ter
  Haar~Romeny}]{DBLP:journals/tip/DashtbozorgZHR18}
\bibinfo{author}{Dashtbozorg, B.}, \bibinfo{author}{Zhang, J.},
  \bibinfo{author}{Huang, F.}, \bibinfo{author}{ter Haar~Romeny, B.M.},
  \bibinfo{year}{2018}.
\newblock \bibinfo{title}{Retinal microaneurysms detection using local
  convergence index features}.
\newblock \bibinfo{journal}{{IEEE} Trans. Image Process.} \bibinfo{volume}{27},
  \bibinfo{pages}{3300--3315}.
\newblock \URLprefix \url{https://doi.org/10.1109/TIP.2018.2815345},
  \DOIprefix\doi{10.1109/TIP.2018.2815345}.
\bibitem[{David et~al.(2016)David, Lou, Ali, Warren, Ryan, Folk and
  Meindert}]{David2016Improved}
\bibinfo{author}{David, A.M.}, \bibinfo{author}{Lou, Y.}, \bibinfo{author}{Ali,
  E.}, \bibinfo{author}{Warren, C.}, \bibinfo{author}{Ryan, A.},
  \bibinfo{author}{Folk, J.C.}, \bibinfo{author}{Meindert, N.},
  \bibinfo{year}{2016}.
\newblock \bibinfo{title}{Improved automated detection of diabetic retinopathy
  on a publicly available dataset through integration of deep learning}.
\newblock \bibinfo{journal}{Investigative Ophthalmology \& Visual Science}
  \bibinfo{volume}{57}, \bibinfo{pages}{5200--}.
\bibitem[{Decenciere et~al.(2013)Decenciere, Cazuguel, Zhang, Thibault, Klein,
  Meyer, Marcotegui, Quellec, Lamard, Danno et~al.}]{decenciere2013teleophta:}
\bibinfo{author}{Decenciere, E.}, \bibinfo{author}{Cazuguel, G.},
  \bibinfo{author}{Zhang, X.}, \bibinfo{author}{Thibault, G.},
  \bibinfo{author}{Klein, J.C.}, \bibinfo{author}{Meyer, F.},
  \bibinfo{author}{Marcotegui, B.}, \bibinfo{author}{Quellec, G.},
  \bibinfo{author}{Lamard, M.}, \bibinfo{author}{Danno, R.}, et~al.,
  \bibinfo{year}{2013}.
\newblock \bibinfo{title}{Teleophta: Machine learning and image processing
  methods for teleophthalmology}.
\newblock \bibinfo{journal}{Irbm} \bibinfo{volume}{34},
  \bibinfo{pages}{196--203}.
\bibitem[{Decenciere et~al.(2014)Decenciere, Zhang, Cazuguel, Lay, Cochener,
  Trone, Gain, Ordonezvarela, Massin, Erginay et~al.}]{decenciere2014feedback}
\bibinfo{author}{Decenciere, E.}, \bibinfo{author}{Zhang, X.},
  \bibinfo{author}{Cazuguel, G.}, \bibinfo{author}{Lay, B.},
  \bibinfo{author}{Cochener, B.}, \bibinfo{author}{Trone, C.},
  \bibinfo{author}{Gain, P.}, \bibinfo{author}{Ordonezvarela, J.},
  \bibinfo{author}{Massin, P.}, \bibinfo{author}{Erginay, A.}, et~al.,
  \bibinfo{year}{2014}.
\newblock \bibinfo{title}{Feedback on a publicly distributed image database:
  The messidor database}.
\newblock \bibinfo{journal}{Image Analysis \& Stereology} \bibinfo{volume}{33},
  \bibinfo{pages}{231--234}.
\bibitem[{Deng et~al.(2009)Deng, Dong, Socher, Li, Li and
  Li}]{DBLP:conf/cvpr/DengDSLL009}
\bibinfo{author}{Deng, J.}, \bibinfo{author}{Dong, W.},
  \bibinfo{author}{Socher, R.}, \bibinfo{author}{Li, L.}, \bibinfo{author}{Li,
  K.}, \bibinfo{author}{Li, F.}, \bibinfo{year}{2009}.
\newblock \bibinfo{title}{Imagenet: {A} large-scale hierarchical image
  database}, in: \bibinfo{booktitle}{2009 {IEEE} Computer Society Conference on
  Computer Vision and Pattern Recognition {(CVPR} 2009), 20-25 June 2009,
  Miami, Florida, {USA}}, \bibinfo{publisher}{{IEEE} Computer Society}. pp.
  \bibinfo{pages}{248--255}.
\newblock \URLprefix \url{https://doi.org/10.1109/CVPR.2009.5206848},
  \DOIprefix\doi{10.1109/CVPR.2009.5206848}.
\bibitem[{Deshmukh and Sivaswamy(2019)}]{DBLP:conf/isbi/DeshmukhS19}
\bibinfo{author}{Deshmukh, A.}, \bibinfo{author}{Sivaswamy, J.},
  \bibinfo{year}{2019}.
\newblock \bibinfo{title}{Synthesis of optical nerve head region of fundus
  image}, in: \bibinfo{booktitle}{16th {IEEE} International Symposium on
  Biomedical Imaging, {ISBI} 2019, Venice, Italy, April 8-11, 2019},
  \bibinfo{publisher}{{IEEE}}. pp. \bibinfo{pages}{583--586}.
\newblock \URLprefix \url{https://doi.org/10.1109/ISBI.2019.8759414},
  \DOIprefix\doi{10.1109/ISBI.2019.8759414}.
\bibitem[{Diaz{-}Pinto et~al.(2019)Diaz{-}Pinto, Colomer, Naranjo, Morales, Xu
  and Frangi}]{DBLP:journals/tmi/Diaz-PintoCNMXF19}
\bibinfo{author}{Diaz{-}Pinto, A.}, \bibinfo{author}{Colomer, A.},
  \bibinfo{author}{Naranjo, V.}, \bibinfo{author}{Morales, S.},
  \bibinfo{author}{Xu, Y.}, \bibinfo{author}{Frangi, A.F.},
  \bibinfo{year}{2019}.
\newblock \bibinfo{title}{Retinal image synthesis and semi-supervised learning
  for glaucoma assessment}.
\newblock \bibinfo{journal}{{IEEE} Trans. Med. Imaging} \bibinfo{volume}{38},
  \bibinfo{pages}{2211--2218}.
\newblock \URLprefix \url{https://doi.org/10.1109/TMI.2019.2903434},
  \DOIprefix\doi{10.1109/TMI.2019.2903434}.
\bibitem[{Edupuganti et~al.(2018)Edupuganti, Chawla and
  Kale}]{DBLP:conf/icip/EdupugantiCK18}
\bibinfo{author}{Edupuganti, V.G.}, \bibinfo{author}{Chawla, A.},
  \bibinfo{author}{Kale, A.}, \bibinfo{year}{2018}.
\newblock \bibinfo{title}{Automatic optic disk and cup segmentation of fundus
  images using deep learning}, in: \bibinfo{booktitle}{2018 {IEEE}
  International Conference on Image Processing, {ICIP} 2018, Athens, Greece,
  October 7-10, 2018}, \bibinfo{publisher}{{IEEE}}. pp.
  \bibinfo{pages}{2227--2231}.
\newblock \URLprefix \url{https://doi.org/10.1109/ICIP.2018.8451753},
  \DOIprefix\doi{10.1109/ICIP.2018.8451753}.
\bibitem[{Feng et~al.(2020)Feng, Zhuo, Pan and
  Tian}]{DBLP:journals/ijon/FengZPT20}
\bibinfo{author}{Feng, S.}, \bibinfo{author}{Zhuo, Z.}, \bibinfo{author}{Pan,
  D.}, \bibinfo{author}{Tian, Q.}, \bibinfo{year}{2020}.
\newblock \bibinfo{title}{Ccnet: {A} cross-connected convolutional network for
  segmenting retinal vessels using multi-scale features}.
\newblock \bibinfo{journal}{Neurocomputing} \bibinfo{volume}{392},
  \bibinfo{pages}{268--276}.
\newblock \URLprefix \url{https://doi.org/10.1016/j.neucom.2018.10.098},
  \DOIprefix\doi{10.1016/j.neucom.2018.10.098}.
\bibitem[{Feng et~al.(2017)Feng, Yang and Yao}]{DBLP:conf/icip/Feng0Y17}
\bibinfo{author}{Feng, Z.}, \bibinfo{author}{Yang, J.}, \bibinfo{author}{Yao,
  L.}, \bibinfo{year}{2017}.
\newblock \bibinfo{title}{Patch-based fully convolutional neural network with
  skip connections for retinal blood vessel segmentation}, in:
  \bibinfo{booktitle}{2017 {IEEE} International Conference on Image Processing,
  {ICIP} 2017, Beijing, China, September 17-20, 2017},
  \bibinfo{publisher}{{IEEE}}. pp. \bibinfo{pages}{1742--1746}.
\newblock \URLprefix \url{https://doi.org/10.1109/ICIP.2017.8296580},
  \DOIprefix\doi{10.1109/ICIP.2017.8296580}.
\bibitem[{Foo et~al.(2020)Foo, Hsu, Lee, Lim and
  Wong}]{DBLP:conf/aaai/FooHLLW20}
\bibinfo{author}{Foo, A.}, \bibinfo{author}{Hsu, W.}, \bibinfo{author}{Lee,
  M.}, \bibinfo{author}{Lim, G.}, \bibinfo{author}{Wong, T.Y.},
  \bibinfo{year}{2020}.
\newblock \bibinfo{title}{Multi-task learning for diabetic retinopathy grading
  and lesion segmentation}, in: \bibinfo{booktitle}{The Thirty-Fourth {AAAI}
  Conference on Artificial Intelligence, {AAAI} 2020, The Thirty-Second
  Innovative Applications of Artificial Intelligence Conference, {IAAI} 2020,
  The Tenth {AAAI} Symposium on Educational Advances in Artificial
  Intelligence, {EAAI} 2020, New York, NY, USA, February 7-12, 2020},
  \bibinfo{publisher}{{AAAI} Press}. pp. \bibinfo{pages}{13267--13272}.
\newblock \URLprefix
  \url{https://aaai.org/ojs/index.php/AAAI/article/view/7035}.
\bibitem[{Frangi et~al.(1998)Frangi, Niessen, Vincken and
  Viergever}]{10.1007/BFb0056195}
\bibinfo{author}{Frangi, A.F.}, \bibinfo{author}{Niessen, W.J.},
  \bibinfo{author}{Vincken, K.L.}, \bibinfo{author}{Viergever, M.A.},
  \bibinfo{year}{1998}.
\newblock \bibinfo{title}{Multiscale vessel enhancement filtering}, in:
  \bibinfo{editor}{Wells, W.M.}, \bibinfo{editor}{Colchester, A.},
  \bibinfo{editor}{Delp, S.} (Eds.), \bibinfo{booktitle}{Medical Image
  Computing and Computer-Assisted Intervention --- MICCAI'98},
  \bibinfo{publisher}{Springer Berlin Heidelberg}, \bibinfo{address}{Berlin,
  Heidelberg}. pp. \bibinfo{pages}{130--137}.
\bibitem[{Fu et~al.(2018a)Fu, Cheng, Xu, Wong, Liu and
  Cao}]{DBLP:journals/tmi/FuCXWLC18}
\bibinfo{author}{Fu, H.}, \bibinfo{author}{Cheng, J.}, \bibinfo{author}{Xu,
  Y.}, \bibinfo{author}{Wong, D.W.K.}, \bibinfo{author}{Liu, J.},
  \bibinfo{author}{Cao, X.}, \bibinfo{year}{2018}a.
\newblock \bibinfo{title}{Joint optic disc and cup segmentation based on
  multi-label deep network and polar transformation}.
\newblock \bibinfo{journal}{{IEEE} Trans. Med. Imaging} \bibinfo{volume}{37},
  \bibinfo{pages}{1597--1605}.
\newblock \URLprefix \url{https://doi.org/10.1109/TMI.2018.2791488},
  \DOIprefix\doi{10.1109/TMI.2018.2791488}.
\bibitem[{Fu et~al.(2018b)Fu, Cheng, Xu, Zhang, Wong, Liu and
  Cao}]{DBLP:journals/tmi/FuCXZWLC18}
\bibinfo{author}{Fu, H.}, \bibinfo{author}{Cheng, J.}, \bibinfo{author}{Xu,
  Y.}, \bibinfo{author}{Zhang, C.}, \bibinfo{author}{Wong, D.W.K.},
  \bibinfo{author}{Liu, J.}, \bibinfo{author}{Cao, X.}, \bibinfo{year}{2018}b.
\newblock \bibinfo{title}{Disc-aware ensemble network for glaucoma screening
  from fundus image}.
\newblock \bibinfo{journal}{{IEEE} Trans. Med. Imaging} \bibinfo{volume}{37},
  \bibinfo{pages}{2493--2501}.
\newblock \URLprefix \url{https://doi.org/10.1109/TMI.2018.2837012},
  \DOIprefix\doi{10.1109/TMI.2018.2837012}.
\bibitem[{Fu et~al.(2019)Fu, Wang, Shen, Cui, Xu, Liu and
  Shao}]{DBLP:conf/miccai/FuWSCX0S19}
\bibinfo{author}{Fu, H.}, \bibinfo{author}{Wang, B.}, \bibinfo{author}{Shen,
  J.}, \bibinfo{author}{Cui, S.}, \bibinfo{author}{Xu, Y.},
  \bibinfo{author}{Liu, J.}, \bibinfo{author}{Shao, L.}, \bibinfo{year}{2019}.
\newblock \bibinfo{title}{Evaluation of retinal image quality assessment
  networks in different color-spaces}, in: \bibinfo{editor}{Shen, D.},
  \bibinfo{editor}{Liu, T.}, \bibinfo{editor}{Peters, T.M.},
  \bibinfo{editor}{Staib, L.H.}, \bibinfo{editor}{Essert, C.},
  \bibinfo{editor}{Zhou, S.}, \bibinfo{editor}{Yap, P.}, \bibinfo{editor}{Khan,
  A.} (Eds.), \bibinfo{booktitle}{Medical Image Computing and Computer Assisted
  Intervention - {MICCAI} 2019 - 22nd International Conference, Shenzhen,
  China, October 13-17, 2019, Proceedings, Part {I}},
  \bibinfo{publisher}{Springer}. pp. \bibinfo{pages}{48--56}.
\newblock \URLprefix \url{https://doi.org/10.1007/978-3-030-32239-7\_6},
  \DOIprefix\doi{10.1007/978-3-030-32239-7\_6}.
\bibitem[{Fu et~al.(2016)Fu, Xu, Lin, Wong and Liu}]{DBLP:conf/miccai/FuXLW016}
\bibinfo{author}{Fu, H.}, \bibinfo{author}{Xu, Y.}, \bibinfo{author}{Lin, S.},
  \bibinfo{author}{Wong, D.W.K.}, \bibinfo{author}{Liu, J.},
  \bibinfo{year}{2016}.
\newblock \bibinfo{title}{Deepvessel: Retinal vessel segmentation via deep
  learning and conditional random field}, in: \bibinfo{editor}{Ourselin, S.},
  \bibinfo{editor}{Joskowicz, L.}, \bibinfo{editor}{Sabuncu, M.R.},
  \bibinfo{editor}{{\"{U}}nal, G.B.}, \bibinfo{editor}{Wells, W.} (Eds.),
  \bibinfo{booktitle}{Medical Image Computing and Computer-Assisted
  Intervention - {MICCAI} 2016 - 19th International Conference, Athens, Greece,
  October 17-21, 2016, Proceedings, Part {II}}, pp. \bibinfo{pages}{132--139}.
\newblock \URLprefix \url{https://doi.org/10.1007/978-3-319-46723-8\_16},
  \DOIprefix\doi{10.1007/978-3-319-46723-8\_16}.
\bibitem[{Fumero et~al.(2011)Fumero, Alay{\'{o}}n, S{\'{a}}nchez, Sigut and
  Gonz{\'{a}}lez{-}Hern{\'{a}}ndez}]{DBLP:conf/cbms/FumeroASSG11}
\bibinfo{author}{Fumero, F.}, \bibinfo{author}{Alay{\'{o}}n, S.},
  \bibinfo{author}{S{\'{a}}nchez, J.L.}, \bibinfo{author}{Sigut, J.F.},
  \bibinfo{author}{Gonz{\'{a}}lez{-}Hern{\'{a}}ndez, M.}, \bibinfo{year}{2011}.
\newblock \bibinfo{title}{{RIM-ONE:} an open retinal image database for optic
  nerve evaluation}, in: \bibinfo{booktitle}{Proceedings of the 24th {IEEE}
  International Symposium on Computer-Based Medical Systems, 27-30 June, 2011,
  Bristol, United Kingdom}, \bibinfo{publisher}{{IEEE} Computer Society}. pp.
  \bibinfo{pages}{1--6}.
\newblock \URLprefix \url{https://doi.org/10.1109/CBMS.2011.5999143},
  \DOIprefix\doi{10.1109/CBMS.2011.5999143}.
\bibitem[{Galdran et~al.(2019)Galdran, Meyer, Costa, Mendon{\c{c}}a and
  Campilho}]{DBLP:conf/isbi/GaldranM0MC19}
\bibinfo{author}{Galdran, A.}, \bibinfo{author}{Meyer, M.I.},
  \bibinfo{author}{Costa, P.}, \bibinfo{author}{Mendon{\c{c}}a, A.M.},
  \bibinfo{author}{Campilho, A.}, \bibinfo{year}{2019}.
\newblock \bibinfo{title}{Uncertainty-aware artery/vein classification on
  retinal images}, in: \bibinfo{booktitle}{16th {IEEE} International Symposium
  on Biomedical Imaging, {ISBI} 2019, Venice, Italy, April 8-11, 2019},
  \bibinfo{publisher}{{IEEE}}. pp. \bibinfo{pages}{556--560}.
\newblock \URLprefix \url{https://doi.org/10.1109/ISBI.2019.8759380},
  \DOIprefix\doi{10.1109/ISBI.2019.8759380}.
\bibitem[{Gargeya and Leng(2017)}]{Gargeya2017Automated}
\bibinfo{author}{Gargeya, R.}, \bibinfo{author}{Leng, T.},
  \bibinfo{year}{2017}.
\newblock \bibinfo{title}{Automated identification of diabetic retinopathy
  using deep learning}.
\newblock \bibinfo{journal}{Ophthalmology} ,
  \bibinfo{pages}{S0161642016317742}.
\bibitem[{Garway-Heath and Hitchings(1998)}]{Garway1998Quantitative}
\bibinfo{author}{Garway-Heath, D.F.}, \bibinfo{author}{Hitchings, R.},
  \bibinfo{year}{1998}.
\newblock \bibinfo{title}{Quantitative evaluation of the optic nerve head in
  early glaucoma}.
\newblock \bibinfo{journal}{British Journal of Ophthalmology}
  \bibinfo{volume}{82}, \bibinfo{pages}{352--361}.
\bibitem[{Ghafoorian et~al.(2017)Ghafoorian, Mehrtash, Kapur, Karssemeijer,
  Marchiori, Pesteie, Guttmann, de~Leeuw, Tempany, van Ginneken, Fedorov,
  Abolmaesumi, Platel and III}]{DBLP:conf/miccai/GhafoorianMKKMP17}
\bibinfo{author}{Ghafoorian, M.}, \bibinfo{author}{Mehrtash, A.},
  \bibinfo{author}{Kapur, T.}, \bibinfo{author}{Karssemeijer, N.},
  \bibinfo{author}{Marchiori, E.}, \bibinfo{author}{Pesteie, M.},
  \bibinfo{author}{Guttmann, C.R.G.}, \bibinfo{author}{de~Leeuw, F.},
  \bibinfo{author}{Tempany, C.M.}, \bibinfo{author}{van Ginneken, B.},
  \bibinfo{author}{Fedorov, A.}, \bibinfo{author}{Abolmaesumi, P.},
  \bibinfo{author}{Platel, B.}, \bibinfo{author}{III, W.M.W.},
  \bibinfo{year}{2017}.
\newblock \bibinfo{title}{Transfer learning for domain adaptation in {MRI:}
  application in brain lesion segmentation}, in: \bibinfo{editor}{Descoteaux,
  M.}, \bibinfo{editor}{Maier{-}Hein, L.}, \bibinfo{editor}{Franz, A.M.},
  \bibinfo{editor}{Jannin, P.}, \bibinfo{editor}{Collins, D.L.},
  \bibinfo{editor}{Duchesne, S.} (Eds.), \bibinfo{booktitle}{Medical Image
  Computing and Computer Assisted Intervention - {MICCAI} 2017 - 20th
  International Conference, Quebec City, QC, Canada, September 11-13, 2017,
  Proceedings, Part {III}}, \bibinfo{publisher}{Springer}. pp.
  \bibinfo{pages}{516--524}.
\newblock \URLprefix \url{https://doi.org/10.1007/978-3-319-66179-7\_59},
  \DOIprefix\doi{10.1007/978-3-319-66179-7\_59}.
\bibitem[{Giancardo et~al.(2012)Giancardo, Meriaudeau, Karnowski, Li, Garg,
  Tobin and Chaum}]{GIANCARDO2012216}
\bibinfo{author}{Giancardo, L.}, \bibinfo{author}{Meriaudeau, F.},
  \bibinfo{author}{Karnowski, T.P.}, \bibinfo{author}{Li, Y.},
  \bibinfo{author}{Garg, S.}, \bibinfo{author}{Tobin, K.W.},
  \bibinfo{author}{Chaum, E.}, \bibinfo{year}{2012}.
\newblock \bibinfo{title}{Exudate-based diabetic macular edema detection in
  fundus images using publicly available datasets}.
\newblock \bibinfo{journal}{Medical Image Analysis} \bibinfo{volume}{16},
  \bibinfo{pages}{216 -- 226}.
\newblock \URLprefix
  \url{http://www.sciencedirect.com/science/article/pii/S1361841511001010},
  \DOIprefix\doi{https://doi.org/10.1016/j.media.2011.07.004}.
\bibitem[{Gondal et~al.(2017)Gondal, K{\"{o}}hler, Grzeszick, Fink and
  Hirsch}]{DBLP:conf/icip/GondalKGFH17}
\bibinfo{author}{Gondal, W.M.}, \bibinfo{author}{K{\"{o}}hler, J.M.},
  \bibinfo{author}{Grzeszick, R.}, \bibinfo{author}{Fink, G.A.},
  \bibinfo{author}{Hirsch, M.}, \bibinfo{year}{2017}.
\newblock \bibinfo{title}{Weakly-supervised localization of diabetic
  retinopathy lesions in retinal fundus images}, in: \bibinfo{booktitle}{2017
  {IEEE} International Conference on Image Processing, {ICIP} 2017, Beijing,
  China, September 17-20, 2017}, \bibinfo{publisher}{{IEEE}}. pp.
  \bibinfo{pages}{2069--2073}.
\newblock \URLprefix \url{https://doi.org/10.1109/ICIP.2017.8296646},
  \DOIprefix\doi{10.1109/ICIP.2017.8296646}.
\bibitem[{Goodfellow et~al.(2014)Goodfellow, Pouget{-}Abadie, Mirza, Xu,
  Warde{-}Farley, Ozair, Courville and
  Bengio}]{DBLP:conf/nips/GoodfellowPMXWOCB14}
\bibinfo{author}{Goodfellow, I.J.}, \bibinfo{author}{Pouget{-}Abadie, J.},
  \bibinfo{author}{Mirza, M.}, \bibinfo{author}{Xu, B.},
  \bibinfo{author}{Warde{-}Farley, D.}, \bibinfo{author}{Ozair, S.},
  \bibinfo{author}{Courville, A.C.}, \bibinfo{author}{Bengio, Y.},
  \bibinfo{year}{2014}.
\newblock \bibinfo{title}{Generative adversarial nets}, in:
  \bibinfo{editor}{Ghahramani, Z.}, \bibinfo{editor}{Welling, M.},
  \bibinfo{editor}{Cortes, C.}, \bibinfo{editor}{Lawrence, N.D.},
  \bibinfo{editor}{Weinberger, K.Q.} (Eds.), \bibinfo{booktitle}{Advances in
  Neural Information Processing Systems 27: Annual Conference on Neural
  Information Processing Systems 2014, December 8-13 2014, Montreal, Quebec,
  Canada}, pp. \bibinfo{pages}{2672--2680}.
\newblock \URLprefix
  \url{http://papers.nips.cc/paper/5423-generative-adversarial-nets}.
\bibitem[{Govindaiah et~al.(2018)Govindaiah, Hussain, Smith and
  Bhuiyan}]{DBLP:conf/isbi/GovindaiahHSB18}
\bibinfo{author}{Govindaiah, A.}, \bibinfo{author}{Hussain, M.A.},
  \bibinfo{author}{Smith, R.T.}, \bibinfo{author}{Bhuiyan, A.},
  \bibinfo{year}{2018}.
\newblock \bibinfo{title}{Deep convolutional neural network based screening and
  assessment of age-related macular degeneration from fundus images}, in:
  \bibinfo{booktitle}{15th {IEEE} International Symposium on Biomedical
  Imaging, {ISBI} 2018, Washington, DC, USA, April 4-7, 2018},
  \bibinfo{publisher}{{IEEE}}. pp. \bibinfo{pages}{1525--1528}.
\newblock \URLprefix \url{https://doi.org/10.1109/ISBI.2018.8363863},
  \DOIprefix\doi{10.1109/ISBI.2018.8363863}.
\bibitem[{Grassmann et~al.(2018)Grassmann, Mengelkamp, Brandl, Harsch and
  Weber}]{Grassmann2018A}
\bibinfo{author}{Grassmann, F.}, \bibinfo{author}{Mengelkamp, J.},
  \bibinfo{author}{Brandl, C.}, \bibinfo{author}{Harsch, S.},
  \bibinfo{author}{Weber, B.H.F.}, \bibinfo{year}{2018}.
\newblock \bibinfo{title}{A deep learning algorithm for prediction of
  age-related eye disease study severity scale for age-related macular
  degeneration from color fundus photography}.
\newblock \bibinfo{journal}{Ophthalmology} \bibinfo{volume}{125},
  \bibinfo{pages}{8280209}.
\bibitem[{van Grinsven et~al.(2016)van Grinsven, van Ginneken, Hoyng, Theelen
  and S{\'{a}}nchez}]{DBLP:journals/tmi/GrinsvenGHTS16}
\bibinfo{author}{van Grinsven, M.J.J.P.}, \bibinfo{author}{van Ginneken, B.},
  \bibinfo{author}{Hoyng, C.B.}, \bibinfo{author}{Theelen, T.},
  \bibinfo{author}{S{\'{a}}nchez, C.I.}, \bibinfo{year}{2016}.
\newblock \bibinfo{title}{Fast convolutional neural network training using
  selective data sampling: Application to hemorrhage detection in color fundus
  images}.
\newblock \bibinfo{journal}{{IEEE} Trans. Med. Imaging} \bibinfo{volume}{35},
  \bibinfo{pages}{1273--1284}.
\newblock \URLprefix \url{https://doi.org/10.1109/TMI.2016.2526689},
  \DOIprefix\doi{10.1109/TMI.2016.2526689}.
\bibitem[{Gulshan et~al.(2016)Gulshan, Peng, Coram, Stumpe, Wu, Narayanaswamy,
  Venugopalan, Widner, Madams, Cuadros, Kim, Raman, Nelson, Mega and
  Webster}]{10.1001/jama.2016.17216}
\bibinfo{author}{Gulshan, V.}, \bibinfo{author}{Peng, L.},
  \bibinfo{author}{Coram, M.}, \bibinfo{author}{Stumpe, M.C.},
  \bibinfo{author}{Wu, D.}, \bibinfo{author}{Narayanaswamy, A.},
  \bibinfo{author}{Venugopalan, S.}, \bibinfo{author}{Widner, K.},
  \bibinfo{author}{Madams, T.}, \bibinfo{author}{Cuadros, J.},
  \bibinfo{author}{Kim, R.}, \bibinfo{author}{Raman, R.},
  \bibinfo{author}{Nelson, P.C.}, \bibinfo{author}{Mega, J.L.},
  \bibinfo{author}{Webster, D.R.}, \bibinfo{year}{2016}.
\newblock \bibinfo{title}{{Development and Validation of a Deep Learning
  Algorithm for Detection of Diabetic Retinopathy in Retinal Fundus
  Photographs}}.
\newblock \bibinfo{journal}{JAMA} \bibinfo{volume}{316},
  \bibinfo{pages}{2402--2410}.
\newblock \URLprefix \url{https://doi.org/10.1001/jama.2016.17216},
  \DOIprefix\doi{10.1001/jama.2016.17216}.
\bibitem[{Gulshan et~al.(2019)Gulshan, Rajan, Widner, Wu, Wubbels, Rhodes,
  Whitehouse, Coram, Corrado, Ramasamy, Raman, Peng and
  Webster}]{10.1001/jamaophthalmol.2019.2004}
\bibinfo{author}{Gulshan, V.}, \bibinfo{author}{Rajan, R.P.},
  \bibinfo{author}{Widner, K.}, \bibinfo{author}{Wu, D.},
  \bibinfo{author}{Wubbels, P.}, \bibinfo{author}{Rhodes, T.},
  \bibinfo{author}{Whitehouse, K.}, \bibinfo{author}{Coram, M.},
  \bibinfo{author}{Corrado, G.}, \bibinfo{author}{Ramasamy, K.},
  \bibinfo{author}{Raman, R.}, \bibinfo{author}{Peng, L.},
  \bibinfo{author}{Webster, D.R.}, \bibinfo{year}{2019}.
\newblock \bibinfo{title}{{Performance of a Deep-Learning Algorithm vs Manual
  Grading for Detecting Diabetic Retinopathy in India}}.
\newblock \bibinfo{journal}{JAMA Ophthalmology} \bibinfo{volume}{137},
  \bibinfo{pages}{987--993}.
\newblock \URLprefix \url{https://doi.org/10.1001/jamaophthalmol.2019.2004},
  \DOIprefix\doi{10.1001/jamaophthalmol.2019.2004}.
\bibitem[{Guo et~al.(2019)Guo, Li, Kang, Li, Zhang and
  Wang}]{DBLP:journals/ijon/GuoLKLZW19}
\bibinfo{author}{Guo, S.}, \bibinfo{author}{Li, T.}, \bibinfo{author}{Kang,
  H.}, \bibinfo{author}{Li, N.}, \bibinfo{author}{Zhang, Y.},
  \bibinfo{author}{Wang, K.}, \bibinfo{year}{2019}.
\newblock \bibinfo{title}{L-seg: An end-to-end unified framework for
  multi-lesion segmentation of fundus images}.
\newblock \bibinfo{journal}{Neurocomputing} \bibinfo{volume}{349},
  \bibinfo{pages}{52--63}.
\newblock \URLprefix \url{https://doi.org/10.1016/j.neucom.2019.04.019},
  \DOIprefix\doi{10.1016/j.neucom.2019.04.019}.
\bibitem[{Guo et~al.(2020a)Guo, Wang, Kang, Liu, Gao and
  Li}]{DBLP:journals/ijon/GuoWKLGL20}
\bibinfo{author}{Guo, S.}, \bibinfo{author}{Wang, K.}, \bibinfo{author}{Kang,
  H.}, \bibinfo{author}{Liu, T.}, \bibinfo{author}{Gao, Y.},
  \bibinfo{author}{Li, T.}, \bibinfo{year}{2020}a.
\newblock \bibinfo{title}{Bin loss for hard exudates segmentation in fundus
  images}.
\newblock \bibinfo{journal}{Neurocomputing} \bibinfo{volume}{392},
  \bibinfo{pages}{314--324}.
\newblock \URLprefix \url{https://doi.org/10.1016/j.neucom.2018.10.103},
  \DOIprefix\doi{10.1016/j.neucom.2018.10.103}.
\bibitem[{Guo et~al.(2020b)Guo, Wang, Zhou, Liu, Wang, Lv, Lv and
  Xie}]{DBLP:conf/isbi/GuoWZLWLLX20}
\bibinfo{author}{Guo, Y.}, \bibinfo{author}{Wang, R.}, \bibinfo{author}{Zhou,
  X.}, \bibinfo{author}{Liu, Y.}, \bibinfo{author}{Wang, L.},
  \bibinfo{author}{Lv, C.}, \bibinfo{author}{Lv, B.}, \bibinfo{author}{Xie,
  G.}, \bibinfo{year}{2020}b.
\newblock \bibinfo{title}{Lesion-aware segmentation network for atrophy and
  detachment of pathological myopia on fundus images}, in:
  \bibinfo{booktitle}{17th {IEEE} International Symposium on Biomedical
  Imaging, {ISBI} 2020, Iowa City, IA, USA, April 3-7, 2020},
  \bibinfo{publisher}{{IEEE}}. pp. \bibinfo{pages}{1242--1245}.
\newblock \URLprefix \url{https://doi.org/10.1109/ISBI45749.2020.9098669},
  \DOIprefix\doi{10.1109/ISBI45749.2020.9098669}.
\bibitem[{Hajabdollahi et~al.(2018)Hajabdollahi, Esfandiarpoor, Najarian,
  Karimi, Samavi and Soroushmehr}]{DBLP:conf/icip/HajabdollahiENK18}
\bibinfo{author}{Hajabdollahi, M.}, \bibinfo{author}{Esfandiarpoor, R.},
  \bibinfo{author}{Najarian, K.}, \bibinfo{author}{Karimi, N.},
  \bibinfo{author}{Samavi, S.}, \bibinfo{author}{Soroushmehr, S.M.R.},
  \bibinfo{year}{2018}.
\newblock \bibinfo{title}{Low complexity convolutional neural network for
  vessel segmentation in portable retinal diagnostic devices}, in:
  \bibinfo{booktitle}{2018 {IEEE} International Conference on Image Processing,
  {ICIP} 2018, Athens, Greece, October 7-10, 2018},
  \bibinfo{publisher}{{IEEE}}. pp. \bibinfo{pages}{2785--2789}.
\newblock \URLprefix \url{https://doi.org/10.1109/ICIP.2018.8451665},
  \DOIprefix\doi{10.1109/ICIP.2018.8451665}.
\bibitem[{He et~al.(2017)He, Gkioxari, Doll{\'{a}}r and
  Girshick}]{DBLP:conf/iccv/HeGDG17}
\bibinfo{author}{He, K.}, \bibinfo{author}{Gkioxari, G.},
  \bibinfo{author}{Doll{\'{a}}r, P.}, \bibinfo{author}{Girshick, R.B.},
  \bibinfo{year}{2017}.
\newblock \bibinfo{title}{Mask {R-CNN}}, in: \bibinfo{booktitle}{{IEEE}
  International Conference on Computer Vision, {ICCV} 2017, Venice, Italy,
  October 22-29, 2017}, \bibinfo{publisher}{{IEEE} Computer Society}. pp.
  \bibinfo{pages}{2980--2988}.
\newblock \URLprefix \url{https://doi.org/10.1109/ICCV.2017.322},
  \DOIprefix\doi{10.1109/ICCV.2017.322}.
\bibitem[{He et~al.(2013)He, Sun and Tang}]{DBLP:journals/pami/He0T13}
\bibinfo{author}{He, K.}, \bibinfo{author}{Sun, J.}, \bibinfo{author}{Tang,
  X.}, \bibinfo{year}{2013}.
\newblock \bibinfo{title}{Guided image filtering}.
\newblock \bibinfo{journal}{{IEEE} Trans. Pattern Anal. Mach. Intell.}
  \bibinfo{volume}{35}, \bibinfo{pages}{1397--1409}.
\newblock \URLprefix \url{https://doi.org/10.1109/TPAMI.2012.213},
  \DOIprefix\doi{10.1109/TPAMI.2012.213}.
\bibitem[{He et~al.(2016)He, Zhang, Ren and Sun}]{DBLP:conf/cvpr/HeZRS16}
\bibinfo{author}{He, K.}, \bibinfo{author}{Zhang, X.}, \bibinfo{author}{Ren,
  S.}, \bibinfo{author}{Sun, J.}, \bibinfo{year}{2016}.
\newblock \bibinfo{title}{Deep residual learning for image recognition}, in:
  \bibinfo{booktitle}{2016 {IEEE} Conference on Computer Vision and Pattern
  Recognition, {CVPR} 2016, Las Vegas, NV, USA, June 27-30, 2016},
  \bibinfo{publisher}{{IEEE} Computer Society}. pp. \bibinfo{pages}{770--778}.
\newblock \URLprefix \url{https://doi.org/10.1109/CVPR.2016.90},
  \DOIprefix\doi{10.1109/CVPR.2016.90}.
\bibitem[{He et~al.(2018)He, Zou, Zhu, Liu, Fu and
  Wang}]{DBLP:conf/icip/HeZZLF018}
\bibinfo{author}{He, Q.}, \bibinfo{author}{Zou, B.}, \bibinfo{author}{Zhu, C.},
  \bibinfo{author}{Liu, X.}, \bibinfo{author}{Fu, H.}, \bibinfo{author}{Wang,
  L.}, \bibinfo{year}{2018}.
\newblock \bibinfo{title}{Multi-label classification scheme based on local
  regression for retinal vessel segmentation}, in: \bibinfo{booktitle}{2018
  {IEEE} International Conference on Image Processing, {ICIP} 2018, Athens,
  Greece, October 7-10, 2018}, \bibinfo{publisher}{{IEEE}}. pp.
  \bibinfo{pages}{2765--2769}.
\newblock \URLprefix \url{https://doi.org/10.1109/ICIP.2018.8451415},
  \DOIprefix\doi{10.1109/ICIP.2018.8451415}.
\bibitem[{He et~al.(2019)He, Zhou, Wang, Cui and
  Shao}]{DBLP:conf/miccai/HeZWC019}
\bibinfo{author}{He, X.}, \bibinfo{author}{Zhou, Y.}, \bibinfo{author}{Wang,
  B.}, \bibinfo{author}{Cui, S.}, \bibinfo{author}{Shao, L.},
  \bibinfo{year}{2019}.
\newblock \bibinfo{title}{Dme-net: Diabetic macular edema grading by auxiliary
  task learning}, in: \bibinfo{editor}{Shen, D.}, \bibinfo{editor}{Liu, T.},
  \bibinfo{editor}{Peters, T.M.}, \bibinfo{editor}{Staib, L.H.},
  \bibinfo{editor}{Essert, C.}, \bibinfo{editor}{Zhou, S.},
  \bibinfo{editor}{Yap, P.}, \bibinfo{editor}{Khan, A.} (Eds.),
  \bibinfo{booktitle}{Medical Image Computing and Computer Assisted
  Intervention - {MICCAI} 2019 - 22nd International Conference, Shenzhen,
  China, October 13-17, 2019, Proceedings, Part {I}},
  \bibinfo{publisher}{Springer}. pp. \bibinfo{pages}{788--796}.
\newblock \URLprefix \url{https://doi.org/10.1007/978-3-030-32239-7\_87},
  \DOIprefix\doi{10.1007/978-3-030-32239-7\_87}.
\bibitem[{Hernandez-Matas et~al.(2017)Hernandez-Matas, Zabulis, Triantafyllou,
  Anyfanti, Douma and Argyros}]{FIRE}
\bibinfo{author}{Hernandez-Matas, C.}, \bibinfo{author}{Zabulis, X.},
  \bibinfo{author}{Triantafyllou, A.}, \bibinfo{author}{Anyfanti, P.},
  \bibinfo{author}{Douma, S.}, \bibinfo{author}{Argyros, A.},
  \bibinfo{year}{2017}.
\newblock \bibinfo{title}{Fire: Fundus image registration dataset}.
\newblock \bibinfo{journal}{Journal for Modeling in Opthalmology (to appear)} .
\bibitem[{Hervella et~al.(2018)Hervella, Rouco, Novo and
  Ortega}]{DBLP:conf/miccai/HervellaRNO18}
\bibinfo{author}{Hervella, {\'{A}}.S.}, \bibinfo{author}{Rouco, J.},
  \bibinfo{author}{Novo, J.}, \bibinfo{author}{Ortega, M.},
  \bibinfo{year}{2018}.
\newblock \bibinfo{title}{Retinal image understanding emerges from
  self-supervised multimodal reconstruction}, in: \bibinfo{editor}{Frangi,
  A.F.}, \bibinfo{editor}{Schnabel, J.A.}, \bibinfo{editor}{Davatzikos, C.},
  \bibinfo{editor}{Alberola{-}L{\'{o}}pez, C.}, \bibinfo{editor}{Fichtinger,
  G.} (Eds.), \bibinfo{booktitle}{Medical Image Computing and Computer Assisted
  Intervention - {MICCAI} 2018 - 21st International Conference, Granada, Spain,
  September 16-20, 2018, Proceedings, Part {I}}, \bibinfo{publisher}{Springer}.
  pp. \bibinfo{pages}{321--328}.
\newblock \URLprefix \url{https://doi.org/10.1007/978-3-030-00928-1\_37},
  \DOIprefix\doi{10.1007/978-3-030-00928-1\_37}.
\bibitem[{Hoover et~al.(2000)Hoover, Kouznetsova and
  Goldbaum}]{DBLP:journals/tmi/HooverKG00}
\bibinfo{author}{Hoover, A.W.}, \bibinfo{author}{Kouznetsova, V.},
  \bibinfo{author}{Goldbaum, M.H.}, \bibinfo{year}{2000}.
\newblock \bibinfo{title}{Locating blood vessels in retinal images by
  piece-wise threshold probing of a matched filter response}.
\newblock \bibinfo{journal}{{IEEE} Trans. Medical Imaging}
  \bibinfo{volume}{19}, \bibinfo{pages}{203--210}.
\newblock \URLprefix \url{https://doi.org/10.1109/42.845178},
  \DOIprefix\doi{10.1109/42.845178}.
\bibitem[{Horta et~al.(2017)Horta, Joshi, Pekala, Pacheco, Kong, Bressler,
  Freund and Burlina}]{DBLP:conf/icmla/HortaJPPKBFB17}
\bibinfo{author}{Horta, A.}, \bibinfo{author}{Joshi, N.},
  \bibinfo{author}{Pekala, M.}, \bibinfo{author}{Pacheco, K.D.},
  \bibinfo{author}{Kong, J.}, \bibinfo{author}{Bressler, N.M.},
  \bibinfo{author}{Freund, D.E.}, \bibinfo{author}{Burlina, P.},
  \bibinfo{year}{2017}.
\newblock \bibinfo{title}{A hybrid approach for incorporating deep visual
  features and side channel information with applications to {AMD} detection},
  in: \bibinfo{editor}{Chen, X.}, \bibinfo{editor}{Luo, B.},
  \bibinfo{editor}{Luo, F.}, \bibinfo{editor}{Palade, V.},
  \bibinfo{editor}{Wani, M.A.} (Eds.), \bibinfo{booktitle}{16th {IEEE}
  International Conference on Machine Learning and Applications, {ICMLA} 2017,
  Cancun, Mexico, December 18-21, 2017}, \bibinfo{publisher}{{IEEE}}. pp.
  \bibinfo{pages}{716--720}.
\newblock \URLprefix \url{https://doi.org/10.1109/ICMLA.2017.00-75},
  \DOIprefix\doi{10.1109/ICMLA.2017.00-75}.
\bibitem[{Hu et~al.(2019)Hu, Chen, Zhong, Ju and
  Yi}]{DBLP:journals/tmi/HuCZJY19}
\bibinfo{author}{Hu, J.}, \bibinfo{author}{Chen, Y.}, \bibinfo{author}{Zhong,
  J.}, \bibinfo{author}{Ju, R.}, \bibinfo{author}{Yi, Z.},
  \bibinfo{year}{2019}.
\newblock \bibinfo{title}{Automated analysis for retinopathy of prematurity by
  deep neural networks}.
\newblock \bibinfo{journal}{{IEEE} Trans. Med. Imaging} \bibinfo{volume}{38},
  \bibinfo{pages}{269--279}.
\newblock \URLprefix \url{https://doi.org/10.1109/TMI.2018.2863562},
  \DOIprefix\doi{10.1109/TMI.2018.2863562}.
\bibitem[{Hu et~al.(2018)Hu, Zhang, Niu, Zhang, Cao, Xiao and
  Gao}]{DBLP:journals/ijon/HuZNZCXG18}
\bibinfo{author}{Hu, K.}, \bibinfo{author}{Zhang, Z.}, \bibinfo{author}{Niu,
  X.}, \bibinfo{author}{Zhang, Y.}, \bibinfo{author}{Cao, C.},
  \bibinfo{author}{Xiao, F.}, \bibinfo{author}{Gao, X.}, \bibinfo{year}{2018}.
\newblock \bibinfo{title}{Retinal vessel segmentation of color fundus images
  using multiscale convolutional neural network with an improved cross-entropy
  loss function}.
\newblock \bibinfo{journal}{Neurocomputing} \bibinfo{volume}{309},
  \bibinfo{pages}{179--191}.
\newblock \URLprefix \url{https://doi.org/10.1016/j.neucom.2018.05.011},
  \DOIprefix\doi{10.1016/j.neucom.2018.05.011}.
\bibitem[{Hu et~al.(2013)Hu, Abr{\`{a}}moff and
  Garvin}]{DBLP:conf/miccai/HuAG13}
\bibinfo{author}{Hu, Q.}, \bibinfo{author}{Abr{\`{a}}moff, M.D.},
  \bibinfo{author}{Garvin, M.K.}, \bibinfo{year}{2013}.
\newblock \bibinfo{title}{Automated separation of binary overlapping trees in
  low-contrast color retinal images}, in: \bibinfo{editor}{Mori, K.},
  \bibinfo{editor}{Sakuma, I.}, \bibinfo{editor}{Sato, Y.},
  \bibinfo{editor}{Barillot, C.}, \bibinfo{editor}{Navab, N.} (Eds.),
  \bibinfo{booktitle}{Medical Image Computing and Computer-Assisted
  Intervention - {MICCAI} 2013 - 16th International Conference, Nagoya, Japan,
  September 22-26, 2013, Proceedings, Part {II}},
  \bibinfo{publisher}{Springer}. pp. \bibinfo{pages}{436--443}.
\newblock \URLprefix \url{https://doi.org/10.1007/978-3-642-40763-5\_54},
  \DOIprefix\doi{10.1007/978-3-642-40763-5\_54}.
\bibitem[{Huang et~al.(2017)Huang, Liu, van~der Maaten and
  Weinberger}]{DBLP:conf/cvpr/HuangLMW17}
\bibinfo{author}{Huang, G.}, \bibinfo{author}{Liu, Z.},
  \bibinfo{author}{van~der Maaten, L.}, \bibinfo{author}{Weinberger, K.Q.},
  \bibinfo{year}{2017}.
\newblock \bibinfo{title}{Densely connected convolutional networks}, in:
  \bibinfo{booktitle}{2017 {IEEE} Conference on Computer Vision and Pattern
  Recognition, {CVPR} 2017, Honolulu, HI, USA, July 21-26, 2017},
  \bibinfo{publisher}{{IEEE} Computer Society}. pp.
  \bibinfo{pages}{2261--2269}.
\newblock \URLprefix \url{https://doi.org/10.1109/CVPR.2017.243},
  \DOIprefix\doi{10.1109/CVPR.2017.243}.
\bibitem[{Huang et~al.(2020)Huang, Lin, Li, Wu, Cheng, Wang, Yuan and
  Tang}]{DBLP:conf/isbi/HuangLLWCWYT20}
\bibinfo{author}{Huang, Y.}, \bibinfo{author}{Lin, L.}, \bibinfo{author}{Li,
  M.}, \bibinfo{author}{Wu, J.}, \bibinfo{author}{Cheng, P.},
  \bibinfo{author}{Wang, K.}, \bibinfo{author}{Yuan, J.},
  \bibinfo{author}{Tang, X.}, \bibinfo{year}{2020}.
\newblock \bibinfo{title}{Automated hemorrhage detection from coarsely
  annotated fundus images in diabetic retinopathy}, in:
  \bibinfo{booktitle}{17th {IEEE} International Symposium on Biomedical
  Imaging, {ISBI} 2020, Iowa City, IA, USA, April 3-7, 2020},
  \bibinfo{publisher}{{IEEE}}. pp. \bibinfo{pages}{1369--1372}.
\newblock \URLprefix \url{https://doi.org/10.1109/ISBI45749.2020.9098319},
  \DOIprefix\doi{10.1109/ISBI45749.2020.9098319}.
\bibitem[{Jiang et~al.(2020)Jiang, Duan, Cheng, Gu, Xia, Fu, Li and
  Liu}]{DBLP:journals/tbe/JiangDCGXFLL20}
\bibinfo{author}{Jiang, Y.}, \bibinfo{author}{Duan, L.},
  \bibinfo{author}{Cheng, J.}, \bibinfo{author}{Gu, Z.}, \bibinfo{author}{Xia,
  H.}, \bibinfo{author}{Fu, H.}, \bibinfo{author}{Li, C.},
  \bibinfo{author}{Liu, J.}, \bibinfo{year}{2020}.
\newblock \bibinfo{title}{Jointrcnn: {A} region-based convolutional neural
  network for optic disc and cup segmentation}.
\newblock \bibinfo{journal}{{IEEE} Trans. Biomed. Engineering}
  \bibinfo{volume}{67}, \bibinfo{pages}{335--343}.
\newblock \URLprefix \url{https://doi.org/10.1109/TBME.2019.2913211},
  \DOIprefix\doi{10.1109/TBME.2019.2913211}.
\bibitem[{Kairouz et~al.(2019)Kairouz, McMahan, Avent, Bellet, Bennis, Bhagoji,
  Bonawitz, Charles, Cormode, Cummings, D'Oliveira, Rouayheb, Evans, Gardner,
  Garrett, Gasc{\'{o}}n, Ghazi, Gibbons, Gruteser, Harchaoui, He, He, Huo,
  Hutchinson, Hsu, Jaggi, Javidi, Joshi, Khodak, Konecn{\'{y}}, Korolova,
  Koushanfar, Koyejo, Lepoint, Liu, Mittal, Mohri, Nock, {\"{O}}zg{\"{u}}r,
  Pagh, Raykova, Qi, Ramage, Raskar, Song, Song, Stich, Sun, Suresh,
  Tram{\`{e}}r, Vepakomma, Wang, Xiong, Xu, Yang, Yu, Yu and
  Zhao}]{DBLP:journals/corr/abs-1912-04977}
\bibinfo{author}{Kairouz, P.}, \bibinfo{author}{McMahan, H.B.},
  \bibinfo{author}{Avent, B.}, \bibinfo{author}{Bellet, A.},
  \bibinfo{author}{Bennis, M.}, \bibinfo{author}{Bhagoji, A.N.},
  \bibinfo{author}{Bonawitz, K.}, \bibinfo{author}{Charles, Z.},
  \bibinfo{author}{Cormode, G.}, \bibinfo{author}{Cummings, R.},
  \bibinfo{author}{D'Oliveira, R.G.L.}, \bibinfo{author}{Rouayheb, S.E.},
  \bibinfo{author}{Evans, D.}, \bibinfo{author}{Gardner, J.},
  \bibinfo{author}{Garrett, Z.}, \bibinfo{author}{Gasc{\'{o}}n, A.},
  \bibinfo{author}{Ghazi, B.}, \bibinfo{author}{Gibbons, P.B.},
  \bibinfo{author}{Gruteser, M.}, \bibinfo{author}{Harchaoui, Z.},
  \bibinfo{author}{He, C.}, \bibinfo{author}{He, L.}, \bibinfo{author}{Huo,
  Z.}, \bibinfo{author}{Hutchinson, B.}, \bibinfo{author}{Hsu, J.},
  \bibinfo{author}{Jaggi, M.}, \bibinfo{author}{Javidi, T.},
  \bibinfo{author}{Joshi, G.}, \bibinfo{author}{Khodak, M.},
  \bibinfo{author}{Konecn{\'{y}}, J.}, \bibinfo{author}{Korolova, A.},
  \bibinfo{author}{Koushanfar, F.}, \bibinfo{author}{Koyejo, S.},
  \bibinfo{author}{Lepoint, T.}, \bibinfo{author}{Liu, Y.},
  \bibinfo{author}{Mittal, P.}, \bibinfo{author}{Mohri, M.},
  \bibinfo{author}{Nock, R.}, \bibinfo{author}{{\"{O}}zg{\"{u}}r, A.},
  \bibinfo{author}{Pagh, R.}, \bibinfo{author}{Raykova, M.},
  \bibinfo{author}{Qi, H.}, \bibinfo{author}{Ramage, D.},
  \bibinfo{author}{Raskar, R.}, \bibinfo{author}{Song, D.},
  \bibinfo{author}{Song, W.}, \bibinfo{author}{Stich, S.U.},
  \bibinfo{author}{Sun, Z.}, \bibinfo{author}{Suresh, A.T.},
  \bibinfo{author}{Tram{\`{e}}r, F.}, \bibinfo{author}{Vepakomma, P.},
  \bibinfo{author}{Wang, J.}, \bibinfo{author}{Xiong, L.}, \bibinfo{author}{Xu,
  Z.}, \bibinfo{author}{Yang, Q.}, \bibinfo{author}{Yu, F.X.},
  \bibinfo{author}{Yu, H.}, \bibinfo{author}{Zhao, S.}, \bibinfo{year}{2019}.
\newblock \bibinfo{title}{Advances and open problems in federated learning}.
\newblock \bibinfo{journal}{CoRR} \bibinfo{volume}{abs/1912.04977}.
\newblock \URLprefix \url{http://arxiv.org/abs/1912.04977}.
\bibitem[{Kanse and Yadav(2019)}]{DBLP:journals/jois/KanseY19}
\bibinfo{author}{Kanse, S.S.}, \bibinfo{author}{Yadav, D.M.},
  \bibinfo{year}{2019}.
\newblock \bibinfo{title}{Retinal fundus image for glaucoma detection: {A}
  review and study}.
\newblock \bibinfo{journal}{J. Intelligent Systems} \bibinfo{volume}{28},
  \bibinfo{pages}{43--56}.
\newblock \URLprefix \url{https://doi.org/10.1515/jisys-2016-0258},
  \DOIprefix\doi{10.1515/jisys-2016-0258}.
\bibitem[{Kauppi et~al.(2007)Kauppi, Kalesnykiene, Kamarainen, Lensu, Sorri,
  Raninen, Voutilainen, Uusitalo, K{\"{a}}lvi{\"{a}}inen and
  Pietil{\"{a}}}]{DBLP:conf/bmvc/KauppiKKLSRVUKP07}
\bibinfo{author}{Kauppi, T.}, \bibinfo{author}{Kalesnykiene, V.},
  \bibinfo{author}{Kamarainen, J.}, \bibinfo{author}{Lensu, L.},
  \bibinfo{author}{Sorri, I.}, \bibinfo{author}{Raninen, A.},
  \bibinfo{author}{Voutilainen, R.}, \bibinfo{author}{Uusitalo, H.},
  \bibinfo{author}{K{\"{a}}lvi{\"{a}}inen, H.}, \bibinfo{author}{Pietil{\"{a}},
  J.}, \bibinfo{year}{2007}.
\newblock \bibinfo{title}{The {DIARETDB1} diabetic retinopathy database and
  evaluation protocol}, in: \bibinfo{editor}{Rajpoot, N.M.},
  \bibinfo{editor}{Bhalerao, A.H.} (Eds.), \bibinfo{booktitle}{Proceedings of
  the British Machine Vision Conference 2007, University of Warwick, UK,
  September 10-13, 2007}, \bibinfo{publisher}{British Machine Vision
  Association}. pp. \bibinfo{pages}{1--10}.
\newblock \URLprefix \url{https://doi.org/10.5244/C.21.15},
  \DOIprefix\doi{10.5244/C.21.15}.
\bibitem[{Kauppi et~al.(2006)Kauppi, Kalesnykiene, kristian Kamarainen, Lensu,
  Sorri, Uusitalo, Kälviäinen and
  Pietilä}]{Kalesnykiene_diaretdb0:evaluation}
\bibinfo{author}{Kauppi, T.}, \bibinfo{author}{Kalesnykiene, V.},
  \bibinfo{author}{kristian Kamarainen, J.}, \bibinfo{author}{Lensu, L.},
  \bibinfo{author}{Sorri, I.}, \bibinfo{author}{Uusitalo, H.},
  \bibinfo{author}{Kälviäinen, H.}, \bibinfo{author}{Pietilä, J.},
  \bibinfo{year}{2006}.
\newblock \bibinfo{title}{Diaretdb0: Evaluation database and methodology for
  diabetic retinopathy algorithms}.
\bibitem[{Keel et~al.(2019)Keel, Wu, Lee, Scheetz and
  He}]{10.1001/jamaophthalmol.2018.6035}
\bibinfo{author}{Keel, S.}, \bibinfo{author}{Wu, J.}, \bibinfo{author}{Lee,
  P.Y.}, \bibinfo{author}{Scheetz, J.}, \bibinfo{author}{He, M.},
  \bibinfo{year}{2019}.
\newblock \bibinfo{title}{{Visualizing Deep Learning Models for the Detection
  of Referable Diabetic Retinopathy and Glaucoma}}.
\newblock \bibinfo{journal}{JAMA Ophthalmology} \bibinfo{volume}{137},
  \bibinfo{pages}{288--292}.
\newblock \URLprefix \url{https://doi.org/10.1001/jamaophthalmol.2018.6035},
  \DOIprefix\doi{10.1001/jamaophthalmol.2018.6035}.
\bibitem[{Khalaf et~al.(2016)Khalaf, Yassine and
  Fahmy}]{DBLP:conf/icip/KhalafYF16}
\bibinfo{author}{Khalaf, A.F.}, \bibinfo{author}{Yassine, I.A.},
  \bibinfo{author}{Fahmy, A.S.}, \bibinfo{year}{2016}.
\newblock \bibinfo{title}{Convolutional neural networks for deep feature
  learning in retinal vessel segmentation}, in: \bibinfo{booktitle}{2016 {IEEE}
  International Conference on Image Processing, {ICIP} 2016, Phoenix, AZ, USA,
  September 25-28, 2016}, \bibinfo{publisher}{{IEEE}}. pp.
  \bibinfo{pages}{385--388}.
\newblock \URLprefix \url{https://doi.org/10.1109/ICIP.2016.7532384},
  \DOIprefix\doi{10.1109/ICIP.2016.7532384}.
\bibitem[{Kr{\"{a}}henb{\"{u}}hl and
  Koltun(2011)}]{DBLP:conf/nips/KrahenbuhlK11}
\bibinfo{author}{Kr{\"{a}}henb{\"{u}}hl, P.}, \bibinfo{author}{Koltun, V.},
  \bibinfo{year}{2011}.
\newblock \bibinfo{title}{Efficient inference in fully connected crfs with
  gaussian edge potentials}, in: \bibinfo{editor}{Shawe{-}Taylor, J.},
  \bibinfo{editor}{Zemel, R.S.}, \bibinfo{editor}{Bartlett, P.L.},
  \bibinfo{editor}{Pereira, F.C.N.}, \bibinfo{editor}{Weinberger, K.Q.} (Eds.),
  \bibinfo{booktitle}{Advances in Neural Information Processing Systems 24:
  25th Annual Conference on Neural Information Processing Systems 2011.
  Proceedings of a meeting held 12-14 December 2011, Granada, Spain}, pp.
  \bibinfo{pages}{109--117}.
\newblock \URLprefix
  \url{http://papers.nips.cc/paper/4296-efficient-inference-in-fully-connected-crfs-with-gaussian-edge-potentials}.
\bibitem[{Krause et~al.(2017)Krause, Gulshan, Rahimy, Karth and
  Webster}]{Krause2017Grader}
\bibinfo{author}{Krause, J.}, \bibinfo{author}{Gulshan, V.},
  \bibinfo{author}{Rahimy, E.}, \bibinfo{author}{Karth, P.},
  \bibinfo{author}{Webster, D.R.}, \bibinfo{year}{2017}.
\newblock \bibinfo{title}{Grader variability and the importance of reference
  standards for evaluating machine learning models for diabetic retinopathy}.
\newblock \bibinfo{journal}{Ophthalmology} \bibinfo{volume}{125}.
\bibitem[{Kromm and Rohr(2020)}]{DBLP:conf/isbi/KrommR20}
\bibinfo{author}{Kromm, C.}, \bibinfo{author}{Rohr, K.}, \bibinfo{year}{2020}.
\newblock \bibinfo{title}{Inception capsule network for retinal blood vessel
  segmentation and centerline extraction}, in: \bibinfo{booktitle}{17th {IEEE}
  International Symposium on Biomedical Imaging, {ISBI} 2020, Iowa City, IA,
  USA, April 3-7, 2020}, \bibinfo{publisher}{{IEEE}}. pp.
  \bibinfo{pages}{1223--1226}.
\newblock \URLprefix \url{https://doi.org/10.1109/ISBI45749.2020.9098538},
  \DOIprefix\doi{10.1109/ISBI45749.2020.9098538}.
\bibitem[{de~La~Torre et~al.(2020)de~La~Torre, Valls and
  Puig}]{DBLP:journals/ijon/TorreVP20}
\bibinfo{author}{de~La~Torre, J.}, \bibinfo{author}{Valls, A.},
  \bibinfo{author}{Puig, D.}, \bibinfo{year}{2020}.
\newblock \bibinfo{title}{A deep learning interpretable classifier for diabetic
  retinopathy disease grading}.
\newblock \bibinfo{journal}{Neurocomputing} \bibinfo{volume}{396},
  \bibinfo{pages}{465--476}.
\newblock \URLprefix \url{https://doi.org/10.1016/j.neucom.2018.07.102},
  \DOIprefix\doi{10.1016/j.neucom.2018.07.102}.
\bibitem[{Li et~al.(2020a)Li, Ye, He, Wang, Qiao and
  Gu}]{DBLP:conf/isbi/LiYHW0G20}
\bibinfo{author}{Li, C.}, \bibinfo{author}{Ye, J.}, \bibinfo{author}{He, J.},
  \bibinfo{author}{Wang, S.}, \bibinfo{author}{Qiao, Y.}, \bibinfo{author}{Gu,
  L.}, \bibinfo{year}{2020}a.
\newblock \bibinfo{title}{Dense correlation network for automated multi-label
  ocular disease detection with paired color fundus photographs}, in:
  \bibinfo{booktitle}{17th {IEEE} International Symposium on Biomedical
  Imaging, {ISBI} 2020, Iowa City, IA, USA, April 3-7, 2020},
  \bibinfo{publisher}{{IEEE}}. pp. \bibinfo{pages}{1--4}.
\newblock \URLprefix \url{https://doi.org/10.1109/ISBI45749.2020.9098340},
  \DOIprefix\doi{10.1109/ISBI45749.2020.9098340}.
\bibitem[{Li et~al.(2020b)Li, Xu, Liu, Li, Wang, Jiang, Wang, Fan and
  Wang}]{DBLP:journals/tmi/LiXLLWJWFW20}
\bibinfo{author}{Li, L.}, \bibinfo{author}{Xu, M.}, \bibinfo{author}{Liu, H.},
  \bibinfo{author}{Li, Y.}, \bibinfo{author}{Wang, X.}, \bibinfo{author}{Jiang,
  L.}, \bibinfo{author}{Wang, Z.}, \bibinfo{author}{Fan, X.},
  \bibinfo{author}{Wang, N.}, \bibinfo{year}{2020}b.
\newblock \bibinfo{title}{A large-scale database and a {CNN} model for
  attention-based glaucoma detection}.
\newblock \bibinfo{journal}{{IEEE} Trans. Med. Imaging} \bibinfo{volume}{39},
  \bibinfo{pages}{413--424}.
\newblock \URLprefix \url{https://doi.org/10.1109/TMI.2019.2927226},
  \DOIprefix\doi{10.1109/TMI.2019.2927226}.
\bibitem[{Li et~al.(2019a)Li, Xu, Wang, Jiang and
  Liu}]{DBLP:conf/cvpr/LiXWJL19}
\bibinfo{author}{Li, L.}, \bibinfo{author}{Xu, M.}, \bibinfo{author}{Wang, X.},
  \bibinfo{author}{Jiang, L.}, \bibinfo{author}{Liu, H.},
  \bibinfo{year}{2019}a.
\newblock \bibinfo{title}{Attention based glaucoma detection: {A} large-scale
  database and {CNN} model}, in: \bibinfo{booktitle}{{IEEE} Conference on
  Computer Vision and Pattern Recognition, {CVPR} 2019, Long Beach, CA, USA,
  June 16-20, 2019}, \bibinfo{publisher}{Computer Vision Foundation / {IEEE}}.
  pp. \bibinfo{pages}{10571--10580}.
\newblock \URLprefix
  \url{http://openaccess.thecvf.com/content\_CVPR\_2019/html/Li\_Attention\_Based\_Glaucoma\_Detection\_A\_Large-Scale\_Database\_and\_CNN\_Model\_CVPR\_2019\_paper.html},
  \DOIprefix\doi{10.1109/CVPR.2019.01082}.
\bibitem[{Li et~al.(2019b)Li, Gao, Wang, Guo, Liu and
  Kang}]{DBLP:journals/isci/LiGWGLK19}
\bibinfo{author}{Li, T.}, \bibinfo{author}{Gao, Y.}, \bibinfo{author}{Wang,
  K.}, \bibinfo{author}{Guo, S.}, \bibinfo{author}{Liu, H.},
  \bibinfo{author}{Kang, H.}, \bibinfo{year}{2019}b.
\newblock \bibinfo{title}{Diagnostic assessment of deep learning algorithms for
  diabetic retinopathy screening}.
\newblock \bibinfo{journal}{Inf. Sci.} \bibinfo{volume}{501},
  \bibinfo{pages}{511--522}.
\newblock \URLprefix \url{https://doi.org/10.1016/j.ins.2019.06.011},
  \DOIprefix\doi{10.1016/j.ins.2019.06.011}.
\bibitem[{Li et~al.(2020c)Li, Hu, Yu, Zhu, Fu and
  Heng}]{DBLP:journals/tmi/LiHYZFH20}
\bibinfo{author}{Li, X.}, \bibinfo{author}{Hu, X.}, \bibinfo{author}{Yu, L.},
  \bibinfo{author}{Zhu, L.}, \bibinfo{author}{Fu, C.}, \bibinfo{author}{Heng,
  P.}, \bibinfo{year}{2020}c.
\newblock \bibinfo{title}{Canet: Cross-disease attention network for joint
  diabetic retinopathy and diabetic macular edema grading}.
\newblock \bibinfo{journal}{{IEEE} Trans. Med. Imaging} \bibinfo{volume}{39},
  \bibinfo{pages}{1483--1493}.
\newblock \URLprefix \url{https://doi.org/10.1109/TMI.2019.2951844},
  \DOIprefix\doi{10.1109/TMI.2019.2951844}.
\bibitem[{Li et~al.(2018a)Li, He, Keel, Meng, Chang and He}]{Li2018Efficacy}
\bibinfo{author}{Li, Z.}, \bibinfo{author}{He, Y.}, \bibinfo{author}{Keel, S.},
  \bibinfo{author}{Meng, W.}, \bibinfo{author}{Chang, R.T.},
  \bibinfo{author}{He, M.}, \bibinfo{year}{2018}a.
\newblock \bibinfo{title}{Efficacy of a deep learning system for detecting
  glaucomatous optic neuropathy based on color fundus photographs}.
\newblock \bibinfo{journal}{Ophthalmology} ,
  \bibinfo{pages}{S0161642017335650}.
\bibitem[{Li et~al.(2018b)Li, Keel, Liu, He, Meng, Scheetz, Lee, Shaw, Ting,
  Wong, Taylor, Chang and He}]{Lidc180147}
\bibinfo{author}{Li, Z.}, \bibinfo{author}{Keel, S.}, \bibinfo{author}{Liu,
  C.}, \bibinfo{author}{He, Y.}, \bibinfo{author}{Meng, W.},
  \bibinfo{author}{Scheetz, J.}, \bibinfo{author}{Lee, P.Y.},
  \bibinfo{author}{Shaw, J.}, \bibinfo{author}{Ting, D.},
  \bibinfo{author}{Wong, T.}, \bibinfo{author}{Taylor, H.},
  \bibinfo{author}{Chang, R.}, \bibinfo{author}{He, M.}, \bibinfo{year}{2018}b.
\newblock \bibinfo{title}{An automated grading system for detection of
  vision-threatening referable diabetic retinopathy on the basis of color
  fundus photographs}.
\newblock \bibinfo{journal}{Diabetes Care} \URLprefix
  \url{https://care.diabetesjournals.org/content/early/2018/09/27/dc18-0147},
  \DOIprefix\doi{10.2337/dc18-0147}.
\bibitem[{Liao et~al.(2020)Liao, Zou, Zhao, Chen, He and
  Zhou}]{DBLP:journals/titb/LiaoZZCHZ20}
\bibinfo{author}{Liao, W.}, \bibinfo{author}{Zou, B.}, \bibinfo{author}{Zhao,
  R.}, \bibinfo{author}{Chen, Y.}, \bibinfo{author}{He, Z.},
  \bibinfo{author}{Zhou, M.}, \bibinfo{year}{2020}.
\newblock \bibinfo{title}{Clinical interpretable deep learning model for
  glaucoma diagnosis}.
\newblock \bibinfo{journal}{{IEEE} J. Biomed. Health Informatics}
  \bibinfo{volume}{24}, \bibinfo{pages}{1405--1412}.
\newblock \URLprefix \url{https://doi.org/10.1109/JBHI.2019.2949075},
  \DOIprefix\doi{10.1109/JBHI.2019.2949075}.
\bibitem[{Lim et~al.(2015)Lim, Cheng, Hsu and Lee}]{DBLP:conf/ictai/LimCHL15}
\bibinfo{author}{Lim, G.}, \bibinfo{author}{Cheng, Y.}, \bibinfo{author}{Hsu,
  W.}, \bibinfo{author}{Lee, M.}, \bibinfo{year}{2015}.
\newblock \bibinfo{title}{Integrated optic disc and cup segmentation with deep
  learning}, in: \bibinfo{booktitle}{27th {IEEE} International Conference on
  Tools with Artificial Intelligence, {ICTAI} 2015, Vietri sul Mare, Italy,
  November 9-11, 2015}, \bibinfo{publisher}{{IEEE} Computer Society}. pp.
  \bibinfo{pages}{162--169}.
\newblock \URLprefix \url{https://doi.org/10.1109/ICTAI.2015.36},
  \DOIprefix\doi{10.1109/ICTAI.2015.36}.
\bibitem[{Lim et~al.(2019)Lim, Lim, Xu, Ting, Wong, Lee and
  Hsu}]{DBLP:conf/aaai/LimLXTWLH19}
\bibinfo{author}{Lim, G.}, \bibinfo{author}{Lim, Z.W.}, \bibinfo{author}{Xu,
  D.}, \bibinfo{author}{Ting, D.S.W.}, \bibinfo{author}{Wong, T.Y.},
  \bibinfo{author}{Lee, M.}, \bibinfo{author}{Hsu, W.}, \bibinfo{year}{2019}.
\newblock \bibinfo{title}{Feature isolation for hypothesis testing in retinal
  imaging: An ischemic stroke prediction case study}, in:
  \bibinfo{booktitle}{The Thirty-Third {AAAI} Conference on Artificial
  Intelligence, {AAAI} 2019, The Thirty-First Innovative Applications of
  Artificial Intelligence Conference, {IAAI} 2019, The Ninth {AAAI} Symposium
  on Educational Advances in Artificial Intelligence, {EAAI} 2019, Honolulu,
  Hawaii, USA, January 27 - February 1, 2019}, \bibinfo{publisher}{{AAAI}
  Press}. pp. \bibinfo{pages}{9510--9515}.
\newblock \URLprefix \url{https://doi.org/10.1609/aaai.v33i01.33019510},
  \DOIprefix\doi{10.1609/aaai.v33i01.33019510}.
\bibitem[{Lin et~al.(2017)Lin, Goyal, Girshick, He and
  Doll{\'{a}}r}]{DBLP:conf/iccv/LinGGHD17}
\bibinfo{author}{Lin, T.}, \bibinfo{author}{Goyal, P.},
  \bibinfo{author}{Girshick, R.B.}, \bibinfo{author}{He, K.},
  \bibinfo{author}{Doll{\'{a}}r, P.}, \bibinfo{year}{2017}.
\newblock \bibinfo{title}{Focal loss for dense object detection}, in:
  \bibinfo{booktitle}{{IEEE} International Conference on Computer Vision,
  {ICCV} 2017, Venice, Italy, October 22-29, 2017}, \bibinfo{publisher}{{IEEE}
  Computer Society}. pp. \bibinfo{pages}{2999--3007}.
\newblock \URLprefix \url{https://doi.org/10.1109/ICCV.2017.324},
  \DOIprefix\doi{10.1109/ICCV.2017.324}.
\bibitem[{Lin et~al.(2018)Lin, Guo, Wang, Wu, Chen, Wang, Chen and
  Wu}]{DBLP:conf/miccai/LinGWWCWCW18}
\bibinfo{author}{Lin, Z.}, \bibinfo{author}{Guo, R.}, \bibinfo{author}{Wang,
  Y.}, \bibinfo{author}{Wu, B.}, \bibinfo{author}{Chen, T.},
  \bibinfo{author}{Wang, W.}, \bibinfo{author}{Chen, D.Z.},
  \bibinfo{author}{Wu, J.}, \bibinfo{year}{2018}.
\newblock \bibinfo{title}{A framework for identifying diabetic retinopathy
  based on anti-noise detection and attention-based fusion}, in:
  \bibinfo{editor}{Frangi, A.F.}, \bibinfo{editor}{Schnabel, J.A.},
  \bibinfo{editor}{Davatzikos, C.}, \bibinfo{editor}{Alberola{-}L{\'{o}}pez,
  C.}, \bibinfo{editor}{Fichtinger, G.} (Eds.), \bibinfo{booktitle}{Medical
  Image Computing and Computer Assisted Intervention - {MICCAI} 2018 - 21st
  International Conference, Granada, Spain, September 16-20, 2018, Proceedings,
  Part {II}}, \bibinfo{publisher}{Springer}. pp. \bibinfo{pages}{74--82}.
\newblock \URLprefix \url{https://doi.org/10.1007/978-3-030-00934-2\_9},
  \DOIprefix\doi{10.1007/978-3-030-00934-2\_9}.
\bibitem[{Liskowski and Krawiec(2016)}]{DBLP:journals/tmi/LiskowskiK16}
\bibinfo{author}{Liskowski, P.}, \bibinfo{author}{Krawiec, K.},
  \bibinfo{year}{2016}.
\newblock \bibinfo{title}{Segmenting retinal blood vessels with deep neural
  networks}.
\newblock \bibinfo{journal}{{IEEE} Trans. Med. Imaging} \bibinfo{volume}{35},
  \bibinfo{pages}{2369--2380}.
\newblock \URLprefix \url{https://doi.org/10.1109/TMI.2016.2546227},
  \DOIprefix\doi{10.1109/TMI.2016.2546227}.
\bibitem[{Liu et~al.(2019a)Liu, Gu and Lu}]{DBLP:conf/miccai/Liu0019}
\bibinfo{author}{Liu, B.}, \bibinfo{author}{Gu, L.}, \bibinfo{author}{Lu, F.},
  \bibinfo{year}{2019}a.
\newblock \bibinfo{title}{Unsupervised ensemble strategy for retinal vessel
  segmentation}, in: \bibinfo{editor}{Shen, D.}, \bibinfo{editor}{Liu, T.},
  \bibinfo{editor}{Peters, T.M.}, \bibinfo{editor}{Staib, L.H.},
  \bibinfo{editor}{Essert, C.}, \bibinfo{editor}{Zhou, S.},
  \bibinfo{editor}{Yap, P.}, \bibinfo{editor}{Khan, A.} (Eds.),
  \bibinfo{booktitle}{Medical Image Computing and Computer Assisted
  Intervention - {MICCAI} 2019 - 22nd International Conference, Shenzhen,
  China, October 13-17, 2019, Proceedings, Part {I}},
  \bibinfo{publisher}{Springer}. pp. \bibinfo{pages}{111--119}.
\newblock \URLprefix \url{https://doi.org/10.1007/978-3-030-32239-7\_13},
  \DOIprefix\doi{10.1007/978-3-030-32239-7\_13}.
\bibitem[{Liu et~al.(2019b)Liu, Wang, Li, Jiang, Han, Ha, Meng and
  He}]{DBLP:conf/miccai/LiuWLJHHMH19}
\bibinfo{author}{Liu, C.}, \bibinfo{author}{Wang, W.}, \bibinfo{author}{Li,
  Z.}, \bibinfo{author}{Jiang, Y.}, \bibinfo{author}{Han, X.},
  \bibinfo{author}{Ha, J.}, \bibinfo{author}{Meng, W.}, \bibinfo{author}{He,
  M.}, \bibinfo{year}{2019}b.
\newblock \bibinfo{title}{Biological age estimated from retinal imaging: {A}
  novel biomarker of aging}, in: \bibinfo{editor}{Shen, D.},
  \bibinfo{editor}{Liu, T.}, \bibinfo{editor}{Peters, T.M.},
  \bibinfo{editor}{Staib, L.H.}, \bibinfo{editor}{Essert, C.},
  \bibinfo{editor}{Zhou, S.}, \bibinfo{editor}{Yap, P.}, \bibinfo{editor}{Khan,
  A.} (Eds.), \bibinfo{booktitle}{Medical Image Computing and Computer Assisted
  Intervention - {MICCAI} 2019 - 22nd International Conference, Shenzhen,
  China, October 13-17, 2019, Proceedings, Part {I}},
  \bibinfo{publisher}{Springer}. pp. \bibinfo{pages}{138--146}.
\newblock \URLprefix \url{https://doi.org/10.1007/978-3-030-32239-7\_16},
  \DOIprefix\doi{10.1007/978-3-030-32239-7\_16}.
\bibitem[{Liu et~al.(2019c)Liu, Li, Wormstone, Qiao, Zhang, Liu, Li, Wang, Mou,
  Pang, Yang, Zangwill, Moghimi, Hou, Bowd, Jiang, Chen, Hu, Xu, Kang, Ji,
  Chang, Tham, Cheung, Ting, Wong, Wang, Weinreb, Xu and
  Wang}]{10.1001/jamaophthalmol.2019.3501}
\bibinfo{author}{Liu, H.}, \bibinfo{author}{Li, L.},
  \bibinfo{author}{Wormstone, I.M.}, \bibinfo{author}{Qiao, C.},
  \bibinfo{author}{Zhang, C.}, \bibinfo{author}{Liu, P.}, \bibinfo{author}{Li,
  S.}, \bibinfo{author}{Wang, H.}, \bibinfo{author}{Mou, D.},
  \bibinfo{author}{Pang, R.}, \bibinfo{author}{Yang, D.},
  \bibinfo{author}{Zangwill, L.M.}, \bibinfo{author}{Moghimi, S.},
  \bibinfo{author}{Hou, H.}, \bibinfo{author}{Bowd, C.},
  \bibinfo{author}{Jiang, L.}, \bibinfo{author}{Chen, Y.}, \bibinfo{author}{Hu,
  M.}, \bibinfo{author}{Xu, Y.}, \bibinfo{author}{Kang, H.},
  \bibinfo{author}{Ji, X.}, \bibinfo{author}{Chang, R.}, \bibinfo{author}{Tham,
  C.}, \bibinfo{author}{Cheung, C.}, \bibinfo{author}{Ting, D.S.W.},
  \bibinfo{author}{Wong, T.Y.}, \bibinfo{author}{Wang, Z.},
  \bibinfo{author}{Weinreb, R.N.}, \bibinfo{author}{Xu, M.},
  \bibinfo{author}{Wang, N.}, \bibinfo{year}{2019}c.
\newblock \bibinfo{title}{{Development and Validation of a Deep Learning System
  to Detect Glaucomatous Optic Neuropathy Using Fundus Photographs}}.
\newblock \bibinfo{journal}{JAMA Ophthalmology} \bibinfo{volume}{137},
  \bibinfo{pages}{1353--1360}.
\newblock \URLprefix \url{https://doi.org/10.1001/jamaophthalmol.2019.3501},
  \DOIprefix\doi{10.1001/jamaophthalmol.2019.3501}.
\bibitem[{Liu et~al.(2019d)Liu, Kong, Li, Zhang and
  Fang}]{DBLP:conf/miccai/LiuKLZF19}
\bibinfo{author}{Liu, P.}, \bibinfo{author}{Kong, B.}, \bibinfo{author}{Li,
  Z.}, \bibinfo{author}{Zhang, S.}, \bibinfo{author}{Fang, R.},
  \bibinfo{year}{2019}d.
\newblock \bibinfo{title}{{CFEA:} collaborative feature ensembling adaptation
  for domain adaptation in unsupervised optic disc and cup segmentation}, in:
  \bibinfo{editor}{Shen, D.}, \bibinfo{editor}{Liu, T.},
  \bibinfo{editor}{Peters, T.M.}, \bibinfo{editor}{Staib, L.H.},
  \bibinfo{editor}{Essert, C.}, \bibinfo{editor}{Zhou, S.},
  \bibinfo{editor}{Yap, P.}, \bibinfo{editor}{Khan, A.} (Eds.),
  \bibinfo{booktitle}{Medical Image Computing and Computer Assisted
  Intervention - {MICCAI} 2019 - 22nd International Conference, Shenzhen,
  China, October 13-17, 2019, Proceedings, Part {V}},
  \bibinfo{publisher}{Springer}. pp. \bibinfo{pages}{521--529}.
\newblock \URLprefix \url{https://doi.org/10.1007/978-3-030-32254-0\_58},
  \DOIprefix\doi{10.1007/978-3-030-32254-0\_58}.
\bibitem[{Liu et~al.(2019e)Liu, Hong, Li, Chen, Zhao and
  Zou}]{DBLP:journals/ijon/LiuHLCZZ19}
\bibinfo{author}{Liu, Q.}, \bibinfo{author}{Hong, X.}, \bibinfo{author}{Li,
  S.}, \bibinfo{author}{Chen, Z.}, \bibinfo{author}{Zhao, G.},
  \bibinfo{author}{Zou, B.}, \bibinfo{year}{2019}e.
\newblock \bibinfo{title}{A spatial-aware joint optic disc and cup segmentation
  method}.
\newblock \bibinfo{journal}{Neurocomputing} \bibinfo{volume}{359},
  \bibinfo{pages}{285--297}.
\newblock \URLprefix \url{https://doi.org/10.1016/j.neucom.2019.05.039},
  \DOIprefix\doi{10.1016/j.neucom.2019.05.039}.
\bibitem[{Liu et~al.(2017)Liu, Cheng, Hu, Wang and
  Bai}]{DBLP:conf/cvpr/LiuCHWB17}
\bibinfo{author}{Liu, Y.}, \bibinfo{author}{Cheng, M.}, \bibinfo{author}{Hu,
  X.}, \bibinfo{author}{Wang, K.}, \bibinfo{author}{Bai, X.},
  \bibinfo{year}{2017}.
\newblock \bibinfo{title}{Richer convolutional features for edge detection},
  in: \bibinfo{booktitle}{2017 {IEEE} Conference on Computer Vision and Pattern
  Recognition, {CVPR} 2017, Honolulu, HI, USA, July 21-26, 2017},
  \bibinfo{publisher}{{IEEE} Computer Society}. pp.
  \bibinfo{pages}{5872--5881}.
\newblock \URLprefix \url{https://doi.org/10.1109/CVPR.2017.622},
  \DOIprefix\doi{10.1109/CVPR.2017.622}.
\bibitem[{Long et~al.(2015)Long, Shelhamer and
  Darrell}]{DBLP:conf/cvpr/LongSD15}
\bibinfo{author}{Long, J.}, \bibinfo{author}{Shelhamer, E.},
  \bibinfo{author}{Darrell, T.}, \bibinfo{year}{2015}.
\newblock \bibinfo{title}{Fully convolutional networks for semantic
  segmentation}, in: \bibinfo{booktitle}{{IEEE} Conference on Computer Vision
  and Pattern Recognition, {CVPR} 2015, Boston, MA, USA, June 7-12, 2015},
  \bibinfo{publisher}{{IEEE} Computer Society}. pp.
  \bibinfo{pages}{3431--3440}.
\newblock \URLprefix \url{https://doi.org/10.1109/CVPR.2015.7298965},
  \DOIprefix\doi{10.1109/CVPR.2015.7298965}.
\bibitem[{Lowell et~al.(2004)Lowell, Hunter, Steel, Basu, Ryder, Fletcher and
  Kennedy}]{DBLP:journals/tmi/LowellHSBRFK04}
\bibinfo{author}{Lowell, J.}, \bibinfo{author}{Hunter, A.},
  \bibinfo{author}{Steel, D.}, \bibinfo{author}{Basu, A.},
  \bibinfo{author}{Ryder, R.}, \bibinfo{author}{Fletcher, E.},
  \bibinfo{author}{Kennedy, L.}, \bibinfo{year}{2004}.
\newblock \bibinfo{title}{Optic nerve head segmentation}.
\newblock \bibinfo{journal}{{IEEE} Trans. Medical Imaging}
  \bibinfo{volume}{23}, \bibinfo{pages}{256--264}.
\newblock \URLprefix \url{https://doi.org/10.1109/TMI.2003.823261},
  \DOIprefix\doi{10.1109/TMI.2003.823261}.
\bibitem[{Luo et~al.(2016)Luo, Li, Urtasun and Zemel}]{DBLP:conf/nips/LuoLUZ16}
\bibinfo{author}{Luo, W.}, \bibinfo{author}{Li, Y.}, \bibinfo{author}{Urtasun,
  R.}, \bibinfo{author}{Zemel, R.S.}, \bibinfo{year}{2016}.
\newblock \bibinfo{title}{Understanding the effective receptive field in deep
  convolutional neural networks}, in: \bibinfo{editor}{Lee, D.D.},
  \bibinfo{editor}{Sugiyama, M.}, \bibinfo{editor}{von Luxburg, U.},
  \bibinfo{editor}{Guyon, I.}, \bibinfo{editor}{Garnett, R.} (Eds.),
  \bibinfo{booktitle}{Advances in Neural Information Processing Systems 29:
  Annual Conference on Neural Information Processing Systems 2016, December
  5-10, 2016, Barcelona, Spain}, pp. \bibinfo{pages}{4898--4906}.
\newblock \URLprefix
  \url{http://papers.nips.cc/paper/6203-understanding-the-effective-receptive-field-in-deep-convolutional-neural-networks}.
\bibitem[{Ma et~al.(2019)Ma, Yu, Ma, Wang, Ding and
  Zheng}]{DBLP:conf/miccai/MaYMWDZ19}
\bibinfo{author}{Ma, W.}, \bibinfo{author}{Yu, S.}, \bibinfo{author}{Ma, K.},
  \bibinfo{author}{Wang, J.}, \bibinfo{author}{Ding, X.},
  \bibinfo{author}{Zheng, Y.}, \bibinfo{year}{2019}.
\newblock \bibinfo{title}{Multi-task neural networks with spatial activation
  for retinal vessel segmentation and artery/vein classification}, in:
  \bibinfo{editor}{Shen, D.}, \bibinfo{editor}{Liu, T.},
  \bibinfo{editor}{Peters, T.M.}, \bibinfo{editor}{Staib, L.H.},
  \bibinfo{editor}{Essert, C.}, \bibinfo{editor}{Zhou, S.},
  \bibinfo{editor}{Yap, P.}, \bibinfo{editor}{Khan, A.} (Eds.),
  \bibinfo{booktitle}{Medical Image Computing and Computer Assisted
  Intervention - {MICCAI} 2019 - 22nd International Conference, Shenzhen,
  China, October 13-17, 2019, Proceedings, Part {I}},
  \bibinfo{publisher}{Springer}. pp. \bibinfo{pages}{769--778}.
\newblock \URLprefix \url{https://doi.org/10.1007/978-3-030-32239-7\_85},
  \DOIprefix\doi{10.1007/978-3-030-32239-7\_85}.
\bibitem[{Mahapatra et~al.(2017)Mahapatra, Bozorgtabar, Hewavitharanage and
  Garnavi}]{DBLP:conf/miccai/MahapatraBHG17}
\bibinfo{author}{Mahapatra, D.}, \bibinfo{author}{Bozorgtabar, B.},
  \bibinfo{author}{Hewavitharanage, S.}, \bibinfo{author}{Garnavi, R.},
  \bibinfo{year}{2017}.
\newblock \bibinfo{title}{Image super resolution using generative adversarial
  networks and local saliency maps for retinal image analysis}, in:
  \bibinfo{editor}{Descoteaux, M.}, \bibinfo{editor}{Maier{-}Hein, L.},
  \bibinfo{editor}{Franz, A.M.}, \bibinfo{editor}{Jannin, P.},
  \bibinfo{editor}{Collins, D.L.}, \bibinfo{editor}{Duchesne, S.} (Eds.),
  \bibinfo{booktitle}{Medical Image Computing and Computer Assisted
  Intervention - {MICCAI} 2017 - 20th International Conference, Quebec City,
  QC, Canada, September 11-13, 2017, Proceedings, Part {III}},
  \bibinfo{publisher}{Springer}. pp. \bibinfo{pages}{382--390}.
\newblock \URLprefix \url{https://doi.org/10.1007/978-3-319-66179-7\_44},
  \DOIprefix\doi{10.1007/978-3-319-66179-7\_44}.
\bibitem[{Maninis et~al.(2016)Maninis, Pont{-}Tuset, Arbel{\'{a}}ez and
  Gool}]{DBLP:conf/miccai/ManinisPAG16}
\bibinfo{author}{Maninis, K.}, \bibinfo{author}{Pont{-}Tuset, J.},
  \bibinfo{author}{Arbel{\'{a}}ez, P.A.}, \bibinfo{author}{Gool, L.V.},
  \bibinfo{year}{2016}.
\newblock \bibinfo{title}{Deep retinal image understanding}, in:
  \bibinfo{editor}{Ourselin, S.}, \bibinfo{editor}{Joskowicz, L.},
  \bibinfo{editor}{Sabuncu, M.R.}, \bibinfo{editor}{{\"{U}}nal, G.B.},
  \bibinfo{editor}{Wells, W.} (Eds.), \bibinfo{booktitle}{Medical Image
  Computing and Computer-Assisted Intervention - {MICCAI} 2016 - 19th
  International Conference, Athens, Greece, October 17-21, 2016, Proceedings,
  Part {II}}, pp. \bibinfo{pages}{140--148}.
\newblock \URLprefix \url{https://doi.org/10.1007/978-3-319-46723-8\_17},
  \DOIprefix\doi{10.1007/978-3-319-46723-8\_17}.
\bibitem[{Massin et~al.(2008)Massin, Chabouis, Erginay, Viens-Bitker,
  Lecleire-Collet, Meas, Guillausseau, Choupot, André and
  Denormandie}]{MASSIN2008227}
\bibinfo{author}{Massin, P.}, \bibinfo{author}{Chabouis, A.},
  \bibinfo{author}{Erginay, A.}, \bibinfo{author}{Viens-Bitker, C.},
  \bibinfo{author}{Lecleire-Collet, A.}, \bibinfo{author}{Meas, T.},
  \bibinfo{author}{Guillausseau, P.J.}, \bibinfo{author}{Choupot, G.},
  \bibinfo{author}{André, B.}, \bibinfo{author}{Denormandie, P.},
  \bibinfo{year}{2008}.
\newblock \bibinfo{title}{Ophdiat: A telemedical network screening system for
  diabetic retinopathy in the ile-de-france}.
\newblock \bibinfo{journal}{Diabetes \& Metabolism} \bibinfo{volume}{34},
  \bibinfo{pages}{227 -- 234}.
\newblock \URLprefix
  \url{http://www.sciencedirect.com/science/article/pii/S1262363608000426},
  \DOIprefix\doi{https://doi.org/10.1016/j.diabet.2007.12.006}.
\bibitem[{Mathis~Antony(2015)}]{oO-Solution}
\bibinfo{author}{Mathis~Antony, d.}, \bibinfo{year}{2015}.
\newblock \bibinfo{title}{Team o\_o solution summary}.
\newblock
  \bibinfo{howpublished}{\url{https://www.kaggle.com/c/diabetic-retinopathy-detection/discussion/15617}}.
\bibitem[{Meng et~al.(2020)Meng, Hashimoto and Satoh}]{DBLP:conf/isbi/MengHS20}
\bibinfo{author}{Meng, Q.}, \bibinfo{author}{Hashimoto, Y.},
  \bibinfo{author}{Satoh, S.}, \bibinfo{year}{2020}.
\newblock \bibinfo{title}{How to extract more information with less burden:
  Fundus image classification and retinal disease localization with
  ophthalmologist intervention}, in: \bibinfo{booktitle}{17th {IEEE}
  International Symposium on Biomedical Imaging, {ISBI} 2020, Iowa City, IA,
  USA, April 3-7, 2020}, \bibinfo{publisher}{{IEEE}}. pp.
  \bibinfo{pages}{1373--1377}.
\newblock \URLprefix \url{https://doi.org/10.1109/ISBI45749.2020.9098600},
  \DOIprefix\doi{10.1109/ISBI45749.2020.9098600}.
\bibitem[{Meyer et~al.(2018)Meyer, Galdran, Mendon{\c{c}}a and
  Campilho}]{DBLP:conf/miccai/MeyerGMC18}
\bibinfo{author}{Meyer, M.I.}, \bibinfo{author}{Galdran, A.},
  \bibinfo{author}{Mendon{\c{c}}a, A.M.}, \bibinfo{author}{Campilho, A.},
  \bibinfo{year}{2018}.
\newblock \bibinfo{title}{A pixel-wise distance regression approach for joint
  retinal optical disc and fovea detection}, in: \bibinfo{editor}{Frangi,
  A.F.}, \bibinfo{editor}{Schnabel, J.A.}, \bibinfo{editor}{Davatzikos, C.},
  \bibinfo{editor}{Alberola{-}L{\'{o}}pez, C.}, \bibinfo{editor}{Fichtinger,
  G.} (Eds.), \bibinfo{booktitle}{Medical Image Computing and Computer Assisted
  Intervention - {MICCAI} 2018 - 21st International Conference, Granada, Spain,
  September 16-20, 2018, Proceedings, Part {II}},
  \bibinfo{publisher}{Springer}. pp. \bibinfo{pages}{39--47}.
\newblock \URLprefix \url{https://doi.org/10.1007/978-3-030-00934-2\_5},
  \DOIprefix\doi{10.1007/978-3-030-00934-2\_5}.
\bibitem[{Mishra et~al.(2020)Mishra, Chen and Hu}]{DBLP:conf/isbi/MishraCH20}
\bibinfo{author}{Mishra, S.}, \bibinfo{author}{Chen, D.Z.},
  \bibinfo{author}{Hu, X.S.}, \bibinfo{year}{2020}.
\newblock \bibinfo{title}{A data-aware deep supervised method for retinal
  vessel segmentation}, in: \bibinfo{booktitle}{17th {IEEE} International
  Symposium on Biomedical Imaging, {ISBI} 2020, Iowa City, IA, USA, April 3-7,
  2020}, \bibinfo{publisher}{{IEEE}}. pp. \bibinfo{pages}{1254--1257}.
\newblock \URLprefix \url{https://doi.org/10.1109/ISBI45749.2020.9098403},
  \DOIprefix\doi{10.1109/ISBI45749.2020.9098403}.
\bibitem[{Mo et~al.(2018)Mo, Zhang and Feng}]{DBLP:journals/ijon/MoZF18}
\bibinfo{author}{Mo, J.}, \bibinfo{author}{Zhang, L.}, \bibinfo{author}{Feng,
  Y.}, \bibinfo{year}{2018}.
\newblock \bibinfo{title}{Exudate-based diabetic macular edema recognition in
  retinal images using cascaded deep residual networks}.
\newblock \bibinfo{journal}{Neurocomputing} \bibinfo{volume}{290},
  \bibinfo{pages}{161--171}.
\newblock \URLprefix \url{https://doi.org/10.1016/j.neucom.2018.02.035},
  \DOIprefix\doi{10.1016/j.neucom.2018.02.035}.
\bibitem[{Moccia et~al.(2018)Moccia, Momi, Hadji and
  Mattos}]{DBLP:journals/cmpb/MocciaMHM18}
\bibinfo{author}{Moccia, S.}, \bibinfo{author}{Momi, E.D.},
  \bibinfo{author}{Hadji, S.E.}, \bibinfo{author}{Mattos, L.S.},
  \bibinfo{year}{2018}.
\newblock \bibinfo{title}{Blood vessel segmentation algorithms - review of
  methods, datasets and evaluation metrics}.
\newblock \bibinfo{journal}{Comput. Methods Programs Biomed.}
  \bibinfo{volume}{158}, \bibinfo{pages}{71--91}.
\newblock \URLprefix \url{https://doi.org/10.1016/j.cmpb.2018.02.001},
  \DOIprefix\doi{10.1016/j.cmpb.2018.02.001}.
\bibitem[{Mohan et~al.(2018)Mohan, Kumar and
  Seelamantula}]{DBLP:conf/icip/MohanKS18}
\bibinfo{author}{Mohan, D.}, \bibinfo{author}{Kumar, J.R.H.},
  \bibinfo{author}{Seelamantula, C.S.}, \bibinfo{year}{2018}.
\newblock \bibinfo{title}{High-performance optic disc segmentation using
  convolutional neural networks}, in: \bibinfo{booktitle}{2018 {IEEE}
  International Conference on Image Processing, {ICIP} 2018, Athens, Greece,
  October 7-10, 2018}, \bibinfo{publisher}{{IEEE}}. pp.
  \bibinfo{pages}{4038--4042}.
\newblock \URLprefix \url{https://doi.org/10.1109/ICIP.2018.8451543},
  \DOIprefix\doi{10.1109/ICIP.2018.8451543}.
\bibitem[{Mohan et~al.(2019)Mohan, Kumar and
  Seelamantula}]{DBLP:conf/icip/MohanKS19}
\bibinfo{author}{Mohan, D.}, \bibinfo{author}{Kumar, J.R.H.},
  \bibinfo{author}{Seelamantula, C.S.}, \bibinfo{year}{2019}.
\newblock \bibinfo{title}{Optic disc segmentation using cascaded
  multiresolution convolutional neural networks}, in: \bibinfo{booktitle}{2019
  {IEEE} International Conference on Image Processing, {ICIP} 2019, Taipei,
  Taiwan, September 22-25, 2019}, \bibinfo{publisher}{{IEEE}}. pp.
  \bibinfo{pages}{834--838}.
\newblock \URLprefix \url{https://doi.org/10.1109/ICIP.2019.8804267},
  \DOIprefix\doi{10.1109/ICIP.2019.8804267}.
\bibitem[{Nasery et~al.(2020)Nasery, Soundararajan and
  Galeotti}]{DBLP:conf/isbi/NaserySG20}
\bibinfo{author}{Nasery, V.}, \bibinfo{author}{Soundararajan, K.B.},
  \bibinfo{author}{Galeotti, J.M.}, \bibinfo{year}{2020}.
\newblock \bibinfo{title}{Learning to segment vessels from poorly illuminated
  fundus images}, in: \bibinfo{booktitle}{17th {IEEE} International Symposium
  on Biomedical Imaging, {ISBI} 2020, Iowa City, IA, USA, April 3-7, 2020},
  \bibinfo{publisher}{{IEEE}}. pp. \bibinfo{pages}{1232--1236}.
\newblock \URLprefix \url{https://doi.org/10.1109/ISBI45749.2020.9098694},
  \DOIprefix\doi{10.1109/ISBI45749.2020.9098694}.
\bibitem[{Natarajan et~al.(2019)Natarajan, Jain, Krishnan, Rogye and
  Sivaprasad}]{10.1001/jamaophthalmol.2019.2923}
\bibinfo{author}{Natarajan, S.}, \bibinfo{author}{Jain, A.},
  \bibinfo{author}{Krishnan, R.}, \bibinfo{author}{Rogye, A.},
  \bibinfo{author}{Sivaprasad, S.}, \bibinfo{year}{2019}.
\newblock \bibinfo{title}{{Diagnostic Accuracy of Community-Based Diabetic
  Retinopathy Screening With an Offline Artificial Intelligence System on a
  Smartphone}}.
\newblock \bibinfo{journal}{JAMA Ophthalmology} \bibinfo{volume}{137},
  \bibinfo{pages}{1182--1188}.
\newblock \URLprefix \url{https://doi.org/10.1001/jamaophthalmol.2019.2923},
  \DOIprefix\doi{10.1001/jamaophthalmol.2019.2923}.
\bibitem[{Nguyen et~al.(2016)Nguyen, Tan, Tapp, Mital, Ting, Wong, Tan, Laude,
  Tai and Tan}]{2016Cost}
\bibinfo{author}{Nguyen, H.V.}, \bibinfo{author}{Tan, G.S.W.},
  \bibinfo{author}{Tapp, R.J.}, \bibinfo{author}{Mital, S.},
  \bibinfo{author}{Ting, D.S.W.}, \bibinfo{author}{Wong, H.T.},
  \bibinfo{author}{Tan, C.S.}, \bibinfo{author}{Laude, A.},
  \bibinfo{author}{Tai, E.S.}, \bibinfo{author}{Tan, N.C.a.},
  \bibinfo{year}{2016}.
\newblock \bibinfo{title}{Cost-effectiveness of a national telemedicine
  diabetic retinopathy screening program in singapore}.
\newblock \bibinfo{journal}{Ophthalmology} , \bibinfo{pages}{2571--2580}.
\bibitem[{Niemeijer et~al.(2010)Niemeijer, van Ginneken, Cree, Mizutani,
  Quellec, S{\'{a}}nchez, Zhang, Hornero, Lamard, Muramatsu, Wu, Cazuguel, You,
  Mayo, Li, Hatanaka, Cochener, Roux, Karray, Garc{\'{\i}}a, Fujita and
  Abr{\`{a}}moff}]{DBLP:journals/tmi/NiemeijerGCMQSZHLMWCYMLHCRKGFA10}
\bibinfo{author}{Niemeijer, M.}, \bibinfo{author}{van Ginneken, B.},
  \bibinfo{author}{Cree, M.J.}, \bibinfo{author}{Mizutani, A.},
  \bibinfo{author}{Quellec, G.}, \bibinfo{author}{S{\'{a}}nchez, C.I.},
  \bibinfo{author}{Zhang, B.}, \bibinfo{author}{Hornero, R.},
  \bibinfo{author}{Lamard, M.}, \bibinfo{author}{Muramatsu, C.},
  \bibinfo{author}{Wu, X.}, \bibinfo{author}{Cazuguel, G.},
  \bibinfo{author}{You, J.}, \bibinfo{author}{Mayo, A.}, \bibinfo{author}{Li,
  Q.}, \bibinfo{author}{Hatanaka, Y.}, \bibinfo{author}{Cochener, B.},
  \bibinfo{author}{Roux, C.}, \bibinfo{author}{Karray, F.},
  \bibinfo{author}{Garc{\'{\i}}a, M.}, \bibinfo{author}{Fujita, H.},
  \bibinfo{author}{Abr{\`{a}}moff, M.D.}, \bibinfo{year}{2010}.
\newblock \bibinfo{title}{Retinopathy online challenge: Automatic detection of
  microaneurysms in digital color fundus photographs}.
\newblock \bibinfo{journal}{{IEEE} Trans. Medical Imaging}
  \bibinfo{volume}{29}, \bibinfo{pages}{185--195}.
\newblock \URLprefix \url{https://doi.org/10.1109/TMI.2009.2033909},
  \DOIprefix\doi{10.1109/TMI.2009.2033909}.
\bibitem[{Niemeijer et~al.(2011)Niemeijer, Xu, Dumitrescu, Gupta, van Ginneken,
  Folk and Abr{\`{a}}moff}]{DBLP:journals/tmi/NiemeijerXDGGFA11}
\bibinfo{author}{Niemeijer, M.}, \bibinfo{author}{Xu, X.},
  \bibinfo{author}{Dumitrescu, A.V.}, \bibinfo{author}{Gupta, P.},
  \bibinfo{author}{van Ginneken, B.}, \bibinfo{author}{Folk, J.C.},
  \bibinfo{author}{Abr{\`{a}}moff, M.D.}, \bibinfo{year}{2011}.
\newblock \bibinfo{title}{Automated measurement of the arteriolar-to-venular
  width ratio in digital color fundus photographs}.
\newblock \bibinfo{journal}{{IEEE} Trans. Med. Imaging} \bibinfo{volume}{30},
  \bibinfo{pages}{1941--1950}.
\newblock \URLprefix \url{https://doi.org/10.1109/TMI.2011.2159619},
  \DOIprefix\doi{10.1109/TMI.2011.2159619}.
\bibitem[{Niu et~al.(2019)Niu, Gu, Lu, Lv, Wang, Sato, Zhang, Xiao, Dai and
  Cheng}]{DBLP:conf/aaai/NiuG0LWSZXDC19}
\bibinfo{author}{Niu, Y.}, \bibinfo{author}{Gu, L.}, \bibinfo{author}{Lu, F.},
  \bibinfo{author}{Lv, F.}, \bibinfo{author}{Wang, Z.}, \bibinfo{author}{Sato,
  I.}, \bibinfo{author}{Zhang, Z.}, \bibinfo{author}{Xiao, Y.},
  \bibinfo{author}{Dai, X.}, \bibinfo{author}{Cheng, T.}, \bibinfo{year}{2019}.
\newblock \bibinfo{title}{Pathological evidence exploration in deep retinal
  image diagnosis}, in: \bibinfo{booktitle}{The Thirty-Third {AAAI} Conference
  on Artificial Intelligence, {AAAI} 2019, The Thirty-First Innovative
  Applications of Artificial Intelligence Conference, {IAAI} 2019, The Ninth
  {AAAI} Symposium on Educational Advances in Artificial Intelligence, {EAAI}
  2019, Honolulu, Hawaii, USA, January 27 - February 1, 2019},
  \bibinfo{publisher}{{AAAI} Press}. pp. \bibinfo{pages}{1093--1101}.
\newblock \URLprefix \url{https://doi.org/10.1609/aaai.v33i01.33011093},
  \DOIprefix\doi{10.1609/aaai.v33i01.33011093}.
\bibitem[{Oliveira et~al.(2018)Oliveira, Pereira and
  Silva}]{DBLP:journals/eswa/OliveiraPS18}
\bibinfo{author}{Oliveira, A.}, \bibinfo{author}{Pereira, S.},
  \bibinfo{author}{Silva, C.A.}, \bibinfo{year}{2018}.
\newblock \bibinfo{title}{Retinal vessel segmentation based on fully
  convolutional neural networks}.
\newblock \bibinfo{journal}{Expert Syst. Appl.} \bibinfo{volume}{112},
  \bibinfo{pages}{229--242}.
\newblock \URLprefix \url{https://doi.org/10.1016/j.eswa.2018.06.034},
  \DOIprefix\doi{10.1016/j.eswa.2018.06.034}.
\bibitem[{Orlando et~al.(2018)Orlando, Breda, van Keer, Blaschko, Blanco and
  Bulant}]{DBLP:conf/miccai/OrlandoBKBBB18}
\bibinfo{author}{Orlando, J.I.}, \bibinfo{author}{Breda, J.B.},
  \bibinfo{author}{van Keer, K.}, \bibinfo{author}{Blaschko, M.B.},
  \bibinfo{author}{Blanco, P.J.}, \bibinfo{author}{Bulant, C.A.},
  \bibinfo{year}{2018}.
\newblock \bibinfo{title}{Towards a glaucoma risk index based on simulated
  hemodynamics from fundus images}, in: \bibinfo{editor}{Frangi, A.F.},
  \bibinfo{editor}{Schnabel, J.A.}, \bibinfo{editor}{Davatzikos, C.},
  \bibinfo{editor}{Alberola{-}L{\'{o}}pez, C.}, \bibinfo{editor}{Fichtinger,
  G.} (Eds.), \bibinfo{booktitle}{Medical Image Computing and Computer Assisted
  Intervention - {MICCAI} 2018 - 21st International Conference, Granada, Spain,
  September 16-20, 2018, Proceedings, Part {II}},
  \bibinfo{publisher}{Springer}. pp. \bibinfo{pages}{65--73}.
\newblock \URLprefix \url{https://doi.org/10.1007/978-3-030-00934-2\_8},
  \DOIprefix\doi{10.1007/978-3-030-00934-2\_8}.
\bibitem[{Orlando et~al.(2020)Orlando, Fu, Breda, van Keer, Bathula,
  Diaz{-}Pinto, Fang, Heng, Kim, Lee, Lee, Li, Liu, Lu, Murugesan, Naranjo,
  Phaye, Shankaranarayana and Bogunovic}]{DBLP:journals/mia/OrlandoFBKBDFHK20}
\bibinfo{author}{Orlando, J.I.}, \bibinfo{author}{Fu, H.},
  \bibinfo{author}{Breda, J.B.}, \bibinfo{author}{van Keer, K.},
  \bibinfo{author}{Bathula, D.R.}, \bibinfo{author}{Diaz{-}Pinto, A.},
  \bibinfo{author}{Fang, R.}, \bibinfo{author}{Heng, P.}, \bibinfo{author}{Kim,
  J.}, \bibinfo{author}{Lee, J.}, \bibinfo{author}{Lee, J.},
  \bibinfo{author}{Li, X.}, \bibinfo{author}{Liu, P.}, \bibinfo{author}{Lu,
  S.}, \bibinfo{author}{Murugesan, B.}, \bibinfo{author}{Naranjo, V.},
  \bibinfo{author}{Phaye, S.S.R.}, \bibinfo{author}{Shankaranarayana, S.M.},
  \bibinfo{author}{Bogunovic, H.}, \bibinfo{year}{2020}.
\newblock \bibinfo{title}{{REFUGE} challenge: {A} unified framework for
  evaluating automated methods for glaucoma assessment from fundus
  photographs}.
\newblock \bibinfo{journal}{Medical Image Anal.} \bibinfo{volume}{59}.
\newblock \URLprefix \url{https://doi.org/10.1016/j.media.2019.101570},
  \DOIprefix\doi{10.1016/j.media.2019.101570}.
\bibitem[{Owen et~al.(2009)Owen, Rudnicka, Mullen, Barman, Monekosso, Whincup,
  Ng and Paterson}]{owen2009measuring}
\bibinfo{author}{Owen, C.G.}, \bibinfo{author}{Rudnicka, A.R.},
  \bibinfo{author}{Mullen, R.J.}, \bibinfo{author}{Barman, S.},
  \bibinfo{author}{Monekosso, D.}, \bibinfo{author}{Whincup, P.H.},
  \bibinfo{author}{Ng, J.}, \bibinfo{author}{Paterson, C.},
  \bibinfo{year}{2009}.
\newblock \bibinfo{title}{Measuring retinal vessel tortuosity in 10-year-old
  children: validation of the computer-assisted image analysis of the retina
  (caiar) program.}
\newblock \bibinfo{journal}{Investigative Ophthalmology \& Visual Science}
  \bibinfo{volume}{50}, \bibinfo{pages}{2004--2010}.
\bibitem[{Pal et~al.(2018)Pal, Moorthy and Shahina}]{DBLP:conf/icip/PalMS18}
\bibinfo{author}{Pal, A.}, \bibinfo{author}{Moorthy, M.R.},
  \bibinfo{author}{Shahina, A.}, \bibinfo{year}{2018}.
\newblock \bibinfo{title}{G-eyenet: {A} convolutional autoencoding classifier
  framework for the detection of glaucoma from retinal fundus images}, in:
  \bibinfo{booktitle}{2018 {IEEE} International Conference on Image Processing,
  {ICIP} 2018, Athens, Greece, October 7-10, 2018},
  \bibinfo{publisher}{{IEEE}}. pp. \bibinfo{pages}{2775--2779}.
\newblock \URLprefix \url{https://doi.org/10.1109/ICIP.2018.8451029},
  \DOIprefix\doi{10.1109/ICIP.2018.8451029}.
\bibitem[{Peng et~al.(2017)Peng, Zhang, Yu, Luo and
  Sun}]{DBLP:conf/cvpr/PengZYLS17}
\bibinfo{author}{Peng, C.}, \bibinfo{author}{Zhang, X.}, \bibinfo{author}{Yu,
  G.}, \bibinfo{author}{Luo, G.}, \bibinfo{author}{Sun, J.},
  \bibinfo{year}{2017}.
\newblock \bibinfo{title}{Large kernel matters - improve semantic segmentation
  by global convolutional network}, in: \bibinfo{booktitle}{2017 {IEEE}
  Conference on Computer Vision and Pattern Recognition, {CVPR} 2017, Honolulu,
  HI, USA, July 21-26, 2017}, \bibinfo{publisher}{{IEEE} Computer Society}. pp.
  \bibinfo{pages}{1743--1751}.
\newblock \URLprefix \url{https://doi.org/10.1109/CVPR.2017.189},
  \DOIprefix\doi{10.1109/CVPR.2017.189}.
\bibitem[{Peng et~al.(2018)Peng, Dharssi, Chen, Keenan, Agrón, Wong, Chew and
  Lu}]{Peng2018DeepSeeNet}
\bibinfo{author}{Peng, Y.}, \bibinfo{author}{Dharssi, S.},
  \bibinfo{author}{Chen, Q.}, \bibinfo{author}{Keenan, T.D.},
  \bibinfo{author}{Agrón, E.}, \bibinfo{author}{Wong, W.T.},
  \bibinfo{author}{Chew, E.Y.}, \bibinfo{author}{Lu, Z.}, \bibinfo{year}{2018}.
\newblock \bibinfo{title}{Deepseenet: A deep learning model for automated
  classification of patient-based age-related macular degeneration severity
  from color fundus photographs}.
\newblock \bibinfo{journal}{Ophthalmology} .
\bibitem[{Phene et~al.(2019)Phene, Dunn, Hammel, Liu, Krause, Kitade,
  Schaekermann, Sayres, Wu, Bora, Semturs, Misra, Huang, Spitze, Medeiros, Maa,
  Gandhi, Corrado, Peng and Webster}]{PHENE20191627}
\bibinfo{author}{Phene, S.}, \bibinfo{author}{Dunn, R.C.},
  \bibinfo{author}{Hammel, N.}, \bibinfo{author}{Liu, Y.},
  \bibinfo{author}{Krause, J.}, \bibinfo{author}{Kitade, N.},
  \bibinfo{author}{Schaekermann, M.}, \bibinfo{author}{Sayres, R.},
  \bibinfo{author}{Wu, D.J.}, \bibinfo{author}{Bora, A.},
  \bibinfo{author}{Semturs, C.}, \bibinfo{author}{Misra, A.},
  \bibinfo{author}{Huang, A.E.}, \bibinfo{author}{Spitze, A.},
  \bibinfo{author}{Medeiros, F.A.}, \bibinfo{author}{Maa, A.Y.},
  \bibinfo{author}{Gandhi, M.}, \bibinfo{author}{Corrado, G.S.},
  \bibinfo{author}{Peng, L.}, \bibinfo{author}{Webster, D.R.},
  \bibinfo{year}{2019}.
\newblock \bibinfo{title}{Deep learning and glaucoma specialists: The relative
  importance of optic disc features to predict glaucoma referral in fundus
  photographs}.
\newblock \bibinfo{journal}{Ophthalmology} \bibinfo{volume}{126},
  \bibinfo{pages}{1627 -- 1639}.
\newblock \URLprefix
  \url{http://www.sciencedirect.com/science/article/pii/S0161642019318755},
  \DOIprefix\doi{https://doi.org/10.1016/j.ophtha.2019.07.024}.
\bibitem[{Playout et~al.(2018)Playout, Duval and
  Cheriet}]{DBLP:conf/miccai/PlayoutDC18}
\bibinfo{author}{Playout, C.}, \bibinfo{author}{Duval, R.},
  \bibinfo{author}{Cheriet, F.}, \bibinfo{year}{2018}.
\newblock \bibinfo{title}{A multitask learning architecture for simultaneous
  segmentation of bright and red lesions in fundus images}, in:
  \bibinfo{editor}{Frangi, A.F.}, \bibinfo{editor}{Schnabel, J.A.},
  \bibinfo{editor}{Davatzikos, C.}, \bibinfo{editor}{Alberola{-}L{\'{o}}pez,
  C.}, \bibinfo{editor}{Fichtinger, G.} (Eds.), \bibinfo{booktitle}{Medical
  Image Computing and Computer Assisted Intervention - {MICCAI} 2018 - 21st
  International Conference, Granada, Spain, September 16-20, 2018, Proceedings,
  Part {II}}, \bibinfo{publisher}{Springer}. pp. \bibinfo{pages}{101--108}.
\newblock \URLprefix \url{https://doi.org/10.1007/978-3-030-00934-2\_12},
  \DOIprefix\doi{10.1007/978-3-030-00934-2\_12}.
\bibitem[{Playout et~al.(2019)Playout, Duval and
  Cheriet}]{DBLP:journals/tmi/PlayoutDC19}
\bibinfo{author}{Playout, C.}, \bibinfo{author}{Duval, R.},
  \bibinfo{author}{Cheriet, F.}, \bibinfo{year}{2019}.
\newblock \bibinfo{title}{A novel weakly supervised multitask architecture for
  retinal lesions segmentation on fundus images}.
\newblock \bibinfo{journal}{{IEEE} Trans. Med. Imaging} \bibinfo{volume}{38},
  \bibinfo{pages}{2434--2444}.
\newblock \URLprefix \url{https://doi.org/10.1109/TMI.2019.2906319},
  \DOIprefix\doi{10.1109/TMI.2019.2906319}.
\bibitem[{Pohlen et~al.(2017)Pohlen, Hermans, Mathias and
  Leibe}]{DBLP:conf/cvpr/PohlenHML17}
\bibinfo{author}{Pohlen, T.}, \bibinfo{author}{Hermans, A.},
  \bibinfo{author}{Mathias, M.}, \bibinfo{author}{Leibe, B.},
  \bibinfo{year}{2017}.
\newblock \bibinfo{title}{Full-resolution residual networks for semantic
  segmentation in street scenes}, in: \bibinfo{booktitle}{2017 {IEEE}
  Conference on Computer Vision and Pattern Recognition, {CVPR} 2017, Honolulu,
  HI, USA, July 21-26, 2017}, \bibinfo{publisher}{{IEEE} Computer Society}. pp.
  \bibinfo{pages}{3309--3318}.
\newblock \URLprefix \url{https://doi.org/10.1109/CVPR.2017.353},
  \DOIprefix\doi{10.1109/CVPR.2017.353}.
\bibitem[{Poplin et~al.(2018)Poplin, Varadarajan, Blumer, Liu, McConnell,
  Corrado, Peng and Webster}]{Poplin2018PredictionOC}
\bibinfo{author}{Poplin, R.}, \bibinfo{author}{Varadarajan, A.V.},
  \bibinfo{author}{Blumer, K.}, \bibinfo{author}{Liu, Y.},
  \bibinfo{author}{McConnell, M.}, \bibinfo{author}{Corrado, G.},
  \bibinfo{author}{Peng, L.}, \bibinfo{author}{Webster, D.R.},
  \bibinfo{year}{2018}.
\newblock \bibinfo{title}{Prediction of cardiovascular risk factors from
  retinal fundus photographs via deep learning}.
\newblock \bibinfo{journal}{Nature Biomedical Engineering} \bibinfo{volume}{2},
  \bibinfo{pages}{158--164}.
\bibitem[{Porwal et~al.(2018)Porwal, Pachade, Kamble, Kokare, Deshmukh,
  Sahasrabuddhe and M{\'{e}}riaudeau}]{DBLP:journals/data/PorwalPKKDSM18}
\bibinfo{author}{Porwal, P.}, \bibinfo{author}{Pachade, S.},
  \bibinfo{author}{Kamble, R.}, \bibinfo{author}{Kokare, M.},
  \bibinfo{author}{Deshmukh, G.}, \bibinfo{author}{Sahasrabuddhe, V.},
  \bibinfo{author}{M{\'{e}}riaudeau, F.}, \bibinfo{year}{2018}.
\newblock \bibinfo{title}{Indian diabetic retinopathy image dataset (idrid):
  {A} database for diabetic retinopathy screening research}.
\newblock \bibinfo{journal}{Data} \bibinfo{volume}{3}, \bibinfo{pages}{25}.
\newblock \URLprefix \url{https://doi.org/10.3390/data3030025},
  \DOIprefix\doi{10.3390/data3030025}.
\bibitem[{Porwal et~al.(2020)Porwal, Pachade, Kokare, Deshmukh, Son, Bae, Liu,
  Wang, Liu, Gao, Wu, Xiao, Wang, Yin, Wang, Danala, He, Choi and
  M{\'{e}}riaudeau}]{DBLP:journals/mia/PorwalPKDSBLWLG20}
\bibinfo{author}{Porwal, P.}, \bibinfo{author}{Pachade, S.},
  \bibinfo{author}{Kokare, M.}, \bibinfo{author}{Deshmukh, G.},
  \bibinfo{author}{Son, J.}, \bibinfo{author}{Bae, W.}, \bibinfo{author}{Liu,
  L.}, \bibinfo{author}{Wang, J.}, \bibinfo{author}{Liu, X.},
  \bibinfo{author}{Gao, L.}, \bibinfo{author}{Wu, T.}, \bibinfo{author}{Xiao,
  J.}, \bibinfo{author}{Wang, F.}, \bibinfo{author}{Yin, B.},
  \bibinfo{author}{Wang, Y.}, \bibinfo{author}{Danala, G.},
  \bibinfo{author}{He, L.}, \bibinfo{author}{Choi, Y.H.},
  \bibinfo{author}{M{\'{e}}riaudeau, F.}, \bibinfo{year}{2020}.
\newblock \bibinfo{title}{Idrid: Diabetic retinopathy - segmentation and
  grading challenge}.
\newblock \bibinfo{journal}{Medical Image Anal.} \bibinfo{volume}{59}.
\newblock \URLprefix \url{https://doi.org/10.1016/j.media.2019.101561},
  \DOIprefix\doi{10.1016/j.media.2019.101561}.
\bibitem[{for Retinopathy~of Prematurity Cooperative~Group(1988)}]{ROPPlus}
\bibinfo{author}{for Retinopathy~of Prematurity Cooperative~Group, T.C.},
  \bibinfo{year}{1988}.
\newblock \bibinfo{title}{Multicenter trial of cryotherapy for retinopathy of
  prematurity: Preliminary results.}
\newblock \bibinfo{journal}{Archives of ophthalmology} \bibinfo{volume}{106},
  \bibinfo{pages}{471--479}.
\bibitem[{Quellec et~al.(2017)Quellec, Charri{\`{e}}re, Boudi, Cochener and
  Lamard}]{DBLP:journals/mia/QuellecCBCL17}
\bibinfo{author}{Quellec, G.}, \bibinfo{author}{Charri{\`{e}}re, K.},
  \bibinfo{author}{Boudi, Y.}, \bibinfo{author}{Cochener, B.},
  \bibinfo{author}{Lamard, M.}, \bibinfo{year}{2017}.
\newblock \bibinfo{title}{Deep image mining for diabetic retinopathy
  screening}.
\newblock \bibinfo{journal}{Medical Image Anal.} \bibinfo{volume}{39},
  \bibinfo{pages}{178--193}.
\newblock \URLprefix \url{https://doi.org/10.1016/j.media.2017.04.012},
  \DOIprefix\doi{10.1016/j.media.2017.04.012}.
\bibitem[{Quellec et~al.(2020)Quellec, Lamard, Conze, Massin and
  Cochener}]{DBLP:journals/mia/QuellecLCMC20}
\bibinfo{author}{Quellec, G.}, \bibinfo{author}{Lamard, M.},
  \bibinfo{author}{Conze, P.}, \bibinfo{author}{Massin, P.},
  \bibinfo{author}{Cochener, B.}, \bibinfo{year}{2020}.
\newblock \bibinfo{title}{Automatic detection of rare pathologies in fundus
  photographs using few-shot learning}.
\newblock \bibinfo{journal}{Medical Image Anal.} \bibinfo{volume}{61},
  \bibinfo{pages}{101660}.
\newblock \URLprefix \url{https://doi.org/10.1016/j.media.2020.101660},
  \DOIprefix\doi{10.1016/j.media.2020.101660}.
\bibitem[{Raghavendra et~al.(2018)Raghavendra, Fujita, Bhandary, Gudigar, Tan
  and Acharya}]{DBLP:journals/isci/RaghavendraFBGT18}
\bibinfo{author}{Raghavendra, U.}, \bibinfo{author}{Fujita, H.},
  \bibinfo{author}{Bhandary, S.V.}, \bibinfo{author}{Gudigar, A.},
  \bibinfo{author}{Tan, J.H.}, \bibinfo{author}{Acharya, U.R.},
  \bibinfo{year}{2018}.
\newblock \bibinfo{title}{Deep convolution neural network for accurate
  diagnosis of glaucoma using digital fundus images}.
\newblock \bibinfo{journal}{Inf. Sci.} \bibinfo{volume}{441},
  \bibinfo{pages}{41--49}.
\newblock \URLprefix \url{https://doi.org/10.1016/j.ins.2018.01.051},
  \DOIprefix\doi{10.1016/j.ins.2018.01.051}.
\bibitem[{Rahimy and Ehsan(2018)}]{Rahimy2018Deep}
\bibinfo{author}{Rahimy}, \bibinfo{author}{Ehsan}, \bibinfo{year}{2018}.
\newblock \bibinfo{title}{Deep learning applications in ophthalmology}.
\newblock \bibinfo{journal}{Current Opinion in Ophthalmology} ,
  \bibinfo{pages}{1}.
\bibitem[{Raj et~al.(2020)Raj, Manjunath, Kumar and
  Seelamantula}]{DBLP:conf/isbi/RajMKS20}
\bibinfo{author}{Raj, P.K.}, \bibinfo{author}{Manjunath, A.},
  \bibinfo{author}{Kumar, J.R.H.}, \bibinfo{author}{Seelamantula, C.S.},
  \bibinfo{year}{2020}.
\newblock \bibinfo{title}{Automatic classification of artery/vein from single
  wavelength fundus images}, in: \bibinfo{booktitle}{17th {IEEE} International
  Symposium on Biomedical Imaging, {ISBI} 2020, Iowa City, IA, USA, April 3-7,
  2020}, \bibinfo{publisher}{{IEEE}}. pp. \bibinfo{pages}{1262--1265}.
\newblock \URLprefix \url{https://doi.org/10.1109/ISBI45749.2020.9098580},
  \DOIprefix\doi{10.1109/ISBI45749.2020.9098580}.
\bibitem[{Redmon and Farhadi(2018)}]{DBLP:journals/corr/abs-1804-02767}
\bibinfo{author}{Redmon, J.}, \bibinfo{author}{Farhadi, A.},
  \bibinfo{year}{2018}.
\newblock \bibinfo{title}{Yolov3: An incremental improvement}.
\newblock \bibinfo{journal}{CoRR} \bibinfo{volume}{abs/1804.02767}.
\newblock \URLprefix \url{http://arxiv.org/abs/1804.02767}.
\bibitem[{Ren et~al.(2015)Ren, He, Girshick and Sun}]{DBLP:conf/nips/RenHGS15}
\bibinfo{author}{Ren, S.}, \bibinfo{author}{He, K.}, \bibinfo{author}{Girshick,
  R.B.}, \bibinfo{author}{Sun, J.}, \bibinfo{year}{2015}.
\newblock \bibinfo{title}{Faster {R-CNN:} towards real-time object detection
  with region proposal networks}, in: \bibinfo{editor}{Cortes, C.},
  \bibinfo{editor}{Lawrence, N.D.}, \bibinfo{editor}{Lee, D.D.},
  \bibinfo{editor}{Sugiyama, M.}, \bibinfo{editor}{Garnett, R.} (Eds.),
  \bibinfo{booktitle}{Advances in Neural Information Processing Systems 28:
  Annual Conference on Neural Information Processing Systems 2015, December
  7-12, 2015, Montreal, Quebec, Canada}, pp. \bibinfo{pages}{91--99}.
\newblock \URLprefix
  \url{http://papers.nips.cc/paper/5638-faster-r-cnn-towards-real-time-object-detection-with-region-proposal-networks}.
\bibitem[{Robinson(2003)}]{article}
\bibinfo{author}{Robinson, B.}, \bibinfo{year}{2003}.
\newblock \bibinfo{title}{Prevalence of asymptomatic eye disease prévalence
  des maladies oculaires asymptomatiques}.
\newblock \bibinfo{journal}{Revue Canadienne D'Optom\'etrie} ,
  \bibinfo{pages}{175--180}.
\bibitem[{Ronneberger et~al.(2015)Ronneberger, Fischer and
  Brox}]{DBLP:conf/miccai/RonnebergerFB15}
\bibinfo{author}{Ronneberger, O.}, \bibinfo{author}{Fischer, P.},
  \bibinfo{author}{Brox, T.}, \bibinfo{year}{2015}.
\newblock \bibinfo{title}{U-net: Convolutional networks for biomedical image
  segmentation}, in: \bibinfo{editor}{Navab, N.}, \bibinfo{editor}{Hornegger,
  J.}, \bibinfo{editor}{III, W.M.W.}, \bibinfo{editor}{Frangi, A.F.} (Eds.),
  \bibinfo{booktitle}{Medical Image Computing and Computer-Assisted
  Intervention - {MICCAI} 2015 - 18th International Conference Munich, Germany,
  October 5 - 9, 2015, Proceedings, Part {III}}, \bibinfo{publisher}{Springer}.
  pp. \bibinfo{pages}{234--241}.
\newblock \URLprefix \url{https://doi.org/10.1007/978-3-319-24574-4\_28},
  \DOIprefix\doi{10.1007/978-3-319-24574-4\_28}.
\bibitem[{Roy et~al.(2017)Roy, Tennakoon, Cao, Sedai, Mahapatra, Maetschke and
  Garnavi}]{DBLP:conf/isbi/RoyTCSMMG17}
\bibinfo{author}{Roy, P.K.}, \bibinfo{author}{Tennakoon, R.B.},
  \bibinfo{author}{Cao, K.}, \bibinfo{author}{Sedai, S.},
  \bibinfo{author}{Mahapatra, D.}, \bibinfo{author}{Maetschke, S.},
  \bibinfo{author}{Garnavi, R.}, \bibinfo{year}{2017}.
\newblock \bibinfo{title}{A novel hybrid approach for severity assessment of
  diabetic retinopathy in colour fundus images}, in: \bibinfo{booktitle}{14th
  {IEEE} International Symposium on Biomedical Imaging, {ISBI} 2017, Melbourne,
  Australia, April 18-21, 2017}, \bibinfo{publisher}{{IEEE}}. pp.
  \bibinfo{pages}{1078--1082}.
\newblock \URLprefix \url{https://doi.org/10.1109/ISBI.2017.7950703},
  \DOIprefix\doi{10.1109/ISBI.2017.7950703}.
\bibitem[{Sabour et~al.(2017)Sabour, Frosst and
  Hinton}]{DBLP:conf/nips/SabourFH17}
\bibinfo{author}{Sabour, S.}, \bibinfo{author}{Frosst, N.},
  \bibinfo{author}{Hinton, G.E.}, \bibinfo{year}{2017}.
\newblock \bibinfo{title}{Dynamic routing between capsules}, in:
  \bibinfo{editor}{Guyon, I.}, \bibinfo{editor}{von Luxburg, U.},
  \bibinfo{editor}{Bengio, S.}, \bibinfo{editor}{Wallach, H.M.},
  \bibinfo{editor}{Fergus, R.}, \bibinfo{editor}{Vishwanathan, S.V.N.},
  \bibinfo{editor}{Garnett, R.} (Eds.), \bibinfo{booktitle}{Advances in Neural
  Information Processing Systems 30: Annual Conference on Neural Information
  Processing Systems 2017, 4-9 December 2017, Long Beach, CA, {USA}}, pp.
  \bibinfo{pages}{3856--3866}.
\newblock \URLprefix
  \url{http://papers.nips.cc/paper/6975-dynamic-routing-between-capsules}.
\bibitem[{Salamat et~al.(2019)Salamat, Missen and
  Rashid}]{DBLP:journals/artmed/SalamatMR19}
\bibinfo{author}{Salamat, N.}, \bibinfo{author}{Missen, M.M.S.},
  \bibinfo{author}{Rashid, A.}, \bibinfo{year}{2019}.
\newblock \bibinfo{title}{Diabetic retinopathy techniques in retinal images:
  {A} review}.
\newblock \bibinfo{journal}{Artif. Intell. Medicine} \bibinfo{volume}{97},
  \bibinfo{pages}{168--188}.
\newblock \URLprefix \url{https://doi.org/10.1016/j.artmed.2018.10.009},
  \DOIprefix\doi{10.1016/j.artmed.2018.10.009}.
\bibitem[{Sandler et~al.(2018)Sandler, Howard, Zhu, Zhmoginov and
  Chen}]{DBLP:journals/corr/abs-1801-04381}
\bibinfo{author}{Sandler, M.}, \bibinfo{author}{Howard, A.G.},
  \bibinfo{author}{Zhu, M.}, \bibinfo{author}{Zhmoginov, A.},
  \bibinfo{author}{Chen, L.}, \bibinfo{year}{2018}.
\newblock \bibinfo{title}{Inverted residuals and linear bottlenecks: Mobile
  networks for classification, detection and segmentation}.
\newblock \bibinfo{journal}{CoRR} \bibinfo{volume}{abs/1801.04381}.
\newblock \URLprefix \url{http://arxiv.org/abs/1801.04381}.
\bibitem[{dos Santos~Ferreira et~al.(2018)dos Santos~Ferreira,
  de~Carvalho~Filho, de~Sousa, Silva and
  Gattass}]{DBLP:journals/eswa/FerreiraFSSG18}
\bibinfo{author}{dos Santos~Ferreira, M.V.},
  \bibinfo{author}{de~Carvalho~Filho, A.O.}, \bibinfo{author}{de~Sousa, A.D.},
  \bibinfo{author}{Silva, A.C.}, \bibinfo{author}{Gattass, M.},
  \bibinfo{year}{2018}.
\newblock \bibinfo{title}{Convolutional neural network and texture
  descriptor-based automatic detection and diagnosis of glaucoma}.
\newblock \bibinfo{journal}{Expert Syst. Appl.} \bibinfo{volume}{110},
  \bibinfo{pages}{250--263}.
\newblock \URLprefix \url{https://doi.org/10.1016/j.eswa.2018.06.010},
  \DOIprefix\doi{10.1016/j.eswa.2018.06.010}.
\bibitem[{Sarhan et~al.(2019)Sarhan, Albarqouni, Yigitsoy, Navab and
  Eslami}]{DBLP:conf/miccai/SarhanAYNE19}
\bibinfo{author}{Sarhan, M.H.}, \bibinfo{author}{Albarqouni, S.},
  \bibinfo{author}{Yigitsoy, M.}, \bibinfo{author}{Navab, N.},
  \bibinfo{author}{Eslami, A.}, \bibinfo{year}{2019}.
\newblock \bibinfo{title}{Multi-scale microaneurysms segmentation using
  embedding triplet loss}, in: \bibinfo{editor}{Shen, D.},
  \bibinfo{editor}{Liu, T.}, \bibinfo{editor}{Peters, T.M.},
  \bibinfo{editor}{Staib, L.H.}, \bibinfo{editor}{Essert, C.},
  \bibinfo{editor}{Zhou, S.}, \bibinfo{editor}{Yap, P.}, \bibinfo{editor}{Khan,
  A.} (Eds.), \bibinfo{booktitle}{Medical Image Computing and Computer Assisted
  Intervention - {MICCAI} 2019 - 22nd International Conference, Shenzhen,
  China, October 13-17, 2019, Proceedings, Part {I}},
  \bibinfo{publisher}{Springer}. pp. \bibinfo{pages}{174--182}.
\newblock \URLprefix \url{https://doi.org/10.1007/978-3-030-32239-7\_20},
  \DOIprefix\doi{10.1007/978-3-030-32239-7\_20}.
\bibitem[{Sayres et~al.(2019)Sayres, Taly, Rahimy, Blumer, Coz, Hammel, Krause,
  Narayanaswamy, Rastegar, Wu, Xu, Barb, Joseph, Shumski, Smith, Sood, Corrado,
  Peng and Webster}]{SAYRES2019552}
\bibinfo{author}{Sayres, R.}, \bibinfo{author}{Taly, A.},
  \bibinfo{author}{Rahimy, E.}, \bibinfo{author}{Blumer, K.},
  \bibinfo{author}{Coz, D.}, \bibinfo{author}{Hammel, N.},
  \bibinfo{author}{Krause, J.}, \bibinfo{author}{Narayanaswamy, A.},
  \bibinfo{author}{Rastegar, Z.}, \bibinfo{author}{Wu, D.},
  \bibinfo{author}{Xu, S.}, \bibinfo{author}{Barb, S.},
  \bibinfo{author}{Joseph, A.}, \bibinfo{author}{Shumski, M.},
  \bibinfo{author}{Smith, J.}, \bibinfo{author}{Sood, A.B.},
  \bibinfo{author}{Corrado, G.S.}, \bibinfo{author}{Peng, L.},
  \bibinfo{author}{Webster, D.R.}, \bibinfo{year}{2019}.
\newblock \bibinfo{title}{Using a deep learning algorithm and integrated
  gradients explanation to assist grading for diabetic retinopathy}.
\newblock \bibinfo{journal}{Ophthalmology} \bibinfo{volume}{126},
  \bibinfo{pages}{552 -- 564}.
\newblock \URLprefix
  \url{http://www.sciencedirect.com/science/article/pii/S0161642018315756},
  \DOIprefix\doi{https://doi.org/10.1016/j.ophtha.2018.11.016}.
\bibitem[{Schmidt-Erfurth et~al.(2018)Schmidt-Erfurth, Sadeghipour, Gerendas,
  Waldstein and Bogunović}]{SCHMIDTERFURTH20181}
\bibinfo{author}{Schmidt-Erfurth, U.}, \bibinfo{author}{Sadeghipour, A.},
  \bibinfo{author}{Gerendas, B.S.}, \bibinfo{author}{Waldstein, S.M.},
  \bibinfo{author}{Bogunović, H.}, \bibinfo{year}{2018}.
\newblock \bibinfo{title}{Artificial intelligence in retina}.
\newblock \bibinfo{journal}{Progress in Retinal and Eye Research}
  \bibinfo{volume}{67}, \bibinfo{pages}{1 -- 29}.
\newblock \URLprefix
  \url{http://www.sciencedirect.com/science/article/pii/S1350946218300119},
  \DOIprefix\doi{https://doi.org/10.1016/j.preteyeres.2018.07.004}.
\bibitem[{Sedai et~al.(2017a)Sedai, Mahapatra, Hewavitharanage, Maetschke and
  Garnavi}]{DBLP:conf/miccai/SedaiMHMG17}
\bibinfo{author}{Sedai, S.}, \bibinfo{author}{Mahapatra, D.},
  \bibinfo{author}{Hewavitharanage, S.}, \bibinfo{author}{Maetschke, S.},
  \bibinfo{author}{Garnavi, R.}, \bibinfo{year}{2017}a.
\newblock \bibinfo{title}{Semi-supervised segmentation of optic cup in retinal
  fundus images using variational autoencoder}, in:
  \bibinfo{editor}{Descoteaux, M.}, \bibinfo{editor}{Maier{-}Hein, L.},
  \bibinfo{editor}{Franz, A.M.}, \bibinfo{editor}{Jannin, P.},
  \bibinfo{editor}{Collins, D.L.}, \bibinfo{editor}{Duchesne, S.} (Eds.),
  \bibinfo{booktitle}{Medical Image Computing and Computer Assisted
  Intervention - {MICCAI} 2017 - 20th International Conference, Quebec City,
  QC, Canada, September 11-13, 2017, Proceedings, Part {II}},
  \bibinfo{publisher}{Springer}. pp. \bibinfo{pages}{75--82}.
\newblock \URLprefix \url{https://doi.org/10.1007/978-3-319-66185-8\_9},
  \DOIprefix\doi{10.1007/978-3-319-66185-8\_9}.
\bibitem[{Sedai et~al.(2017b)Sedai, Tennakoon, Roy, Cao and
  Garnavi}]{DBLP:conf/isbi/SedaiTRCG17}
\bibinfo{author}{Sedai, S.}, \bibinfo{author}{Tennakoon, R.B.},
  \bibinfo{author}{Roy, P.K.}, \bibinfo{author}{Cao, K.},
  \bibinfo{author}{Garnavi, R.}, \bibinfo{year}{2017}b.
\newblock \bibinfo{title}{Multi-stage segmentation of the fovea in retinal
  fundus images using fully convolutional neural networks}, in:
  \bibinfo{booktitle}{14th {IEEE} International Symposium on Biomedical
  Imaging, {ISBI} 2017, Melbourne, Australia, April 18-21, 2017},
  \bibinfo{publisher}{{IEEE}}. pp. \bibinfo{pages}{1083--1086}.
\newblock \URLprefix \url{https://doi.org/10.1109/ISBI.2017.7950704},
  \DOIprefix\doi{10.1109/ISBI.2017.7950704}.
\bibitem[{Selvaraju et~al.(2017)Selvaraju, Cogswell, Das, Vedantam, Parikh and
  Batra}]{DBLP:conf/iccv/SelvarajuCDVPB17}
\bibinfo{author}{Selvaraju, R.R.}, \bibinfo{author}{Cogswell, M.},
  \bibinfo{author}{Das, A.}, \bibinfo{author}{Vedantam, R.},
  \bibinfo{author}{Parikh, D.}, \bibinfo{author}{Batra, D.},
  \bibinfo{year}{2017}.
\newblock \bibinfo{title}{Grad-cam: Visual explanations from deep networks via
  gradient-based localization}, in: \bibinfo{booktitle}{{IEEE} International
  Conference on Computer Vision, {ICCV} 2017, Venice, Italy, October 22-29,
  2017}, \bibinfo{publisher}{{IEEE} Computer Society}. pp.
  \bibinfo{pages}{618--626}.
\newblock \URLprefix \url{https://doi.org/10.1109/ICCV.2017.74},
  \DOIprefix\doi{10.1109/ICCV.2017.74}.
\bibitem[{Sengupta et~al.(2020)Sengupta, Singh, Leopold, Gulati and
  Lakshminarayanan}]{DBLP:journals/artmed/SenguptaSLGL20}
\bibinfo{author}{Sengupta, S.}, \bibinfo{author}{Singh, A.},
  \bibinfo{author}{Leopold, H.A.}, \bibinfo{author}{Gulati, T.},
  \bibinfo{author}{Lakshminarayanan, V.}, \bibinfo{year}{2020}.
\newblock \bibinfo{title}{Ophthalmic diagnosis using deep learning with fundus
  images - {A} critical review}.
\newblock \bibinfo{journal}{Artif. Intell. Medicine} \bibinfo{volume}{102},
  \bibinfo{pages}{101758}.
\newblock \URLprefix \url{https://doi.org/10.1016/j.artmed.2019.101758},
  \DOIprefix\doi{10.1016/j.artmed.2019.101758}.
\bibitem[{Shah et~al.(2019)Shah, Kasukurthi and
  Pande}]{DBLP:conf/isbi/ShahKP19}
\bibinfo{author}{Shah, S.}, \bibinfo{author}{Kasukurthi, N.},
  \bibinfo{author}{Pande, H.}, \bibinfo{year}{2019}.
\newblock \bibinfo{title}{Dynamic region proposal networks for semantic
  segmentation in automated glaucoma screening}, in: \bibinfo{booktitle}{16th
  {IEEE} International Symposium on Biomedical Imaging, {ISBI} 2019, Venice,
  Italy, April 8-11, 2019}, \bibinfo{publisher}{{IEEE}}. pp.
  \bibinfo{pages}{578--582}.
\newblock \URLprefix \url{https://doi.org/10.1109/ISBI.2019.8759171},
  \DOIprefix\doi{10.1109/ISBI.2019.8759171}.
\bibitem[{Shankaranarayana et~al.(2019)Shankaranarayana, Ram, Mitra and
  Sivaprakasam}]{DBLP:journals/titb/Shankaranarayana19}
\bibinfo{author}{Shankaranarayana, S.M.}, \bibinfo{author}{Ram, K.},
  \bibinfo{author}{Mitra, K.}, \bibinfo{author}{Sivaprakasam, M.},
  \bibinfo{year}{2019}.
\newblock \bibinfo{title}{Fully convolutional networks for monocular retinal
  depth estimation and optic disc-cup segmentation}.
\newblock \bibinfo{journal}{{IEEE} J. Biomed. Health Informatics}
  \bibinfo{volume}{23}, \bibinfo{pages}{1417--1426}.
\newblock \URLprefix \url{https://doi.org/10.1109/JBHI.2019.2899403},
  \DOIprefix\doi{10.1109/JBHI.2019.2899403}.
\bibitem[{Shen et~al.(2020)Shen, Sheng, Fang, Li, Dai, Stolte, Qin, Jia and
  Shen}]{DBLP:journals/mia/ShenSFLDSQJS20}
\bibinfo{author}{Shen, Y.}, \bibinfo{author}{Sheng, B.}, \bibinfo{author}{Fang,
  R.}, \bibinfo{author}{Li, H.}, \bibinfo{author}{Dai, L.},
  \bibinfo{author}{Stolte, S.}, \bibinfo{author}{Qin, J.},
  \bibinfo{author}{Jia, W.}, \bibinfo{author}{Shen, D.}, \bibinfo{year}{2020}.
\newblock \bibinfo{title}{Domain-invariant interpretable fundus image quality
  assessment}.
\newblock \bibinfo{journal}{Medical Image Anal.} \bibinfo{volume}{61},
  \bibinfo{pages}{101654}.
\newblock \URLprefix \url{https://doi.org/10.1016/j.media.2020.101654},
  \DOIprefix\doi{10.1016/j.media.2020.101654}.
\bibitem[{Silva et~al.(2009)Silva, Liew, Wong, Chang and Wong}]{2009Retinal}
\bibinfo{author}{Silva, D.A.D.}, \bibinfo{author}{Liew, G.},
  \bibinfo{author}{Wong, M.C.}, \bibinfo{author}{Chang, H.M.},
  \bibinfo{author}{Wong, T.Y.}, \bibinfo{year}{2009}.
\newblock \bibinfo{title}{Retinal vascular caliber and extracranial carotid
  disease in patients with acute ischemic stroke: the multi-centre retinal
  stroke (mcrs) study.}
\newblock \bibinfo{journal}{Stroke; a journal of cerebral circulation}
  \bibinfo{volume}{40}, \bibinfo{pages}{3695--9}.
\bibitem[{Simonyan and Zisserman(2015)}]{DBLP:journals/corr/SimonyanZ14a}
\bibinfo{author}{Simonyan, K.}, \bibinfo{author}{Zisserman, A.},
  \bibinfo{year}{2015}.
\newblock \bibinfo{title}{Very deep convolutional networks for large-scale
  image recognition}, in: \bibinfo{editor}{Bengio, Y.}, \bibinfo{editor}{LeCun,
  Y.} (Eds.), \bibinfo{booktitle}{3rd International Conference on Learning
  Representations, {ICLR} 2015, San Diego, CA, USA, May 7-9, 2015, Conference
  Track Proceedings}.
\newblock \URLprefix \url{http://arxiv.org/abs/1409.1556}.
\bibitem[{Sivaprasad et~al.(2014)Sivaprasad, Arden, Prevost, Crosby-Nwaobi,
  Holmes, Kelly, Murphy, Rubin, Vasconcelos and Hykin}]{2014A}
\bibinfo{author}{Sivaprasad, S.}, \bibinfo{author}{Arden, G.},
  \bibinfo{author}{Prevost, A.T.}, \bibinfo{author}{Crosby-Nwaobi, R.},
  \bibinfo{author}{Holmes, H.}, \bibinfo{author}{Kelly, J.},
  \bibinfo{author}{Murphy, C.}, \bibinfo{author}{Rubin, G.},
  \bibinfo{author}{Vasconcelos, J.}, \bibinfo{author}{Hykin, P.},
  \bibinfo{year}{2014}.
\newblock \bibinfo{title}{A multicentre phase iii randomised controlled
  single-masked clinical trial evaluating the clinical efficacy and safety of
  light-masks at preventing dark-adaptation in the treatment of early diabetic
  macular oedema (cleopatra): study protocol for a randomised controlled
  trial}.
\newblock \bibinfo{journal}{Trials} .
\bibitem[{Sivaswamy et~al.(2014)Sivaswamy, Krishnadas, Joshi, Jain and
  Tabish}]{DBLP:conf/isbi/SivaswamyKJJT14}
\bibinfo{author}{Sivaswamy, J.}, \bibinfo{author}{Krishnadas, S.R.},
  \bibinfo{author}{Joshi, G.D.}, \bibinfo{author}{Jain, M.},
  \bibinfo{author}{Tabish, A.U.S.}, \bibinfo{year}{2014}.
\newblock \bibinfo{title}{Drishti-gs: Retinal image dataset for optic nerve
  head(onh) segmentation}, in: \bibinfo{booktitle}{{IEEE} 11th International
  Symposium on Biomedical Imaging, {ISBI} 2014, April 29 - May 2, 2014,
  Beijing, Chin, Beijing, China}, \bibinfo{publisher}{{IEEE}}. pp.
  \bibinfo{pages}{53--56}.
\newblock \URLprefix \url{https://doi.org/10.1109/ISBI.2014.6867807},
  \DOIprefix\doi{10.1109/ISBI.2014.6867807}.
\bibitem[{Soomro et~al.(2019)Soomro, Afifi, Gao, Hellwich, Zheng and
  Paul}]{DBLP:journals/eswa/SoomroAGHZP19}
\bibinfo{author}{Soomro, T.A.}, \bibinfo{author}{Afifi, A.J.},
  \bibinfo{author}{Gao, J.}, \bibinfo{author}{Hellwich, O.},
  \bibinfo{author}{Zheng, L.}, \bibinfo{author}{Paul, M.},
  \bibinfo{year}{2019}.
\newblock \bibinfo{title}{Strided fully convolutional neural network for
  boosting the sensitivity of retinal blood vessels segmentation}.
\newblock \bibinfo{journal}{Expert Syst. Appl.} \bibinfo{volume}{134},
  \bibinfo{pages}{36--52}.
\newblock \URLprefix \url{https://doi.org/10.1016/j.eswa.2019.05.029},
  \DOIprefix\doi{10.1016/j.eswa.2019.05.029}.
\bibitem[{Springenberg et~al.(2015)Springenberg, Dosovitskiy, Brox and
  Riedmiller}]{DBLP:journals/corr/SpringenbergDBR14}
\bibinfo{author}{Springenberg, J.T.}, \bibinfo{author}{Dosovitskiy, A.},
  \bibinfo{author}{Brox, T.}, \bibinfo{author}{Riedmiller, M.A.},
  \bibinfo{year}{2015}.
\newblock \bibinfo{title}{Striving for simplicity: The all convolutional net},
  in: \bibinfo{editor}{Bengio, Y.}, \bibinfo{editor}{LeCun, Y.} (Eds.),
  \bibinfo{booktitle}{3rd International Conference on Learning Representations,
  {ICLR} 2015, San Diego, CA, USA, May 7-9, 2015, Workshop Track Proceedings}.
\newblock \URLprefix \url{http://arxiv.org/abs/1412.6806}.
\bibitem[{Staal et~al.(2004)Staal, Abr{\`{a}}moff, Niemeijer, Viergever and van
  Ginneken}]{DBLP:journals/tmi/StaalANVG04}
\bibinfo{author}{Staal, J.}, \bibinfo{author}{Abr{\`{a}}moff, M.D.},
  \bibinfo{author}{Niemeijer, M.}, \bibinfo{author}{Viergever, M.A.},
  \bibinfo{author}{van Ginneken, B.}, \bibinfo{year}{2004}.
\newblock \bibinfo{title}{Ridge-based vessel segmentation in color images of
  the retina}.
\newblock \bibinfo{journal}{{IEEE} Trans. Medical Imaging}
  \bibinfo{volume}{23}, \bibinfo{pages}{501--509}.
\newblock \URLprefix \url{https://doi.org/10.1109/TMI.2004.825627},
  \DOIprefix\doi{10.1109/TMI.2004.825627}.
\bibitem[{Szegedy et~al.(2017)Szegedy, Ioffe, Vanhoucke and
  Alemi}]{DBLP:conf/aaai/SzegedyIVA17}
\bibinfo{author}{Szegedy, C.}, \bibinfo{author}{Ioffe, S.},
  \bibinfo{author}{Vanhoucke, V.}, \bibinfo{author}{Alemi, A.A.},
  \bibinfo{year}{2017}.
\newblock \bibinfo{title}{Inception-v4, inception-resnet and the impact of
  residual connections on learning}, in: \bibinfo{editor}{Singh, S.P.},
  \bibinfo{editor}{Markovitch, S.} (Eds.), \bibinfo{booktitle}{Proceedings of
  the Thirty-First {AAAI} Conference on Artificial Intelligence, February 4-9,
  2017, San Francisco, California, {USA}}, \bibinfo{publisher}{{AAAI} Press}.
  pp. \bibinfo{pages}{4278--4284}.
\newblock \URLprefix
  \url{http://aaai.org/ocs/index.php/AAAI/AAAI17/paper/view/14806}.
\bibitem[{Szegedy et~al.(2015)Szegedy, Liu, Jia, Sermanet, Reed, Anguelov,
  Erhan, Vanhoucke and Rabinovich}]{DBLP:conf/cvpr/SzegedyLJSRAEVR15}
\bibinfo{author}{Szegedy, C.}, \bibinfo{author}{Liu, W.}, \bibinfo{author}{Jia,
  Y.}, \bibinfo{author}{Sermanet, P.}, \bibinfo{author}{Reed, S.E.},
  \bibinfo{author}{Anguelov, D.}, \bibinfo{author}{Erhan, D.},
  \bibinfo{author}{Vanhoucke, V.}, \bibinfo{author}{Rabinovich, A.},
  \bibinfo{year}{2015}.
\newblock \bibinfo{title}{Going deeper with convolutions}, in:
  \bibinfo{booktitle}{{IEEE} Conference on Computer Vision and Pattern
  Recognition, {CVPR} 2015, Boston, MA, USA, June 7-12, 2015},
  \bibinfo{publisher}{{IEEE} Computer Society}. pp. \bibinfo{pages}{1--9}.
\newblock \URLprefix \url{https://doi.org/10.1109/CVPR.2015.7298594},
  \DOIprefix\doi{10.1109/CVPR.2015.7298594}.
\bibitem[{Szegedy et~al.(2016)Szegedy, Vanhoucke, Ioffe, Shlens and
  Wojna}]{DBLP:conf/cvpr/SzegedyVISW16}
\bibinfo{author}{Szegedy, C.}, \bibinfo{author}{Vanhoucke, V.},
  \bibinfo{author}{Ioffe, S.}, \bibinfo{author}{Shlens, J.},
  \bibinfo{author}{Wojna, Z.}, \bibinfo{year}{2016}.
\newblock \bibinfo{title}{Rethinking the inception architecture for computer
  vision}, in: \bibinfo{booktitle}{2016 {IEEE} Conference on Computer Vision
  and Pattern Recognition, {CVPR} 2016, Las Vegas, NV, USA, June 27-30, 2016},
  \bibinfo{publisher}{{IEEE} Computer Society}. pp.
  \bibinfo{pages}{2818--2826}.
\newblock \URLprefix \url{https://doi.org/10.1109/CVPR.2016.308},
  \DOIprefix\doi{10.1109/CVPR.2016.308}.
\bibitem[{Tan et~al.(2018)Tan, Bhandary, Sivaprasad, Hagiwara, Bagchi,
  Raghavendra, Rao, Raju, Shetty, Gertych, Chua and
  Acharya}]{DBLP:journals/fgcs/TanBSHBRRRSGCA18}
\bibinfo{author}{Tan, J.H.}, \bibinfo{author}{Bhandary, S.V.},
  \bibinfo{author}{Sivaprasad, S.}, \bibinfo{author}{Hagiwara, Y.},
  \bibinfo{author}{Bagchi, A.}, \bibinfo{author}{Raghavendra, U.},
  \bibinfo{author}{Rao, A.K.}, \bibinfo{author}{Raju, B.},
  \bibinfo{author}{Shetty, N.S.}, \bibinfo{author}{Gertych, A.},
  \bibinfo{author}{Chua, K.C.}, \bibinfo{author}{Acharya, U.R.},
  \bibinfo{year}{2018}.
\newblock \bibinfo{title}{Age-related macular degeneration detection using deep
  convolutional neural network}.
\newblock \bibinfo{journal}{Future Gener. Comput. Syst.} \bibinfo{volume}{87},
  \bibinfo{pages}{127--135}.
\newblock \URLprefix \url{https://doi.org/10.1016/j.future.2018.05.001},
  \DOIprefix\doi{10.1016/j.future.2018.05.001}.
\bibitem[{Tan et~al.(2017)Tan, Fujita, Sivaprasad, Bhandary, Rao, Chua and
  Acharya}]{DBLP:journals/isci/TanFSBRCA17}
\bibinfo{author}{Tan, J.H.}, \bibinfo{author}{Fujita, H.},
  \bibinfo{author}{Sivaprasad, S.}, \bibinfo{author}{Bhandary, S.V.},
  \bibinfo{author}{Rao, A.K.}, \bibinfo{author}{Chua, K.C.},
  \bibinfo{author}{Acharya, U.R.}, \bibinfo{year}{2017}.
\newblock \bibinfo{title}{Automated segmentation of exudates, haemorrhages,
  microaneurysms using single convolutional neural network}.
\newblock \bibinfo{journal}{Inf. Sci.} \bibinfo{volume}{420},
  \bibinfo{pages}{66--76}.
\newblock \URLprefix \url{https://doi.org/10.1016/j.ins.2017.08.050},
  \DOIprefix\doi{10.1016/j.ins.2017.08.050}.
\bibitem[{Tasman et~al.(2006)Tasman, Patz, Mcnamara, Kaiser, Trese and
  Smith}]{Tasman2006Retinopathy}
\bibinfo{author}{Tasman, W.}, \bibinfo{author}{Patz, A.},
  \bibinfo{author}{Mcnamara, J.A.}, \bibinfo{author}{Kaiser, R.S.},
  \bibinfo{author}{Trese, M.T.}, \bibinfo{author}{Smith, B.T.},
  \bibinfo{year}{2006}.
\newblock \bibinfo{title}{Retinopathy of prematurity: The life of a lifetime
  disease}.
\newblock \bibinfo{journal}{American Journal of Ophthalmology}
  \bibinfo{volume}{141}, \bibinfo{pages}{0--174}.
\bibitem[{Taylor et~al.(2019)Taylor, Brown, Gupta, Campbell, Ostmo, Chan, Dy,
  Erdogmus, Ioannidis, Kim, Kalpathy-Cramer, Chiang, for~the Imaging and
  in~Retinopathy~of Prematurity~Consortium}]{10.1001/jamaophthalmol.2019.2433}
\bibinfo{author}{Taylor, S.}, \bibinfo{author}{Brown, J.M.},
  \bibinfo{author}{Gupta, K.}, \bibinfo{author}{Campbell, J.P.},
  \bibinfo{author}{Ostmo, S.}, \bibinfo{author}{Chan, R.V.P.},
  \bibinfo{author}{Dy, J.}, \bibinfo{author}{Erdogmus, D.},
  \bibinfo{author}{Ioannidis, S.}, \bibinfo{author}{Kim, S.J.},
  \bibinfo{author}{Kalpathy-Cramer, J.}, \bibinfo{author}{Chiang, M.F.},
  \bibinfo{author}{for~the Imaging}, \bibinfo{author}{in~Retinopathy~of
  Prematurity~Consortium, I.}, \bibinfo{year}{2019}.
\newblock \bibinfo{title}{{Monitoring Disease Progression With a Quantitative
  Severity Scale for Retinopathy of Prematurity Using Deep Learning}}.
\newblock \bibinfo{journal}{JAMA Ophthalmology} \bibinfo{volume}{137},
  \bibinfo{pages}{1022--1028}.
\newblock \URLprefix \url{https://doi.org/10.1001/jamaophthalmol.2019.2433},
  \DOIprefix\doi{10.1001/jamaophthalmol.2019.2433}.
\bibitem[{Tham et~al.(2014)Tham, Li, Wong, Quigley, Aung and
  Cheng}]{tham2014global}
\bibinfo{author}{Tham, Y.C.}, \bibinfo{author}{Li, X.}, \bibinfo{author}{Wong,
  T.Y.}, \bibinfo{author}{Quigley, H.A.}, \bibinfo{author}{Aung, T.},
  \bibinfo{author}{Cheng, C.}, \bibinfo{year}{2014}.
\newblock \bibinfo{title}{Global prevalence of glaucoma and projections of
  glaucoma burden through 2040 a systematic review and meta-analysis}.
\newblock \bibinfo{journal}{Ophthalmology} \bibinfo{volume}{121},
  \bibinfo{pages}{2081--2090}.
\bibitem[{{The Age-Related Eye Disease Study Research Group}(1999)}]{1999573}
\bibinfo{author}{{The Age-Related Eye Disease Study Research Group}},
  \bibinfo{year}{1999}.
\newblock \bibinfo{title}{The age-related eye disease study (areds): Design
  implications areds report no. 1}.
\newblock \bibinfo{journal}{Controlled Clinical Trials} \bibinfo{volume}{20},
  \bibinfo{pages}{573 -- 600}.
\newblock \URLprefix
  \url{http://www.sciencedirect.com/science/article/pii/S0197245699000318},
  \DOIprefix\doi{https://doi.org/10.1016/S0197-2456(99)00031-8}.
\bibitem[{Ting et~al.(2018a)Ting, Pasquale, Peng, Campbell, Lee, Raman, Tan,
  Schmetterer, Keane and Wong}]{Artificial2018}
\bibinfo{author}{Ting, D.}, \bibinfo{author}{Pasquale, L.},
  \bibinfo{author}{Peng, L.}, \bibinfo{author}{Campbell, J.},
  \bibinfo{author}{Lee, A.}, \bibinfo{author}{Raman, R.}, \bibinfo{author}{Tan,
  G.}, \bibinfo{author}{Schmetterer, L.}, \bibinfo{author}{Keane, P.},
  \bibinfo{author}{Wong, T.}, \bibinfo{year}{2018}a.
\newblock \bibinfo{title}{Artificial intelligence and deep learning in
  ophthalmology}.
\newblock \bibinfo{journal}{British Journal of Ophthalmology}
  \bibinfo{volume}{103}, \bibinfo{pages}{bjophthalmol--2018}.
\newblock \DOIprefix\doi{10.1136/bjophthalmol-2018-313173}.
\bibitem[{Ting et~al.(2019)Ting, Peng, Varadarajan, Keane, Burlina, Chiang,
  Schmetterer, Pasquale, Bressler, Webster, Abramoff and Wong}]{TING2019100759}
\bibinfo{author}{Ting, D.S.}, \bibinfo{author}{Peng, L.},
  \bibinfo{author}{Varadarajan, A.V.}, \bibinfo{author}{Keane, P.A.},
  \bibinfo{author}{Burlina, P.M.}, \bibinfo{author}{Chiang, M.F.},
  \bibinfo{author}{Schmetterer, L.}, \bibinfo{author}{Pasquale, L.R.},
  \bibinfo{author}{Bressler, N.M.}, \bibinfo{author}{Webster, D.R.},
  \bibinfo{author}{Abramoff, M.}, \bibinfo{author}{Wong, T.Y.},
  \bibinfo{year}{2019}.
\newblock \bibinfo{title}{Deep learning in ophthalmology: The technical and
  clinical considerations}.
\newblock \bibinfo{journal}{Progress in Retinal and Eye Research}
  \bibinfo{volume}{72}, \bibinfo{pages}{100759}.
\newblock \URLprefix
  \url{http://www.sciencedirect.com/science/article/pii/S1350946218300909},
  \DOIprefix\doi{https://doi.org/10.1016/j.preteyeres.2019.04.003}.
\bibitem[{Ting et~al.(2017)Ting, Cheung, Lim, Tan, Quang, Gan, Hamzah,
  Garcia-Franco, San~Yeo, Lee, Wong, Sabanayagam, Baskaran, Ibrahim, Tan,
  Finkelstein, Lamoureux, Wong, Bressler, Sivaprasad, Varma, Jonas, He, Cheng,
  Cheung, Aung, Hsu, Lee and Wong}]{10.1001/jama.2017.18152}
\bibinfo{author}{Ting, D.S.W.}, \bibinfo{author}{Cheung, C.Y.L.},
  \bibinfo{author}{Lim, G.}, \bibinfo{author}{Tan, G.S.W.},
  \bibinfo{author}{Quang, N.D.}, \bibinfo{author}{Gan, A.},
  \bibinfo{author}{Hamzah, H.}, \bibinfo{author}{Garcia-Franco, R.},
  \bibinfo{author}{San~Yeo, I.Y.}, \bibinfo{author}{Lee, S.Y.},
  \bibinfo{author}{Wong, E.Y.M.}, \bibinfo{author}{Sabanayagam, C.},
  \bibinfo{author}{Baskaran, M.}, \bibinfo{author}{Ibrahim, F.},
  \bibinfo{author}{Tan, N.C.}, \bibinfo{author}{Finkelstein, E.A.},
  \bibinfo{author}{Lamoureux, E.L.}, \bibinfo{author}{Wong, I.Y.},
  \bibinfo{author}{Bressler, N.M.}, \bibinfo{author}{Sivaprasad, S.},
  \bibinfo{author}{Varma, R.}, \bibinfo{author}{Jonas, J.B.},
  \bibinfo{author}{He, M.G.}, \bibinfo{author}{Cheng, C.Y.},
  \bibinfo{author}{Cheung, G.C.M.}, \bibinfo{author}{Aung, T.},
  \bibinfo{author}{Hsu, W.}, \bibinfo{author}{Lee, M.L.},
  \bibinfo{author}{Wong, T.Y.}, \bibinfo{year}{2017}.
\newblock \bibinfo{title}{{Development and Validation of a Deep Learning System
  for Diabetic Retinopathy and Related Eye Diseases Using Retinal Images From
  Multiethnic Populations With Diabetes}}.
\newblock \bibinfo{journal}{JAMA} \bibinfo{volume}{318},
  \bibinfo{pages}{2211--2223}.
\newblock \URLprefix \url{https://doi.org/10.1001/jama.2017.18152},
  \DOIprefix\doi{10.1001/jama.2017.18152}.
\bibitem[{Ting et~al.(2018b)Ting, Pasquale, Peng, Campbell, Lee, Raman, Tan,
  Schmetterer, Keane and Wong}]{Ting2018Artificial}
\bibinfo{author}{Ting, D.S.W.}, \bibinfo{author}{Pasquale, L.R.},
  \bibinfo{author}{Peng, L.}, \bibinfo{author}{Campbell, J.P.},
  \bibinfo{author}{Lee, A.Y.}, \bibinfo{author}{Raman, R.},
  \bibinfo{author}{Tan, G.S.W.}, \bibinfo{author}{Schmetterer, L.},
  \bibinfo{author}{Keane, P.A.}, \bibinfo{author}{Wong, T.Y.},
  \bibinfo{year}{2018}b.
\newblock \bibinfo{title}{Artificial intelligence and deep learning in
  ophthalmology}.
\newblock \bibinfo{journal}{British Journal of Ophthalmology} .
\bibitem[{Tu et~al.(2020)Tu, Gao, Zhou, Chen, Fu, Gu, Cheng, Yu and
  Liu}]{DBLP:conf/isbi/TuGZCFGCYL20}
\bibinfo{author}{Tu, Z.}, \bibinfo{author}{Gao, S.}, \bibinfo{author}{Zhou,
  K.}, \bibinfo{author}{Chen, X.}, \bibinfo{author}{Fu, H.},
  \bibinfo{author}{Gu, Z.}, \bibinfo{author}{Cheng, J.}, \bibinfo{author}{Yu,
  Z.}, \bibinfo{author}{Liu, J.}, \bibinfo{year}{2020}.
\newblock \bibinfo{title}{Sunet: {A} lesion regularized model for simultaneous
  diabetic retinopathy and diabetic macular edema grading}, in:
  \bibinfo{booktitle}{17th {IEEE} International Symposium on Biomedical
  Imaging, {ISBI} 2020, Iowa City, IA, USA, April 3-7, 2020},
  \bibinfo{publisher}{{IEEE}}. pp. \bibinfo{pages}{1378--1382}.
\newblock \URLprefix \url{https://doi.org/10.1109/ISBI45749.2020.9098673},
  \DOIprefix\doi{10.1109/ISBI45749.2020.9098673}.
\bibitem[{V and Sivaswamy(2019)}]{DBLP:conf/isbi/VS19}
\bibinfo{author}{V, S.A.}, \bibinfo{author}{Sivaswamy, J.},
  \bibinfo{year}{2019}.
\newblock \bibinfo{title}{Matching the characteristics of fundus and smartphone
  camera images}, in: \bibinfo{booktitle}{16th {IEEE} International Symposium
  on Biomedical Imaging, {ISBI} 2019, Venice, Italy, April 8-11, 2019},
  \bibinfo{publisher}{{IEEE}}. pp. \bibinfo{pages}{569--572}.
\newblock \URLprefix \url{https://doi.org/10.1109/ISBI.2019.8759381},
  \DOIprefix\doi{10.1109/ISBI.2019.8759381}.
\bibitem[{Varadarajan et~al.(2017)Varadarajan, Poplin, Blumer,
  Angerm{\"{u}}ller, Ledsam, Chopra, Keane, Corrado, Peng and
  Webster}]{DBLP:journals/corr/abs-1712-07798}
\bibinfo{author}{Varadarajan, A.V.}, \bibinfo{author}{Poplin, R.},
  \bibinfo{author}{Blumer, K.}, \bibinfo{author}{Angerm{\"{u}}ller, C.},
  \bibinfo{author}{Ledsam, J.}, \bibinfo{author}{Chopra, R.},
  \bibinfo{author}{Keane, P.A.}, \bibinfo{author}{Corrado, G.},
  \bibinfo{author}{Peng, L.}, \bibinfo{author}{Webster, D.R.},
  \bibinfo{year}{2017}.
\newblock \bibinfo{title}{Deep learning for predicting refractive error from
  retinal fundus images}.
\newblock \bibinfo{journal}{CoRR} \bibinfo{volume}{abs/1712.07798}.
\newblock \URLprefix \url{http://arxiv.org/abs/1712.07798}.
\bibitem[{Vos et~al.(2016)Vos, Allen, Arora, Barber, Bhutta, Brown, Carter,
  Casey and et~al.}]{20161545}
\bibinfo{author}{Vos, T.}, \bibinfo{author}{Allen, C.}, \bibinfo{author}{Arora,
  M.}, \bibinfo{author}{Barber, R.M.}, \bibinfo{author}{Bhutta, Z.A.},
  \bibinfo{author}{Brown, A.}, \bibinfo{author}{Carter, A.},
  \bibinfo{author}{Casey, D.C.}, \bibinfo{author}{et~al., F.J.C.},
  \bibinfo{year}{2016}.
\newblock \bibinfo{title}{Global, regional, and national incidence, prevalence,
  and years lived with disability for 310 diseases and injuries, 1990–2015: a
  systematic analysis for the global burden of disease study 2015}.
\newblock \bibinfo{journal}{The Lancet} \bibinfo{volume}{388},
  \bibinfo{pages}{1545 -- 1602}.
\newblock \URLprefix
  \url{http://www.sciencedirect.com/science/article/pii/S0140673616316786},
  \DOIprefix\doi{https://doi.org/10.1016/S0140-6736(16)31678-6}.
\bibitem[{Wang et~al.(2019a)Wang, Qiu and He}]{DBLP:conf/miccai/Wang0H19}
\bibinfo{author}{Wang, B.}, \bibinfo{author}{Qiu, S.}, \bibinfo{author}{He,
  H.}, \bibinfo{year}{2019}a.
\newblock \bibinfo{title}{Dual encoding u-net for retinal vessel segmentation},
  in: \bibinfo{editor}{Shen, D.}, \bibinfo{editor}{Liu, T.},
  \bibinfo{editor}{Peters, T.M.}, \bibinfo{editor}{Staib, L.H.},
  \bibinfo{editor}{Essert, C.}, \bibinfo{editor}{Zhou, S.},
  \bibinfo{editor}{Yap, P.}, \bibinfo{editor}{Khan, A.} (Eds.),
  \bibinfo{booktitle}{Medical Image Computing and Computer Assisted
  Intervention - {MICCAI} 2019 - 22nd International Conference, Shenzhen,
  China, October 13-17, 2019, Proceedings, Part {I}},
  \bibinfo{publisher}{Springer}. pp. \bibinfo{pages}{84--92}.
\newblock \URLprefix \url{https://doi.org/10.1007/978-3-030-32239-7\_10},
  \DOIprefix\doi{10.1007/978-3-030-32239-7\_10}.
\bibitem[{Wang et~al.(2020)Wang, Zhang, Huang, Wang and
  Chen}]{DBLP:conf/isbi/00210HWC20}
\bibinfo{author}{Wang, K.}, \bibinfo{author}{Zhang, X.},
  \bibinfo{author}{Huang, S.}, \bibinfo{author}{Wang, Q.},
  \bibinfo{author}{Chen, F.}, \bibinfo{year}{2020}.
\newblock \bibinfo{title}{Ctf-net: Retinal vessel segmentation via deep
  coarse-to-fine supervision network}, in: \bibinfo{booktitle}{17th {IEEE}
  International Symposium on Biomedical Imaging, {ISBI} 2020, Iowa City, IA,
  USA, April 3-7, 2020}, \bibinfo{publisher}{{IEEE}}. pp.
  \bibinfo{pages}{1237--1241}.
\newblock \URLprefix \url{https://doi.org/10.1109/ISBI45749.2020.9098742},
  \DOIprefix\doi{10.1109/ISBI45749.2020.9098742}.
\bibitem[{Wang and Deng(2018)}]{DBLP:journals/ijon/WangD18}
\bibinfo{author}{Wang, M.}, \bibinfo{author}{Deng, W.}, \bibinfo{year}{2018}.
\newblock \bibinfo{title}{Deep visual domain adaptation: {A} survey}.
\newblock \bibinfo{journal}{Neurocomputing} \bibinfo{volume}{312},
  \bibinfo{pages}{135--153}.
\newblock \URLprefix \url{https://doi.org/10.1016/j.neucom.2018.05.083},
  \DOIprefix\doi{10.1016/j.neucom.2018.05.083}.
\bibitem[{Wang et~al.(2019b)Wang, Yu, Li, Yang, Fu and
  Heng}]{DBLP:conf/miccai/WangYLYFH19}
\bibinfo{author}{Wang, S.}, \bibinfo{author}{Yu, L.}, \bibinfo{author}{Li, K.},
  \bibinfo{author}{Yang, X.}, \bibinfo{author}{Fu, C.}, \bibinfo{author}{Heng,
  P.}, \bibinfo{year}{2019}b.
\newblock \bibinfo{title}{Boundary and entropy-driven adversarial learning for
  fundus image segmentation}, in: \bibinfo{editor}{Shen, D.},
  \bibinfo{editor}{Liu, T.}, \bibinfo{editor}{Peters, T.M.},
  \bibinfo{editor}{Staib, L.H.}, \bibinfo{editor}{Essert, C.},
  \bibinfo{editor}{Zhou, S.}, \bibinfo{editor}{Yap, P.}, \bibinfo{editor}{Khan,
  A.} (Eds.), \bibinfo{booktitle}{Medical Image Computing and Computer Assisted
  Intervention - {MICCAI} 2019 - 22nd International Conference, Shenzhen,
  China, October 13-17, 2019, Proceedings, Part {I}},
  \bibinfo{publisher}{Springer}. pp. \bibinfo{pages}{102--110}.
\newblock \URLprefix \url{https://doi.org/10.1007/978-3-030-32239-7\_12},
  \DOIprefix\doi{10.1007/978-3-030-32239-7\_12}.
\bibitem[{Wang et~al.(2019c)Wang, Yu, Yang, Fu and
  Heng}]{DBLP:journals/tmi/WangYYFH19}
\bibinfo{author}{Wang, S.}, \bibinfo{author}{Yu, L.}, \bibinfo{author}{Yang,
  X.}, \bibinfo{author}{Fu, C.}, \bibinfo{author}{Heng, P.},
  \bibinfo{year}{2019}c.
\newblock \bibinfo{title}{Patch-based output space adversarial learning for
  joint optic disc and cup segmentation}.
\newblock \bibinfo{journal}{{IEEE} Trans. Med. Imaging} \bibinfo{volume}{38},
  \bibinfo{pages}{2485--2495}.
\newblock \URLprefix \url{https://doi.org/10.1109/TMI.2019.2899910},
  \DOIprefix\doi{10.1109/TMI.2019.2899910}.
\bibitem[{Wang et~al.(2019d)Wang, Ju, Zhao and Ge}]{DBLP:conf/miccai/WangJZG19}
\bibinfo{author}{Wang, X.}, \bibinfo{author}{Ju, L.}, \bibinfo{author}{Zhao,
  X.}, \bibinfo{author}{Ge, Z.}, \bibinfo{year}{2019}d.
\newblock \bibinfo{title}{Retinal abnormalities recognition using regional
  multitask learning}, in: \bibinfo{editor}{Shen, D.}, \bibinfo{editor}{Liu,
  T.}, \bibinfo{editor}{Peters, T.M.}, \bibinfo{editor}{Staib, L.H.},
  \bibinfo{editor}{Essert, C.}, \bibinfo{editor}{Zhou, S.},
  \bibinfo{editor}{Yap, P.}, \bibinfo{editor}{Khan, A.} (Eds.),
  \bibinfo{booktitle}{Medical Image Computing and Computer Assisted
  Intervention - {MICCAI} 2019 - 22nd International Conference, Shenzhen,
  China, October 13-17, 2019, Proceedings, Part {I}},
  \bibinfo{publisher}{Springer}. pp. \bibinfo{pages}{30--38}.
\newblock \URLprefix \url{https://doi.org/10.1007/978-3-030-32239-7\_4},
  \DOIprefix\doi{10.1007/978-3-030-32239-7\_4}.
\bibitem[{Wang et~al.(2019e)Wang, Xu, Li, Wang and
  Guan}]{DBLP:conf/miccai/WangXLWG19}
\bibinfo{author}{Wang, X.}, \bibinfo{author}{Xu, M.}, \bibinfo{author}{Li, L.},
  \bibinfo{author}{Wang, Z.}, \bibinfo{author}{Guan, Z.},
  \bibinfo{year}{2019}e.
\newblock \bibinfo{title}{Pathology-aware deep network visualization and its
  application in glaucoma image synthesis}, in: \bibinfo{editor}{Shen, D.},
  \bibinfo{editor}{Liu, T.}, \bibinfo{editor}{Peters, T.M.},
  \bibinfo{editor}{Staib, L.H.}, \bibinfo{editor}{Essert, C.},
  \bibinfo{editor}{Zhou, S.}, \bibinfo{editor}{Yap, P.}, \bibinfo{editor}{Khan,
  A.} (Eds.), \bibinfo{booktitle}{Medical Image Computing and Computer Assisted
  Intervention - {MICCAI} 2019 - 22nd International Conference, Shenzhen,
  China, October 13-17, 2019, Proceedings, Part {I}},
  \bibinfo{publisher}{Springer}. pp. \bibinfo{pages}{423--431}.
\newblock \URLprefix \url{https://doi.org/10.1007/978-3-030-32239-7\_47},
  \DOIprefix\doi{10.1007/978-3-030-32239-7\_47}.
\bibitem[{Wang et~al.(2019f)Wang, Dong, Rosario, Xu, Xie and
  Xing}]{DBLP:conf/isbi/WangDR0XX19}
\bibinfo{author}{Wang, Z.}, \bibinfo{author}{Dong, N.},
  \bibinfo{author}{Rosario, S.D.}, \bibinfo{author}{Xu, M.},
  \bibinfo{author}{Xie, P.}, \bibinfo{author}{Xing, E.P.},
  \bibinfo{year}{2019}f.
\newblock \bibinfo{title}{Ellipse detection of optic disc-and-cup boundary in
  fundus images}, in: \bibinfo{booktitle}{16th {IEEE} International Symposium
  on Biomedical Imaging, {ISBI} 2019, Venice, Italy, April 8-11, 2019},
  \bibinfo{publisher}{{IEEE}}. pp. \bibinfo{pages}{601--604}.
\newblock \URLprefix \url{https://doi.org/10.1109/ISBI.2019.8759173},
  \DOIprefix\doi{10.1109/ISBI.2019.8759173}.
\bibitem[{Wang et~al.(2017)Wang, Yin, Shi, Fang, Li and
  Wang}]{DBLP:conf/miccai/WangYSFLW17}
\bibinfo{author}{Wang, Z.}, \bibinfo{author}{Yin, Y.}, \bibinfo{author}{Shi,
  J.}, \bibinfo{author}{Fang, W.}, \bibinfo{author}{Li, H.},
  \bibinfo{author}{Wang, X.}, \bibinfo{year}{2017}.
\newblock \bibinfo{title}{Zoom-in-net: Deep mining lesions for diabetic
  retinopathy detection}, in: \bibinfo{editor}{Descoteaux, M.},
  \bibinfo{editor}{Maier{-}Hein, L.}, \bibinfo{editor}{Franz, A.M.},
  \bibinfo{editor}{Jannin, P.}, \bibinfo{editor}{Collins, D.L.},
  \bibinfo{editor}{Duchesne, S.} (Eds.), \bibinfo{booktitle}{Medical Image
  Computing and Computer Assisted Intervention - {MICCAI} 2017 - 20th
  International Conference, Quebec City, QC, Canada, September 11-13, 2017,
  Proceedings, Part {III}}, \bibinfo{publisher}{Springer}. pp.
  \bibinfo{pages}{267--275}.
\newblock \URLprefix \url{https://doi.org/10.1007/978-3-319-66179-7\_31},
  \DOIprefix\doi{10.1007/978-3-319-66179-7\_31}.
\bibitem[{Wilkinson et~al.(2003)Wilkinson, Iii, Klein, Lee, Agardh, Davis,
  Dills, Kampik, Pararajasegaram and Verdaguer}]{Wilkinson2003Proposed}
\bibinfo{author}{Wilkinson, C.P.}, \bibinfo{author}{Iii, F.L.F.},
  \bibinfo{author}{Klein, R.E.}, \bibinfo{author}{Lee, P.P.},
  \bibinfo{author}{Agardh, C.D.}, \bibinfo{author}{Davis, M.},
  \bibinfo{author}{Dills, D.}, \bibinfo{author}{Kampik, A.},
  \bibinfo{author}{Pararajasegaram, R.}, \bibinfo{author}{Verdaguer, J.T.},
  \bibinfo{year}{2003}.
\newblock \bibinfo{title}{Proposed international clinical diabetic retinopathy
  and diabetic macular edema disease severity scales}.
\newblock \bibinfo{journal}{Ophthalmology} \bibinfo{volume}{110},
  \bibinfo{pages}{0--1682}.
\bibitem[{Wu et~al.(2019)Wu, Xia, Song, Zhang, Liu, Zhang and
  Cai}]{DBLP:conf/miccai/WuX0ZLZC19}
\bibinfo{author}{Wu, Y.}, \bibinfo{author}{Xia, Y.}, \bibinfo{author}{Song,
  Y.}, \bibinfo{author}{Zhang, D.}, \bibinfo{author}{Liu, D.},
  \bibinfo{author}{Zhang, C.}, \bibinfo{author}{Cai, W.}, \bibinfo{year}{2019}.
\newblock \bibinfo{title}{Vessel-net: Retinal vessel segmentation under
  multi-path supervision}, in: \bibinfo{editor}{Shen, D.},
  \bibinfo{editor}{Liu, T.}, \bibinfo{editor}{Peters, T.M.},
  \bibinfo{editor}{Staib, L.H.}, \bibinfo{editor}{Essert, C.},
  \bibinfo{editor}{Zhou, S.}, \bibinfo{editor}{Yap, P.}, \bibinfo{editor}{Khan,
  A.} (Eds.), \bibinfo{booktitle}{Medical Image Computing and Computer Assisted
  Intervention - {MICCAI} 2019 - 22nd International Conference, Shenzhen,
  China, October 13-17, 2019, Proceedings, Part {I}},
  \bibinfo{publisher}{Springer}. pp. \bibinfo{pages}{264--272}.
\newblock \URLprefix \url{https://doi.org/10.1007/978-3-030-32239-7\_30},
  \DOIprefix\doi{10.1007/978-3-030-32239-7\_30}.
\bibitem[{Wu et~al.(2018)Wu, Xia, Song, Zhang and
  Cai}]{DBLP:conf/miccai/WuXSZC18}
\bibinfo{author}{Wu, Y.}, \bibinfo{author}{Xia, Y.}, \bibinfo{author}{Song,
  Y.}, \bibinfo{author}{Zhang, Y.}, \bibinfo{author}{Cai, W.},
  \bibinfo{year}{2018}.
\newblock \bibinfo{title}{Multiscale network followed network model for retinal
  vessel segmentation}, in: \bibinfo{editor}{Frangi, A.F.},
  \bibinfo{editor}{Schnabel, J.A.}, \bibinfo{editor}{Davatzikos, C.},
  \bibinfo{editor}{Alberola{-}L{\'{o}}pez, C.}, \bibinfo{editor}{Fichtinger,
  G.} (Eds.), \bibinfo{booktitle}{Medical Image Computing and Computer Assisted
  Intervention - {MICCAI} 2018 - 21st International Conference, Granada, Spain,
  September 16-20, 2018, Proceedings, Part {II}},
  \bibinfo{publisher}{Springer}. pp. \bibinfo{pages}{119--126}.
\newblock \URLprefix \url{https://doi.org/10.1007/978-3-030-00934-2\_14},
  \DOIprefix\doi{10.1007/978-3-030-00934-2\_14}.
\bibitem[{Wu et~al.(2020)Wu, Xia, Song, Zhang and
  Cai}]{DBLP:journals/nn/WuXSZC20}
\bibinfo{author}{Wu, Y.}, \bibinfo{author}{Xia, Y.}, \bibinfo{author}{Song,
  Y.}, \bibinfo{author}{Zhang, Y.}, \bibinfo{author}{Cai, W.},
  \bibinfo{year}{2020}.
\newblock \bibinfo{title}{Nfn +: {A} novel network followed network for retinal
  vessel segmentation}.
\newblock \bibinfo{journal}{Neural Networks} \bibinfo{volume}{126},
  \bibinfo{pages}{153--162}.
\newblock \URLprefix \url{https://doi.org/10.1016/j.neunet.2020.02.018},
  \DOIprefix\doi{10.1016/j.neunet.2020.02.018}.
\bibitem[{{Xie} and {Tu}(2015)}]{7410521}
\bibinfo{author}{{Xie}, S.}, \bibinfo{author}{{Tu}, Z.}, \bibinfo{year}{2015}.
\newblock \bibinfo{title}{Holistically-nested edge detection}, in:
  \bibinfo{booktitle}{2015 IEEE International Conference on Computer Vision
  (ICCV)}, pp. \bibinfo{pages}{1395--1403}.
\bibitem[{Xu et~al.(2015)Xu, Ba, Kiros, Cho, Courville, Salakhutdinov, Zemel
  and Bengio}]{DBLP:conf/icml/XuBKCCSZB15}
\bibinfo{author}{Xu, K.}, \bibinfo{author}{Ba, J.}, \bibinfo{author}{Kiros,
  R.}, \bibinfo{author}{Cho, K.}, \bibinfo{author}{Courville, A.C.},
  \bibinfo{author}{Salakhutdinov, R.}, \bibinfo{author}{Zemel, R.S.},
  \bibinfo{author}{Bengio, Y.}, \bibinfo{year}{2015}.
\newblock \bibinfo{title}{Show, attend and tell: Neural image caption
  generation with visual attention}, in: \bibinfo{editor}{Bach, F.R.},
  \bibinfo{editor}{Blei, D.M.} (Eds.), \bibinfo{booktitle}{Proceedings of the
  32nd International Conference on Machine Learning, {ICML} 2015, Lille,
  France, 6-11 July 2015}, \bibinfo{publisher}{JMLR.org}. pp.
  \bibinfo{pages}{2048--2057}.
\newblock \URLprefix \url{http://proceedings.mlr.press/v37/xuc15.html}.
\bibitem[{Xu et~al.(2020)Xu, Zhang, Li, Guan and
  Zhang}]{DBLP:journals/titb/XuZLGZ20}
\bibinfo{author}{Xu, X.}, \bibinfo{author}{Zhang, L.}, \bibinfo{author}{Li,
  J.}, \bibinfo{author}{Guan, Y.}, \bibinfo{author}{Zhang, L.},
  \bibinfo{year}{2020}.
\newblock \bibinfo{title}{A hybrid global-local representation {CNN} model for
  automatic cataract grading}.
\newblock \bibinfo{journal}{{IEEE} J. Biomed. Health Informatics}
  \bibinfo{volume}{24}, \bibinfo{pages}{556--567}.
\newblock \URLprefix \url{https://doi.org/10.1109/JBHI.2019.2914690},
  \DOIprefix\doi{10.1109/JBHI.2019.2914690}.
\bibitem[{Xue et~al.(2019)Xue, Yan, Qu, Qi, Qiu, Zhang, Chen, Liu, Li and
  Liu}]{DBLP:journals/kbs/XueYQQQZCLLL19}
\bibinfo{author}{Xue, J.}, \bibinfo{author}{Yan, S.}, \bibinfo{author}{Qu, J.},
  \bibinfo{author}{Qi, F.}, \bibinfo{author}{Qiu, C.}, \bibinfo{author}{Zhang,
  H.}, \bibinfo{author}{Chen, M.}, \bibinfo{author}{Liu, T.},
  \bibinfo{author}{Li, D.}, \bibinfo{author}{Liu, X.}, \bibinfo{year}{2019}.
\newblock \bibinfo{title}{Deep membrane systems for multitask segmentation in
  diabetic retinopathy}.
\newblock \bibinfo{journal}{Knowl. Based Syst.} \bibinfo{volume}{183}.
\newblock \URLprefix \url{https://doi.org/10.1016/j.knosys.2019.104887},
  \DOIprefix\doi{10.1016/j.knosys.2019.104887}.
\bibitem[{Yan et~al.(2018a)Yan, Cui, Wang, Liu, Liu, Wei, Yin and
  Zheng}]{DBLP:conf/miccai/YanCWLLWYZ18}
\bibinfo{author}{Yan, F.}, \bibinfo{author}{Cui, J.}, \bibinfo{author}{Wang,
  Y.}, \bibinfo{author}{Liu, H.}, \bibinfo{author}{Liu, H.},
  \bibinfo{author}{Wei, B.}, \bibinfo{author}{Yin, Y.}, \bibinfo{author}{Zheng,
  Y.}, \bibinfo{year}{2018}a.
\newblock \bibinfo{title}{Deep random walk for drusen segmentation from fundus
  images}, in: \bibinfo{editor}{Frangi, A.F.}, \bibinfo{editor}{Schnabel,
  J.A.}, \bibinfo{editor}{Davatzikos, C.},
  \bibinfo{editor}{Alberola{-}L{\'{o}}pez, C.}, \bibinfo{editor}{Fichtinger,
  G.} (Eds.), \bibinfo{booktitle}{Medical Image Computing and Computer Assisted
  Intervention - {MICCAI} 2018 - 21st International Conference, Granada, Spain,
  September 16-20, 2018, Proceedings, Part {II}},
  \bibinfo{publisher}{Springer}. pp. \bibinfo{pages}{48--55}.
\newblock \URLprefix \url{https://doi.org/10.1007/978-3-030-00934-2\_6},
  \DOIprefix\doi{10.1007/978-3-030-00934-2\_6}.
\bibitem[{Yan et~al.(2019a)Yan, Han, Wang, Qiu, Xiong and
  Cui}]{DBLP:conf/isbi/YanHWQXC19}
\bibinfo{author}{Yan, Z.}, \bibinfo{author}{Han, X.}, \bibinfo{author}{Wang,
  C.}, \bibinfo{author}{Qiu, Y.}, \bibinfo{author}{Xiong, Z.},
  \bibinfo{author}{Cui, S.}, \bibinfo{year}{2019}a.
\newblock \bibinfo{title}{Learning mutually local-global u-nets for
  high-resolution retinal lesion segmentation in fundus images}, in:
  \bibinfo{booktitle}{16th {IEEE} International Symposium on Biomedical
  Imaging, {ISBI} 2019, Venice, Italy, April 8-11, 2019},
  \bibinfo{publisher}{{IEEE}}. pp. \bibinfo{pages}{597--600}.
\newblock \URLprefix \url{https://doi.org/10.1109/ISBI.2019.8759579},
  \DOIprefix\doi{10.1109/ISBI.2019.8759579}.
\bibitem[{Yan et~al.(2018b)Yan, Yang and Cheng}]{DBLP:journals/tbe/YanYC18}
\bibinfo{author}{Yan, Z.}, \bibinfo{author}{Yang, X.}, \bibinfo{author}{Cheng,
  K.}, \bibinfo{year}{2018}b.
\newblock \bibinfo{title}{Joint segment-level and pixel-wise losses for deep
  learning based retinal vessel segmentation}.
\newblock \bibinfo{journal}{{IEEE} Trans. Biomed. Engineering}
  \bibinfo{volume}{65}, \bibinfo{pages}{1912--1923}.
\newblock \URLprefix \url{https://doi.org/10.1109/TBME.2018.2828137},
  \DOIprefix\doi{10.1109/TBME.2018.2828137}.
\bibitem[{Yan et~al.(2019b)Yan, Yang and Cheng}]{DBLP:journals/titb/YanYC19}
\bibinfo{author}{Yan, Z.}, \bibinfo{author}{Yang, X.}, \bibinfo{author}{Cheng,
  K.}, \bibinfo{year}{2019}b.
\newblock \bibinfo{title}{A three-stage deep learning model for accurate
  retinal vessel segmentation}.
\newblock \bibinfo{journal}{{IEEE} J. Biomed. Health Informatics}
  \bibinfo{volume}{23}, \bibinfo{pages}{1427--1436}.
\newblock \URLprefix \url{https://doi.org/10.1109/JBHI.2018.2872813},
  \DOIprefix\doi{10.1109/JBHI.2018.2872813}.
\bibitem[{Yang et~al.(2017)Yang, Li, Li, Wu, Fan and
  Zhang}]{DBLP:conf/miccai/YangLLWFZ17}
\bibinfo{author}{Yang, Y.}, \bibinfo{author}{Li, T.}, \bibinfo{author}{Li, W.},
  \bibinfo{author}{Wu, H.}, \bibinfo{author}{Fan, W.}, \bibinfo{author}{Zhang,
  W.}, \bibinfo{year}{2017}.
\newblock \bibinfo{title}{Lesion detection and grading of diabetic retinopathy
  via two-stages deep convolutional neural networks}, in:
  \bibinfo{editor}{Descoteaux, M.}, \bibinfo{editor}{Maier{-}Hein, L.},
  \bibinfo{editor}{Franz, A.M.}, \bibinfo{editor}{Jannin, P.},
  \bibinfo{editor}{Collins, D.L.}, \bibinfo{editor}{Duchesne, S.} (Eds.),
  \bibinfo{booktitle}{Medical Image Computing and Computer Assisted
  Intervention - {MICCAI} 2017 - 20th International Conference, Quebec City,
  QC, Canada, September 11-13, 2017, Proceedings, Part {III}},
  \bibinfo{publisher}{Springer}. pp. \bibinfo{pages}{533--540}.
\newblock \URLprefix \url{https://doi.org/10.1007/978-3-319-66179-7\_61},
  \DOIprefix\doi{10.1007/978-3-319-66179-7\_61}.
\bibitem[{Yin et~al.(2019)Yin, Wu, Xu, Min, Yang, Zhang and
  Tan}]{DBLP:conf/miccai/YinWXMYZT19}
\bibinfo{author}{Yin, P.}, \bibinfo{author}{Wu, Q.}, \bibinfo{author}{Xu, Y.},
  \bibinfo{author}{Min, H.}, \bibinfo{author}{Yang, M.},
  \bibinfo{author}{Zhang, Y.}, \bibinfo{author}{Tan, M.}, \bibinfo{year}{2019}.
\newblock \bibinfo{title}{Pm-net: Pyramid multi-label network for joint optic
  disc and cup segmentation}, in: \bibinfo{editor}{Shen, D.},
  \bibinfo{editor}{Liu, T.}, \bibinfo{editor}{Peters, T.M.},
  \bibinfo{editor}{Staib, L.H.}, \bibinfo{editor}{Essert, C.},
  \bibinfo{editor}{Zhou, S.}, \bibinfo{editor}{Yap, P.}, \bibinfo{editor}{Khan,
  A.} (Eds.), \bibinfo{booktitle}{Medical Image Computing and Computer Assisted
  Intervention - {MICCAI} 2019 - 22nd International Conference, Shenzhen,
  China, October 13-17, 2019, Proceedings, Part {I}},
  \bibinfo{publisher}{Springer}. pp. \bibinfo{pages}{129--137}.
\newblock \URLprefix \url{https://doi.org/10.1007/978-3-030-32239-7\_15},
  \DOIprefix\doi{10.1007/978-3-030-32239-7\_15}.
\bibitem[{Yu et~al.(2019)Yu, Zhao, Gong, Wang, Li, Yang, Dong, Li and
  Zhang}]{DBLP:conf/miccai/Yu0GWLYDLZ19}
\bibinfo{author}{Yu, F.}, \bibinfo{author}{Zhao, J.}, \bibinfo{author}{Gong,
  Y.}, \bibinfo{author}{Wang, Z.}, \bibinfo{author}{Li, Y.},
  \bibinfo{author}{Yang, F.}, \bibinfo{author}{Dong, B.}, \bibinfo{author}{Li,
  Q.}, \bibinfo{author}{Zhang, L.}, \bibinfo{year}{2019}.
\newblock \bibinfo{title}{Annotation-free cardiac vessel segmentation via
  knowledge transfer from retinal images}, in: \bibinfo{editor}{Shen, D.},
  \bibinfo{editor}{Liu, T.}, \bibinfo{editor}{Peters, T.M.},
  \bibinfo{editor}{Staib, L.H.}, \bibinfo{editor}{Essert, C.},
  \bibinfo{editor}{Zhou, S.}, \bibinfo{editor}{Yap, P.}, \bibinfo{editor}{Khan,
  A.} (Eds.), \bibinfo{booktitle}{Medical Image Computing and Computer Assisted
  Intervention - {MICCAI} 2019 - 22nd International Conference, Shenzhen,
  China, October 13-17, 2019, Proceedings, Part {II}},
  \bibinfo{publisher}{Springer}. pp. \bibinfo{pages}{714--722}.
\newblock \URLprefix \url{https://doi.org/10.1007/978-3-030-32245-8\_79},
  \DOIprefix\doi{10.1007/978-3-030-32245-8\_79}.
\bibitem[{Yu et~al.(2020)Yu, Qin, Zhuang, Ding, Qin and
  Choo}]{DBLP:journals/ijon/YuQZDQC20}
\bibinfo{author}{Yu, L.}, \bibinfo{author}{Qin, Z.}, \bibinfo{author}{Zhuang,
  T.}, \bibinfo{author}{Ding, Y.}, \bibinfo{author}{Qin, Z.},
  \bibinfo{author}{Choo, K.R.}, \bibinfo{year}{2020}.
\newblock \bibinfo{title}{A framework for hierarchical division of retinal
  vascular networks}.
\newblock \bibinfo{journal}{Neurocomputing} \bibinfo{volume}{392},
  \bibinfo{pages}{221--232}.
\newblock \URLprefix \url{https://doi.org/10.1016/j.neucom.2018.11.113},
  \DOIprefix\doi{10.1016/j.neucom.2018.11.113}.
\bibitem[{Zhang et~al.(2019a)Zhang, Fu, Yan, Zhang, Wu, Yang, Tan and
  Xu}]{DBLP:conf/miccai/ZhangFYZWYTX19}
\bibinfo{author}{Zhang, S.}, \bibinfo{author}{Fu, H.}, \bibinfo{author}{Yan,
  Y.}, \bibinfo{author}{Zhang, Y.}, \bibinfo{author}{Wu, Q.},
  \bibinfo{author}{Yang, M.}, \bibinfo{author}{Tan, M.}, \bibinfo{author}{Xu,
  Y.}, \bibinfo{year}{2019}a.
\newblock \bibinfo{title}{Attention guided network for retinal image
  segmentation}, in: \bibinfo{editor}{Shen, D.}, \bibinfo{editor}{Liu, T.},
  \bibinfo{editor}{Peters, T.M.}, \bibinfo{editor}{Staib, L.H.},
  \bibinfo{editor}{Essert, C.}, \bibinfo{editor}{Zhou, S.},
  \bibinfo{editor}{Yap, P.}, \bibinfo{editor}{Khan, A.} (Eds.),
  \bibinfo{booktitle}{Medical Image Computing and Computer Assisted
  Intervention - {MICCAI} 2019 - 22nd International Conference, Shenzhen,
  China, October 13-17, 2019, Proceedings, Part {I}},
  \bibinfo{publisher}{Springer}. pp. \bibinfo{pages}{797--805}.
\newblock \URLprefix \url{https://doi.org/10.1007/978-3-030-32239-7\_88},
  \DOIprefix\doi{10.1007/978-3-030-32239-7\_88}.
\bibitem[{Zhang et~al.(2019b)Zhang, Zhong, Yang, Gao, Hu, Chen and
  Yi}]{DBLP:journals/kbs/ZhangZYGHCY19}
\bibinfo{author}{Zhang, W.}, \bibinfo{author}{Zhong, J.},
  \bibinfo{author}{Yang, S.}, \bibinfo{author}{Gao, Z.}, \bibinfo{author}{Hu,
  J.}, \bibinfo{author}{Chen, Y.}, \bibinfo{author}{Yi, Z.},
  \bibinfo{year}{2019}b.
\newblock \bibinfo{title}{Automated identification and grading system of
  diabetic retinopathy using deep neural networks}.
\newblock \bibinfo{journal}{Knowl. Based Syst.} \bibinfo{volume}{175},
  \bibinfo{pages}{12--25}.
\newblock \URLprefix \url{https://doi.org/10.1016/j.knosys.2019.03.016},
  \DOIprefix\doi{10.1016/j.knosys.2019.03.016}.
\bibitem[{Zhang and Chung(2018)}]{DBLP:conf/miccai/ZhangC18}
\bibinfo{author}{Zhang, Y.}, \bibinfo{author}{Chung, A.C.S.},
  \bibinfo{year}{2018}.
\newblock \bibinfo{title}{Deep supervision with additional labels for retinal
  vessel segmentation task}, in: \bibinfo{editor}{Frangi, A.F.},
  \bibinfo{editor}{Schnabel, J.A.}, \bibinfo{editor}{Davatzikos, C.},
  \bibinfo{editor}{Alberola{-}L{\'{o}}pez, C.}, \bibinfo{editor}{Fichtinger,
  G.} (Eds.), \bibinfo{booktitle}{Medical Image Computing and Computer Assisted
  Intervention - {MICCAI} 2018 - 21st International Conference, Granada, Spain,
  September 16-20, 2018, Proceedings, Part {II}},
  \bibinfo{publisher}{Springer}. pp. \bibinfo{pages}{83--91}.
\newblock \URLprefix \url{https://doi.org/10.1007/978-3-030-00934-2\_10},
  \DOIprefix\doi{10.1007/978-3-030-00934-2\_10}.
\bibitem[{{Zhang} et~al.(2013){Zhang}, {Liu}, {Yin}, {Lee}, {Wong} and
  {Sung}}]{6566371}
\bibinfo{author}{{Zhang}, Z.}, \bibinfo{author}{{Liu}, J.},
  \bibinfo{author}{{Yin}, F.}, \bibinfo{author}{{Lee}, B.},
  \bibinfo{author}{{Wong}, D.W.K.}, \bibinfo{author}{{Sung}, K.R.},
  \bibinfo{year}{2013}.
\newblock \bibinfo{title}{Achiko-k: Database of fundus images from glaucoma
  patients}, in: \bibinfo{booktitle}{2013 IEEE 8th Conference on Industrial
  Electronics and Applications (ICIEA)}, pp. \bibinfo{pages}{228--231}.
\bibitem[{Zhang et~al.(2014)Zhang, Srivastava, Liu, Chen, Duan, Wong, Kwoh,
  Wong and Liu}]{DBLP:journals/midm/ZhangSLCDWKW014}
\bibinfo{author}{Zhang, Z.}, \bibinfo{author}{Srivastava, R.},
  \bibinfo{author}{Liu, H.}, \bibinfo{author}{Chen, X.}, \bibinfo{author}{Duan,
  L.}, \bibinfo{author}{Wong, D.W.K.}, \bibinfo{author}{Kwoh, C.K.},
  \bibinfo{author}{Wong, T.Y.}, \bibinfo{author}{Liu, J.},
  \bibinfo{year}{2014}.
\newblock \bibinfo{title}{A survey on computer aided diagnosis for ocular
  diseases}.
\newblock \bibinfo{journal}{{BMC} Med. Inf. {\&} Decision Making}
  \bibinfo{volume}{14}, \bibinfo{pages}{80}.
\newblock \URLprefix \url{https://doi.org/10.1186/1472-6947-14-80},
  \DOIprefix\doi{10.1186/1472-6947-14-80}.
\bibitem[{{Zhang} et~al.(2010){Zhang}, {Yin}, {Liu}, {Wong}, {Tan}, {Lee},
  {Cheng} and {Wong}}]{5626137}
\bibinfo{author}{{Zhang}, Z.}, \bibinfo{author}{{Yin}, F.S.},
  \bibinfo{author}{{Liu}, J.}, \bibinfo{author}{{Wong}, W.K.},
  \bibinfo{author}{{Tan}, N.M.}, \bibinfo{author}{{Lee}, B.H.},
  \bibinfo{author}{{Cheng}, J.}, \bibinfo{author}{{Wong}, T.Y.},
  \bibinfo{year}{2010}.
\newblock \bibinfo{title}{Origa-light: An online retinal fundus image database
  for glaucoma analysis and research}, in: \bibinfo{booktitle}{2010 Annual
  International Conference of the IEEE Engineering in Medicine and Biology},
  pp. \bibinfo{pages}{3065--3068}.
\bibitem[{Zhao et~al.(2020a)Zhao, Li and Cheng}]{DBLP:journals/pr/ZhaoLC20}
\bibinfo{author}{Zhao, H.}, \bibinfo{author}{Li, H.}, \bibinfo{author}{Cheng,
  L.}, \bibinfo{year}{2020}a.
\newblock \bibinfo{title}{Improving retinal vessel segmentation with joint
  local loss by matting}.
\newblock \bibinfo{journal}{Pattern Recognit.} \bibinfo{volume}{98}.
\newblock \URLprefix \url{https://doi.org/10.1016/j.patcog.2019.107068},
  \DOIprefix\doi{10.1016/j.patcog.2019.107068}.
\bibitem[{Zhao et~al.(2018)Zhao, Li, Maurer{-}Stroh and
  Cheng}]{DBLP:journals/mia/ZhaoLMC18}
\bibinfo{author}{Zhao, H.}, \bibinfo{author}{Li, H.},
  \bibinfo{author}{Maurer{-}Stroh, S.}, \bibinfo{author}{Cheng, L.},
  \bibinfo{year}{2018}.
\newblock \bibinfo{title}{Synthesizing retinal and neuronal images with
  generative adversarial nets}.
\newblock \bibinfo{journal}{Medical Image Anal.} \bibinfo{volume}{49},
  \bibinfo{pages}{14--26}.
\newblock \URLprefix \url{https://doi.org/10.1016/j.media.2018.07.001},
  \DOIprefix\doi{10.1016/j.media.2018.07.001}.
\bibitem[{Zhao et~al.(2019a)Zhao, Li, Maurer{-}Stroh, Guo, Deng and
  Cheng}]{DBLP:journals/tmi/ZhaoLMGDC19}
\bibinfo{author}{Zhao, H.}, \bibinfo{author}{Li, H.},
  \bibinfo{author}{Maurer{-}Stroh, S.}, \bibinfo{author}{Guo, Y.},
  \bibinfo{author}{Deng, Q.}, \bibinfo{author}{Cheng, L.},
  \bibinfo{year}{2019}a.
\newblock \bibinfo{title}{Supervised segmentation of un-annotated retinal
  fundus images by synthesis}.
\newblock \bibinfo{journal}{{IEEE} Trans. Medical Imaging}
  \bibinfo{volume}{38}, \bibinfo{pages}{46--56}.
\newblock \URLprefix \url{https://doi.org/10.1109/TMI.2018.2854886},
  \DOIprefix\doi{10.1109/TMI.2018.2854886}.
\bibitem[{Zhao et~al.(2019b)Zhao, Yang, Cao and
  Li}]{DBLP:conf/miccai/ZhaoYCL19}
\bibinfo{author}{Zhao, H.}, \bibinfo{author}{Yang, B.}, \bibinfo{author}{Cao,
  L.}, \bibinfo{author}{Li, H.}, \bibinfo{year}{2019}b.
\newblock \bibinfo{title}{Data-driven enhancement of blurry retinal images via
  generative adversarial networks}, in: \bibinfo{editor}{Shen, D.},
  \bibinfo{editor}{Liu, T.}, \bibinfo{editor}{Peters, T.M.},
  \bibinfo{editor}{Staib, L.H.}, \bibinfo{editor}{Essert, C.},
  \bibinfo{editor}{Zhou, S.}, \bibinfo{editor}{Yap, P.}, \bibinfo{editor}{Khan,
  A.} (Eds.), \bibinfo{booktitle}{Medical Image Computing and Computer Assisted
  Intervention - {MICCAI} 2019 - 22nd International Conference, Shenzhen,
  China, October 13-17, 2019, Proceedings, Part {I}},
  \bibinfo{publisher}{Springer}. pp. \bibinfo{pages}{75--83}.
\newblock \URLprefix \url{https://doi.org/10.1007/978-3-030-32239-7\_9},
  \DOIprefix\doi{10.1007/978-3-030-32239-7\_9}.
\bibitem[{Zhao et~al.(2020b)Zhao, Chen, Liu, Chen, Guo and
  Li}]{DBLP:journals/titb/ZhaoCLCGL20}
\bibinfo{author}{Zhao, R.}, \bibinfo{author}{Chen, X.}, \bibinfo{author}{Liu,
  X.}, \bibinfo{author}{Chen, Z.}, \bibinfo{author}{Guo, F.},
  \bibinfo{author}{Li, S.}, \bibinfo{year}{2020}b.
\newblock \bibinfo{title}{Direct cup-to-disc ratio estimation for glaucoma
  screening via semi-supervised learning}.
\newblock \bibinfo{journal}{{IEEE} J. Biomed. Health Informatics}
  \bibinfo{volume}{24}, \bibinfo{pages}{1104--1113}.
\newblock \URLprefix \url{https://doi.org/10.1109/JBHI.2019.2934477},
  \DOIprefix\doi{10.1109/JBHI.2019.2934477}.
\bibitem[{Zhao et~al.(2019c)Zhao, Chen, Liu, Zou and
  Li}]{DBLP:conf/miccai/ZhaoCLZ019}
\bibinfo{author}{Zhao, R.}, \bibinfo{author}{Chen, Z.}, \bibinfo{author}{Liu,
  X.}, \bibinfo{author}{Zou, B.}, \bibinfo{author}{Li, S.},
  \bibinfo{year}{2019}c.
\newblock \bibinfo{title}{Multi-index optic disc quantification via multitask
  ensemble learning}, in: \bibinfo{editor}{Shen, D.}, \bibinfo{editor}{Liu,
  T.}, \bibinfo{editor}{Peters, T.M.}, \bibinfo{editor}{Staib, L.H.},
  \bibinfo{editor}{Essert, C.}, \bibinfo{editor}{Zhou, S.},
  \bibinfo{editor}{Yap, P.}, \bibinfo{editor}{Khan, A.} (Eds.),
  \bibinfo{booktitle}{Medical Image Computing and Computer Assisted
  Intervention - {MICCAI} 2019 - 22nd International Conference, Shenzhen,
  China, October 13-17, 2019, Proceedings, Part {I}},
  \bibinfo{publisher}{Springer}. pp. \bibinfo{pages}{21--29}.
\newblock \URLprefix \url{https://doi.org/10.1007/978-3-030-32239-7\_3},
  \DOIprefix\doi{10.1007/978-3-030-32239-7\_3}.
\bibitem[{Zhao and Li(2020)}]{DBLP:journals/mia/ZhaoL20}
\bibinfo{author}{Zhao, R.}, \bibinfo{author}{Li, S.}, \bibinfo{year}{2020}.
\newblock \bibinfo{title}{Multi-indices quantification of optic nerve head in
  fundus image via multitask collaborative learning}.
\newblock \bibinfo{journal}{Medical Image Anal.} \bibinfo{volume}{60}.
\newblock \URLprefix \url{https://doi.org/10.1016/j.media.2019.101593},
  \DOIprefix\doi{10.1016/j.media.2019.101593}.
\bibitem[{Zhao et~al.(2019d)Zhao, Liao, Zou, Chen and
  Li}]{DBLP:conf/aaai/ZhaoLZC019}
\bibinfo{author}{Zhao, R.}, \bibinfo{author}{Liao, W.}, \bibinfo{author}{Zou,
  B.}, \bibinfo{author}{Chen, Z.}, \bibinfo{author}{Li, S.},
  \bibinfo{year}{2019}d.
\newblock \bibinfo{title}{Weakly-supervised simultaneous evidence
  identification and segmentation for automated glaucoma diagnosis}, in:
  \bibinfo{booktitle}{The Thirty-Third {AAAI} Conference on Artificial
  Intelligence, {AAAI} 2019, The Thirty-First Innovative Applications of
  Artificial Intelligence Conference, {IAAI} 2019, The Ninth {AAAI} Symposium
  on Educational Advances in Artificial Intelligence, {EAAI} 2019, Honolulu,
  Hawaii, USA, January 27 - February 1, 2019}, \bibinfo{publisher}{{AAAI}
  Press}. pp. \bibinfo{pages}{809--816}.
\newblock \URLprefix \url{https://doi.org/10.1609/aaai.v33i01.3301809},
  \DOIprefix\doi{10.1609/aaai.v33i01.3301809}.
\bibitem[{Zhao et~al.(2019e)Zhao, Zhang, Hao, Tian, Chua, Chen and
  Xu}]{DBLP:conf/icip/ZhaoZHTCCX19}
\bibinfo{author}{Zhao, Z.}, \bibinfo{author}{Zhang, K.}, \bibinfo{author}{Hao,
  X.}, \bibinfo{author}{Tian, J.}, \bibinfo{author}{Chua, M.C.H.},
  \bibinfo{author}{Chen, L.}, \bibinfo{author}{Xu, X.}, \bibinfo{year}{2019}e.
\newblock \bibinfo{title}{Bira-net: Bilinear attention net for diabetic
  retinopathy grading}, in: \bibinfo{booktitle}{2019 {IEEE} International
  Conference on Image Processing, {ICIP} 2019, Taipei, Taiwan, September 22-25,
  2019}, \bibinfo{publisher}{{IEEE}}. pp. \bibinfo{pages}{1385--1389}.
\newblock \URLprefix \url{https://doi.org/10.1109/ICIP.2019.8803074},
  \DOIprefix\doi{10.1109/ICIP.2019.8803074}.
\bibitem[{Zheng et~al.(2013)Zheng, Cheng, Lamoureux, Chiang, Anuar, Wang,
  Mitchell, Saw and Wong}]{zheng2013how}
\bibinfo{author}{Zheng, Y.}, \bibinfo{author}{Cheng, C.},
  \bibinfo{author}{Lamoureux, E.L.}, \bibinfo{author}{Chiang, P.P.},
  \bibinfo{author}{Anuar, A.R.}, \bibinfo{author}{Wang, J.J.},
  \bibinfo{author}{Mitchell, P.}, \bibinfo{author}{Saw, S.},
  \bibinfo{author}{Wong, T.Y.}, \bibinfo{year}{2013}.
\newblock \bibinfo{title}{How much eye care services do asian populations need?
  projection from the singapore epidemiology of eye disease (seed) study}.
\newblock \bibinfo{journal}{Investigative Ophthalmology \& Visual Science}
  \bibinfo{volume}{54}, \bibinfo{pages}{2171--2177}.
\bibitem[{Zhou et~al.(2016)Zhou, Khosla, Lapedriza, Oliva and
  Torralba}]{DBLP:conf/cvpr/ZhouKLOT16}
\bibinfo{author}{Zhou, B.}, \bibinfo{author}{Khosla, A.},
  \bibinfo{author}{Lapedriza, {\`{A}}.}, \bibinfo{author}{Oliva, A.},
  \bibinfo{author}{Torralba, A.}, \bibinfo{year}{2016}.
\newblock \bibinfo{title}{Learning deep features for discriminative
  localization}, in: \bibinfo{booktitle}{2016 {IEEE} Conference on Computer
  Vision and Pattern Recognition, {CVPR} 2016, Las Vegas, NV, USA, June 27-30,
  2016}, \bibinfo{publisher}{{IEEE} Computer Society}. pp.
  \bibinfo{pages}{2921--2929}.
\newblock \URLprefix \url{https://doi.org/10.1109/CVPR.2016.319},
  \DOIprefix\doi{10.1109/CVPR.2016.319}.
\bibitem[{Zhou et~al.(2019)Zhou, He, Cui, Zhu, Liu and
  Shao}]{DBLP:conf/miccai/ZhouHCZL019}
\bibinfo{author}{Zhou, Y.}, \bibinfo{author}{He, X.}, \bibinfo{author}{Cui,
  S.}, \bibinfo{author}{Zhu, F.}, \bibinfo{author}{Liu, L.},
  \bibinfo{author}{Shao, L.}, \bibinfo{year}{2019}.
\newblock \bibinfo{title}{High-resolution diabetic retinopathy image synthesis
  manipulated by grading and lesions}, in: \bibinfo{editor}{Shen, D.},
  \bibinfo{editor}{Liu, T.}, \bibinfo{editor}{Peters, T.M.},
  \bibinfo{editor}{Staib, L.H.}, \bibinfo{editor}{Essert, C.},
  \bibinfo{editor}{Zhou, S.}, \bibinfo{editor}{Yap, P.}, \bibinfo{editor}{Khan,
  A.} (Eds.), \bibinfo{booktitle}{Medical Image Computing and Computer Assisted
  Intervention - {MICCAI} 2019 - 22nd International Conference, Shenzhen,
  China, October 13-17, 2019, Proceedings, Part {I}},
  \bibinfo{publisher}{Springer}. pp. \bibinfo{pages}{505--513}.
\newblock \URLprefix \url{https://doi.org/10.1007/978-3-030-32239-7\_56},
  \DOIprefix\doi{10.1007/978-3-030-32239-7\_56}.
\bibitem[{Zhou et~al.(2020)Zhou, Li and Li}]{DBLP:journals/tmi/ZhouLL20}
\bibinfo{author}{Zhou, Y.}, \bibinfo{author}{Li, G.}, \bibinfo{author}{Li, H.},
  \bibinfo{year}{2020}.
\newblock \bibinfo{title}{Automatic cataract classification using deep neural
  network with discrete state transition}.
\newblock \bibinfo{journal}{{IEEE} Trans. Med. Imaging} \bibinfo{volume}{39},
  \bibinfo{pages}{436--446}.
\newblock \URLprefix \url{https://doi.org/10.1109/TMI.2019.2928229},
  \DOIprefix\doi{10.1109/TMI.2019.2928229}.
\bibitem[{Zhou(2018)}]{zhou2018a}
\bibinfo{author}{Zhou, Z.}, \bibinfo{year}{2018}.
\newblock \bibinfo{title}{A brief introduction to weakly supervised learning}.
\newblock \bibinfo{journal}{National Science Review} \bibinfo{volume}{5},
  \bibinfo{pages}{44--53}.
\bibitem[{{Zhu} et~al.(2017){Zhu}, {Park}, {Isola} and {Efros}}]{8237506}
\bibinfo{author}{{Zhu}, J.}, \bibinfo{author}{{Park}, T.},
  \bibinfo{author}{{Isola}, P.}, \bibinfo{author}{{Efros}, A.A.},
  \bibinfo{year}{2017}.
\newblock \bibinfo{title}{Unpaired image-to-image translation using
  cycle-consistent adversarial networks}, in: \bibinfo{booktitle}{2017 IEEE
  International Conference on Computer Vision (ICCV)}, pp.
  \bibinfo{pages}{2242--2251}.
\bibitem[{Zou et~al.(2020)Zou, He, Zhao, Zhu, Liao and
  Li}]{DBLP:journals/ijon/ZouHZZLL20}
\bibinfo{author}{Zou, B.}, \bibinfo{author}{He, Z.}, \bibinfo{author}{Zhao,
  R.}, \bibinfo{author}{Zhu, C.}, \bibinfo{author}{Liao, W.},
  \bibinfo{author}{Li, S.}, \bibinfo{year}{2020}.
\newblock \bibinfo{title}{Non-rigid retinal image registration using an
  unsupervised structure-driven regression network}.
\newblock \bibinfo{journal}{Neurocomputing} \bibinfo{volume}{404},
  \bibinfo{pages}{14--25}.
\newblock \URLprefix \url{https://doi.org/10.1016/j.neucom.2020.04.122},
  \DOIprefix\doi{10.1016/j.neucom.2020.04.122}.

\end{thebibliography}

\end{document}